\documentclass[12pt,twoside]{book}

\usepackage{graphpap,geometry}

\usepackage{setspace}
\usepackage{times}
\newcommand{\Tr}{{\rm Tr}}                                      
\newcommand{\tr}{{\rm tr}\;}                                    
\newcommand{\Ln}{{\rm Ln}}  
\newcommand{\mqr}{m^{\text q}_R}
\newcommand{\slashit}[1]{#1 \kern-.45em\slash}
\newcommand{\slashp}{\slashit p}
\newcommand{\slashP}{P \kern-.65em\slash }

\usepackage{amsmath}
\usepackage{amsfonts,amssymb,amscd}

\usepackage{graphicx}
\usepackage{openone}
\usepackage{array}
\usepackage{title}

\begin{document}

\pagenumbering{roman}
\thispagestyle{empty}
\title{Topics in quantum field theory: Renormalization groups in Hamiltonian 
framework and baryon structure in a non-local QCD model}
\author{Amir H. Rezaeian}
\dept{School of Physics and Astronomy, The University of Manchester, UK}
\institute{University of Manchester Institute of Science and Technology}
\degree{Doctor of Philosophy}
\date{August 2004}
\maketitle

\chapter*{}

{\em{"We are at the very beginning of time for the human race. It is not
unreasonable that we grapple with problems. But there are tens of
thousands of years in the future. Our responsibility is to do what we
can, learn what we can, improve the solutions, and pass them
on."}}\newline 
$~~~\hspace{1.5cm}$Richard Feynman


\chapter*{Acknowledgement}
I must thank my wife, Shiva, for all her patience and love. Without her company and support this work would never have been
completed.
 
I would like to thank my supervisor Niels Walet for his wise advice,
encouragement, support, and his friendship over the duration of my
PhD. Niels has taught me how to become an independent researcher as I
should. His democratic attitude toward me allowed me to learn many
different aspects of physics during my PhD, thanks for all your help
Niels.

I would also like to thank Raymond F. Bishop, the head of our
theoretical physics group, for all his continuous support and encouragement which made my
PhD possible.

It has been my great pleasure to work with Mike Birse, his warm
character and sense of humour makes it very easy to learn a lot,
I am very grateful to him for all his inspiration and help.

My special thanks go to Hans-Juergen Pirner and Franz Wegner for their
hospitality and fruitful discussions and remarks during my visit in Heidelberg.

I would like to thank Bob Plant for useful correspondence and Boris Krippa for interesting discussions.

I thank my fellow colleagues at the department, Theodoros, Bernardo,
Toby, Neill, Daniel for their friendship.

Finally, I would like to acknowledge support from British Government
Oversea Research Award and UMIST scholarship.

\newcommand\dekname{Declaration}
\chapter*{\dekname}
No portion of the work referred to in this thesis has been submitted
in support of an application for another degree or qualification of
this or any other university, or other institution of learning.
\chapter*{Abstract}

In this thesis I investigate aspects of two problems.  In the first
part of this thesis, I will investigate how an effective field theory
can be constructed.  One of the most fundamental questions in physics
is how new degrees of freedom emerge from a fundamental theory. In the
Hamiltonian framework this can be rephrased as finding the correct
representation for the Hamiltonian matrix. The similarity (not
essentially unitary) renormalization group provides us with an
intuitive framework, where a transition from a perturbative region to
a non-perturbative one can be realised and physical properties can be
computed in a unified way.  In this context, we have shown that the
well-known coupled-cluster many-body theory techniques can be
incorporated in the Wilsonian renormalization group to provide a very
powerful framework for construction of effective Hamiltonian field
theories. Eventhough the formulation is intrinsically
non-perturbative, we have shown that a loop-expansion can be
implemented.

The second part of my thesis is rather phenomenologically
orientated. In this part, I will employ an effective field-theoretical
model as can be constructed by means of the techniques of the first part of my thesis, a
quark-confining non-local Nambu-Jona-Lasinio model and study the
nucleon and diquarks in this model. For certain parameters the model
exhibits quark confinement, in the form of a propagator without real
poles. After truncation of the two-body channels to the scalar and
axial-vector diquarks, a relativistic Faddeev equation for nucleon
bound states is solved in the covariant diquark-quark picture.  The
dependence of the nucleon mass on diquark masses is studied in
detail. We find parameters that lead to a simultaneous reasonable
description of pions and nucleons. Both the diquarks contribute
attractively to the nucleon mass. Axial-vector diquark correlations
are seen to be important, especially in the confining phase of the
model. We study the possible implications of quark confinement for the
description of the diquarks and the nucleon. In particular, we find
that it leads to a more compact nucleon.

\chapter*{General remarks}
This thesis is organized as follows: in the first three chapters, we
concentrate on renormalization group  methods in Hamiltonian framework. In chapter 1, we
introduce the coupled-cluster theory. In chapter 2, we show how
renormalization group can be employed in the context of the
coupled-cluster theory. In order to highlight the merits and
the shortcomings of our approach over previous ones, in Sec.~2.2, we review
different RG methods in Hamiltonian framework. Different aspects of our approach is introduced in sections
2.4-2.8. In sections 2.9 and 2.10, as illustrative examples, we apply
our formulation on the $\Phi^{4}$ theory and an extended Lee model. In
chapter 3, we pursue a different approach for the renormalization of
the many-body problem. We show that a combination of the coupled-cluster
theory and the Feshbach projection techniques leads to a renormalized
generalized Brueckner theory. 

In the second part of this thesis, we investigate the baryon structure
in a chiral quantum chromodynamics model based on the relativistic Faddeev approach. In
sections 4.1 and 4.2, we introduce the most important properties of
QCD which are needed for the modelling of hadrons. In sections 4.3 and
4.4, we show how an effective low-energy field theory can be
constructed from the underlying QCD theory. In chapter 5, we introduce
alternative  field theoretical approaches for describing baryons, such as
Skyrme models, bag models and diquark-quark models in the context of
the relativistic Faddeev approach. Finally, in chapter 6, we study
baryons based on the diquark-quark picture in a quark-confining
non-local NJL model. The non-local NJL model is introduced in
Sec.~6.2. In Sec.~6.3, we discuss the pionic sector of the model. In
Sec.~6.4 the diquark problem is solved and discussed. In Sec. 6.5 the
three-body problem of baryons is investigated. The numerical technique involved
in solving the effective Bethe-Salpeter equation is given and the
results for three-body sector are presented.

\bibliographystyle{unsrt}

\tableofcontents
\cleardoublepage
\pagenumbering{arabic}
\chapter{Basic structure of the coupled-cluster formalism}
\section{Introduction} 
In order to understand fully the properties of
quantum many-body systems, various methods have been developed which
aim to go beyond perturbation theory. One of the simplest approaches has
been the so-called configuration-interaction method which diagonalises
the Hamiltonian in a finite subspace of the full many-body Hilbert
space. An extension of this method has been introduced via various
versions of coupled-cluster methods
\cite{coester-k,nccm,ccm1,ccm3}. The coupled cluster
method (CCM) in its simplest form originated in nuclear physics around
forty years ago in the work of Coester and K\"ummel
\cite{coester-k}. The configuration-interaction method (CIM), and various version of coupled-cluster methods: normal
coupled cluster method (NCCM)\cite{coester-k,nccm} and the extended
coupled cluster method (ECCM)\cite{ccm1,ccm3} form a hierarchy of
many-body formulations for describing quantum systems of
interacting particles or fields
\cite{h1}. They are denoted generically as independent-cluster (IC)
parametrizations, in the sense that they incorporate the many-body
correlations via sets of amplitudes that describe the various
correlated clusters within the interacting system as mutually
independent entities. The intrinsic non-perturbative nature of the
methods is considered to be one of their advantages which make them
almost universally applicable in many-body physics. The IC methods
differ from each other in the way they incorporate the locality and
separability properties; in a diagrammatic language they differ in
their linking properties. Each of the IC methods has been shown
\cite{ccm3} to provide an exact mapping of the original quantum
mechanical problem to a corresponding classical mechanics in terms of
a set of multiconfigurational canonical field amplitudes.  The merit
of IC has been outlined in Ref.~\cite{h1} and literature cited
therein.

\section{Formalism}
 In this section we concentrate on the NCCM and the ECCM from a formal viewpoint.

Exponential structures arise frequently in physics for
 similar underlying fundamental reasons. For example, in
the Ursell-Mayer theory in statistical mechanics, in the Goldstone
linked-cluster theorem \cite{h8} and the Gell-Mann and Low theorem
\cite{h9}. The complexity of the vacuum (the ground state) of an arbitrary many-body system in
the NCCM parametrization \cite{coester-k,nccm, h1} is expressed by an infinite set of correlation
amplitudes $\{s_{I},\tilde{s}_{I}\}$ which have to be determined by the dynamics,
\begin{eqnarray}
&&|\psi\rangle=K(t)e^{S}|\psi_{0}\rangle, \hspace{2cm}\langle\tilde{\psi}|=\frac{1}{K(t)}\langle\psi_{0}|\tilde{S}e^{-S}, \label{1}\nonumber\\
&&S=\sum_{I\neq 0} s_{I}C_{I}^{\dagger}, \hspace{2cm} \tilde{S}=1+\sum_{I\neq 0}\tilde{s_{I}}C_{I}. \
\end{eqnarray}
Here $K(t)$ is a time-dependent scale factor. The coefficients $s_{I}$
and $\tilde{s_{I}}$ are time dependent.  The intermediate
normalization condition $\langle\tilde{\psi}|\psi\rangle=1$ is
explicit for all times $t$. We restrict ourselves to the
non-degenerate system, so that the exact states of the system may
sensibly be refereed to some suitably chosen single reference state
denoted as $ |\psi_{0}\rangle$. The state $ |\psi_{0}\rangle$ is a
ground state, e.g. a special (Hartree) bare vacuum in quantum field
theory (QFT). The function $|\psi_{0}\rangle$ can be chosen rather
generally, but is tied to the choice of generalized creation operators
$\{C^{\dag}_{I}\}$; the state $|\psi_{0}\rangle$ is annihilated by
$\{C_{I}\}$ $\forall I\neq 0$ (where $C_{0}\equiv 1$, the identity
operator) and is a cyclic vector in the sense that the algebra of all
possible operators in the many-body Hilbert space $\mathcal{H}$ is
spanned by the two Abelian subalgebras of creation and annihilation
operators defined with respect to it. In this way we can define proper
complete orthonormal sets of mutually commuting configuration creations
operators $\{C^{\dag}_{I}\}$ and their Hermitian adjoint counterparts
$\{C_{I}\}$, defined in terms of a complete set of many-body
configuration $\{I\}$. These are, in turn, defined by a set-index $I$,
which labels the cluster configuration created by $C^{\dag}_{I}$ with
respect to the reference state $|\psi_{0}\rangle$. Therefore, $\{I\}$
defines a subsystem or cluster within the full system of a given
configuration and the actual choice of these clusters depends upon the
particular system under consideration. We assume that the creation and
annihilation subalgebras and the state $|\psi_{0}\rangle$ are cyclic,
so that all ket states in the Hilbert space $\mathcal{H}$ can be
constructed from linear combinations of states
$\{C^{\dag}_{I}|\psi_{0}\rangle\}$; and for the bra states with
respect to states $\{\langle\psi_{0}|C_{I}\}$. It is well-known in the
many-body application that the above parametrization Eq.~(\ref{1}),
guarantees automatically proper size-extensivity (see section 1.3) and
conformity with the Goldstone linked-cluster theorem \cite{h8} to all
levels of truncation. In contrast, the configuration-interaction
method is size extensive and linked only in the full,
infinite-dimensional space
\cite{h1}. The NCCM parametrization of bra- and ket-states Eq.~(\ref{1}), in its
asymmetrical (independent) form, does not manifestly preserve their
Hermitian congugacy, hence we have here a biorthogonal formulation of
the many-body problem. However, this is the most reasonable
parametrization if one is to preserve the canonical form of the equations of
motion with respect to phase space $\{ s_{I},\tilde{s}_{I}\}$ and the
Hellmann-Feynman theorem \footnote{According to Hellmann-Feynman
theorem, if we perturb the Hamiltonian, $H\to H'=H+\lambda A$ (where
$\lambda $ is infinitesimally small quantity) then the ground state
energy changes as $E_{0}\to E_{0}+\lambda
dE_{0}/d\lambda+O(\lambda^{2})$ with
$dE_{0}/d\lambda=\langle\psi|dH/d\lambda |\psi\rangle$.}
\cite{ccm3,h1}. Nevertheless, non-hermiticity is negligible if the
reference state and its complement are not strongly correlated
\cite{16}. We may hope that $S$ and $\tilde{S}$ are small (in a
somewhat ill-defined non-perturbative sense), in other words, we may
require that some of the coherence has already been obtained by
optimizing the reference state. (This can be done, for example, by a
Hartree-Bogolubov transformation). Then the remaining correlations can
be added via the CCM\footnote{ The application of this procedure to
two-dimensional $\phi^{4}$ theory has been shown to give a stable
result outside the critical region \cite{17}.}. Therefore, defining a
good reference state can in principle control the accuracy of CCM.

In the CIM, one defines the ket and bra states as follows,
\begin{eqnarray}
&&|\psi\rangle=F|\psi_{0}\rangle; \hspace{2cm} \langle\tilde{\psi}|=\langle\psi_{0}|\tilde{F},\nonumber\\
&&F=\sum_{I}f_{I}C^{\dag}_{I}; \hspace{2cm} \tilde{F}=\sum_{I}\tilde{f}_{I}C_{I},\label{cim}\
\end{eqnarray}
where the normalization condition $\langle\tilde{\psi}|\psi\rangle=1$
can not be imposed trivially. Although the CIM has a simpler
parametrization than the CCM, it does not satisfy the
size-extensivity after truncation and contains unlinked pieces emerging from the products of noninteracting subclusters. A formal
relation between the CIM and coupled cluster theory will be demonstrated
in section 1.3.

While all ground-state expectation values $\langle\tilde{\psi}|
A|\psi\rangle =\bar{A}(s_{I},\tilde{s}_{I})$ and amplitudes
$\{s_{I}\}$ are linked in NCCM, the amplitudes $\{\tilde{s_{I}}\}$
contain unlinked terms. This is resolved in the ECCM
\cite{ccm1,h1} where we reparametrize the Hilbert space such that
all basic amplitudes are linked,
\begin{eqnarray}
&&|\psi\rangle=K(t)e^{S}|\psi_{0}\rangle; \hspace{2cm} \langle\tilde{\psi}|=\frac{1}{K(t)}\langle\psi_{0}|e^{\tilde{\Sigma}}e^{-S}, \label{2} \nonumber\\
&& \Sigma|\psi_{0}\rangle=Q e^{\tilde{\Sigma}}e^{-S} S|\psi_{0}\rangle; \hspace{1.4cm} Q=1-|\psi_{0}\rangle\langle\psi_{0}|,\nonumber\\
&&\Sigma=\sum_{I\neq 0} \sigma_{I}C^{\dag}_{I}; \hspace{2.75cm} \tilde{\Sigma}=\sum_{I\neq 0}\tilde{\sigma}_{I}C_{I}.\
\end{eqnarray}
The inverse relationships between the ECCM and the NCCM counterparts are given by
\begin{equation}
\sigma_{I}\equiv\langle\psi_{0}|C_{I}\tilde{S}S|\psi_{0}\rangle, \hspace{1.5cm}
s_{I}\equiv\langle\psi_{0}|C_{I}e^{\tilde{\Sigma}}\Sigma|\psi_{0}\rangle.
\end{equation}
The ECCM amplitudes $\{\sigma_{I},\tilde{\sigma}_{I}\}$ are
canonically conjugate (which comes from time dependent
variation)\footnote{The equation of motion for the ECCM amplitudes can be obtained from a variational principle \cite{ccm3} by requiring the action-like functional 
\begin{equation}
\mathcal{A}=\int dt \langle \tilde{\psi}|i\partial/\partial t- H(t)|\psi\rangle,
\end{equation}
to be stationary against small variations of amplitudes. After some straightforward algebra, one can obtain
\begin{equation}
\mathcal{A}=\int dt \Big[-i\sum_{I\neq 0}\dot{\tilde{\sigma}}_{I}\sigma_{I}-\langle H\rangle\Big].
\end{equation}
The stationary conditions lead us to the pair of equations of motion,
\begin{equation}
i \dot{\sigma}_{I}=\frac{\delta \langle H\rangle}{\delta \tilde{\sigma}_{I}} \hspace{2cm} i\dot{\tilde{\sigma}}_{I}=-\frac{\delta\langle H\rangle}{\delta\sigma_{I}}.
\end{equation}
Therefore, $\{\sigma_{I},\tilde{\sigma}_{I}\}$ are obviously canonically conjugate to each other in the usual terminology of classical Hamiltonian mechanics. 
We will elaborate more on this property of the coupled-cluster theory
in the context of the renormalization group in the chapter 2.7.}. It
must be clear that this parametrization for the ECCM is not unique,
however it is complete and sufficient to specify the ECCM phase
space. It has been pointed out earlier that this choice of
parametrization is the most convenient one
\cite{ccm1}. The ECCM is believed to be the unique formulation of
quantum many-body with full locality and separability at all levels of
approximation. The individual amplitudes $\{s_{I},\tilde{s}_{I}\}$ (or
$\{\sigma_{I},\tilde{\sigma}_{I}\}$) in both parametrizations are
determined independently by solving an infinite set of non-linear
equations which emerge from the dynamics of the quantum system. In
practice, one needs to truncate both sets of coefficients. A
consistent truncation scheme is the so called SUB($n$) scheme, where
the $n$-body partition of the operator $\{s_{I},\tilde{s}_{I}\}$ (or
$\{\sigma_{I},\tilde{\sigma}_{I}\}$) is truncated so that the general set-index $\{I\}$  
contains up to $n$-tuple excitation (e.g., of single particle for
bosonic systems with a reference state). Determination of the amplitudes corresponds to summing
infinite sets of diagrams which in perturbation language take into
account arbitrary high-order contributions in the coupling constant,
therefore the NCCM and ECCM are not an expansion in this coupling
constant. This property demands a precise truncation scheme for
hierarchies without losing the renormalizability of the theory. We
will consider this problem in the next chapter.

In the following, we confine our consideration to a
real Klein-Gordon field, as an example. The configuration operators are specified
by $\{C_{I}\to a_{k_{1}}...a_{k_{I}} , C_{I}^{\dag}\to
a^{\dag}_{k_{1}}...a^{\dag}_{k_{I}}\}$ with subalgebra
\begin{equation}
 [a_{k},a^{\dag}_{k'}]=\delta_{kk'}, \hspace{2cm}  [a_{k},a_{k'}]=0. \label{0.2}
\end{equation}
The corresponding $S$ and $\tilde{S}$ operators are
\begin{eqnarray}
&&S=\sum_{n=1}S_{n}, \hspace{2cm} S_{n}=\sum_{q_{1},..q_{n}}\frac{1}{n!}s_{n}(q_{1},..q_{n})a^{\dag}_{q_{1}}..a^{\dag}_{q_{n}}, \label{a0.3}\nonumber\\
&&\tilde{S}=1+\sum_{n=1}\tilde{S}_{n},
\hspace{1.3cm}\tilde{S}_{n}=\sum_{q_{1},..q_{n}}\frac{1}{n!}\tilde{s}_{n}(q_{1},..q_{n})a_{q_{1}}..a_{q_{n}},\
\end{eqnarray}
 and $|\psi_{0}\rangle$ is the Fock vacuum. The individual amplitudes
 $\{s_{n},\tilde{s}_{n}\}$ which describe excitations of $n$ Fock
 particles have to be fixed by the dynamics of the quantum
 system. Using Fock states in QFT has often been ambiguous due to
 problems connected with Haag's theorem \cite{h13} (Haag's theorem
 says that there can be no interaction picture - that we cannot use
 the Fock space of noninteracting particles as a Hilbert space - in
 the sense that we would identify Hilbert spaces via field polynomials
 acting on a vacuum at a certain time).
It is well-known that the algebraic structure, does not, in general, fix the Hilbert space representation and
 therefore dynamical considerations are required. The IC formalism
 invokes dynamics rigorously. The dynamical principle to fix the physical vacuum is
 Poincar\'e invariance,
\begin{equation}
H|\psi\rangle=P|\psi\rangle=L|\psi\rangle=K|\psi\rangle=0,  \label{3}
\end{equation}
where $H, P, L$ and $K$ are generators of the Poincar\'e group (Hamiltonian, momentum, angular momentum and boost operators, respectively). The
excited states are no longer invariant under these symmetry operations
and the spectrum can be obtained in an extended version of IC
\cite{h14}.  By putting the Ansatz Eqs.~(\ref{1},\ref{a0.3}) into the
conditions Eq.~(\ref{3}) one can obtain:
\begin{eqnarray}
&&\langle q_{1},...,q_{n}|e^{-S}Pe^{S}|\psi_{0}\rangle=\langle q_{1},...,q_{n}|(P+[P,S])|\psi_{0}\rangle=\Big[\sum_{i=1}^{n}q_{i}\Big]s_{n}(q_{1},...,q_{n})=0, \label{4}\nonumber\\
&&\\
&&\langle\psi_{0}|\tilde{S}e^{-S}Pe^{S}|q_{1},...,q_{n}\rangle=\langle\psi_{0}|\tilde{S}\big(P+[P,S]\big)|q_{1},...,q_{n}\rangle= 
\Big[\sum_{i=1}^{n}q_{i}\Big]\tilde{s}_{n}(q_{1},...,q_{n})\nonumber\\
&&+\sum_{m}\frac{\tilde{s}_{n+m}(q_{1},...,q_{n+m})}{m!}\Big[[\sum_{i=1}^{m}q_{i}]s_{m}(q_{1},...,q_{m})\Big]=0,\label{new-added4}\\
&&\langle q_{1},...,q_{n}|e^{-S}Le^{S}|\psi_{0}\rangle=\sum_{\alpha}^{n}\epsilon_{ijl}(q_{\alpha})_{j}\frac{\partial}{\partial(q_{\alpha})_{l}}s_{n}(q_{1},...,q_{n})=0.  \label{5}\
\end{eqnarray}
In the same fashion one can impose the condition
$\langle\psi_{0}|\tilde{S}e^{-S}Le^{S}|q_{1},...,q_{n}\rangle=0$, which leads to the following condition, having made used of the equation (\ref{5}): 
\begin{equation}
\sum_{\alpha=1}^{n}\epsilon_{ijl}(q_{\alpha})_{j}\frac{\partial}{\partial(q_{\alpha})_{l}}\tilde{s}_{n}(q_{1},...,q_{n})=0 \label{5-ray}
\end{equation}
Similarly for Hamiltonian and boost operators we have 
\begin{eqnarray}
&&\langle q_{1},...,q_{n}|e^{-S}He^{S}|\psi_{0}\rangle=\langle\psi_{0}|\tilde{S}e^{-S}He^{S}|q_{1},...,q_{n}\rangle=0,
\label{6}\\ &&\langle
q_{1},...,q_{n}|e^{-S}Ke^{S}|\psi_{0}\rangle=\langle\psi_{0}|\tilde{S}e^{-S}Ke^{S}|q_{1},...,q_{n}\rangle=0. \label{7}\
\end{eqnarray}
The equations (\ref{4},\ref{new-added4},\ref{5},\ref{5-ray}) lead us to
\begin{equation}
s_{n}(q_{1},...,q_{n})=\delta\Big[\sum_{i=1}^{n}q_{i}\Big]s_{n}(\{q_{i}.q_{j}\}), \hspace{1cm} \tilde{s}_{n}(q_{1},...,q_{n})=\delta\Big[\sum_{i=1}^{n}q_{i}\Big]\tilde{s}_{n}(\{q_{i}.q_{j}\}),
\end{equation}
which means that $\{s_{I},\tilde{s}_{I}\}$ depend on scalar quantities only and
momentum is preserved. The same result can be obtained for the ECCM
parametrization. To complete the determination of phase space, we use
the energy hierarchy Eq.~(\ref{6}), where $H$ is a normal ordered
Hamiltonian and the vacuum energy vanishes. Conceptually, it is evident
that Eq.~(\ref{6}) suffices to fix all amplitudes without invoking
Eq.~(\ref{7}), but explicit verification to all orders seems to be
impossible.

The fully linked feature of the $e^{-S}He^{S}$ term in Eq.~(\ref{6}) can be made
explicit by denoting it as $\{He^{S}\}_{\mathcal{L}}$, which can be written as a set of nested commutators,
\begin{equation}
e^{-S}He^{S}=\{He^{S}\}_{\mathcal{L}}=H+[H,S]+\frac{1}{2!}[[H,S],S]+...\hspace{1cm} .\label{10}
\end{equation}
This procedure is still rigorous. Lorentz symmetry, stability and
causality are examples of features normally expected to hold in
physical quantum field theories. In renormalized QFT stability and
causality are closely intertwined with Lorentz invariance. For
example, stability includes the need for energy positivity of Fock
states of ordinary momenta, while causality is implemented
microscopically by the requirement that observables commute at
spacelike separation \cite{h15}, so-called microcausality. In
addition, both are expected to hold in all inertial frames. A stable
and causal theory without Lorentz symmetry could in principle still be
acceptable \cite{h16}. In the framework of many-body theory, medium
contributions always affect the local and global properties of hadrons
especially at high density. In dense matter, the Pauli principle and
cluster properties can affect causality since they restrict
the permissible process in a scattering reaction. To understand these effects we need firstly to consider
if our formalism itself can in principle preserve the causal structure
of given physical system. Let us introduce a new set
$\{b_{k},b_{k}^{\dag}\}$ which are connected to previous
$\{a_{k},a_{k'}^{\dag}\}$ defined in Eq.~(\ref{0.2}) via a generalized
Bogolubov transformation,
\begin{equation}
b^{\dag}_{k}=A_{ki}a^{\dag}_{i}+B_{ki}a_{i}+D_{i},\label{b}
\end{equation}
This is the most general linear transformation, which preserves
commutator relations ($b_{k}$ is an annihilation operator and $b^{\dag}_{k}$ is
the Hermitian conjugate of $b_{k}$),
\begin{equation}
 [b_{k},b^{\dag}_{k'}]=\delta_{kk'}, \hspace{2cm}  [b_{k},b_{k'}]=0, \label{11}
\end{equation}
provided that $AA^{\dag}-BB^{\dag}=1$. This obviously preserves the
commutator relations for boson fields and preserves microcausality
explicitly. The bare ``$a$ vacuum'' defined by
$a_{k}|\psi_{0}\rangle_{a}=0$,  is replaced by a bare ``$b$ vacuum''
which satisfies $b_{k}|\psi_{0}\rangle_{b}=0$. Using Thouless theorem \cite{h18}, $|\psi\rangle_{b}$ can be written as
\begin{equation}
|\psi\rangle_{b}=N^{-1/2}e^{S_{1}+S_{2}}|\psi\rangle_{a}, \label{11-2}
\end{equation}
with
\begin{equation}
S_{1}=\sum_{k}S^{1}_{k}(A,B,D)a^{\dag}_{k}, \hspace{1.5cm} S_{2}=\sum_{kk'}\frac{1}{2!}S^{2}_{kk'}(A,B,D)a^{\dag}_{k} a^{\dag}_{k'},
\end{equation}
where $S^{1}$ and $S^{2}$ are known functions of the matrices $A$ and
$B$ and the vector $D$ \cite{h18}. It is obvious that $|\psi\rangle_{b}$ is a
low-order approximation to the CCM wave function Eq.~(\ref{1}) or
(\ref{2}). The above-mentioned parametrization of vacuum wave function can be generalized to the CCM
wave function, however it may require a nonlinear transformation from
which it can be constructed.  The inspiration for this transformation
can be taken from the extension IC formulation for excited states
\cite{h14}
\begin{equation}
b^{\dag}_{k}=e^{S}F(k), \hspace{2cm}
F(k)=a_{k}^{\dag}+\sum_{n=3}^{\infty}F_{n}(p_{1},...,p_{n-1},k)a^{\dag}_{p_{1}}...a^{\dag}_{p_{n-1}}, \label{14}
\end{equation}
where the correlation operator $S$ is known from Eq.~(\ref{a0.3}) and $F$ is a
new amplitude which includes momentum conservation.  This new
amplitudes has to be determined and it changes the momentum of the bosons before
creation. It is not hard to derive equations for the energy spectrum
which lead us to $N$-body effective Hamiltonians
\cite{h7} and yields the folded diagrams of degenerate many-body
perturbation theory \cite{h19}. This nonlinear transformation
manifestly invalidates the commutator relation Eq.~(\ref{11}) and
accordingly the microcausal commutation relations of the boson
field. It should be noted that causality of the underlying theory can
not be fully determined at this level and one needs to take into account the
dynamics of the underlying quantum system. Causality in the context of
IC formulation can be ensured by requiring
\begin{equation}
\langle\tilde{\psi}|[\phi(X),\phi(Y)]|\psi\rangle=0, \hspace{2cm} (X-Y)^2<0 , \label{a}
\end{equation}
where $Y$ denotes space-time coordinates $(y,y_{0})$ and $\phi(Y)$ is Klein-Gordon field operator which is expressed in terms of Fock space operators by
\begin{equation}
\phi(Y)=\sum_{k}\xi(Y)_{k}a_{k}+\xi^{\dag}_{k}(Y)a^{\dag}_{k}, \label{a-new-fock}
\end{equation}
where $\xi_{k}(Y)$ form a complete orthonormal set of states. Eq.~(\ref{a}) can be evaluated explicitly by using the following identity \cite{ccm3,h1}
\begin{equation}
\langle\tilde{\psi}|[A,B]|\psi\rangle=\overline{[A,B]}=\sum_{I}\frac{\partial\bar{A}}{\partial{x_{I}}}\frac{\partial\bar{B}}{\partial{\tilde{x}_{I}}}-\frac{\partial\bar{A}}{\partial{\tilde{x}_{I}}}\frac{\partial\bar{B}}{\partial{x_{I}}}, \label{17}
\end{equation}
where $\{x_{I},\tilde{x}_{I}\}$ are the canonical coordinate and
momenta of the NCCM or the ECCM parametrization. By making use of
Eqs.~(\ref{a0.3},\ref{10}) and (\ref{a-new-fock}) one can find 
\begin{equation}
\langle\tilde{\psi}|\phi(Y)|\psi\rangle=\overline{\phi(Y)}=\sum_{n=1}\sum_{k}\xi_{k}(Y)
\tilde{x}_{n-1}(q_{1},...,q_{n-1})x_{n}(q_{1},...,q_{n-1},k)+\xi^{\dag}_{k}(Y)\tilde{x}_{1}(k),
\end{equation}
where $\{x_{n}\to s_{n},\sigma_{n},\tilde{x}_{n}\to
\tilde{s}_{n},\tilde{\sigma}_{n}\}$ and we define
$\tilde{x}_{0}=1$. By exploiting Eq. (\ref{17}) one can show
\begin{equation}
\overline{[\phi(X),\phi(Y)]}=\sum_{k,k'}[\xi_{k}(X)\xi^{\dag}_{k'}(Y)-\xi^{\dag}_{k}(X)\xi_{k'}(Y)]=\Delta(X-Y)-\Delta(Y-X).
\end{equation}
The function $\Delta(X-Y)$ is the Pauli-Jordan function \cite{h15} defined for
field operators expanded in plane-wave basis. When $(X-Y)^2<0$, we can
perform a Lorentz transformation on the second term, taking $(X-Y)\to
-(X-Y)$. The two terms are therefore equal and cancel to give zero, hence, Eq.~({\ref{a}) is satisfied. It
should be noted that for a general quantum system by considering just
the dynamics of the underlying system, determining the amplitudes
$\{x_{I},\tilde{x}_{I}\}$ and verifying Eq.~(\ref{a}) at every level
of truncation one can ensure causality. Obviously this might introduce
a lower limit of truncation in a consistent SUB(n) scheme which
contains causality.

In relativistic quantum mechanics there is another
distinct type of causality, the fact that there is a well-posed initial
value problem, so-called Cauchy causality. In the local field theory the Poincar\'e
invariance implies the existence of a unitary representation of the
Poincar\'e group that acts on the Hilbert space. In other words, the
transformed final state is uniquely determined by time evolving the
transformed initial state. Therefore, Cauchy causality is a
consequence of Poincar\'e invariance and is independent of any
consideration concerning microcausality.

One of the standard approaches in the nuclear many-body theory is to
introduce an effective interaction and reduce the full many-body
problem to a problem in a small model-space. As an example for the
meson-nucleon system below threshold, it was shown that one has either
hermitian effective operators with all desirable transformation
properties under the Poincar\'e group or non-hermitian ones obtained by the
CCM with desirable simplicity \cite{h17}.  Thus effective operators
introduced by the CCM can not manifestly preserve Cauchy causality, at
any level of truncation. However, it economically disentangles and
reduces the complexities of the many-body problem.  It is notable that
in some cases, the violation of relativistic invariance due to
truncation is consistent with errors introduced via the approximation
\cite{h18}. In practice one should not expect that Poincar\'e invariance
holds for an approximation, and forcing the approximated wave
function to obey Lorentz invariance might lead to inconsistent results. Having
said that, we will show that in the context of coupled-cluster renormalization group 
framework, one can in principle define a truncation scheme where
Poincar\'e invariance is preserved.

\section{Linked-cluster theorem and Wightman functionals}
In this section we proceed from a formal view to show the relation
between the CIM and the coupled cluster method. We will show that the
CCM is a natural reparametrization of the CIM which incorporates the
full size-extensivity. For this purpose we exploit the reconstruction
theorem \cite{app1}, well known in axiomatic Quantum field theory
\cite{app2}:

Denote the state vector of the vacuum by $|\Omega\rangle$ (in the language of the CCM $|\Omega\rangle$ is the full interacting ground state $|\psi\rangle$). The physical vacuum expectation value of products of local fields
\begin{equation}
w^{n}(x_{1}...x_{n})=\langle\Omega|\phi(x_{1})...\phi(x_{n})|\Omega\rangle,   \label{w-int}
\end{equation}
are tempered distributions\footnote{Tempered distributions generalize
the bounded (or slow-growing) locally integrable functions; all
distributions with compact support and all square-integrable functions
can be viewed as tempered distributions.} over
$\mathbb{R}^{4n}$. These $w^{n}$ are called Wightman distributions. If
the hierarchy of distributions $w^{n}(n=0,...)$ is known then the
Hilbert space can be constructed. This is the so-called reconstruction
theorem \cite{app1}.

Now consider a configuration $(x_{1},...,x_{n})$ consisting of several
clusters. A cluster here is a subset of points so as all the points in
one cluster have a large space-like separation from all the points in
any other clusters. One may expect that in infinite separation between
the clusters, one has
\begin{equation}
w^{n}(x_{1},...,x_{n})\sim \prod_{r}w^{n_{r}}(y_{r,1},...y_{r,n_{r}}), \label{he1-1}
\end{equation}
the index $r$ and $n_{r} $ denote the cluster and number of points in
the $r$-th cluster, respectively. $y_{r,n_{k}}$ indicates a subset of
the $x_ {k}$ belonging to this cluster. Thus we require that in the
vacuum state the correlation of quantities relating to different
regions decreases to zero as the space-like separation of the regions
increases to infinity.  Therefore it makes sense to introduce another
hierarchy of functions $w^{n}_{T} $,
\begin{equation}
 w^{n}(x_{1},...,x_{n})=\sum_{p}
\prod_{r}w^{n_{r}}_{T}(y_{r,k}) \label{a2}.
\end{equation}
Here $p$ denotes a partition of the set of points $x_{i}$ into subsets
labeled by the index $r$. The sum is over all possible partitions, The
objects $w^{n}_{T}$ are called truncated functions or correlated
functions\footnote{For free fields all truncated functions with $n\neq2$ vanish, thus it suffices to know the two-point functions.}.  Within
the hierarchy $\{w^{n}\}$ the truncated functions can be computed
recursively:
\begin{eqnarray}
&&w^{1}_{T}(x)=w^{1}(x),\nonumber\\
&&w^{2}_{T}(x_{1},x_{2})=w^{2}(x_{1},x_{2})-w^{1}(x_{1})w^{1}(x_{2}),\\
&&\vdots\nonumber\
\end{eqnarray}
The asymptotic property Eq.~(\ref{he1-1}) of the $w^{n}$ is converted to the simpler property
\begin{equation}
w^{n}_{T}(x_{1},...x_{n})\to 0,
\end{equation}
if any space like separation $x_{i}-x_{j}$ goes to infinity.

Let us now consider a hierarchy of functions $P^{n}(k_{1},...k_{n})$
(for simplicity totally symmetric under permutation of their
arguments). We define a generating functional $\mathcal{P}\{f\}$ for an arbitrary
function $f(k)$:
\begin{equation}
 \mathcal{P}\{f\}=\sum_{n}\frac{1}{n!}\int  P^{n}(k_{1},...k_{n})f(k_{1})...f(k_{n}). \label{a3}
\end{equation}
The $P^{n}$ can be found from
\begin{equation}
P^{n}(k_{1},...k_{n})=\frac{\delta^{n}}{\delta f(x_{1})..\delta f(x_{n})}\mathcal{P}\{f\}|_{f=0}.
\end{equation}
Now one can simply use the definition Eq.~(\ref{a2}) to obtain a relation between the hierarchy
$\{P^{n}\}$ and the hierarchy of the truncated functions $\{P^{n}_{T}\}$ in terms of respective generating function, namely
\begin{equation}
\mathcal{P}\{f\}=e^{ \mathcal{P}_{T}\{f\}}.\label{he4}
\end{equation}
This relation is called the Linked cluster theorem
\cite{app2}. Having in mind the relations Eqs.~(\ref{a3}) and
(\ref{he4}), one observes that the CCM parametrization introduced in
Eqs.~(\ref{1}) and (\ref{2}) is a natural reparametrization of the CIM defined in Eq.~(\ref{cim}) which
incorporates the size-extensivity \cite{h1}, at finite levels of
truncation. This leads to a cluster decomposition and linked
decomposition property of CCM at any level of truncation. Thus for
extensive variables as the energy, the linked terms lead to
contributions which obey the proper linear scaling in the particle
number. The CIM contains unlinked diagrams for the ground-state
energy expectation value, and thereby suffers from the so-called
size-extensivity problems.

The cluster decomposition condition Eq.~(\ref{he1-1}) is fundamental
to a quantum field theory. It may break down partly when, in a
statistical-mechanical sense, the theory is in a mixed phase
\cite{clu-dec}. This implies that there is more than one possible vacuum
state, and therefore the cluster decomposition should be
restored in principle if one builds the Hilbert space on one of the vacua.

Notice that in QCD, the cluster decomposition for colour singlet
objects can break down due to the confinement. In axiomatic local
quantum field theory with an indefinite metric space $\nu$ (e.g.,
Minkowski space) the following theorem holds; for the vacuum
expectation values of two (smeared local) operators $A$ and $B$ with
spacelike distance $R$ from each other \cite{germany,germany0}:
\begin{eqnarray}
|\langle\Omega|A(x)B(0)|\Omega\rangle&-&\langle\Omega|A(x)|\Omega\rangle\langle\Omega|B(0)|\Omega\rangle|
\nonumber\\ &\leq&\left\{\begin{array}{ll}\text{const}\times
R^{-3/2+2N}e^{-MR}& \text{if there is a mass gap M},\\
\text{const}\times R^{-2+2N}&\text{if there is no mass
gap},\end{array}\right.\label{conf}\
\end{eqnarray}
where $M$ is the mass gap. (Herein, we assume $N=0$, however, a
positive $N$ is possible for the indefinite inner product
structure in $\nu$.) In order to avoid the decomposition property
for product of unobservable operators $A$ and $B$ which together
with Kugo-Ojima criterion\footnote{The Kugo-Ojima confinement
scenario describes a mechanism by which the physical state space
contains only colourless states, the colored states are not
BRS-singlets and therefore do not appear in $S$-matrix elements,
since they are confined \cite{germany,germany1}.}
\cite{germany,germany1} for the confinement is equivalent to
failure of the decomposition property for coloured clusters, there
can not be a mass gap in the indefinite space $\nu$. This would
thus eliminate the possibility of scattering a physical state into
color singlet states consisting of widely separated colored
clusters (the ``behind-the-moon'' problem). However, this has no
information for the physical spectrum of the mass operator in
$\mathcal{H}$ which indeed does have a mass gap. In another words,
there is a mass gap (but not in the full indefinite space $\nu$)
in the semi-definite physical subspace induced by the BRST charge
operator $Q_{B}$, $\nu_{\text{phy}}=\text{Ker} Q_{B}$, where
states within are annihilated by $Q_{B}$,
\begin{eqnarray}
v_{\text{phys}}&=&\{|\psi\rangle\in v: Q_{B}|\psi\rangle=0\}=\text{Ker} Q_{B},\nonumber\\
\mathcal{H}(Q_{B},v)&=& \text{Ker} Q_{B}/\text{Im} Q_{B},\
\end{eqnarray}
where states in $\text{Im} Q_{B}= (\text{Ker} Q_{B})^{\bot}$ are
called BRST-coboundaries in the terminology of de Rham cohomology and
do not contribute in $\nu_{\text{phys}}$ \cite{geom}. It has been shown
\cite{germany2} that there is a connection between Kugo-Ojima
criterion and an infrared enhancement of the ghost propagator. In
Landau gauge, the gluon-ghost vertex function offers a convenient
possibility to define a non-perturbative running coupling. The
infrared fixed point obtained from this running coupling determines
the two-point color-octet interactions and leads to the existence of
unphysical massless states which are necessary to escape the cluster
decomposition of colored clusters \cite{germany}. Notice that a
dynamical mechanism, responsible for breaking of cluster decomposition
for colored objects, is yet to be discovered, since our knowledge about
color confinement is still preliminary.

\chapter{Hamiltonian renormalization Groups}
\section{Introduction}
The main goal of traditional renormalization theory is to determine
when and how the cancellation of divergences originating from the
locality of quantum field theory may occur. This is essential if one
wants to have meaningful quantitative results. However, it is by no
means obvious how quantum fluctuations associated with short distance
scales can be incorporated and controlled through the choice of only a few
parameters, typically the bare masses and coupling constants, or by
the counterterms in renormalized perturbation theory.

The development of Wilson's renormalization group (RG) formalism
\cite{18} allowed physicists to produce a logically consistent
picture of renormalization in which perturbation theory at
arbitrary high energy scale can be matched with the perturbation
expansion at another scale, without invoking the details of
intermediate scales. In the Wilsonian approach, all of the
parameters of a renormalizable field theory can be thought of as
scale-dependent objects and their flows are governed by the
so-called RG differential equations.

Motivated by Wilson's picture, effective field theory (EFT) approaches have
been introduced to replace complicated fundamental theories with
simpler theories based only on the relevant degrees of freedom at the physical
scale of interest \cite{eff1}. The basis of the EFT concept is the recognition of
the importance of different typical energy scales in nature, where each
scale has its own characteristic degrees of freedom. In strong
interactions the transition from the fundamental to the effective level is
induced by a phase transition that takes place around
$\Lambda_{\text{QCD}}\simeq 1 \text{GeV}$ via the spontaneous breaking of
chiral symmetry, which generates pseudoscalar Goldstone bosons.  This
coincides, of course, with the emergence of nuclei and nuclear mather,
as opposed to the quark-gluon plasma and quark matter expected to
occur at high temperature and high density. Therefore, at low energies
($E<\Lambda_{\text{QCD}}$), the relevant degrees of freedom are not quarks
and gluons, but pseudoscalar mesons and other hadrons. The
resulting description is a chiral EFT, which has much in common with
traditional potential models. In particular, there might be an
intermediate regime where a non-relativistic model is inadequate but
where relatively few hadronic degrees of freedom can be used to
faithfully describe both nuclear structure and response. If this is
the case, one should be able to describe strongly interacting hadronic
systems in this effective model.

The power of Hamiltonian methods is well known from the study of
non-relativistic many-body systems and from strongly-interacting
few particle systems, even though a Lagrangian approach is usually
chosen for relativistic theories. One might prefer to obtain the
effective interaction using the covariant Lagrangian formalism and
then consider the ground state and collectively excited states in
the non-covariant Hamiltonian formalism by exploiting many-body
techniques. For the description of physical states and in
particular bound states, the Hamiltonian formalism is preferable
over the Lagrangian one. This is due to the fact that such
problems are not naturally defined covariantly. For a bound state
the interaction time scale is infinite, and thus a
time-independent approach is better suited. However, the
Lagrangian can not generally be converted to a Hamiltonian if the
effective Lagrangian contains higher-order time-derivatives, since
no Legendre transformation exits for such a case. Therefore one
may wish to obtain the effective interaction in a unified
self-consistent way within the Hamiltonian formalism. The
renormalization group transformation is used to derive the
physical Hamiltonian that can describe experiment. To implement a
RG calculation, one should define a space of Hamiltonians and find
a certain RG transformation which maps this space into itself.
Then one should study the topology of the Hamiltonian space, by
searching for the fixed points and studying the trajectories of
the Hamiltonian with respect to these fixed points. At the fixed
points we have a scale-invariant quantum field theory. Near the
fixed point, the irrelevant operators in the original Hamiltonian
have small coefficients, and we are only left with the relevant
and marginal operators. The coefficients of these operators
correspond to the parameters of the renormalized field theory. In
this way, in a renormalizable field theory, we disregard
information about the evolution of irrelevant terms. The fact that
irrelevant terms can be dropped near the fixed point does not
necessarily imply that they are unimportant at low-energy scales
of experimental interest. This depends on how sensitive the
physical observables are to physics near the scale of the cutoff.
Therefore, it is of interest to develop a RG method in which the
irrelevant variables are treated on an equal footing with the
relevant and marginal variables (the schemes presented here
have this advantage).

Hamiltonian methods for strongly-interacting systems are intrinsically
non-perturbative and usually contain a Tamm-Dancoff type
approximation, in the sense that one expands the bound state in states containing a small number of particles. This
truncation of the Fock space gives rise to a new class of
non-perturbative divergences, since the truncation does not allow us
to take into account all diagrams for any given order in perturbation
theory. Therefore renormalization issues have to be considered
carefully. Two very different remedies for this issue are the use of
light-front Tamm-Dancoff field theory (LFFT)
\cite{3} and the application of the coupled cluster method (CCM)
\cite{h1,5}. However, both methods are too complicated to attack the issue
in a self-consistent way. In the last decade extensive attempts have
been made to give a workable prescription for renormalization within the
Hamiltonian formalism \cite{aps1,aps2,10,me}. Commonly unitary
transformations are used to decouple the high- and low-energy modes
aiming at the partial diagonalization of the Hamiltonian.  One of the
most elegant approaches in this context is the similarity
renormalization group (SRG) proposed by Glazek and Wilson
\cite{aps1} (and by Wegner \cite{aps2} independently).
The SRG \cite{aps1,aps2} is designed to be free of small energy
denominators and to eliminate interactions which change the energies
of the unperturbed states by a large amount.  However, there are
several problems with this approach: it is hard to incorporate loop
expansions within the method, the SRG can not systematically remove
interactions which change the number of particles (i.e, when the
Hamiltonian is not diagonal in particle number space), and most
importantly, the computations are complex and there is no efficient
non-perturbative calculation scheme.

In this chapter we introduce a new method \cite{me,me1} for obtaining the low-energy
effective operators in the framework of a CCM approach. The
transformation constructed avoids the small denominators that plague
old-fashioned perturbation theory. Neither perturbation theory nor
unitarity of the transformation are essential for this method. The
method is non-perturbative, since there is no expansion in the
coupling constant; nonetheless, the CCM can be conceived as a
topological expansion in the number of correlated excitations. We show
that introducing a double similarity transformation using
linked-cluster amplitudes will simplify the partial diagonalization
underlying renormalization in Hamiltonian approaches. However, a price
must be paid: due to the truncation the similarity transformations are
not unitary, and accordingly the hermiticity of the resultant
effective Hamiltonian is not manifest. This is related to the fact
that we have a biorthogonal representation of the given many-body
problem. There is a long tradition of such approaches. The first we
are aware of are Dyson-type bosonization schemes \cite{11}. {[}Here
one chooses to map the generators of a Lie algebra, such that the
raising generators have a particularly simple representation.{]} The
space of states is mapped onto a larger space where the physically
realizable states are obtained by constrained dynamics. This is
closely related to CCM formalism, where the extended phase space is a
complex manifold, the physical subspace constraint function is of
second class and the physical shell itself is a K\"{a}hler manifold
\cite{12}. The second is the Suzuki-Lee method in the nuclear
many-body (NMT) problem \cite{13,Ni}, which reduces the full many-body
problem to a problem in a small configuration space and introduces a
related effective interaction. The effective interaction is naturally
understood as the result of certain transformations which decouple a
model space from its complement. As is well known in the theory of
effective interactions, unitarity of the transformation used for
decoupling or diagonalization is not necessary. Actually, the
advantage of a non-unitarity approach is that it can give a very
simple description for both diagonalization and ground state.
This has been discussed by many authors \cite{14} and, although it
might lead to a non-hermitian effective Hamiltonian, it has been shown
that hermiticity can be recovered \cite{12,15}. Notice that defining a good
model space can in principle control the accuracy of CCM \cite{16,17}.

To solve the relativistic bound state problem one needs to
systematically and simultaneously decouple 1) the high-energy from
low-energy modes and 2) the many- from the few-particle states. We
emphasize in this chapter that CCM can in principle be an adequate
method to attack both these requirements. Our hope is to fully utilize
Wilsonian Exact renormalization group \cite{18} within the CCM
formalism. Here the high energy modes will be integrated out leading
to a modified low-energy Hamiltonian in an effective many-body space.
Notice that our formulation does not depend on the form of dynamics
and can be used for any quantization scheme, e.g., equal time or
light-cone.

\section{Traditional approaches and their problems}
The Tamm-Dancoff approximation \cite{tamm} was developed in the 1950's
to describe a relativistic bound state in terms of a small number of
particles. It was soon revealed that the Tamm-Dancoff truncation gives
rise to a new class of non-perturbative divergences, since the
truncation does not allow us to take into account all diagrams at a
given order in perturbation theory. On the other hand, any naive
renormalization violates Poincar\'{e} symmetry and the cluster
decomposition property (the cluster decomposition means that if two
subsystems at very large space-like separation cease to interact then
the wave function becomes multiplicatively separable). One of the
simplest example of the Tamm-Dancoff approximation is the constituent
picture of QCD where one describes a QCD bound state
$|\psi\rangle$ within a truncated Fock-space,
\begin{equation}
|\psi\rangle=\phi_{1}|q\bar{q}\rangle+\phi_{2}|q\bar{q}g\rangle+\phi_{3}|q\bar{q}q\bar{q}\rangle+...\nonumber
\end{equation}
we use a shorthand notation for the Fock-space where $q$ is a
quark, $\bar{q}$ an antiquark, and $g$ stands for a gluon. Now the
bound-state problem is solved via the Schr\"odinger equation
$H^{QCD}|\psi\rangle=E|\psi\rangle$. It is well known that in the
complicated equal-time vacuum bound states contain an infinite
number of particles that are part of the physical vacuum on which
hadrons are built. On the other hand, interactions in a field
theory couple states with arbitrarily large difference in both
free energy and numbers of particles. Thus any Fock-space
expansion can hardly be justified without being supplemented with
a prescription for decoupling of the high-energy from the
low-energy modes and the many-body from few-particle states. In
the context of the Hamiltonian formulation this problem can be
expresed by asking how the Hamiltonian matrix is diagonalized in
particle- and momentum-space.

In his earliest work Wilson\cite{aps3} exploited a Bloch type
transformation \cite{aps4} to reduce the Hamiltonian matrix by
lowering a cutoff which was initially imposed on the individual
states.  Later Wilson abandoned this formulation in favour of a
Lagrangian one. The most important reason was that the Bloch
transformation is ill-defined and produces unphysical
divergences. These divergences emerge from denominators which contain
a small energy difference between states retained and states removed
by the transformation, and appear across the boundary line at
$\lambda$ .
\begin{figure}[!tp]
\vspace{1cm}
\includegraphics[height=.23\textheight]{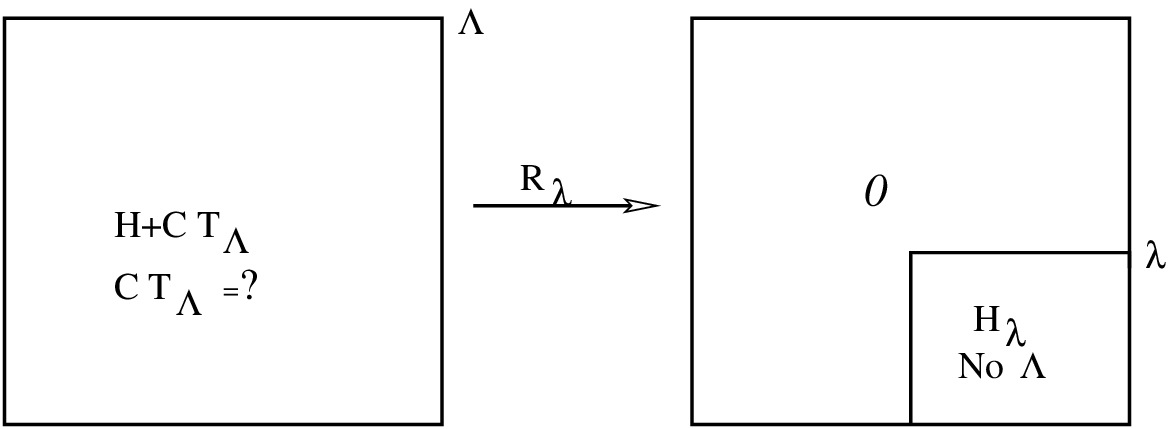}
\includegraphics[height=.23\textheight]{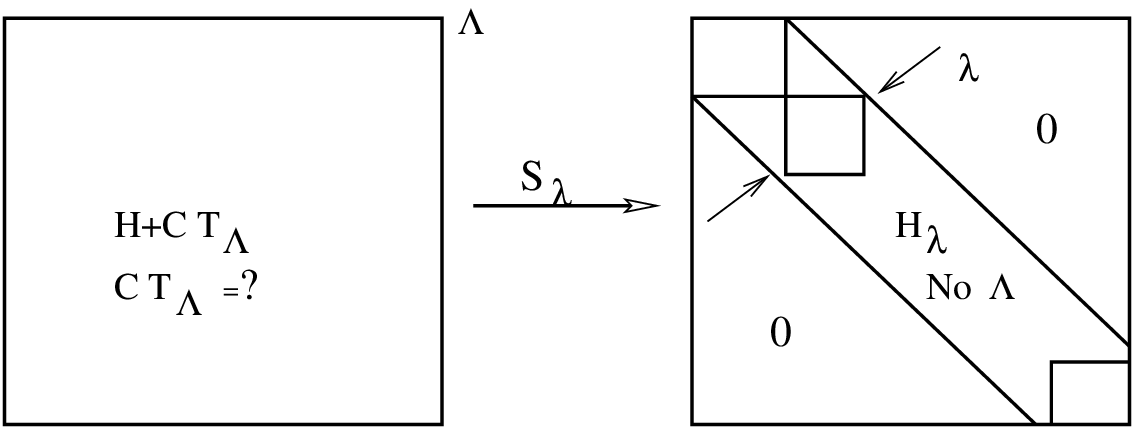}
\caption{Two ways to run a cutoff on free energy. In top a cutoff on the magnitude of the energy is lowered from $\Lambda$ to $\lambda$, leading to a small matrix (this scheme corresponds to the original Wilson RG). In down we show similarity renormalization group scheme where a cutoff is imposed on the off-diagonal part of the matrix and diagonalization is led to a narrow matrix with a band width $\lambda$. $CT_{\Lambda}$ denotes the counterterms.}
\end{figure}

Two remedies for this issue are the use of light front coordinates
\cite{3} and application of the CCM \cite{5}. In the light-front
Tamm-Dancoff field theory (LFFT) the quantization plane is chosen to
coincide with the light front, therefore the divergences that plagued
the original theories seem to disappear \cite{6} since here vacuum
remains trivial. Furthermore, not having to include interactions in
boost operators allows a renormalizable truncation scheme
\cite{7}. One of the most important difficulties in LFFT is the
complicated structure of the renormalization process \cite{8}. In
principle, ad-hoc counterterms can not be prevented if one is to
preserve the underlying symmetry.  In the standard form of CCM, on the
other hand, the amplitudes obey a system of coupled non-linear
equations which contain some ill-defined terms because of ultraviolet
divergences. It has been shown \cite{9} that the ill-defined
amplitudes, which are also called critical topologies, can be
systematically removed, by exploiting the linked-cluster property of
the ground state.  This can be done by introducing a mapping which
transfers them into a finite representations without making any
approximation such as a coupling expansion. Thus far this resummation
method has been restricted to superrenormalizable theories due to its
complexity.

Recently Wilson and Glazek \cite{aps1} and independently Wegner
\cite{aps2} have re-investigated this issue and introduced a new
scheme, the similarity renormalization group (SRG). The SRG
resembles the original Wilsonian renormalization group formulation
\cite{aps3}, since a transformation that explicitly runs the
cutoff is developed. However, here one runs a cutoff on energy
differences rather than on individual states.  In the SRG
framework one has to calculate a narrow matrix instead of a small
matrix, and the cutoff can be conceived as a band width (see Fig.~[2.1]). Therefore
by construction the perturbation expansion for transformed
Hamiltonians contains no small-energy denominators. Here we review
the general formulation of the SRG.

\subsection{Glazek-Wilson formulation}
The detailed description of Glazek-Wilson RG method can be found in
Ref.~\cite{aps1}. Here we only concentrate on the key elements of their
method. We introduce a unitary transformation aiming at partially
diagonalizing the Hamiltonian so that no couplings between states with
energy differences larger than $\lambda$ are present. The unitary
transformation defines a set of Hamiltonians $H_{\lambda}$ which
interpolate between the initial Hamiltonian ($\lambda=\Delta$ ) and
the effective Hamiltonian $H_{\lambda}$. This $\lambda$ plays the role
of a flow parameter. We will assume that $H_{\lambda}$ is dominated by
its diagonal part which we will denote $H_{0\lambda}$ with eigenvalues
$E_{i\lambda}$ and eigenstates $|i\rangle$:
\begin{equation}
\langle f|H_{0\lambda}|i\rangle=E_{i\lambda}\langle f|i\rangle.
\end{equation}
Therefore the Hamiltonian at scale
$\lambda$ can be written $H_{\lambda}=H_{0\lambda}+H_{I\lambda}$,
where $H_{I\lambda}$ is non-diagonal part. Notice that $H_{0\lambda}$ is not necessarily the bare free Hamiltonian which is independent of $\lambda$.  
In order to introduce the infinitesimal unitary transformations which
induce the infinitesimal changes in $H_{\lambda}$ when $\lambda$
changes by an infinitesimal amount, we need to define various zones of the operators We introduce an
auxiliary function $x_{ij\lambda}=\langle i|x_{\lambda}|j\rangle$ of
the states with labels $i$ and $j$ for a given $\lambda$. If we denote
the eigenvalue of $H_{0\lambda}$ with $E_{i\lambda}$, $x$ is defined
as
\begin{equation}
x_{ij\lambda}=\frac{|E_{i\lambda}-E_{j\lambda}|}{E_{i\lambda}+E_{j\lambda}+\lambda}.
\end{equation}
The modulus of the function $x_{ij\lambda}$ is close to $1$ when
one of the energies is much larger than the other and also much
larger than the cutoff $\lambda$ and $x$ approaches $0$ when the
energies are similar or small in comparison to the cutoff. One
then introduces smooth projectors 
\begin{equation}
u_{ij\lambda}=\langle i|u_{\lambda}|j\rangle=u(x_{ij\lambda}),
\end{equation} 
and 
\begin{equation}
r_{ij\lambda}=1-u_{ij\lambda}=r(x_{ij\lambda}).
\end{equation}
By means of $u$ and $r$ one can separate each matrix into two parts
$M=D(M)+R(M)$, where we define
\begin{equation}
D(M)_{ij}=u_{ij\lambda}M_{ij},
\end{equation}
and corresponding
\begin{equation}
R(M)_{ij}=r_{ij\lambda}M_{ij}.
\end{equation}
The functions $u$ and $r$ are
needed in order to ensure smoothness and differentiability, and
consequently a continuous transition between different parts of
the Hamiltonian matrix. They are the key elements in making the
diagonalization free of small-energy denominators. These functions
are implemented as a ``form factor`` in every vertex of the
interaction. We construct an infinitesimal unitary transformation
eliminating the part of Hamiltonian which has only non-zero
elements far away from the diagonal, thus they can not produce
small-energy denominators.  This continuous unitary transformation
satisfies,
\begin{equation}
\frac{dH_{\lambda}}{d\lambda}=[T_{\lambda},H_{\lambda}].\label{q1}
\end{equation}
The generator $T_{\lambda}$ is anti-Hermitian and is chosen in such a way
that
\begin{equation}
 H_{\lambda}=D(Q_{\lambda})=Q_{\lambda}u_{\lambda},
\end{equation}
where $Q$ is arbitrary in the
far-off-diagonal region. In terms of $Q$, the matrix elements of
Eq. ({\ref{q1}) satisfy
\begin{equation}
\frac{du_{ij\lambda}}{d\lambda}Q_{ij\lambda}+u_{ij\lambda}\frac{dQ_{ij\lambda}}{d\lambda}=T_{ij\lambda}(E_{j\lambda}-E_{i\lambda})+[T_{\lambda},H_{I\lambda}]_{ij}.\end{equation}
Notice that $Q_{\lambda}$ can be arbitrary in the far-off diagonal
region where $u_{\lambda}=\frac{du_{\lambda}}{d\lambda}=0$ if it is finite and
its derivative is finite. In the above equation we have two unknowns,
$\frac{dQ_{\lambda}}{d\lambda}$ and $T_{\lambda}$. As additional input we use the fact
that, for a given $\lambda$, the Hamiltonian $H_{\lambda}$ can be
additionally unitary transformed without violating the relation
$D(Q_{\lambda})=H_{\lambda}$. We regroup the above equation in the
form,
\begin{equation}
u_{ij\lambda}\frac{dQ_{ij\lambda}}{d\lambda}-T_{ij\lambda}(E_{j\lambda}-E_{i\lambda})=[T_{\lambda},H_{I\lambda}]_{ij}-\frac{du_{ij\lambda}}{d\lambda}Q_{ij\lambda}=G_{ij\lambda}. \label{n-1}
\end{equation}
The unknowns on the left hand-side are determined by first solving
Eq.~(\ref{n-1}) neglecting the commutator on the right-hand side; the
result is then substituted into the right-hand side, and solved
iteratively until convergence is reached. 
The $D$ and $R$ parts of the operator $G$
can be defined,
\begin{eqnarray}
u_{ij\lambda}\frac{dQ_{ij\lambda}}{d\lambda}&\equiv&D(G_{\lambda})_{ij},\nonumber\\
T_{ij\lambda}(E_{j\lambda}-E_{i\lambda})&\equiv&-R(G_{\lambda})_{ij}.\label{q2}\
\end{eqnarray}
By evaluating the matrix elements of both sides of the above equations in
different zones of the operators, one obtains differential equations
for matrix elements of $Q_{\lambda}$. Therefore, from Eqs.~(\ref{n-1},
\ref{q2}) one can immediately obtain the generator $T$ and the
Hamiltonian flow equation in terms of the matrix elements $\langle
f|H_{\lambda}|i\rangle=H_{\lambda fi}$ and $\langle
f|T_{\lambda}|i\rangle=T_{\lambda fi}$,
\begin{eqnarray}
T_{ij\lambda}&=&\frac{r_{ij\lambda}}{E_{i\lambda}-E_{j\lambda}}\left([T_{\lambda},H_{I\lambda}]_{ij}-\frac{d\ln u_{ij\lambda}}{d\lambda}H_{ij\lambda}\right),\label{q3}\\
\frac{dH_{ij\lambda}}{d\lambda}&=&u_{ij\lambda}[T_{\lambda},H_{I\lambda}]_{ij}+r_{ij\lambda}\frac{d\ln u_{ij\lambda}}{d\lambda}H_{ij\lambda}.\label{q4}\
\end{eqnarray}
It is obvious from the above equations that no small-energy
denominators $E_{i\lambda}-E_{j\lambda}$ arises out-side of the
band (off-diagonal region), and within the band zone, we have
$T=0$. The equations (\ref{q3},\ref{q4}) can be solved
iteratively.

\subsection{Wegner Formulation}
Wegner's formulation of the SRG is defined in a very elegant way
aiming at diagonalization of the Hamiltonian in a block-diagonal
form with the number of particles conserved in each block. Again,
a unitary transformation is used with flow parameter $s$ that
range from $0$ to $\infty$,
\begin{equation}
\frac{dH(s)}{ds}=[T(s),H(s)].\label{q5}
\end{equation}
We separate the Hamiltonian in a diagonal part $H_{d}$ and the
remainder $H_{r}$. We use the fact that $\tr H^{2}$ is invariant
under the unitary transformation, therefore we have 
\begin{equation}
\tr H^{2}_{d}+\tr H^{2}_{r}=\tr H^{2}=\text{const}.
\end{equation} 
This means that $\tr H^{2}_{r}$ falls monotonically if $\tr H^{2}_{d}$
increases.  One can use Eq.~(\ref{q5}) to obtain
\begin{equation}
\frac{d\tr H^{2}_{d}}{ds}=\frac{d}{ds}\sum_{i}H^{2}_{ii}=2\sum_{i}H_{ii}\sum_{j}\left(T_{ij}H_{ji}-H_{ij}T_{ji}\right)=2\sum_{ij}T_{ji}H_{ji}(H_{jj}-H_{ii}).
\end{equation}
In order to ensure that $\sum_{i\neq j}H^{2}_{ij}$ falls monotonically, one can simply choose the generator as $T_{ji}=H_{ji}(H_{jj}-H_{ii})$ or
\begin{equation}
T(s)=[H_{d}(s),H_{r}(s)].\label{n1-correct}
\end{equation}
One can show by making use of this definition that,
\begin{eqnarray}
\frac{dH_{ij}(s)}{ds}&=&\sum_{j}\left(H_{ii}(s)+H_{jj}(s)-2H_{kk}(s)\right)H_{ik}(s)H_{kj}(s),\nonumber\\
\frac{d}{ds}\sum_{i\neq j}H^{2}_{ij}&=&-\frac{d}{ds}\sum_{k}H^{2}_{kk}=-2\sum_{ij}T^{2}_{ij}.\
\end{eqnarray}
Because $\sum_{i\neq j}H^{2}_{ij}$ falls monotonously and is restricted from below, 
When $s\to\infty$ the derivative vanishes and we have $T_{s}\to
0$, at this limit the procedure of block-diagonalization is
completed. At this point, the unitary transformation
Eqs~(\ref{q5},\ref{n1-correct}) is completely defined. The only
freedom is in the choice of separation of the Hamiltonian into a
``diagonal'' and a ``rest'' part. Of course this depends on the
given physical problem.  As a illustration of the method, we show
how perturbation theory can be applied in this formalism. For a
given values of $s$ we have,
\begin{eqnarray}
H(s)&=&H^{(0)}_{d}+H^{(1)}_{r}+H^{(2)}_{r}+...,\\
T(s)&=&[H_{d}(s),H_{r}(s)]=[H^{(0)}_{d},H^{(1)}_{r}]+[H^{(0)}_{d},H^{(2)}_{r}]+...~ .\nonumber\\
&=&T^{(1)}+T^{(2)}+...,\
\end{eqnarray}
where the superscript denotes the order in the bare coupling
constant. The part $H^{(0)}_{d}$ is the free Hamiltonian. The
index $r$ denotes the rest of the Hamiltonian. Note that generally
the diagonal part in the flow equation is the full particle number
conserving part of the effective Hamiltonian. The choice of only
$H^{(0)}_{d}$ as the diagonal part leads to the simplest
band-diagonal structure where the particle number is conserved. We
use the basis of the eigenfunctions of the free Hamiltonian
$H^{(0)}_{d}|i\rangle=E_{i}|i\rangle$ to obtain the matrix
elements of Eqs~(\ref{q5},\ref{n1-correct}),
\begin{eqnarray}
\frac{dH_{ij}}{ds}&=&-(E_{i}-E_{j})^{2}H^{(1)}_{rij}+[T^{(1)},H^{(1)}_{rij}]-(E_{i}-E_{j})^{2}H^{(2)}_{rij}+...,\\
T_{ij}&=&(E_{i}-E_{j})H^{(1)}_{rij}+(E_{i}-E_{j})H^{(2)}_{rij}+..., \
\end{eqnarray}
where the energy differences are given by
\begin{equation}
E_{i}-E_{j}=\sum_{k=1}^{n_{1}}E_{ik}-\sum_{k=1}^{n_{2}}E_{jk},
\end{equation}
and $E_{ik}$ and $E_{jk}$ are the energies of the creation and
annihilation particles, respectively. To leading order in
perturbation theory, one finds
\begin{eqnarray}
\frac{dH^{(1)}_{rij}}{ds}&=&-(E_{i}-E_{j})^{2}H^{(1)}_{rij},\nonumber\\
H^{(1)}_{rij}(s)&=&H^{(1)}_{rij}(s=0)u_{ij}(s),\nonumber\\
u_{ij}(s)&=&e^{-(E_{i}-E_{j})^{2}s}.\
\end{eqnarray}
The flow-parameter $s$ has dimension $1/(\text{energy})^{2}$ and
is related to the similarity width $\lambda$ (ultraviolet cutoff)
by $s=1/\lambda^{2}$. This implies that matrix elements of the
interaction which change the number of particles are strongly
suppressed, since we have $|E_{i}-E_{j}|>\lambda$. The similarity
generator in Wegner's formulation corresponds to the choice of a
gaussian similarity function with uniform width.  In the next
leading order, one has to deal separately with the diagonal and
rest parts.  In analogy to the Glazek-Wilson method we introduce $
H^{(2)}_{r}=u(s)\bar{H}^{(2)}_{r}(s)$, and the solution reads,
\begin{eqnarray}
\bar{H}^{(2)}_{rij}(s)&=&\bar{H}^{(2)}_{rij}(s=0)+\int_{0}^{s}ds' u(s')[T^{(1)},H^{(1)}]_{rij}(s'),\nonumber\\
H^{(2)}_{dij}(s)&=&H^{(2)}_{dij}(s=0)+\int_{0}^{s}ds' [T^{(1)},H^{(1)}]_{dij}(s').\
\end{eqnarray}
It is obvious that for the non-diagonal term a smooth form factor
appears in a natural way to suppress the off-diagonal interaction.
In other words, the particle number changing interactions are
eliminated while a new terms are produced. The procedure can be
extended to arbitrarily high orders. The counterterms can be
determined order-by-order using the idea of coupling coherence,
namely that under similarity transformation Hamiltonian remains
form invariant (see next section).

\section{Coupling coherence condition}
One of the most severe problems for the traditional light-front RG
is that an infinite number of relevant and marginal operators are
required \cite{cohp}. This is due to the fact that light-front
cutoff violates the underlying symmetry, e.g., Lorentz invariance
and gauge symmetries. Since these are continuous symmetries, their
violation in principle leads to infinite number of symmetry
violating counterterms (with new couplings), in order to maintain
the symmetry of the effective Hamiltonian. In terms of the
effective field theory approach, some sort of fine tuning is
required to fix the strength of the new couplings so that the
underlying symmetry is restored. One should note that in order to
reduce the number of momentum degrees of freedom, one must
introduce a real cutoff, such as a momentum cutoff or a lattice
cutoff. However, dealing with divergences do not require
necessarily a decrease in degrees of freedom. In fact, one may
even increase the degrees of freedom, e.g., the Pauli-Villars or
the dimensional regularization methods \cite{ph4}. The main idea
behind the coupling coherence renormalization condition
\cite{coh,coh1} is that the Hamiltonian is form-invariant on the
RG trajectory. This condition isolates and repairs the hidden
symmetries \cite{coh,coh1}.
\begin{equation}
H(\Lambda)\equiv \mathcal{H}(\mu),
\end{equation}
In order words, rewriting the Hamiltonian in different degrees of
freedom does not change the operator itself. One may think of $\mathcal{H}(\mu)$
as QCD written in terms of constituent quarks and gluons and
$H(\Lambda)$ as the same QCD Hamiltonian written in terms of canonical
quarks and gluons, associated with partons and current quarks.  The
SRG and the coupled-cluster RG with coupling coherence allows one to
construct effective theories with the same number of couplings as the
underlying fundamental theory, even when the cutoff violate symmetries
of the theory. This does not preclude the emergence of the new
couplings, however, they depend on the original coupling and will
vanish if the fundamental couplings are turned off. This boundary
condition together with the RG equations determines their dependence on the fundamental coupling.

As an illustrative example \cite{coh1}, we consider following interaction
\begin{equation}
V(\phi)=\frac{\lambda_{1}}{4!}\phi^{4}_{1}+\frac{\lambda_{2}}{4!}\phi^{2}_{2}+\frac{\lambda_{3}}{4!}\phi^{2}_{1}\phi^{2}_{2},
\end{equation}
where $\phi_{1}$ and $\phi_{2}$ are scalar fields. We want to
investigate that under what conditions the couplings are independent
of each other. Consider the Gell-Mann-Low equations \cite{gel} up to one-loop in perturbation theory, ignoring the masses,
\begin{eqnarray}
\frac{\partial \lambda_{1}}{\partial t}&=&3\xi\lambda^{2}_{1}+\frac{1}{12}\xi\lambda^{2}_{3},\nonumber\\
\frac{\partial \lambda_{2}}{\partial t}&=&3\xi\lambda^{2}_{2}+\frac{1}{12}\xi\lambda^{2}_{3},\nonumber\\
\frac{\partial \lambda_{3}}{\partial t}&=&\frac{2}{3}\xi\lambda^{3}_{2}+\xi\lambda_{3}(\lambda_{1}+\lambda_{2}),\label{c1}\
\end{eqnarray}
where $t=\log(\Lambda/\mu)$ and $\xi=\hbar/(16\pi^{2})$. Assume
that there is only one independent coupling
$\bar{\lambda}=\lambda_{1}$, and $\lambda_{2}$, $\lambda_{3}$ are
functions of $\bar{\lambda}$. Now one can simplify Eq.~(\ref{c1}),
\begin{eqnarray}
\left(3\bar{\lambda}^{2}+\frac{1}{12}\lambda^{2}_{3}\right)\frac{\partial \lambda_{2}}{\partial \bar{\lambda}}&=&3\lambda^{2}_{2}+\frac{1}{12}\lambda^{2}_{3},\nonumber\\
\left(3\bar{\lambda}^{2}+\frac{1}{12}\lambda^{2}_{3}\right)\frac{\partial \lambda_{3}}{\partial \bar{\lambda}}&=&\frac{2}{3}\lambda^{2}_{3}+\bar{\lambda}\lambda_{3}+\lambda_{2}\lambda_{3}.\label{c2}\
\end{eqnarray}
In the leading order, the above equations have two distinct
solutions, one is when $\lambda_{2}=\bar{\lambda}$ and
$\lambda_{3}=2\bar{\lambda}$. In this case we have
\begin{equation}
V(\phi)=\frac{\bar{\lambda}}{4!}\left(\phi^{2}_{1}+\phi^{2}_{2}\right)^{2}.
\end{equation}
Therefore we recover the $O(2)$ symmetric theory. The other
solution is $\lambda_{2}=\bar{\lambda}$ and
$\lambda_{3}=6\bar{\lambda}$ which leads to two decoupled scalar
fields,
\begin{equation}
V(\phi)=\frac{\bar{\lambda}}{2.4!}\left((\phi_{1}+\phi_{2})^{4}+(\phi_{1}-\phi_{2})^{4}\right),
\end{equation}
One can conclude that $\lambda_{2}$ and $\lambda_{3}$ do not run
independently with the cutoff if and only if there is a symmetry
which connects their strength to $\lambda_{1}$. The condition that
a limited number of couplings run with cutoff independently
reveals the symmetries broken by the regulator and repairs them.
More interesting, it may be used as well to uncover symmetries
that are broken by the vacuum. This may reconcile the trivial
vacuum in light-front field theory and vacuum symmetry breaking
problem.
\section{General formulation of the similarity renormalization group}
In this section we pave the way for an introduction of the coupled-cluster RG, and
consider the similarity renormalization group in a more general
framework without requiring the unitarity. The discussion in this
section is partially based on the work of Suzuki and Okamoto
\cite{Ni}. Let us consider a system described by a Hamiltonian
$H(\Lambda)$ which has, at the very beginning, a large cut-off
$\Lambda$.  We assume that the renormalized Hamiltonian
$H^{\text{eff}}(\Lambda)$ up to scale $\Lambda$ can be written as the
sum of the canonical Hamiltonian and a {}``counterterm''
$H_{C}(\Lambda)$,
\begin{equation}
  H^{\text{eff}}(\Lambda)=H(\Lambda)+H_{C}(\Lambda)\,.
\end{equation}
Our aim is to construct the renormalized Hamiltonian by obtaining this
counterterm. Now imagine that we restrict the Hamiltonian to a lower
energy scale $(\mu)$, where we want to find an effective Hamiltonian
$H^{\text{eff}}(\mu)$ which has the same energy spectrum as the
original Hamiltonian in the smaller space. The cut-off $\mu$ can be
conceived as a flow parameter. The value of $\mu=\Lambda$ corresponds
to the initial bare regulated Hamiltonian. Formally, we wish to
transform the Hamiltonian to a new basis, where the medium-energy
modes $\mu<k<\Lambda$, decouple from the low-energy ones, while the
low-energy spectrum remains unchanged. We split the Hilbert space by
means of flow-parameter $\mu$ into two subspaces, the
intermediate-energy space $Q$ containing modes with $\mu<k<\Lambda$
and a low-energy space $P$ with $k\leq\mu$. Our renormalization
approach is based on decoupling of the complement space $Q$ from the
model space $P$. Thereby the decoupling transformation generates a new
effective interaction $\delta H(\mu,\Lambda)$ containing the effects
of physics between the scales $\Lambda$ and $\mu$. One can then
determine the counterterm by requiring coupling
coherence~\cite{10,coh,coh1}, namely that the transformed Hamiltonian
has the same form as the original one but with $\Lambda$ replaced by
$\mu$ everywhere. (This is in contrast to the popular Effective Field
Theory approach, where one includes all permissible couplings of a
given order and fixes them by requiring observable computed be both
cutoff-independent and Lorentz covariant.)

We define two projector operators, also called $P$ and $Q$, which
project a state onto the model space and its complement, satisfy
$P^{2}=P$, $Q^{2}=Q$, $PQ=0$ and $P+Q=1$. We introduce an isometry
operator $G$ which maps states in the $P$- onto the $Q$- space,
\begin{equation}
|q\rangle=G|p\rangle\hspace{1cm}(|q\rangle\in Q,|p\rangle\in P)\,.
\end{equation}
 The operator $G$ is the basic ingredient in a family of {}``integrating-out
operators'', and passes information about the correlations of the
high energy modes to the low-energy space. The operator $G$ obeys
$G=QGP$, $GQ=0$, $PG=0$ and $G^{n}=0$ for $n\geqslant2$. The counterintuitive choice that $G$ maps from model to complement space is due to the definition
Eq.~(\ref{eq4}) below (c.f.\ the relation between the active and passive
view of rotations). In order to give a general form of the effective
low-energy Hamiltonian, we define another operator $X(n,\mu,\Lambda)$,
\begin{equation}
X(n,\mu,\Lambda)=(1+G)(1+G^{\dag}G+GG^{\dag})^{n}\,.
\label{eq2}
\end{equation}
 ($n$ is a real number.) The inverse of $X(n,\mu,\Lambda)$ can be
obtained explicitly,
\begin{equation}
X^{-1}(n,\mu,\Lambda)=(1+G^{\dag}G+GG^{\dag})^{-n}(1-G)\,.
\label{eq3}
\end{equation}
The special case $n=0$ is equivalent to the transformation introduced
in ref.~\cite{25} to relate the hermitian and non-hermitian effective
operators in the energy-independent Suzuki-Lee approach. We now
consider the transformation of $H(\Lambda)$ defined as
\begin{equation}
\overline{H}(n,\mu,\Lambda)=X^{-1}(n,\mu,\Lambda)H(\Lambda)X(n,\mu,\Lambda)\,,
\label{eq4}
\end{equation}
where we have
\begin{equation}
H(\Lambda)\rightarrow\overline{H}(n,\mu,\Lambda)\equiv H(\mu)+\delta H(\mu,\Lambda)\,.
\end{equation}
One can prove that if $\overline{H}(n,\mu,\Lambda)$ satisfies the
desirable decoupling property,
\begin{equation}
Q\overline{H}(n,\mu,\Lambda)P=0\,,
\label{eq5}
\end{equation}
or more explicitly, by substituting the definition of
$X(n,\mu,\Lambda)$ and $X^{-1}(n,\mu,\Lambda)$ from
Eqs.~(\ref{eq2})--(\ref{eq3}),
\begin{equation}
QH(\Lambda)P+QH(\Lambda)QG-GPH(\Lambda)P-GPH(\Lambda)QG=0\,,
\label{eq6}
\end{equation}
that $H^{\text{eff}}(\mu) \equiv {\mathcal{H}}(n,\mu) =
P\overline{H}(n,\mu,\Lambda)P$ is an effective Hamiltonian for the low
energy degrees of freedom.  In other words, it should have the same
low-energy eigenvalues as the original Hamiltonian. The proof is as
follows:

Consider an eigenvalue equation in the $P$ space for a state
$|\phi(k)\rangle\in P$,
\begin{equation}
P\overline{H}(n,\mu,\Lambda)P|\phi(k)\rangle=
E_{k}PX^{-1}(n,\mu,\Lambda)X(n,\mu,\Lambda)P|\phi(k)\rangle\,.
\label{eq7}
\end{equation}
By multiplying both sides by $X(n,\mu,\Lambda)$ and making use of the
decoupling property Eq. (\ref{eq5}), we obtain
\begin{equation}
H(\Lambda)X(n,\mu,\Lambda)P|\phi(k)\rangle=
E_{k}X(n,\mu,\Lambda)P|\phi(k)\rangle\,.
\label{eq8}
\end{equation}
This equation means that $E_{k}$ in Eq.~(\ref{eq7}) agrees with one of
the eigenvalue of $H(\Lambda)$ and $X(n,\mu,\Lambda)P|\phi(k)\rangle$
is the corresponding eigenstate. Now we demand that
\begin{equation}
H^{\text{eff}}(\mu)\equiv H(\mu)+H_{C}(\mu)\,.
\end{equation}
This requirement uniquely determines the counterterm $H_{C}$.

In the same way we can also obtain the $Q$-space effective
Hamiltonian, from the definition of $\overline{H}(n,\mu,\Lambda)$. It
can be seen that if $G$ satisfies the requirement in Eq.~(\ref{eq6}),
then we have additional decoupling condition
\begin{equation}
P\overline{H}(n,\mu,\Lambda)Q=0\,.
\label{decoupling}
\end{equation}
Although the above condition is not independent of the decoupling
condition previously introduced in Eq.~(\ref{eq5}) in the exact form,
but this is not the case after involvement of some approximation (truncation). We
will argue later that \emph{both} of the decoupling conditions
Eqs.~(\ref{eq5}) and~(\ref{decoupling}) are necessary in order to have
a sector-independent renormalization scheme. The word ``sector'' here
means the given truncated Fock space. Let us now clarify the meaning
of this concept. To maintain the generality of the previous
discussion, we use here the well known Bloch-Feshbach
formalism~\cite{aps4,b,fesh}.  The Bloch-Feshbach method exploits
projection operators in the Hilbert space in order to determine
effective operators in some restricted model space. This technique
seems to be more universal than Wilson's renormalization formulated in
a Lagrangian framework. This is due to the fact that in the
Bloch-Feshbach formalism, other irrelevant degrees of freedom (such as
high angular momentum, spin degrees of freedom, number of particles,
etc.) can be systematically eliminated in the same fashion.

Assume that the full space Schr\"{o}dinger equation is
$H|\psi\rangle=E_{\psi}|\psi\rangle$ and for simplicity $|\psi\rangle$ has been normalized to one. The similarity transformed
Schr\"{o}dinger equation now reads
$\overline{H}|\psi_{X^{-1}}\rangle=E_{\psi}|\psi_{X^{-1}}\rangle$, where we
defined $|\psi_{X^{-1}}\rangle=X^{-1}|\psi\rangle$ and $\overline{H}$ is a similarity transformed Hamiltonian. This equation is
now separated into two coupled equations for $P$- and $Q$-space. 
\begin{eqnarray}
&&(E_{\psi}-P\overline{H}P)P|\psi_{X^{-1}}\rangle=P\overline{H}Q|\psi_{X^{-1}}\rangle,\label{ray-de1}\\
&&(E_{\psi}-Q\overline{H}Q)Q|\psi_{X^{-1}}\rangle=Q\overline{H}P|\psi_{X^{-1}}\rangle.\label{ray-de2}\
\end{eqnarray}
We may formally solve Eq.~(\ref{ray-de2}) as
\begin{equation}
Q|\psi_{X^{-1}}\rangle=\frac{Q\overline{H}P}{E_{\psi}-Q\overline{H}Q}P|\psi_{X^{-1}}\rangle, \label{ray-non1}
\end{equation}
and substitute this equation into Eq.~(\ref{ray-de1}) to obtain a formally exact uncoupled equation in $P$-space,
\begin{equation}
H^{\text{eff}}P|\psi_{X^{-1}}\rangle=E_{\psi}P|\psi_{X^{-1}}\rangle,\label{F0-new}
\end{equation}
where we have
\begin{equation}
H^{\text{eff}} =P\overline{H}P+P\overline{H}Q\frac{1}{E_{\psi}-Q\overline{H}Q}Q\overline{H}P.\label{F1}
\end{equation}
The effective Hamiltonian $H^{\text{eff}}$ constructed in this fashion is explicitly 
energy dependent. 
This equation resembles the one for Brueckner's reaction matrix (or
``G''-matrix) equation in nuclear many-body theory (NMT). In the same
way for arbitrary operator $O$ (after a potential similarity
transformation), we construct the effective operator. Let us define a similarity transformed operator $\overline{O}$,
\begin{eqnarray}
&&\overline{O}=\sum\sum|\phi_{x^{-1}}\rangle\langle\phi_{x^{-1}}|\overline{O}|\psi_{x^{-1}}\rangle\langle \psi_{x^{-1}}|,\nonumber\\
&& \overline{O}=\sum\sum|\phi_{x^{-1}}\rangle\langle\phi_{x^{-1}}|(P+Q)\overline{O}(P+Q)|\psi_{x^{-1}}\rangle\langle \psi_{x^{-1}}|,\label{ray-non2}\
\end{eqnarray}
where $|\psi\rangle$ and $|\phi\rangle$ are eigen functions of
Hamiltonian and we have
$\sum|\phi_{x^{-1}}\rangle\langle\phi_{x^{-1}}|=\sum
|\psi_{x^{-1}}\rangle\langle \psi_{x^{-1}}|=1$ and $P+Q=1$. We now write
$\langle\phi_{x^{-1}}|Q$ and $Q|\psi_{x^{-1}}\rangle$ in terms of their
solution in the $P$-space obtained in Eq.~(\ref{ray-non1}). One can
immediately show
\begin{eqnarray}
&&\langle\phi_{x^{-1}}|P\overline{O}Q|\psi_{x^{-1}}\rangle= \langle\phi_{x^{-1}}|P\overline{O}Q\frac{Q\overline{H}P}{E_{\psi}-Q\overline{H}Q}P|\psi_{x^{-1}}\rangle, \nonumber\\
&&\langle\phi_{x^{-1}}|Q\overline{O}P|\psi_{x^{-1}}\rangle=\langle\phi_{x^{-1}}|P\frac{P\overline{H}Q}{E_{\phi}-Q\overline{H}Q}Q\overline{O}P|\psi_{x^{-1}}\rangle,\nonumber\\
&&\langle\phi_{x^{-1}}|Q\overline{O}Q|\psi_{x^{-1}}\rangle=\langle\phi_{x^{-1}}|P\frac{P\overline{H}Q}{E_{\phi}-Q\overline{H}Q}Q\overline{O}Q
\frac{Q\overline{H}P}{E_{\psi}-Q\overline{H}Q}P|\psi_{x^{-1}}\rangle.\
\end{eqnarray}
By plugging the above equations into Eq.~(\ref{ray-non2}), one can obtain the effective operator in the $P$-space
\begin{equation}
\overline{O}=\sum\sum|\phi_{x^{-1}}\rangle\langle\phi_{x^{-1}}|PO^{\text{eff}}P|\psi_{x^{-1}}\rangle\langle \psi_{x^{-1}}|=PO^{\text{eff}}P,\label{f-final1}
\end{equation}
where we have  
\begin{eqnarray}
O^{\text{eff}} & = & P\overline{O}P+P\overline{H}Q\frac{1}{E_{\phi}-Q\overline{H}Q}Q\overline{O}P+ P\overline{O}Q\frac{1}{E_{\psi}-Q\overline{H}Q}Q\overline{H}P\nonumber\\
&+& P\overline{H}Q\frac{1}{E_{\phi}-Q\overline{H}Q}Q\overline{O}Q\frac{1}{E_{\psi}-Q\overline{H}Q} Q\overline{H}P.\label{F2}\nonumber\\
\end{eqnarray}
Notice that Eq.~(\ref{F2}) can be converted into the form of Eq.~(\ref{F1}) with the effective Schr\"{o}dinger equation Eq.~(\ref{F0-new}) when $\overline{O}\to\overline{H}$
\footnote{To this end, we organize Eq.~(\ref{F2}) when $\overline{O}\to\overline{H}$, 
\begin{eqnarray}
O^{\text{eff}} & = & P\overline{O}P+P\overline{H}Q\Big(\frac{1}{E_{\phi}-Q\overline{H}Q}+ \frac{1}{E_{\psi}-Q\overline{H}Q}
+\frac{1}{E_{\phi}-Q\overline{H}Q}
Q\overline{H}Q\frac{1}{E_{\psi}-Q\overline{H}Q}\Big)Q\overline{H}P,
\nonumber\\
&=& P\overline{O}P+P\overline{H}Q\Big(\frac{1}{E_{\psi}-Q\overline{H}Q}+\frac{E_{\psi}}{(E_{\phi}-Q\overline{H}Q)(E_{\psi}-Q\overline{H}Q)}\Big)Q\overline{H}P.\label{F2-new1}\
\end{eqnarray}
One now can use Eq.~(\ref{ray-non1}) to rewrite the last part of the above equation into the form of 
\begin{eqnarray}
\langle \phi_{x^{-1}}|P\overline{H}Q\Big(\frac{E_{\psi}}{(E_{\phi}-Q\overline{H}Q)(E_{\psi}-Q\overline{H}Q)}\Big)Q\overline{H}P|\psi_{x^{-1}}\rangle
&=&E_{\psi}\langle\phi_{x^{-1}}|Q^{2}|\psi_{x^{-1}}\rangle \label{last-new}\\
&=&E_{\psi}\left(\langle\phi_{x^{-1}}|\psi_{x^{-1}}\rangle-\langle\phi_{x^{-1}}|P|\psi_{x^{-1}}\rangle\right),\nonumber\
\end{eqnarray}
where we used $Q+P=1$ and $Q^{2}=Q$. Having made used of Eqs.~(\ref{f-final1},\ref{F2-new1},\ref{last-new}), one can immediately obtain
Eq.~(\ref{F0-new}) with the effective Hamiltonian in the form of Eq.~(\ref{F1}).}.

The $E$-dependence in Eqs.~(\ref{F1}) and~(\ref{F2}) emerges from the
fact that the effective interaction in the reduced space is not
assumed to be decoupled from the excluded space. However, by using the
decoupling conditions introduced in Eqs.~(\ref{eq5})
and~(\ref{decoupling}), we observe that energy dependence can be
removed, and the effective operators become
\begin{eqnarray}
H^{\text{eff}} & = & P\overline{H}P={\mathcal{H}}(n,\mu)\,,
\nonumber \\
O^{\text{eff}} & = & P\overline{O}P={\mathcal{O}}(n,\mu)\,.
\label{final}\
\end{eqnarray}
The decoupling property makes the operators in one sector independent
of the other sector. The effects of the excluded sector is taken into
account by imposing the decoupling conditions. This is closely related
to the folded diagram method in NMT for removing
energy-dependence~\cite{folded}. (It is well-known in NMT that
$E$-dependence in the $G$-matrix emerges from non-folded diagrams
which can be systematically eliminated using the effective interaction
approach). The above argument was given without assuming an explicit
form for $X$ and thus the decoupling conditions are more fundamental
than the prescription used to derive these conditions. We will show
later that one can choose a transformation $X$, together with the
model space and its complement, which avoids the occurrence of ``small
energy denominators''.  We now show that Lorentz covariance in a given
sector does not hinge on a special form of similarity
transformation. Assume the existence of ten Poincar\'{e} generators $L_{i}$
satisfying
\begin{equation}
[L_{i},L_{j}]=\sum a_{ij}^{k}L_{k}\,,
\end{equation}
where $a_{ij}^{k}$ are the structure coefficients. One can show
that if the operators $L_{i}$ satisfy the decoupling conditions~
$Q\bar{L}_{i}P=0$ and $P\bar{L}_{i}Q=0$ it follows that
\begin{equation}
[L_{i}^{\text{eff}},L_{j}^{\text{eff}}]=\sum
a_{ij}^{k}L_{k}^{\text{eff}}\,.
\end{equation}
This leads to a relativistic description even after simultaneously
integrating out the high-frequency modes and reducing the number of
particles. Indeed we conjecture that requiring the decoupling
conditions makes the effective Hamiltonian free of
Lorentz-noninvariant operators for a given truncated sector regardless
of the regularization scheme.  However, one may still be faced with an
effective Hamiltonian which violates gauge invariance (for e.g., when
sharp cutoff is employed).

Note that the solution to Eq.~(\ref{eq6}) is independent of the number
$n$. One can make use of Eq.~(\ref{eq6}) and its complex conjugate to
show that for any real number $n$, the following relation for the
effective low-energy Hamiltonian
\begin{equation}
{\mathcal{H}}(n,\mu)=\mathcal{H}^{\dag}(-n-1,\mu)\,.
\label{eq10}
\end{equation}
The case $n=-1/2$ is special since the effective Hamiltonian is
hermitian,
\begin{equation}
{\mathcal{H}}(-1/2,\mu)=(P+G^{\dag}G)^{1/2}H(\Lambda)(P+G)(P+G^{\dag}G)^{-1/2}\,.
\label{hermi}
\end{equation}
Hermiticity can be verified from the relation~\cite{26}
\begin{equation}
e^{T}P=(1+G)(P+G^{\dag}G)^{-1/2}\,,
\end{equation}
where,
\begin{equation}
T=\arctan(G-G^{\dag})=\sum_{n=0}^{\infty}
\frac{(-1)^{n}}{2n+1}\big(G(G^{\dag}G)^{n}-\textrm{h.c}.)\,.
\end{equation}
Since the operator $T$ is anti-hermitian, $e^{T}$ is a unitary
operator. From the above expression Eq.~(\ref{hermi}) can be written
in the explicitly hermitian form
\begin{equation}
{\mathcal{H}}\left(-\frac12,\mu\right)=Pe^{-T}H(\Lambda)e^{T}P\,.
\end{equation}
As was already emphasized, renormalization based on unitary
transformations is more complicated and non-economical. Thus we will
explore a non-unitary approach. An interesting non-hermitian effective
low-energy Hamiltonian can be obtained for $n=0$,
\begin{equation}
{\mathcal{H}}(0,\mu,\Lambda)=PH(\Lambda)(P+QG)\,.
\label{eq11}
\end{equation}
This form resembles the Bloch and Horowitz type of effective
Hamiltonian as used in NMT~\cite{b}, and was the one used by Wilson in his original work on quantum
field theory~\cite{aps4}(see section II). In the context of the CCM, this form leads
to the folded diagram expansion well known in many-body
theory~\cite{h19}. It is of interest that various effective low-energy
Hamiltonians can be constructed according to Eq.~(\ref{final}) by the
use of the mapping operator $G$ which all obey the decoupling property
Eq.~(\ref{eq5}) and Eq.~(\ref{decoupling}). Neither perturbation
theory nor hermiticity is essential for this large class of effective
Hamiltonians.

\section{The coupled-cluster renormalization group}
The CCM approach is, of course, just one of the ways of describing the
relevant spectrum by means of non-unitary transformations. According
to our prescription the model space is
$P:\{|L\rangle\otimes|0,b\rangle_{h},L\leq\mu\}$, where
$|0,b\rangle_{h}$ is a bare high energy vacuum (the ground state of
high-momentum of free-Hamiltonian) which is annihilated by all the
high frequency annihilation operators $\{ C_{I}\}$ (for a given
quantization scheme, e.g., equal time or light-cone) , the set of
indices $\{ I\}$ therefore defines a subsystem, or cluster, within the
full system of a given configuration. The actual choice depends upon
the particular system under consideration. In general, the operators
$\{ C_{I}\}$ will be products (or sums of products) of suitable
single-particle operators. We assume that the annihilation and its
corresponding creation $\{ C_{I}^{\dag}\}$ subalgebras and the state
$|0,b\rangle_{h}$ are cyclic, so that the linear combination of state
$\{ C_{I}^{\dag}|0,b\rangle_{h}\}$ and $\{~_{h}\langle b,0|C_{I}\}$
span the decoupled Hilbert space of the high-momentum modes,
$\{|H\rangle\}$, where $\mu<H<\Lambda$.  It is also convenient, but
not necessary, to impose the orthogonality condition,
$\langle0|C_{I}C_{J}^{\dag}|0\rangle=\delta(I,J)$, where $\delta(I,J)$
is a Kronecker delta symbol. The complement space is
$Q:\{|L\rangle\otimes\left(|H\rangle-|0,b\rangle_{h}\right)\}$.
Our main goal is to decouple the $P$-space from the $Q$-space. This
gives sense to the partial diagonalization of the high-energy part of
the Hamiltonian. The states in full Hilbert space are constructed by
adding correlated clusters of high-energy modes onto the $P$-space, or
equivalently integrating out the high-energy modes from the
Hamiltonian,
\begin{eqnarray}
|f\rangle&=&X(\mu,\Lambda)|p\rangle=e^{\hat{S}}e^{-\hat{S}'}|0,b\rangle_{h}
\otimes|L\rangle=e^{\hat{S}}|0,b\rangle_{h}\otimes|L\rangle\,,
\label{eq13}\\
\langle\widetilde{f}|&=&\langle L|\otimes~_{h}\langle b,0|X^{-1}
(\mu,\Lambda)=\langle L|\otimes~_{h}\langle0|e^{\hat{S}'}e^{-\hat{S}}\,,
\label{eq14}
\end{eqnarray}
where the operators $X(\mu,\Lambda)$ and $X^{-1}(\mu,\Lambda)$ have
been expanded in terms of independent coupled cluster excitations $I$,
\begin{equation}
\begin{array}[b]{rclrcl}
\hat{S}&=&\displaystyle\sum_{m=0}\hat{S}_{m}\left(\frac{\mu}{\Lambda}\right)^{m}\,,
\qquad&
\hat{S}_{m}&=&\displaystyle\sideset{}{'}\sum_{I}\hat{s}^{m}_{I}C_{I}^{\dag}\,,
\\[15pt]
\hat{S}'&=&\displaystyle\sum_{m=0}\hat{S}'_{m}\left(\frac{\mu}{\Lambda}\right)^{m},
\qquad&
\hat{S}'_{m}&=&\displaystyle\sideset{}{'}\sum_{I}{}\hat{s}'^{m}_{I}C_{I}\, \label{def-s}.
\end{array}
\end{equation}
\begin{figure}[!tp]
\centerline{\includegraphics[clip,width=10 cm]{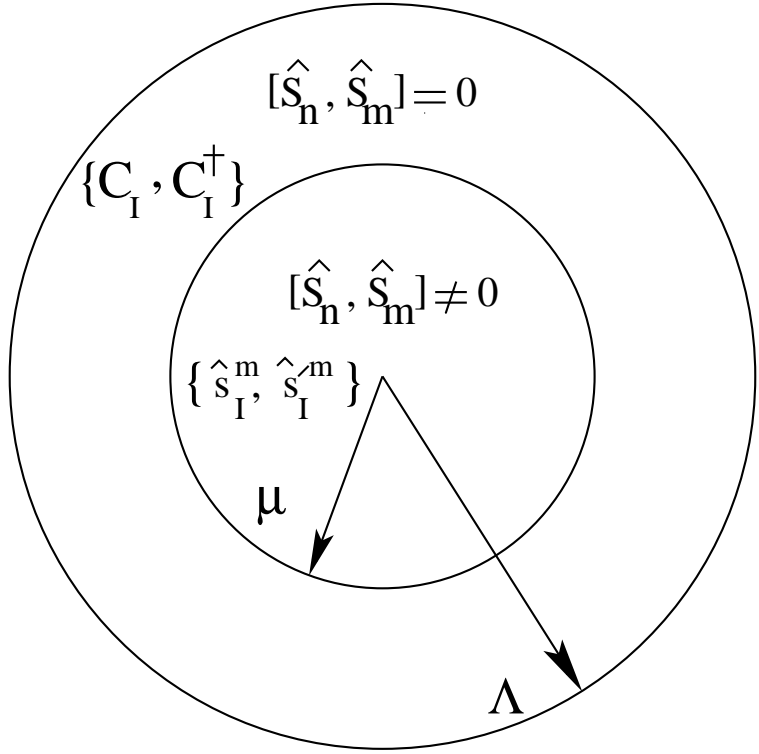}}
\caption{Shows the Wilsonian shells and the low-energy phase space $\{\hat{s}^{m}_{I},\hat{s}^{'m}_{I}\}$ which is induced 
by integrating out the high-energy modes $\{C_{I},C^{\dag}_{I}\}$, please see the text for details. }
\end{figure}
Here the primed sum means that at least one fast particle is created
or destroyed $(I\neq0)$, and momentum conservation in $P\oplus Q$
is included in $\hat{s}_{I}$ and
$\hat{s}'_{I}$. $\hat{S}_{m}(\hat{S}'_{m})$ are not generally
commutable in the low-energy Fock space, whereas they are by
construction commutable in the high-energy Fock space.  
One of the crucial difference between our approach and the traditional
CCM \cite{ccm1} is that, here $\hat{s}^{m}_{I}$ and
$\hat{s}'^{m}_{I}$ are only $c$-number in the high-energy Fock space,
but they are operators in the low-energy Fock space (unless when
$\mu\to 0$). We have in the Wilsonian high-energy shell
$[\hat{S},C^{\dag}_{I}]=[\hat{S}',C_{I}]=0$ (see Fig.~2.2). Therefore, one can
guarantee the proper size-extensivity and consequently conformity with
the linked-cluster theorem at any level of approximation in the
Wilsonian high-energy shell. Nevertheless one can still apply the
standard CCM to the induced effective low-energy Hamiltonian to extend
the proper size-extensivity to the entire Fock space.

It is immediately clear that states in the interacting Hilbert space
are normalized, $\langle\widetilde{f}|f\rangle=~_{h}\langle
b,0|0,b\rangle_{h}=1$.  We have two types of parameters in this
procedure: One is the coupling constant of the theory ($\lambda$), and
the other is the ratio of cutoffs ($\mu/\Lambda$). The explicit power
counting makes the degree of divergence of each order smaller than the
previous one. According to our logic, Eq.~(\ref{eq11}) can be written
as
\begin{equation} {\mathcal{H}}(\mu)=\,_{h}\langle
b,0|X^{-1}(\mu,\Lambda)H(\Lambda)X(\mu,\Lambda)|0,b\rangle_{h}\,,
\label{eq15}
\end{equation}
with $X(\mu,\Lambda)$ and $X^{-1}(\mu,\Lambda)$ defined in
Eq.~(\ref{eq13}) and Eq.~(\ref{eq14}). We require that effective
Hamiltonian ${\mathcal{H}}(\mu)$ obtained in this way remains form
invariant or coherent~\cite{10,coh,coh1}.  This requirement satisfies on an
infinitely long renormalization group trajectory and thus does
constitutes a renormalized Hamiltonian. Thereby one can readily
identify the counter terms produced from expansion of
Eq.~(\ref{eq15}).

The individual amplitudes for a given $m$,
$\{\hat{s}_{I}^{m},\hat{s}'{}_{I}^{m}\}\equiv\{\hat{s}_{I},\hat{s}'_{I}\}_{m}$,
have to be fixed by the dynamics of quantum system. This is a
complicated set of requirements. However, we require less than that.
Suppose that after a similar transformation of Hamiltonian,
$\overline{H}$, we obtain an effective Hamiltonian of the form
\begin{equation}
\overline{H}=H(\text{low})+H_{\text{free}}(\text{high})+C_{I}^{\dag}V_{IJ}C_{J}\,,
\label{eqH}
\end{equation}
where $V_{IJ}$ is an arbitrary operator in the low frequency space.
The $I$ and $J$ indices should be chosen such that the last term in
Eq.~(\ref{eqH}) contains at least one creation- operator and one
annihilation-operator of high frequency. By using
Rayleigh-Schr\"{o}dinger perturbation theory, it can be shown that the
free high-energy vacuum state of $H_{\text{free}}(\text{high})$ is
annihilated by Eq.~(\ref{eqH}) and remains without correction at any
order of perturbation theory.  Having said that, we will now consider
how to find the individual amplitudes
$\{\hat{s}_{I},\hat{s}'_{I}\}_{m}$ that transfer the Hamiltonian into
the form Eq.~(\ref{eqH}). We split the Hamiltonian in five parts:
\begin{equation}
H=H_{1}+H_{2}^{\text{free}}(\text{high})+
V_{C}(C_{I}^{\dag})+V_{A}(C_{I})+V_{B}\,,
\label{eq16}
\end{equation}
where $H_{1}$ contains only the low frequency modes with $k\leq\mu$,
$H_{2}$ is the free Hamiltonian for all modes with $\mu<k<\Lambda$,
$V_{C}$ contains low frequency operators and products of the high
frequency creation operators $C_{I}^{\dag}$ and $V_{A}$ is the hermitian
conjugate of $V_{C}$. The remaining terms are contained in $V_{B}$,
these terms contain at least one annihilation and creation operators
of the high energy modes. Our goal is to eliminate $V_{C}$ and $V_{A}$
since $V_{B}$ annihilates the vacuum. The ket-state coefficients
$\{\hat{s}_{I}\}_{m}$ are worked out via the ket-state Schr\"{o}dinger
equation $H(\Lambda)|f\rangle=E|f\rangle$ written in the form
\begin{equation}
\langle0|C_{I}e^{-\hat{S}}He^{\hat{S}}|0\rangle=0\,,\qquad
\forall I\neq0\,.
\label{eq17}
\end{equation}
The bra-state coefficients $\{\hat{s}_{I},\hat{s}'_{I}\}_{m}$ are
obtained by making use of the Schr\"{o}dinger equation defined for the
bra-state,~$\langle\widetilde{f}|H(\Lambda)=\langle\widetilde{f}|E$.
First we project both sides on $C_{I}^{\dag}|0\rangle$, then we
eliminate $E$ by making use of the ket-state equation projection with
the state $\langle0|e^{\hat{S}'}C_{I}^{\dag}$ to yield the equations
\begin{equation}
\langle0|e^{\hat{S}'}e^{-\hat{S}}[H,C_{I}^{\dag}]e^{\hat{S}}
e^{-\hat{S}'}|0\rangle=0\,,\qquad\forall I\neq0\,.
\label{eq18}
\end{equation}
Alternatively one can in a unified way apply
$e^{\hat{S}}e^{-\hat{S}'}C_{I}^{\dag}|0\rangle$ on the Schr\"{o}dinger
equation for the bra-state and obtain
\begin{equation}
\langle0|e^{\hat{S}'}e^{-\hat{S}}He^{\hat{S}}e^{-\hat{S}'}
C_{I}^{\dag}|0\rangle=0\,,\qquad\forall I\neq0\,.
\label{eq19}
\end{equation}
Equation~(\ref{eq17}) and Eqs.~(\ref{eq18}) or~(\ref{eq19}) provide
two sets of formally exact, microscopic, operatorial coupled
non-linear equations for the ket and bra. One can solve the coupled
equations in Eq.~(\ref{eq17}) to work out $\{\hat{s}_{I}\}_{m}$ and
then use them as an input in Eqs.~(\ref{eq18}) or~(\ref{eq19}).

It is important to notice that Eqs.~(\ref{eq17}) and~(\ref{eq18}) can
be also derived by requiring that the effective low-energy Hamiltonian
defined in Eq.~(\ref{eq15}), be stationary (i.e.\
$\delta\mathcal{H}(\mu)=0$) with respect to all variations in each of
the independent functional $\{\hat{s}_{I},\hat{s}'_{I}\}_{m}$. One can
easily verify that the requirements
$\delta\mathcal{H}(\mu)/\delta\hat{s}^{m}_{I}=0$ and
$\delta\mathcal{H}(\mu)/\delta\hat{s}'^{m}_{I}=0$ yield Eqs.~(\ref{eq18})
and~(\ref{eq17}). The combination of Eqs.~(\ref{eq17})
and~(\ref{eq18}) does not manifestly satisfy the decoupling property
as set out in Eqs.~(\ref{eq5}) and~(\ref{decoupling}).  On the other
hand Eqs.~(\ref{eq17}) and~(\ref{eq19}) satisfy these
conditions. Equations~(\ref{eq17}) and~(\ref{eq19}) imply that all
interactions including creation and annihilation of fast particles
(``$I$'') are eliminated from the transformed Hamiltonian
$\mathcal{H(\mu)}$ in Eq.~(\ref{eq15}). In other words, these are
decoupling conditions leading to the elimination of $V_{C}$ and
$V_{A}$ from Eq.~(\ref{eq16}), which is, in essence, a
block-diagonalization. Therefore it makes sense for our purpose to use
Eqs.~(\ref{eq17}) and~(\ref{eq19}) for obtaining the unknown
coefficients. We postpone the discussion of the connection between decoupling and variational equations in next section.

So far everything has been introduced rigorously without invoking any
approximation. In practice one needs to truncate both sets of
coefficients $\{\hat{s}_{I},\hat{s}'_{I}\}_{m}$ at a given order of
$m$. A consistent truncation scheme is the so-called SUB($\mathcal{N},\mathcal{M}$)
scheme, where the $n$-body partition of the operator
$\{\hat{S},\hat{S}'\}$ is truncated so that one sets the higher
partition with $I>\mathcal{N}$ to zero up to a given accuracy $m=\mathcal{M}$. Notice that,
Eqs.~(\ref{eq18}) and~(\ref{eq19}) provide two equivalent sets of
equations in the exact form, however after the truncation they can in
principle be different.  Eqs.~(\ref{eq17}) and~(\ref{eq19}) are
compatible with the decoupling property at any level of the
truncation, whereas Eqs.~(\ref{eq17}) combined with~(\ref{eq18}) are
fully consistent with HFT at any level of truncation. Thus the
low-energy effective form of an arbitrary operator can be computed
according to Eq.~(\ref{final}) in the same truncation scheme used for
the renormalization of the Hamiltonian.  In particular, we will show
that only in the lowest order ($m=0$), equations~(\ref{eq18})
and~(\ref{eq19}) are equivalent, independent of the physical system
and the truncation scheme.

Although our method is non-perturbative, perturbation theory can be
recovered from it. In this way, its simple structure for loop expansion
will be obvious and we will observe that at lower order hermiticity
is preserved. Now we illustrate how this is realizable in our approach.
Assume that $V_{C}$ and $V_{A}$ are of order $\lambda$, we will
diagonalize the Hamiltonian, at leading order in $\lambda$ up to
the desired accuracy in $\mu/\Lambda$. We use the commutator-expansion Eq.~(\ref{10}) to organize
Eq.~(\ref{eq17}) perturbatively in order of $m$,
aiming at elimination of the high momenta degree of freedom up to the
first order in the coupling constant, thus yields
\begin{eqnarray}
m=0:&&\langle0|C_{I}(V_{C}+[H_{2},\hat S_{0}])|0\rangle=0\,,
\nonumber \\
m=1:&&\langle0|C_{I}([H_{1},\hat S_{0}]+[H_{2},\hat S_{1}]+
[V_{A},\hat S_{1}]+[V_{C},\hat S_{1}])|0\rangle=0\,,
\nonumber \\
&&\qquad\vdots
\nonumber \\
m=n:&&\langle0|C_{I}([H_{1},\hat S_{n-1}]+[H_{2},\hat S_{n}]+
[V_{A},\hat S_{n}]+[V_{C},\hat S_{n}])|0\rangle=0\,,
\label{eq21}
\end{eqnarray}
where $I\neq0$. Notice that $\hat S_{0}$ is chosen to cancel $V_{C}$
in the effective Hamiltonian, hence it is at least of order of
$\lambda$, consequently it generates a new term $[H_{1},\hat S_{0}]$
which is of higher order in $\mu/\Lambda$ and can be canceled out on
the next orders by $\hat S_{1}$. The logic for obtaining the equations
above is based on the fact that $\hat S_{n}$ should be smaller than
$\hat S_{n-1}$ (for sake of convergence) and that the equations should
be consistent with each other. Since $H_{2},V_{A},V_{C}\approx\Lambda$
and $H_{1}\approx\mu$, from Eq.~(\ref{eq21}) we have the desired
relation $\hat S_{n}\approx\frac{\mu}{\Lambda}\hat S_{n-1}$.  Notice
that if $\mu\sim\Lambda$ then one needs to keep all coupled equations
in Eq.~(\ref{eq21}) up to $m=n$. In other words, there is no
perturbative expansion in ratio of cutoffs. This may introduce
small-energy denominator i. e, $1/(\Lambda-\mu)$, since one may obtain
the solution of these equations, e. g., $S_{0}$ in form of a
geometrical series which can be resummed and leads to a small-energy
denominator \footnote{Thanks to Prof. F. Wegner, for bringing this
point to my attention.}. Having said that as far as the
renormalization is concern we are always interested on condition that
$\Lambda>>\mu$. Furthermore, for $\mu\sim\Lambda$ one should resort to
the non-perturbative decoupling equations which are by construction free of
small-energy denominator. The same procedure can be applied for
Eq.~(\ref{eq19}) which leads to the introduction of a new series of
equations in order of $m$,
\begin{eqnarray}
m=0:&&\langle0|(V_{A}-[H_{2},\hat S'_{0}])C_{I}^{\dag}|0\rangle=0\,,
\nonumber \\
m=1:&&\langle0|([H_{1},\hat S'_{0}]+[H_{2},\hat S'_{1}]+[V_{C},\hat S'_{1}]+
[V_{A},\hat S'_{1}]-[V_{A},\hat S_{1}])C_{I}^{\dag}|0\rangle=0\,,
\nonumber \\
&&\qquad\vdots
\nonumber \\
m=n:&&\langle0|([H_{1},\hat S'_{n-1}]+[H_{2},\hat S'_{n}]+[V_{C},\hat S'_{n}]+
[V_{A},\hat S'_{n}]-[V_{A},\hat S_{n}])C_{I}^{\dag}|0\rangle=0\,.\qquad
\label{eq23}
\end{eqnarray}
Alternatively, we can use Eq.~(\ref{eq18}) to yield the equations
\begin{eqnarray}
m=0:&&\langle0|\big([V_{A},C_{I}^{\dag}]-\big[[H_{2},C_{I}^{\dag}],
\hat S'_{0}\big]\big)|0\rangle=0\,,
\nonumber \\
m=1:&&\langle0|\big(\big[[V_{A},C_{I}^{\dag}],\hat S_{1}\big]-
\big[[H_{2},C_{I}^{\dag}],\hat S'_{1}\big]-\big[[V_{A},C_{I}^{\dag}],
\hat S'_{1}\big]\big)|0\rangle=0\,,
\nonumber \\
&&\qquad\vdots
\nonumber \\
m=n:&&\langle0|\big(\big[V_{A},C_{I}^{\dag}],\hat S_{n}\big]-
\big[[H_{2},C_{I}^{\dag}],\hat S'_{n}\big]-\big[[V_{A},C_{I}^{\dag}],
\hat S'_{n}\big]\big)|0\rangle=0\,.\qquad
\label{eq24}
\end{eqnarray}
It is obvious that at order $m=0$, Eqs.~(\ref{eq23}) and~(\ref{eq24})
are the same and $\hat S'_{0}=\hat S_{0}^{\dag}$, which indicates that the
similarity transformation at this level remains unitary. It should be
noted that diagonalization at first order in the coupling constant
introduces a low-energy effective Hamiltonian in Eq.~(\ref{eq15})
which is valid up to the order $\lambda^{3}$. In the same way,
diagonalization at second order in $\lambda$ modifies the Hamiltonian
at order $\lambda^{4}$ and leads generally to a non-unitarity
transformation. In this way one can proceed to diagonalize the
Hamiltonian at a given order in $\lambda$ with desired accuracy in
$\mu/\Lambda$ . Finally, the renormalization process is completed by
introducing the correct $Z(\Lambda)$ factors which redefine the
divergences emerging from Eq.~(\ref{eq15}).

\section{The decoupling conditions versus the variational  principle}
In section 2.3 we introduced the effective low-energy phase space
$\{\hat{s}_{I},\hat{s}'_{I}\}_{m}$ induced by integrating out the
high-energy modes. We argued later that the induced low-energy
phase space can be obtained either from Eqs.~(\ref{eq17},
\ref{eq19}) which are manifestly compatible with the decoupling
conditions or from Eqs.~(\ref{eq17}, \ref{eq18}) compatible with
the variational principle and the Hellmann-Feynman theorem. Here
we will show that in fact this two formulations are related
through an isomorphic transformation (or in a strict sense by a
symplectomorphism). In what follows, for simplicity, we absorb the
factor $(\frac{\mu}{\Lambda})^{m}$ in Eq.~(\ref{def-s}) into the
definition of $\hat{S}_{m}$ and $\hat{S}'_{m}$, hence we assume
$\hat{S}_{m}(\hat{S}'_{m})\simeq (\frac{\mu}{\Lambda})^{m}$,
therefore we have desirable convergence relation
$\hat{S}_{m+1}(\hat{S}'_{m+1})\simeq
\frac{\mu}{\Lambda}\hat{S}_{m}(\hat{S}'_{m})$. In the spirit of
Schr\"odinger representation in quantum field theory, one may now
define a generalized many-body ket and bra wave functional at a
given scale $\mu$,
\begin{equation}
|\hat{\psi_{\mu}}\rangle=e^{\hat{S}(\mu,\Lambda)}e^{-\hat{S}'(\mu,\Lambda)}|0\rangle=e^{\hat{S}(\mu,\Lambda)}|0\rangle,
\hspace{2cm}
\langle\hat{\tilde{\psi}}_{\mu}|=\langle 0|e^{\hat{S}'(\mu,\Lambda)}e^{-\hat{S}(\mu,\Lambda)}, \label{wave}
\end{equation}
It is clear that by construction, we have
$\langle\hat{\tilde{\psi}}_{\mu}|\hat{\psi_{\mu}}\rangle=1$ at any
level of approximation. Notice that the bra and ket are parametrised
independently, since they are not hermitian-adjoint of each
other. Therefore, we have a biorthogonal representation of the
many-body system. We will illuminate below the main underlying reasons
behind this parametrization. With this definition the effective
low-energy Hamiltonian Eq.~(\ref{eq15}) is rewritten as,
\begin{equation}
\hat{\mathcal{H}}(\mu)=\langle\hat{\tilde{\psi}}_{\mu}|H|\hat{\psi_{\mu}}\rangle. \label{rg1-1}
\end{equation}
Here, the bra and ket wave functional are built by adding independent
clusters of high-energy correlation (where $\mu<k<\Lambda$) in the
vacuum of the free high-energy Hamiltonian, or by integrating out the high-energy modes
from the Hamiltonian. The unknown low-energy operators
$\{\hat{s}^{m}_{I},\hat{s}'^{m}_{I}\}$ are obtained by solving the Schr\"{o}dinger
equation in the high-energy shell ($\mu<k<\Lambda$) and are decoupling equations,
\begin{eqnarray}
D_{1}&=&Q\bar{H}P=\langle0|C_{I}e^{-\hat{S}}He^{\hat{S}}|0\rangle=0\,,\qquad \forall I\neq0\, \label{rg2}\\
D_{2}&=&P\bar{H}Q=\langle0|e^{\hat{S}'}e^{-\hat{S}}He^{\hat{S}}e^{-\hat{S}'}
C_{I}^{\dag}|0\rangle=0,\qquad\forall I\neq0\,.\label{rg3}\
\end{eqnarray}
 The decoupling equations Eqs.~(\ref{rg2},\ref{rg3}) imply that all
 interactions containing creation and annihilation of ``$I$''
 high-momentum particles are eliminated from the transformed
 Hamiltonian while it generates new low-momentum interactions through
 $\{\hat{s}^{m}_{I},\hat{s}'^{m}_{I}\}$. These equations show the
 changes of the generalized wave functional and accordingly the
 effective Hamiltonian with the flow-parameter $\mu$.  Alternatively
 one can obtain $\{\hat{s}^{m}_{I},\hat{s}'^{m}_{I}\}$ by requiring
 that the effective low-energy be stationary
 $\delta\hat{\mathcal{H}}(\mu)=0$ with respect to all variations in
 each of the independent functional
 $\{\hat{s}^{m}_{I},\hat{s}'^{m}_{I}\}$,
\begin{eqnarray}
&&\frac{\delta\hat{\mathcal{H}}(\mu)}{\delta \hat{s}^{m}_{I}}=0 \to
\langle0|e^{\hat{S}'}e^{-\hat{S}}[H,C_{I}^{\dag}]e^{\hat{S}}
e^{-\hat{S}'}|0\rangle=0\,,\qquad\forall I\neq0\,\label{rg4}\\
&&\frac{\delta\hat{\mathcal{H}}(\mu)}{\delta \hat{s}'^{m}_{I}}=0 \to
D_{1}=0\label{rg5}\
\end{eqnarray}
where $D_{1}$ is introduced in Eq.~(\ref{rg2}). Notice that we have
already shown that Eq.~(\ref{rg4}) is as well derivable from dynamics
. The decoupling conditions Eqs.~(\ref{rg2}, \ref{rg3}) seems generally
to be in conflict with variational equations
(\ref{rg4}, \ref{rg5}). Here we will show that in fact this two
formulations are related via an isomorphic transformation (or in a
strict sense by a symplectomorphism). Let us introduce a new set of
variables $\{\hat{\sigma}^{m}_{I},\hat{\bar{\sigma}}^{m}_{I}\}$ within
the low-energy phase space, by transforming the induced low-energy
phase space operators set $\{\hat{s}^{m}_{I},\hat{s}'^{m}_{I}\}$ into
a new set $\{\hat{\sigma}^{m}_{I},\hat{\bar{\sigma}}^{m}_{I}\}$,
\begin{eqnarray}
&&\left\{\begin{array}{ll}\hat{\sigma}^{m}_{I}=\langle
0|C_{I}e^{\hat{S}'}S|0\rangle=\displaystyle\sideset{}{'}\sum_{J}\hat{s}^{m}_{J}\hat{\omega}_{JI},\\
\hat{\bar{\sigma}}^{m}_{I}=\hat{s}'^{m}_{I},\end{array}\right.\label{ge1}\
\end{eqnarray}
where the low-energy operator $\hat{\omega}_{JI}$ is defined as
\begin{equation}
\hat{\omega}_{JI}=\langle 0|C_{I}e^{\hat{S}'}C^{\dag}_{J}|0\rangle.
\end{equation}
In the same way, one may conversely write $\hat{s}$ in terms of $ \hat{\sigma}$,
\begin{equation}
\hat{s}^{m}_{I}(\hat{\sigma},\hat{\bar{\sigma}})=\displaystyle\sideset{}{'}\sum_{J}\hat{\sigma}^{m}_{J}\hat{\bar{\omega}}_{JI},
\end{equation}
where we have used the following orthogonality property,
\begin{eqnarray}
&&\hat{\bar{\omega}}_{JI}=\langle 0|C_{I}e^{-\hat{S}'}C^{\dag}_{J}|0\rangle,\nonumber\\
&& \sum_{K} \hat{\omega}_{JK}\hat{\bar{\omega}}_{KI}= \sum _{K}\hat{\bar{\omega}}_{JK}\hat{\omega}_{KI}=\delta(J,I).\label{or}\
\end{eqnarray}
We use the canonical transformation Eq.~(\ref{ge1}) to rewrite the
decoupling equations $D_{1}$ and $D_{2}$ in terms of derivative with respect to new
variables, one can immediately show,
\begin{equation}
D_{1}=\langle
0|C_{I}e^{\hat{S}'}e^{-\hat{S}}H(\Lambda)e^{\hat{S}}e^{-\hat{S}'}|0\rangle=\frac{\delta\hat{\mathcal{H}}(\mu)}{\delta\hat{s}'^{m}_{I}}
=\frac{\delta\hat{\mathcal{H}}}{\delta\hat{\bar{\sigma}}^{m}_{I}}+\displaystyle\sideset{}{'}\sum_{J}\hat{\sigma}^{m}_{I+J}\frac{\delta\hat{\mathcal{H}}}{\delta
\hat{\sigma}^{m}_{J}}. \label{va1}
\end{equation}
In the same fashion, having made use of the closure identity $I$ in the high-energy Fock space $\mathcal{G}$, and Eq.~(\ref{or}),
one can rewrite the other decoupling equation $D_{2}$ in the form of
\begin{eqnarray}
D_{2}&=&\langle
0|e^{\hat{S}'}e^{-\hat{S}}H(\Lambda)e^{\hat{S}}(I)e^{-\hat{S}'}C^{\dag}_{I}|0\rangle,\nonumber\\
&=&\hat{\mathcal{H}}(\mu)\langle 0|e^{-\hat{S}'}C_{I}^{\dag}|0\rangle+\displaystyle\sideset{}{'}\sum_{J}\hat{\bar{\omega}}_{IJ}\{\langle
0|e^{\hat{S}'}e^{-\hat{S}}[H(\Lambda),C_{I}^{\dag}]e^{\hat{S}}|0\rangle,\nonumber\\
&+&\langle 0|e^{\hat{S}'}C^{\dag}_{I}e^{-\hat{S}'}(I)e^{\hat{S}'}e^{-\hat{S}}H(\Lambda)e^{\hat{S}}e^{-\hat{S}'}|0\rangle\},\nonumber\\
&=&\frac{\delta\hat{\mathcal{H}}(\mu)}{\delta\hat{\sigma}^{m}_{I}}
+\displaystyle\sideset{}{'}\sum_{J}
\frac{\delta\hat{\mathcal{H}}(\mu)}{\delta\hat{\bar{\sigma}}^{m}_{J}}\hat{L}_{JI}+
\displaystyle\sideset{}{'}\sum_{J,K}\frac{\delta\hat{\mathcal{H}}(\mu)}{\delta\hat{\sigma}^{m}_{J}}\hat{\sigma}^{m}_{J+K}\hat{L}_{KI},\label{va2} \
\end{eqnarray}
where in final step we have made use of the orthogonality relation
Eq.~(\ref{or}) and the relation in Eq.~(\ref{va1}). The operator $\hat{L}_{IJ}$ is defined as
\begin{equation}
\hat{L}_{IJ}=\sum_{K}\langle 0|e^{\hat{S}'}C^{\dag}_{K}e^{-\hat{S}'}C^{\dag}_{J}|0\rangle\langle 0|C_{K}e^{-\hat{S}'}C^{\dag}_{I}|0\rangle .
\end{equation}
The operator $\hat{L}_{IJ}$ is symmetric and doubly linked. It is
obvious from Eqs.~(\ref{va1}, \ref{va2}) that the requirement of
$\delta\hat{\mathcal{H}}(\mu)/\delta\hat{\sigma}^{m}_{I}=0$ and
$\delta\hat{\mathcal{H}}(\mu)/\delta\hat{\bar{\sigma}}^{m}_{I}=0$ lead directly
to the decoupling conditions $D_{1}=0$ and $D_{2}=0$. However the
reverse is not always correct. Therefore, the decoupling conditions are
weaker than the variational equations. A set of similar canonical variables
as in Eq.~(\ref{ge1}) (where the variables are $c$-numbers), was introduced by
Arponen, Bishop and Pajanne \cite{ccm1} in the context of the traditional CCM and
turned out to be quite practical.

We have already proved (in section 2.4) that the effective low-energy operators Eq.~(\ref{eq15})
or Eq.~(\ref{rg1-1}) equipped with decoupling conditions
Eqs.~(\ref{rg2},\ref{rg3}) have the same low-energy eigenvalues
as the original Hamiltonian. The decoupling conditions are thus sufficient
requirements to ensure partial diagonalization of the Hamiltonian in
the particle and momentum space.

Having obtained the effective bra and ket Eq.~(\ref{wave}) at a given scale $\mu$, one can compute the effective low-energy of an arbitrary operator $A$,
\begin{equation}
\hat{\mathcal{A}}(\hat{s}_{I},\hat{s}'_{I})=\langle\hat{\tilde{\psi}}_{\mu}|A|\hat{\psi_{\mu}}\rangle, \label{ava}
\end{equation}
One may now pose the question if the approximation (truncation) used to
obtain the induced bra- and ket-state, and accordingly the effective
Hamiltonian, is sufficient to obtain the effective low-energy of
a given operator by Eq.~(\ref{ava}). On the other hand, one might be
curious about the necessity of the double similarity transformation
and as well the introduction of two independent sets of variables
$\{\hat{s}_{I}^{m},\hat{s}'^{m}_{I}\}$, since one could have defined a
single transformation $H\to e^{-\hat{S}}He^{\hat{S}}$ where $\hat{S}$
is defined in Eq.~(\ref{def-s}) (without introducing a new set of variables
$\{\hat{s}'^{m}_{I}\}$, and invoking the definition of the bar
state). In this case, the variables set $\{\hat{s}^{m}_{I}\}$
would be determined by Eq.~(\ref{rg2}) alone. Let us pursue this idea and apply
the Hellmann-Feynman theorem\footnote{The
Hellmann-Feynman theorem is originally introduced for the ground
state, however its proof is more general and can be applied
here.} to such a parametrization. In our framework, the
Hellmann-Feynman theorem  states that if we perturb
the Hamiltonian $H\to H'=H+J A$, where $J$ is an infinitesimally
small (a source term) and $A$ is an arbitrary operator, such that the
effective Hamiltonian changes as $\hat{\mathcal{H}}\to
\hat{\mathcal{H}}'=\hat{\mathcal{H}}+J d\hat{\mathcal{H}}/dJ+O(J^{2})$ then we have,
\begin{equation}
d\hat{\mathcal{H}}/dJ=\hat{\mathcal{A}}=\langle \psi_{\mu}|dH/dJ|\psi_{\mu}\rangle,
\end{equation}
where we define an effective low-energy operator
$\hat{\mathcal{A}}=\langle\psi_{\mu}|A|\psi_{\mu}\rangle$. We can find the effective low-energy operator
$\hat{\mathcal{A}}$ by using the Hellmann-Feynman theorem,
\begin{equation}
\hat{\mathcal{A}}=\frac{d}{dJ}\langle 0|e^{-\hat{S}}(H+JA)e^{\hat{S}}|0\rangle=\langle 0|e^{-\hat{S}}Ae^{\hat{S}}|0\rangle+\langle 0|e^{-\hat{S}}He^{\hat{S}}C^{\dag}_{I}|0\rangle\frac{\delta \hat{s}^{m}_{I}}{\delta J}, \label{h-f}
\end{equation}
where we used $[\hat{S},C_{I}^{\dag}]=0$. If we now calculate $\delta \hat{s}^{m}_{I}$ from the first decoupling equation
(\ref{rg2}) which involves only $\hat{S}$, having retained only $O(J,\delta
S)$ terms, we find
\begin{equation}
\langle 0|C_{I}e^{-(\hat{S}+\delta \hat{S})}(H+JA)e^{(\hat{S}+\delta \hat{S})}|0\rangle=\langle 0|C_{I}e^{-\hat{S}}JAe^{\hat{S}}|0\rangle+\langle 0|C_{I}e^{-\hat{S}}[H, C^{\dag}_{J}]e^{\hat{S}}|0\rangle\delta \hat{s}^{m}_{J}=0.
\end{equation}
Therefore, one can show that,
\begin{equation}
C^{\dag}_{I}|0\rangle \frac{\delta \hat{s}^{m}_{I}}{\delta J}=\mathcal{Q}(\hat{\mathcal{H}}-\mathcal{Q}e^{-\hat{S}}He^{\hat{S}}\mathcal{Q})^{-1}\mathcal{Q}e^{-\hat{S}}Ae^{\hat{S}}|0\rangle, \label{ds}
\end{equation}
where the operator $\mathcal{Q}=1-|0\rangle\langle
0|=\displaystyle\sideset{}{'}\sum_{J}C^{\dag}_{J}|0\rangle\langle 0|C_{J}$
is introduced. Now one can make use of Eq.~(\ref{ds}) to show that the right-hand side of Eq.~(\ref{h-f}) can be recast in the following form,
\begin{equation}
\hat{\mathcal{A}}=\langle 0| e^{\hat{S}'}e^{-\hat{S}}Ae^{\hat{S}}|0\rangle, \label{ava-n}
\end{equation}
where we introduced the notation $e^{\hat{S}'}$ by,
\begin{equation}
\langle 0|e^{\hat{S}'}=\langle 0|+\langle 0|e^{-\hat{S}}He^{\hat{S}}\mathcal{Q}(\hat{\mathcal{H}}-\mathcal{Q}e^{-\hat{S}}He^{S}\mathcal{Q})^{-1}\mathcal{Q}. \label{eccm}
\end{equation}
Interestingly, Eq.~(\ref{eccm}) satisfies the second decoupling
condition Eq.~(\ref{rg3}). Thereby Eq.~(\ref{ava-n}) becomes exactly
equivalent to Eq.~(\ref{ava}). Therefore the use of the
the Hellmann-Feynman theorem and assuming the coupled-cluster
parametrization leads naturally to the definition of the bra-state in
Eq.~(\ref{wave}) and emergence of a new set of variable $\hat{S}'$. In other
words, no other bra-state parametrization (including hermitian-adjoint
of ket-state) is compatible with the Hellmann-Feynman theorem.

It is well known that the traditional multiplicative renormalization
is not sufficient for the renormalization of more than one composite
operator inserted into the renormalized Green functions. To avoid ad
hoc subtractions, one can introduce the composite operators into the
Lagrangian with a space-time dependent source (coupling). It has been
shown that the renormalization of the source produces requires counter
terms to render all Green functions containing the insertion of
composite operators renormalized \cite{rc}. These new counterterms do
not affect the renormalization of the original theory.

The main advantage of the compatibility of our parametrization with the Hellmann-Feynman
theorem underlies that here the effective low-energy of an arbitrary
operator, can be obtained in the same truncation scheme
SUB$(\mathcal{N},\mathcal{M})$ used for the Hamiltonian matrix
\footnote{As Thouless pointed out, the Hellmann-Feynman theorem
immediately implies that an expectation value $\langle A\rangle $ of
an arbitrary operator is computed diagrammatically from the same set
of Goldstone diagrams as for the energy $\langle H\rangle $, where all
interaction is replaced by the operator $A$.}. Moreover, the
above-mentioned technique used in Lagrangian formalism, can be
employed in our framework by means of the Hellmann-Feynman
theorem. Therefore, the renormalization of an arbitrary operator can
be calculated in a unified way. This is indeed, one of advantages of the
non-unitary parametrization of the similarity renormalization group.

\section{The symplectic structure}
Despite extensive progress in development of various RG techniques,
little is still known about the geometrical interpretation of the RG
\cite{g}. Here, we introduce the geometrical structure emerging from
our approach.  We define a low-energy action-like functional
$\hat{\mathcal{A}}$, having integrated out fast modes and making use
of the reparametrization Eq.~(\ref{ge1}),
\begin{eqnarray}
\hat{\mathcal{A}}(\mu)&=&\int dt \langle 0|e^{\hat{S}'(t)}e^{-\hat{S}(t)}(i\frac{\delta}{\delta t}-\hat{H}(t,\Lambda))e^{\hat{S}(t)}e^{-\hat{S}'(t)}|0\rangle,\nonumber\\
&=&\int dt (i \displaystyle\sideset{}{'}\sum_{I}
\hat{\bar{\sigma}}_{I}\hat{\dot{\sigma}}_{I}-\hat{\mathcal{H}}(\mu,\sigma,\bar{\sigma})),\nonumber\\
&=&\int dt (-i \displaystyle\sideset{}{'}\sum_{I}
\hat{\dot{\bar{\sigma}}}_{I}\hat{\sigma}_{I}-\hat{\mathcal{H}}(\mu,\sigma,\bar{\sigma})),\
\end{eqnarray}
where in the final step we employed integration by parts. The operator $\mathcal{H}(\mu,\sigma,\bar{\sigma})$ is the effective
low-energy Hamiltonian defined in Eq.~(\ref{eq15}) and $\hat{\sigma}_{I}(\hat{\bar{\sigma}}_{I})$ are defined,
\begin{equation}
\hat{\sigma}_{I}=\sum_{m=0}^{\mathcal{M}}\hat{\sigma}^{m}_{I} \hspace{2cm}
\hat{\bar{\sigma}}_{I}=\sum_{m=0}^{\mathcal{M}}\hat{\bar{\sigma}}^{m}_{I} \label{news}
\end{equation}
 The stationary of
$\hat{\mathcal{A}}$ with respect to the complete set of variables
$\{\hat{\sigma}_{I},\hat{\bar{\sigma}}_{I}\}$ for a given truncation
SUB$(\mathcal{N},\mathcal{M})$ yields
\begin{equation}
\delta\hat{\mathcal{A}}(\mu)=0 \longrightarrow \left(i\frac{\delta\hat{\sigma}_{I}}{\delta t}=\frac{\delta\hat{\mathcal{H}}(\mu)}{\delta\hat{\bar{\sigma}}_{I}};\hspace{1cm}
-i\frac{\delta\hat{\bar{\sigma}}_{I}}{\delta t}=\frac{\delta\hat{\mathcal{H}}(\mu)}{\delta\hat{\sigma}_{I}}\right).\label{motion}
\end{equation}
These equations can be obtained for the set
$\{\hat{s}_{I},\hat{s}'_{I}\}$ as well, without invoking the canonical
transformation Eq.~(\ref{ge1}). However, if one wants to obtain the
above equations in a straightforward manner from time-dependent
Schr\"odinger equation, then the canonical transformation
Eq.~(\ref{ge1}) is necessary.

We note that although our transformation is not unitary and
the parametrization of $\hat{S}$ and $\hat{S}'$ have
been introduced in a very asymmetric fashion, their fundamental dynamics in
the low-energy phase space follows the canonical equation of motion. Therefore the induced low-energy operators
$\hat{\sigma}_{I}$ and $\hat{\bar{\sigma}}_{I}$ are canonically conjugate in the
terminology of the classical Hamiltonian mechanics. This can be made even more suggestive by
defining a new set of variables, the generalized field $\hat{\phi}_{I}$ and
their canonically conjugate generalized momentum densities
$\hat{\pi}_{I}$,
\begin{equation}
\hat{\phi}_{I}=\frac{1}{2}(\hat{\sigma}_{I}+\hat{\bar{\sigma}}_{I}), \hspace{1cm} \hat{\pi}_{I}=\frac{i}{2}(\hat{\bar{\sigma}}_{I}-\hat{\sigma}_{I}).
\end{equation}
In terms of the new operators the equation of motion Eq.~(\ref{motion}) can be rewritten into the form
\begin{eqnarray}
&&\frac{d\hat{\phi}_{I}}{dt}=\{\hat{\phi}_{I},\hat{\mathcal{H}}(\mu)\}=\frac{\delta\hat{\mathcal{H}}(\mu)}{\delta\hat{\pi}_{I}},\nonumber\\
&&\frac{d\hat{\pi}_{I}}{dt}=\{\hat{\pi}_{I},\hat{\mathcal{H}}(\mu)\}=-\frac{\delta\hat{\mathcal{H}}(\mu)}{\delta\hat{\phi}_{I}},\label{motion2}\
\end{eqnarray}
where a generalized Poisson bracket $\{A,B\}$ for two arbitrary operators is defined as
\begin{equation}
\{A,B\}=\displaystyle\sideset{}{'}\sum_{I}\left(\frac{\delta A}{\delta\hat{\phi}_{I}}\frac{\delta B}{\delta\hat{\pi}_{I}}-\frac{\delta B}{\delta\hat{\phi}_{I}}\frac{\delta A}{\delta\hat{\pi}_{I}}\right).\label{po}
\end{equation}
For the nonzero mutual Poisson brackets of canonical coordinates and
momentums we have $\{\hat{\phi}_{I},\hat{\pi}_{J}\}=\delta(I,J)$.  We
can also look at the behaviour of the low-energy effective operator
for a product of operators. One can write the product of operators
after a double similarity transformation as a product of
transformed operators,
\begin{eqnarray}
\langle\hat{\tilde{\psi}}_{\mu}|AB|\hat{\psi_{\mu}}\rangle&=&\langle 0|e^{\hat{S}'}e^{-\hat{S}}ABe^{\hat{S}}e^{-\hat{S}'}|0\rangle=\langle 0|\bar{A}\bar{B}|0\rangle,\label{class}\\
&=&\hat{\mathcal{A}}\hat{\mathcal{B}}+\displaystyle\sideset{}{'}\sum_{I}\langle 0|e^{\hat{S}'}e^{-\hat{S}}Ae^{\hat{S}}e^{-\hat{S}'}C^{\dag}_{I}|0\rangle\langle 0|C_{I}e^{\hat{S}'}e^{-\hat{S}}Be^{\hat{S}}e^{-\hat{S}'}|0\rangle, \nonumber\
\end{eqnarray}
where we have used the closure identity in $Q$-space. The
effective operator $\hat{\mathcal{A}}$ and $\hat{\mathcal{B}}$ are defined
by Eq.~(\ref{ava}). Now we express the last part of Eq.~(\ref{class})
in terms of a derivatives of the effective operators
$\hat{\mathcal{A}}(\hat{\mathcal{B}})$ with respect to low-energy phase space
$\{\hat{\sigma}_{I},\hat{\bar{\sigma}}_{I}\}$. In fact, we have
already shown such relations in Eqs.~(\ref{va1}, \ref{va2}). The same
relationships are valid for any arbitrary operator for new variables
$\{\hat{\sigma}_{I},\hat{\bar{\sigma}}_{I}\}$. Therefore with a
replacement of $H\to A, B$ and $\mathcal{H} \to
\mathcal{A},\mathcal{B}$ in Eqs.~(\ref{va1}, \ref{va2}) we have the
desired equations. The same procedure can be carried out for evaluating of $\langle\hat{\tilde{\psi}}_{\mu}|BA|\hat{\psi_{\mu}}\rangle$. After some straightforward algebra, one finds that,
\begin{equation}
\langle\hat{\tilde{\psi}}_{\mu}|AB-BA|\hat{\psi_{\mu}}\rangle=\langle\hat{\tilde{\psi}}_{\mu}|[A,B]|\hat{\psi_{\mu}}\rangle=i\{\hat{\mathcal{A}},\hat{\mathcal{B}}\},
\end{equation}
where the Poisson bracket is as defined in Eq.~(\ref{po}).

Therefore our parametrization is clearly suggestive that the effective
low-energy phase space induced by integrating out the high-energy
particles can be described in terms of highly non-linear classical
dynamics with the canonical coordinates
$(\hat{\phi}_{I},\hat{\pi}_{I})$, without losing any
quantum-field-theoretical information. The canonical equation of
motions Eq.~(\ref{motion2}) are valid at any level of approximation
SUB$(\mathcal{N},\mathcal{M})$ on the renormalization group
trajectory. In this way, we are led to a manifestation of the
correspondence principle in a more generalized form, that is the
induced low-energy Hamiltonian obtained by integrating short-distance
modes are governed by a hierarchy of non-linear classical mechanical
equations for quasi-local fields $\hat{\phi}_{I}$ and $\hat{\pi}_{I}$.
One may hope to recover the full classical mechanics (with c-numbers
variables as coordinates) at the other extreme of the RG trajectory.

The general form for the action completely determines the symplectic
structure of our low-energy phase space $
(\hat{\phi}_{I},\hat{\pi}_{I})$. The phase space is the cotangent
bundle\footnote{In differential geometry, the cotangent bundle is the
union of all cotangent spaces of a manifold \cite{geom}.} of the configuration
space $\mathcal{C}$, $\Gamma=T^{*}(\mathcal{C})$ (the coordinates
$\hat{\phi}_{I}$ label the points of configuration space
$\mathcal{C}$) \cite{geom}. The Poisson bracket Eq.~(\ref{po}) induces
a symplectic structure
\begin{equation}
\omega=\frac{1}{2}\omega_{ab}dx^{a}\wedge dx^{b},
\end{equation}
where the coordinates on the symplectic manifold $\Gamma$ are denoted by
$x^{a}\in \{\hat{\phi}_{I}\}$ and $\omega_{ab}$ is the inverse of the symplectic matrix
\begin{equation}
\omega^{ab}=\{x^{a},x^{b}\}\iff \{A,B\}=\omega^{ab}A_{,a}B_{,b}.
\end{equation}
Moreover, a symplectic form defines an isomorphism between the tangent
and cotangent spaces of $\Gamma$. One may associate a vector field $X_{f}$ to every function $f\in \mathcal{C}^{\infty}(\Gamma)$ by
\begin{equation}
i(X_{f})\omega=-df,
\end{equation}
where $i(X) $ and $X_{f}$ denote the interior product and the symplectic
gradient of $f$, respectively. $X_{f}$ is the so-called the Hamiltonian vector field of
$f$, and  it generates a flow on $\Gamma$ which leaves $\omega$ invariant,
since the Lie derivative of $\omega$ along $X_{f} $ is zero. In this
way, one can rewrite the Poisson bracket in the form
\begin{equation}
\{f,g\}=i(X_{f})i(X_{g})\omega=i(X_{f})dg=\omega(X_{f},X_{g})\in\mathcal{C}^{\infty}(\Gamma),
\end{equation}
which shows the change of $g$ along $X_{f}$. We require that $\omega$
to be closed ($d\omega=0$)\footnote{A symplectic manifold
($\Gamma,\omega$) is a smooth real $N$-dimensional manifold without
boundary, equipped with a closed non-degenerate two-form $\omega$,
i. e., $d\omega=0$ where $d$ is the exterior differential
\cite{geom}.}, which implies the Jacobi identities
$(d\omega)(X_{f},X_{g},X_{h})=0$. Therefore, the existence of Poisson
bracket in our definition of the coordinates of the phase-space introduces a
symplectic manifold for the phase-space.

\section{The constrained induced phase space}
In this section we closely follow Ref.~\cite{12} (one should pay
some extra care here, since the phase space here is operatorial rather
than $c$-number).  Let us assume that the renormalized form of an
arbitrary operator $A$, Eq.~(\ref{ava}) can be obtained by,
\begin{equation}
\mathcal{A} =\langle 0|e^{\hat{S}'}e^{-\hat{S}}Ae^{\hat{S}}e^{-\hat{S}'}|0\rangle\equiv\frac{\langle 0|e^{\hat{S}^{\dag}}Ae^{\hat{S}}|0\rangle}{\langle 0|e^{\hat{S}^{\dag}}e^{\hat{S}}|0\rangle}.\label{he1-2}\
\end{equation}
Generally the left-hand side does not agree with right-hand side (it is by no means clear that this relation will be held after
a truncation). In order to ensure the unitarity of the similarity transformation we require,
\begin{equation}
\langle 0|e^{\hat{S}'}\equiv\frac{\langle
0|e^{\hat{S}^{\dag}}e^{\hat{S}}}{\langle
0|e^{\hat{S}^{\dag}}e^{\hat{S}}|0\rangle},\hspace{2cm}
e^{\hat{S}'^{\dag}}|0\rangle\equiv\frac{e^{\hat{S}^{\dag}}e^{\hat{S}}|0\rangle}{{\langle
0|e^{\hat{S}^{\dag}}e^{\hat{S}}|0\rangle}}.\label{he2}
\end{equation}
Thereby, the ket-state and bra-state defined in Eq.~(\ref{wave})
become hermitian-adjoint of each another. We assume the induced low-energy phase
space to be a complex manifold
$\{(\hat{s}_{I},\hat{s}'_{I}),(\hat{s}^{*}_{I},\hat{s}'^{*}_{I})\}$. The
hermiticity conditions are introduced by constraint functions
$\hat{\chi}_{I}(\mu)$ and $\hat{\chi}^{*}_{I}(\mu)$
\begin{eqnarray}
&&\hat{\chi}_{I}(\mu)\equiv (\langle
0|e^{\hat{S}^{\dag}}e^{\hat{S}}|0\rangle)^{-1}\langle
0|e^{\hat{S}^{\dag}}C^{\dag}_{I}e^{\hat{S}}|0\rangle-\langle
0|\hat{S}'C^{\dag}_{I}|0\rangle, \nonumber\\
&&\hat{\chi}^{*}_{I}(\mu)\equiv (\langle
0|e^{\hat{S}^{\dag}}e^{\hat{S}}|0\rangle)^{-1}\langle
0|e^{\hat{S}^{\dag}}C_{I}e^{\hat{S}}|0\rangle-\langle
0|C_{I}\hat{S}'^{*}|0\rangle. \
\end{eqnarray}
Therefore the physical submanifold shell is defined through:
\begin{eqnarray}
&&\hat{\chi}_{I}= 0 \longrightarrow  PX^{-1}(\mu,\Lambda)C^{\dag}_{I}X(\mu,\Lambda)P=\sum_{m=0}^{\mathcal{M}}\hat{s}'^{m}_{I}\equiv\hat{s}'_{I},\nonumber\\
&&\hat{\chi}_{I}^{*}= 0\longrightarrow PX^{-1}(\mu,\Lambda)C_{I}X(\mu,\Lambda)P=\sum_{m=0}^{\mathcal{M}}\hat{s}'^{*m}_{I}\equiv \hat{s}'^{*}_{I},\
\end{eqnarray}
where $X(\mu,\Lambda)$ is defined in Eq.~(\ref{eq13}, \ref{eq14}). This implies that
in the physical submanifold where we have exact hermiticity, and there is
an isomorphic mapping between a cluster of high-energy creation
$C^{\dag}_{I}$ (annihilation $C_{I}$) operators and a low-energy
operators $\hat{s}'_{I}(\hat{s}'^{*}_{I})$.  This isomorphism is
invariant under the renormalization group transformation. We introduce a
Poisson bracket for the complex representation of the phase space,
\begin{equation}
\{A,B\}=\displaystyle\sideset{}{'}\sum_{I}\left(\frac{\delta A}{\delta \hat{s}_{I}}\frac{\delta B}{\delta \hat{s}'_{I}}-
\frac{\delta A}{\delta \hat{s}'_{I}}\frac{\delta B}{\delta \hat{s}_{I}}+
\frac{\delta A}{\delta \hat{s}'^{*}_{I}}\frac{\delta B}{\delta \hat{s}^{*}_{I}}-\frac{\delta B}{\delta \hat{s}^{*}_{I}}\frac{\delta B}{\delta \hat{s}'^{*}_{I}}\right).
\end{equation}
The non-zero commutators of the canonical coordinates follow the
canonical symplectic structure
$\{\hat{s}_{I},\hat{s}'_{I}\}=
\{\hat{s}^{*}_{I},\hat{s}'^{*}_{J}\}=\delta(I,J)$.
The nature of the constraint can be revealed by considering the
commutators between the constraints functional $\hat{\chi}_{I}$ and
$\hat{\chi}^{*}_{I}$. After some tedious but straightforward
algebra, one obtains,
\begin{eqnarray}
&&\{\hat{\chi}_{I},\hat{\chi}_{J}\}=\{\hat{\chi}^{*}_{I},\hat{\chi}^{*}_{J}\}=0,\nonumber\\
&&\{\hat{\chi}_{I},\hat{\chi}^{*}_{J}\}=
2\left(\langle 0|e^{\hat{S}'}e^{S}C_{I}C^{\dag}_{J}e^{\hat{S}}e^{-\hat{S}'}|0\rangle -\hat{\bar{\sigma}}_{I}\hat{\bar{\sigma}}^{*}_{J}\right).\nonumber\
\end{eqnarray}
This implies that the constraints are of second class and do not
correspond to any gauge symmetry degrees of freedom. However, these
superfluous degrees of freedom can be eliminated by Dirac bracket
technique \cite{di}. Of course, since the constrained manifold $N$ has dimension less than the full
manifold $M$, one may define a pullback map $f^{*}:T^{*}(M)\to
T^{*}(N)$ to obtain the induced symplectic structure of the constraint
surface $\hat{\omega}^{0}$, $\hat{\omega}^{0}=f^{*}\hat{\omega}$. In analogy to ordinary CCM \cite{12}, we define a symplectic two-form in the full manifold,
\begin{equation}
\hat{\omega}=\displaystyle\sideset{}{'}\sum_{I}(d\hat{s}_{I}\wedge d\hat{s}'_{I}+d\hat{s}'^{*}_{I}\wedge d\hat{s}^{*}_{I}).
\end{equation}
The induced symplectic two-form on the
physical shell can be found by substituting the value of
$d\hat{s}'_{I}$ and $d\hat{s}'^{*}_{I}$ in terms of $d\hat{s}_{I}$ and
$d\hat{s}^{*}_{I}$ by using on-shell condition
$d\hat{\chi}_{I}=d\hat{\chi}^{*}_{I}=0$, hence we find
\begin{equation}
\hat{\omega}^{0}=2\displaystyle\sideset{}{'}\sum_{I,J}\hat{w}^{IJ}d\hat{\sigma}_{I}\wedge d\hat{\sigma}^{*}_{J},
\end{equation}
where
$\hat{w}^{IJ}=\frac{\delta\hat{\sigma}^{*}_{I}}{\delta\hat{\bar{\sigma}}_{J}}$,
consequently $\hat{\omega}^{0}$ is closed and $\hat{\omega}$ is positive
matrix. Therefore the induced physical low-energy phase space is a
K\"ahler manifold \footnote{A K\"ahler manifold is a Hermitian
manifold (M,g) whose K\"ahler form $\Omega$ is closed:
$d\Omega=0$. The metric $g$ is called the K\"ahler metric of $M$ \cite{geom}.}\cite{geom}. This defines a positive
hermitian metric in the physical shell \footnote{ Notice that it is well known
that the geometrical quantization can be applied on K\"ahler manifold
since it has a natural polarization.}.

In the following next two sections, we apply our RG formalism to
compute the effective Hamiltonian for $\phi^{4}$ and extended
Lee theory up to two- and one-loop order, respectively.

\section{Example I: $\Phi^{4}$ theory}
In this section, we obtain the effective Hamiltonian of $\phi^{4}$
theory up to two-loop order in equal-time quantization. In following we will quote from Ref.~\cite{me1}. The bare
$\phi^{4}$ theory Hamiltonian is
\cite{ph4}
\begin{equation}
H=\int d^{3}x\left(
\frac{1}{2}\pi^{2}(x)+\frac{1}{2}\phi(x)\big(-\nabla^{2}+m^{2}\big)\phi(x)+g\phi^{4}(x)\right).
\end{equation}
According to our logic the ultraviolet-finite Hamiltonian is
obtained by introducing counterterms, which depend on the UV
cutoff $\Lambda$ and some arbitrary renormalization scale. This
redefines the parameters of the theory and defines the effective
low-energy Hamiltonian. The renormalized Hamiltonian has the form
\begin{equation}
H=\int d^{3}x\left(
\frac{Z_{\pi}}{2}\pi^{2}(x)+\frac{1}{2}\sqrt{Z_{\phi}}\phi(x)\big(-\nabla^{2}+Z_{m}m^{2}\big)\sqrt{Z_{\phi}}\phi(x)+Z_{g}Z_{\phi}^{2}g
\phi^{4}(x)+...\right).
\end{equation}
Each of the $Z$-factors has an expansion of the form.
\begin{equation}
Z=1+f_{1}(\Lambda)\lambda +f_{2}(\Lambda)\lambda^{2}+\hdots,
\end{equation}
where $\lambda$ is a generic coupling constant of theory and has been
defined at a given renormalization scale $M$. The functions $f_{n}$
will be obtained order-by-order, by summing up contributions of
the fast modes between $\mu$ and $\Lambda$, in the sense that $Z(\Lambda)\to Z(\mu)$ and $f_{n}(\Lambda)\to
f_{n}(\mu)$. This means that the low-energy correlation functions are
invariants of the renormalization group flow. One can therefore assume
that the $Z$'s are initially $1$ and choose the corresponding $f$'s
from the condition that the cut-off dependence be cancelled out after
computing the effective Hamiltonian in the desired loop order.
Even though the newly generated interactions are sensitive to the
regularization scheme (as
 is well known \cite{local}, a sharp cutoff may lead to new
non-local interaction terms), nevertheless one can ignore these if
they are finite and do not produce any divergence as the cutoff $\Lambda$ approaches to infinity. We now
split field operators into high- and low-momentum modes;
$\phi(x)=\phi_{L}(x)+\phi_{H}(x)$, where $\phi_{L}(x)$ denotes
modes of low-frequency with momentum $k\leq \mu$ and $\phi_{H}(x)$
denotes modes of high-frequency with momentum constrained to a
shell $\mu<k\leq\Lambda$. The field $\phi_{L}(x)$ can be conceived
as a background to which the $\phi_{H}(x)$-modes are coupled.
Therefore, in the standard diagrammatic language, integrating out
the high-frequency modes $\phi_{H}(x)$ implies that only
high-frequency modes appear in internal lines. The field
$\phi_{H}(x)$ is represented in Fock space as
\begin{equation}
\phi_{H}(x)=\sum_{\mu<k\leq\Lambda}\frac{1}{\sqrt{2\omega_{k}}}(a_{k}e^{ikx}+a^{\dag}_{k}e^{-ikx}),\label{a4}
\end{equation}
where $\omega_{k}=\sqrt{k^{2}+m^{2}}$ and the operators $a_{k}$
and $a_{k}^{\dag}$ satisfy the standard boson commutation rules.
From now on all summations are implicitly over the high-frequency
modes $\mu<k\leq\Lambda$. The Hamiltonian in terms of high- and
low-frequency modes can be written as, after normal ordering with
respect to high-frequency modes,
\begin{equation}
 H=H_{1}+H_{2}+V_{B}+V_{C}+V_{A},
\end{equation}
where we define,
\begin{eqnarray}\label{a5}
 H_{1}&=&\int\left(
\frac{1}{2}\pi_{L}^{2}(x)+\frac{1}{2}\phi_{L}(x)\big(-\nabla^{2}+m^{2}\big)\phi_{L}(x)+g
\phi_{L}^{4}(x)\right),\nonumber\\
H_{2}&=&\sum\omega_{k}a^{\dag}_{k}a_{k},\nonumber\\
V_{B}&=&g\sum\int\frac{e^{i(p+q+r-k)x}}{\sqrt{\omega_{k}\omega_{p}\omega{q}\omega_{r}}}a^{\dag}_{k}a_{p}a_{q}a_{r}
+\frac{3e^{i(p+q-r-k)x}}{4\sqrt{\omega_{k}\omega_{p}\omega{q}\omega_{r}}}a^{\dag}_{k}a^{\dag}_{p}a_{q}a_{r}
\nonumber\\
&&+6\phi_{L}(x)\frac{e^{i(p+q-k)x}}{\sqrt{2\omega_{k}\omega_{p}\omega_{q}}}a^{\dag}_{k}a_{p}a_{q}+
3\big(\phi^{2}_{L}(x)+\frac{1}{2\omega_{r}}\big)\frac{e^{i(k-p)x}}{\sqrt{\omega_{k}\omega_{p}}}a^{\dag}_{p}a_{k}\nonumber\\
&&+\frac{3\phi_{L}^{2}(x)}{2\omega_{r}}+\text{h.c.},\nonumber\\
V_{C}&=&g\sum\int
V^{4}_{C}~a^{\dag}_{k}a^{\dag}_{p}a^{\dag}_{q}a^{\dag}_{r}+V^{3}_{C}
~a^{\dag}_{k}a^{\dag}_{p}a^{\dag}_{q}+V^{2}_{C}~a^{\dag}_{k}a^{\dag}_{p}+V^{1}_{C}~a^{\dag}_{k},\nonumber\\
V_{A}&=&V_{C}^{\dag},\nonumber\\
V^{1}_{C}&=&\Big(\frac{6\phi_{L}(x)}{\omega_{p}}+4\phi_{L}^{3}(x)\Big)\frac{e^{-ikx}}
{\sqrt{2\omega_{k}}},\hspace{2.5cm}
V^{2}_{C}=3\Big(\phi^{2}_{L}(x)+\frac{1}{2\omega_{r}}\Big)\frac{e^{-i(k+p)x}}{\sqrt{\omega_{k}\omega_{p}}},\nonumber\\
V^{3}_{C}&=&2\phi_{L}(x)\frac{e^{-i(k+p+q)x}}{\sqrt{2\omega_{k}\omega_{p}\omega_{k}}},\hspace{3.7cm}
V^{4}_{C}=\frac{e^{-i(k+p+q+r)x}}{4\sqrt{\omega_{k}\omega_{p}\omega_{q}\omega_{r}}}.\
\end{eqnarray}
The high-energy configurations in the Fock space are
specified by
\(\{C_{I}\to  \prod_{i=1} a_{k_{i}}\}\) and \(\{C^{\dag}_{I}\to
 \prod_{i=1}a_{k_{i}}^{\dag})\}\). Up to two-loop expansion, our
 renormalization scheme requires to keep $S(S')$ at least to order $n=4$,
 which allows us to eliminate the pure terms $V_{C}$ and $V_{A}$ at a lower level of expansion. The
 $\hat{S}(\hat{S}')$ operators consistent with a $SUB(4,m)$ truncation
 scheme are,
\begin{eqnarray}
\hat{S}_{m}&=&\int\sum\left(\hat{S}^{1}_{m}~a^{\dag}_{k}+\hat{S}^{2}_{m}~a^{\dag}_{k}a^{\dag}_{p}+ \hat{S}^{3}_{m}~a^{\dag}_{k}a^{\dag}_{p}a^{\dag}_{q}+\hat{S}^{4}_{m}~a^{\dag}_{k}a^{\dag}_{p}a^{\dag}_{q}a^{\dag}_{r}\right),\nonumber\\
\hat{S}'_{m}&=& \int\sum \left(\hat{S}'^{1}_{m}~a_{k}+\hat{S}'^{2}_{m}~a_{k}a_{p}+\hat{S}'^{3}_{m}~a_{k}a_{p}a_{q}+\hat{S}^{4}_{m}~a_{k}a_{p}a_{q}
a_{r}\right).\label{a6}\
\end{eqnarray}
 We split the diagonalization of the
Hamiltonian matrix in an upper and lower triangle part, by using the
double similarity transformation. One may notice that the ``most
non-diagonal'' terms in the Hamiltonian are $V_{C}$ and $V_{A}$ (in
the light-front Hamiltonian such terms do not exist because modes with
longitudinal momentum identically zero are not allowed). The potential
$V_{B}$ is already partially diagonalized and does not change the
vacuum of the high-energy states.  Therefore, here we employ a minimal
scheme, aiming at removal of $V_{A}$ and $V_{C}$ only.

We restrict ourselves to the elimination
of the high-energy degrees of freedom up to the first order in the
coupling constant $g$ and second order in the ratio of cutoffs
\(\mu/\Lambda\). Therefore, our truncation scheme is called
$SUB(4,2)$. For $m=0$ one finds,
\begin{eqnarray}
  S^{1}_{0}&=&-g\frac{V^{1}_{C}}{\omega_{k}}, \hspace{4.8cm}
  S^{2}_{0}=-g\frac{V^{2}_{C}}{\omega_{k}+\omega_{p}},\nonumber\\
  S^{3}_{0}&=&-g\frac{V^{3}_{C}}{\omega_{k}+\omega_{p}+
  \omega_{q}}, \hspace{3cm}
S^{4}_{0}=-g\frac{V^{4}_{C}}{\omega_{k}+\omega_{p}+
  \omega_{q}+\omega_{r}},\label{a7}\
\end{eqnarray}
where the $V^{1-4}_{C}$ are defined in Eq.~(\ref{a5}). Here, one
has $S'_{0}=S_{0}^{\dag}$. At this stage the results for
the one-loop renormalization can be computed. We evaluate the
effective Hamiltonian by substituting $S(S')$ from Eqs.~(\ref{a6})
and (\ref{a7}) into Eq.~(\ref{eq15}). In order to achieve
renormalization, one should identify the potentially divergent
terms ( when $ \Lambda\to \infty$) in the expansion of
$H^{\text{eff}}(\mu)$. Such a process generally can be done by
inventing a power-counting rule, using the property
$S_{n}\simeq\frac{\mu}{\Lambda }S_{n-1} $. Here we take
$\omega_{k}\simeq|k|$ for $\mu\gg m$ and replace \(\sum_{k}\) by
\(\int \frac{d^{3}k}{(2\pi)^{3}}\). The standard tadpole one-loop
mass renormalization arises from $V_{B}$ due to normal-ordering.
We add this divergent term to $H_{1}$ and renormalize the bare
mass
\begin{eqnarray}
\delta H^{\text{1-loop}}&=& \langle 0|V_{B}|0\rangle=6g\sum\int \frac{\phi^{2}(x)}{2\omega_{k}}=\frac{3g}{4\pi^{2}}(\Lambda^{2}-\mu^{2})\int d^{3}x \phi^{2}(x),\nonumber\\
Z_{m}&=& 1-\frac{3g}{2\pi^{2}}(\Lambda^{2}-\mu^{2}).\
\end{eqnarray}
In this order the contribution of the terms \([V_{C},S],
[V_{A},S']\) and \([H_{1},S(S')]\) are zero, after projection on
to the high-energy vacuum. The only divergent contributions come
from \([V_{A}^{2(3)},S^{2(3)}_{0}]\) due to a double and third
contraction of the high-frequency fields respectively. There are
two other divergent terms, \(([V_{C}^{2(3)},S'^{2(3)}_{0}]\),
however they are harmless and are cancelled out by the divergence
of \([[H_{2},S_{0}],S'^{2(3)}_{0}]\). One thus obtains,
\begin{eqnarray}
\delta H&=& -\frac{18g^{2}}{(2\pi)^{6}}\int\frac{\phi^{2}(x)\phi^{2}(y)}{\omega_{k}\omega_{p}(\omega_{k}+\omega_{p})}e^{i(k+p)(x-y)}\nonumber\\
&-&\frac{12g^{2}}{(2\pi)^{9}}\int\frac{\phi(x)\phi(y)}{\omega_{k}\omega_{p}\omega_{q}(\omega_{k}+\omega_{p}+\omega_{q})}e^{i(k+p+q)(x-y)}.\label{integral}\
\end{eqnarray}
In general evaluation of integrals like Eq.~(\ref{integral}) may
produce non-localities. This is due to the fact that the total
momentum in integrands of Eq.~(\ref{integral}), namely $r_{1}=p+q$
and $r_{2}=k+p+q $ are in the low-momentum space. To evaluate such
integrations, one can firstly reduce the potential divergent
integrals by a change of variable, for example for the first
integrand we use $p,q\to p,r_{1}$, and then expand the integrand
in $r_{1}/p$. Therefore, after expansion and evaluating the
momentum integrals, one may be faced with non-analytic terms in
the low-momentum space. However here these are non-divergent and will
thus be ignored. We find
\begin{eqnarray}
\delta H^{\text{1-loop}}&=&-\frac{9g^{2}}{2\pi^{2}}\ln\left(\frac{\Lambda}{\mu}\right)\int d^{3}x \phi^{4}(x)-\frac{3g^{2}}{2\pi^{4}}(2\ln 2-1)\Lambda^{2}\int d^{3}x \phi^{2}(x)\nonumber\\
&+&\frac{3g^{2}}{16\pi^{4}}\ln\left(\frac{\Lambda}{\mu}\right)\int d^{3}x(\nabla\phi(x))^{2}+\text{finite terms}.
\
\end{eqnarray}
One can immediately deduce the renormalization factors $Z_{g}$ and $Z_{\phi}$ from above expression
\begin{eqnarray}
Z_{g}&=&1+\frac{9g^{2}}{2\pi^{2}}\ln\left(\frac{\Lambda}{\mu}\right),\\
Z_{\phi}&=&1-\frac{3g^{2}}{8\pi^{4}}\ln\left(\frac{\Lambda}{\mu}\right).\label{z}\
\end{eqnarray}
The unknown coefficients in expression $S_{1}$  is computed by
making use of Eq.~(\ref{a7}) and solving coupled equations
(\ref{eq17}), therefore one may yield,
\begin{eqnarray}
S^{1}_{1}&=&\frac{6g
e^{-ikx}}{\omega_{k}^{2}\sqrt{2\omega_{k}}}\Big(
2\phi_{L}(x)-2i\pi_{L}(x)\phi_{L}^{2}(x)-\frac{i\pi_{L}(x)}{\omega_{p}}\Big)-\frac{g}{\omega_{k}}\sum_{\nu=1}^{3}\frac{1}{\nu!}V_{A}^{\nu}S_{1}^{\nu+1},\nonumber\\
S^{2}_{1}&=&\frac{3g
e^{-i(k+p)x}}{(\omega_{k}+\omega_{p})^{2}\sqrt{\omega_{k}\omega_{p}}}\Big(1-i2\pi_{L}(x)
\phi_{L}(x)\Big)-\frac{g}{\omega_{k}+\omega_{p}}\Big([V^{1}_{C},S^{1}_{1}]+ \sum_{\nu=1}^{2}\frac{1}{\nu!}V_{A}^{\nu}S_{1}^{\nu+2}\Big)\nonumber\\
S^{3}_{1}&=&-\frac{2ig
e^{-i(k+p+q)x}}{(\omega_{k}+\omega_{p}+
\omega_{q})^{2}\sqrt{2\omega_{k}\omega_{p}\omega_{q}}}\pi_{L}(x)-\frac{g}
{(\omega_{k}+\omega_{p}+\omega_{q})}\Big(V^{1}_{A}S^{4}_{1}
+\sum_{\nu=1}^{2}[V^{\nu}_{C},S^{3-\nu}_{1}]\Big),\nonumber\\
S^{4}_{1}&=&-\frac{g}{(\omega_{k}+\omega_{p}+\omega_{q}+\omega_{r})}\sum_{\nu=1}^{3}[V^{4-\nu}_{C},S^{\nu}_{1}].\label{a8}\
\end{eqnarray}
In the above expression summation over dummy momentum indices is
assumed. One can find $\hat{S}'_{1}$ in the same manner by
exploiting  Eq.~(\ref{eq19}) and using Eq.~(\ref{a8}) as an input,
which leads to
\begin{equation}
S'^{\nu}_{1}=(S^{\nu}_{1})^{\dag}+S'^{\nu a}_{1} \hspace{2cm} \nu=1,...,4,\label{s'}
\end{equation}
with the notations,
\begin{eqnarray}
S'^{1a}_{1}&=&\frac{g}{\omega_{k}}\Big(\sum_{\nu=1}^{3}\frac{1}{\nu!}S'^{(\nu+1)a}_{1}V^{\nu}_{C}-\sum_{\nu=1}^{3}\frac{1}{\nu!}V^{\nu+1}_{A}S^{\nu}_{1}\Big),\nonumber\\
S'^{2a}_{1}
&=&\frac{g}{\omega_{k}+\omega_{p}}\Big(\sum_{\nu=1}^{2}\frac{1}{\nu!}S'^{(\nu+2)a}_{1}V^{\nu}_{C}(q)-\sum_{\nu=1}^{2}\frac{1}{\nu!}V^{\nu+2}_{A}S^{\nu}_{1}+[V^{1}_{A},S'^{1a}_{1}]\Big),\nonumber\\
S'^{3a}_{1}&=&\frac{g}{\omega_{k}+\omega_{p}+\omega_{q}}\Big(S'^{4a}_{1}V^{1}_{C}-V^{4}_{A}S^{1}_{1}+\sum_{\nu=1}^{2}
[V^{\nu}_{A},S'^{(3-\nu)a}_{1}]\Big),\nonumber\\
S'^{4a}_{1}&=&\frac{g}{(\omega_{k}+\omega_{p}+\omega_{q}+\omega_{r})}\sum_{\nu=1}^{3}[V^{\nu}_{A},S'^{(4-\nu)a}_{1}].\label{a9}\
\end{eqnarray}
The only divergent contribution up to order $g^{2}$ arises from,
\begin{equation}
\delta H=-\langle 0|[H_{1},S_{1}],S'_{0}]|0\rangle,
\end{equation}
After the evaluation of the leading divergent part, we find that
\begin{equation}
\delta H=-\frac{3g^{2}}{16\pi^{4}}\ln\left(\frac{\Lambda}{\mu}\right)\int d^{3}x~ \pi^{2}(x),\label{pi}
\end{equation}
which contributes to the two-loop wave-function renormalization
$Z_{\pi}$. By comparing Eqs.~(\ref{z}) and (\ref{pi}), one may
conclude that $Z_{\pi}=Z_{\phi}^{-1}$, as it should be. To finish
the renormalization up to two-loop order, one should also take
into account the contribution at order $g^{3}$. The divergent
terms at this level originate from
\begin{equation}
\delta H=-\langle 0|\Big[[\big(V_{A}+1/2V_{C}+V_{B}\big),S_{0}],S'_{0}\Big]|0\rangle.
\end{equation}
 After a straightforward but lengthy computation one can obtain the leading divergent parts,
\begin{equation}
\delta H=\frac{27g^{3}}{2\pi^{4}}\Big[\big[\ln\left(\frac{\Lambda}{\mu}\right)\big]^{2}+\ln\left(\frac{\Lambda}{\mu}\right)\Big]\int d^{3}x  \phi^{4}(x),
\end{equation}
this term should be added to Eq.~(\ref{integral}), therefore one
can immediately deduce the correct total renormalization factor
$Z_{g}$ up to two-loop order,
\begin{equation}
Z_{g}=1+\frac{9g^{2}}{2\pi^{2}}\ln\left(\frac{\Lambda}{\mu}\right)+\frac{g^{3}}{4\pi^{4}}\left(81\left(\ln\left(\frac{\Lambda}{\mu}\right)\right)^{2}-51\ln\left(\frac{\Lambda}{\mu}\right)\right). \label{f1}
\end{equation}
One can now immediately obtain the well-known \cite{ph4} two-loop
$\beta$-function and anomalous dimension by making use of
Eqs.~(\ref{z}, \ref{f1}).
\begin{eqnarray}
&&\beta(g)=\frac{\partial g}{\partial \log \mu}|_{\Lambda}=\frac{9}{2\pi^{2}}g^{2}-\frac{51}{4\pi^{4}}g^{3},\\
&&\gamma(g)=\frac{1}{2}\frac{\partial \log Z_{\phi}}{\partial \log \mu}|_{\Lambda}=\frac{3}{16\pi^{4}}g^{2}.\
\end{eqnarray}
It is important to point out that the diagonalization at first
order in the coupling constant defines a correct low-energy
effective Hamiltonian which is valid up to order $g^{3}$. Having
said that, from Eq.~(\ref{s'}) one can observe that the
non-hermiticity of the $\hat{S}$ operator appears at order $g^{2}$
and in a lower order of $\mu/\Lambda$. As we have shown,
non-hermiticity is negligible up to two-loop order (asymmetric
terms appear in irrelevant contributions (which are non-divergent and vanish as $\Lambda$ goes to infinity). We conjecture that,
for the present model, non-hermitian terms only appear in irrelevant
contributions, whatever the order of truncation.

\section{Example II: Extended Lee Model}
As another illustrative example, we will now apply coupled-cluster RG to determine
the effective Hamiltonian for an extended Lee model (ELM) up to the one-loop order.

We define four kinds of particles, the $V$-particle and $N$-particle as two different fermions and the $\theta$ and $\bar{\theta}$ as a scalar boson and anti-boson respectively. Here $a(k)$, $a^{\dag}(k)$ and $b(k)$, $b^{\dag}(k)$ are the annihilation and creation operators which satisfy boson commutator rules. The $V(p)$, $V^{\dag}(p)$ and $N(p)$, $N^{\dag}(p)$ define the fermion sector and obey the usual anticommutator rules. The bare ELM Hamiltonian then reads
\begin{eqnarray}
H&=&H_{0}+H_{I},\label{eq25}\nonumber\\
H_{0}&=&\int{d^{3}p
~\omega_{V}(p)V^{\dag}(p)V(p)}+\int{d^{3}p
~\omega_{N}(p)N^{\dag}(p)N(p)}\nonumber\\ &+&\int{d^{3}k
~\omega_{\theta}(k)a^{\dag}(k)a(k)}+\int{d^{3}k
~\omega_{\bar{\theta}}(k)b^{\dag}(k)b(k)},\nonumber\\
H_{I}&=&\lambda_{1}(2\pi)^{-3/2}\int{\frac{d^{3}kd^{3}p}{(2\omega_{\theta}(k))^{1/2}}V^{\dag}(p)N(p-k)a(k)}\nonumber\\
&+&\lambda_{2}(2\pi)^{-3/2}\int{\frac{d^{3}kd^{3}p}{(2\omega_{\bar{\theta}}(k))^{1/2}}N^{\dag}(p)V(p-k)b(k)}+\text{h.c.}.\
\end{eqnarray}
 The kinetic energy generically is defined
 \(\omega_{O}(k)=\sqrt{k^{2}+m^{2}_{O}}\) where the indice $O$ can be either
 $(V,N,\theta,\bar{\theta})$. The interaction term in $H_{I}$
 describes the processes;
\begin{eqnarray}
&& V\rightleftarrows N+\theta,\label{eq26}\\
&& N\rightleftarrows V+\bar{\theta}. \label{eq27}\
\end{eqnarray}
The crossing symmetry become manifest if we take
$\lambda_{1}=\lambda_{2}$ and equal masses for boson and anti-boson.
For sake of generality we will ignore crossing symmetry at the
moment. The Lee model \cite{19} can be recovered if we decouple the  anti-boson
$\bar{\theta}$, $\lambda_{2}\to 0$. In the Lee model the virtual
process Eq.~(\ref{eq27}) is not included and thus the $N$-particle state
remains unrenormalized and the model become exactly solvable.

It is believed that the Lee model is asymptotically free for
space-time dimension $D$ less than four \cite{22}. With on-shell
renormalization one can show that the Lee model for $D>4$ (odd $D$) is
ultraviolet stable and not asymptotically free \cite{23}. It is well
known that such a model in four dimension exhibit a ghost state as the
cutoff is removed. The Hamiltonian Eq.~(\ref{eq25}) exhibits two
symmetries; it is straightforward to verify that following operators
commute with $H$
\begin{eqnarray}
B&=&\int{d^{3}p~V^{\dag}(p)V(p)}+\int{d^{3}p~N^{\dag}(p)N(p)},\label{sym}\nonumber\\
Q&=&\int{d^{3}p~N^{\dag}(p)N(p)}+\int{d^{3}k~b^{\dag}(p)b(p)}-\int{d^{3}k~a^{\dag}(p)a(p)}.\
\end{eqnarray}
Clearly $B$ is a baryon number operator and $Q$ is a charge
operator. We assign the charges $1,0,-1$ and $1$ to the
$N,V,\theta$ and $\bar{\theta}$, respectively. The sectors of the ELM
are labeled by the eigenvalue $(b,q)$ of the operators $(B,Q)$. According
to our formulation the ultraviolet-finite Hamiltonian is obtained
by introducing $Z$-factors, which depend on the UV cutoff $\Lambda$
and some arbitrary renormalization scale $M$ in such way that
effective Hamiltonian does not depend on $\Lambda$. The bare
Hamiltonian can be rewritten
\begin{eqnarray}
H&=&\int{d^{3}p~Z^{2}_{V}Z_{M_{V}}\omega_{V}(p)V^{\dag}(p)V(p)}+\int{d^{3}p~Z^{2}_{N}Z_{M_{N}}\omega_{N}(p)N^{\dag}(p)N(p)}\nonumber\\
&+&\int{d^{3}k~Z^{2}_{\theta}Z_{M_{\theta}}\omega_{\theta}(k)a^{\dag}(k)a(k)}+\int{d^{3}k~Z^{2}_{\bar{\theta}}Z_{M_{\bar{\theta}}}\omega_{\bar{\theta}}(k)b^{\dag}(k)b(k)}\nonumber\\
&+&\int{\frac{\lambda_{1}d^{3}kd^{3}p}{(2(2\pi)^{3}\omega_{\theta}(k))^{1/2}}}Z_{\lambda_{1}}Z_{V}Z_{N}Z_{\theta}V^{\dag}(p)N(p-k)a(k)\nonumber\\
&+&\int{\frac{\lambda_{2}d^{3}kd^{3}p}{(2(2\pi)^{3}\omega_{\bar{\theta}}(k))^{1/2}}}Z_{\lambda_{2}}Z_{V}Z_{N}Z_{\bar{\theta}} N^{\dag}(p)V(p-k)b(k)\label{eq28}\nonumber\\
&+&\text{h.c.}~.\
\end{eqnarray}
 We split the original Hamiltonian in the form of Eq.~(\ref{eq16});
\begin{eqnarray}
H_{1}&=&H(\int^{\mu}_{0}),\nonumber\\
H_{2}&=&H_{0}(\int^{\Lambda}_{\mu}),\nonumber\\
V_{C}&=&\int^{\mu}_{0}\int^{\Lambda}_{\mu}\frac{d^{3}p'd^{3}k}{(2(2\pi)^{3}\omega_{\theta}(k))^{1/2}}\lambda_{1} N^{\dag}(p'-k)V(p')a^{\dag}(k)\nonumber\\
&+&\int^{\mu}_{0}\int^{\Lambda}_{\mu}\frac{d^{3}p'd^{3}k}{(2(2\pi)^{3}\omega_{\bar{\theta}}(k))^{1/2}}\lambda_{2}V^{\dag}(p'-k)N(p')b^{\dag}(k), \nonumber\\
V_{A}&=&V^{\dag}_{C},\nonumber\\
V_{B}&=&\int^{\mu}_{0}\int^{\Lambda}_{\mu}\frac{d^{3}pd^{3}k'}{(2(2\pi)^{3}\omega_{\theta}(k'))^{1/2}}\lambda_{1}V^{\dag}(p)N(p-k')a(k')\nonumber\\
&+&\int^{\mu}_{0}\int^{\Lambda}_{\mu}\frac{d^{3}pd^{3}k'}{(2(2\pi)^{3}\omega_{\bar{\theta}}(k'))^{1/2}}\lambda_{2}N^{\dag}(p)V(p-k')b(k')\nonumber\\
&+&\int^{\Lambda}_{\mu}\int^{\Lambda}_{\mu}\frac{d^{3}pd^{3}k}{(2(2\pi)^{3}\omega_{{\theta}}(k))^{1/2}}\lambda_{1}V^{\dag}(p)N(p-k)a(k)\nonumber\\
&+&\int^{\Lambda}_{\mu}\int^{\Lambda}_{\mu}\frac{d^{3}pd^{3}k}{(2(2\pi)^{3}\omega_{\bar{\theta}}(k))^{1/2}}\lambda_{2}N^{\dag}(p)V(p-k)b(k)+\text{h.c.}~ .\
\end{eqnarray}
Here $p'$ and $k'$ stand for low momenta ($p',k'<\mu$). If the
arguments of an operator are all low momenta ($p'$ or $k'$), this
indicates low momentum operators. The arguments in $H(\int^{\mu}_{0})$
and $H_{0}(\int^{\Lambda}_{\mu})$ means that all the momentum
integrations involved in Eq.~(\ref{eq25}) are running between
$0<p'<\mu$ for the former and $\mu<p<\Lambda$ for the latter,
respectively. The configuration space of the high momentum operators
are specified by \(\{C_{I}\to V^{n_{1}}N^{n_{2}}a^{n_{3}}b^{n_{4}},
C^{\dag}_{I}\to{C^{\dag}_{I}\to
(V^{\dag})^{n_{1}}(N^{\dag})^{n_{2}}(a^{\dag})^{n_{3}}(b^{\dag})^{n_{4}}}\}\)
with $ n_{1}+n_{2}+n_{3}+n_{4}=I $. Aiming at a one-loop expansion the
corresponding $S$ and $S'$ operators which preserve the symmetry
property Eq.~(\ref{sym}), can be chosen as
\begin{eqnarray}
S_{m}&=&\int{d^{3}p'd^{3}k~S^{1}_{m}(p')V^{\dag}(p'-k)b^{\dag}(k)}+\int{d^{3}p'd^{3}k~S^{2}_{m}(p')N^{\dag}(p'-k)a^{\dag}(k)},\nonumber\\
S^{1}_{m}&=&S^{N}_{m}N(p')+S^{Vb}_{m}V(p'-k')b(k')+S^{Va}_{m}V(p'+k')a^{\dag}(k'),\nonumber\\
S^{2}_{m}&=&S^{V}_{m}V(p')+S^{Na}_{m}N(p'-k')a(k')+S^{Nb}_{m}N(p'+k')b^{\dag}(k').\label{eq29}\
\end{eqnarray}
We have ignored the $I=1$ configuration, since there are no tadpole type
diagrams.( The truncation of $S_{I}$ in configuration space should be
consistent with our loop expansion.) Here we confine our
attention to the elimination of the high-momentum degrees of freedom up to the
first order in coupling constant and second order in $\mu/\Lambda$. The
unknown coefficients in Eq. (\ref{eq29}) can be obtained by making use of
Eq. (\ref{eq21}),

\begin{eqnarray}
S^{V}_{0}&=&\frac{
\lambda_{1}}{(2(2\pi)^{3}\omega_{\theta}(k))^{1/2}(\omega_{N}(p'-k)+\omega_{\theta}(k))},\nonumber\\
S^{N}_{0}&=&\frac{\lambda_{2}}{(2(2\pi)^{3}\omega_{\bar{\theta}}(k))^{1/2}(\omega_{V}(p'-k)+\omega_{\bar{\theta}}(k))},\nonumber\\
S^{Vb}_{0}&=&S^{Va}_{0}=0,\nonumber\\
S^{V}_{1}&=&\frac{\lambda_{1}\omega_{V}(p')}{(2(2\pi)^{3}\omega_{\theta}(k))^{1/2}(\omega_{N}(p'-k)+\omega_{\theta}(k))^{2}},\nonumber\\
S^{N}_{1}&=&\frac{\lambda_{2}\omega_{N}(p')}{(2(2\pi)^{3}\omega_{\bar{\theta}}(k))^{1/2}(\omega_{V}(p'-k)+\omega_{\bar{\theta}}(k))^{2}},\nonumber\\
S^{Va}_{1}&=&\frac{\lambda_{1}\lambda_{2}}{2(2\pi)^{3}(\omega_{\bar{\theta}}(k)\omega_{\theta}(k'))^{1/2}(\omega_{V}(p'-k)+\omega_{\bar{\theta}}(k))^{2}},
\nonumber\\
S^{Nb}_{1}&=&\frac{\lambda_{1}\lambda_{2}}{2(2\pi)^{3}(\omega_{\bar{\theta}}(k')\omega_{\theta}(k))^{1/2}(\omega_{N}(p'-k)+\omega_{\theta}(k))^{2}},\nonumber\\
S^{Na}_{1}&=&\frac{\lambda_{1}^{2}}{2(2\pi)^{3}(\omega_{\theta}(k')\omega_{\theta}(k))^{1/2}(\omega_{N}(p'-k)+\omega_{\theta}(k))^{2}},\nonumber\\
S^{Vb}_{1}&=&\frac{\lambda_{2}^{2}}{2(2\pi)^{3}(\omega_{\bar{\theta}}(k')\omega_{\bar{\theta}}(k))^{1/2}(\omega_{V}(p'-k)+\omega_{\bar{\theta}}(k))^{2}}.\
\end{eqnarray}
It is easy to observe that Eq.~(\ref{eq23}) will be satisfied if we
require $ S'_{m}=S^{\dag}_{m}$, since, up to
the first in $\lambda$ the similarity transformation introduced in
Eq.~(\ref{eq15}) remains unitary. Equally one could use Eq.~(\ref{eq24}) to
obtain $S'$, it is obtained that $S'_{0}=S_{0}^{\dag}$ and $S'_{1}=0$,
we will show that the renormalization feature of our model up to this
order will remain unchanged, however the effective low-energy
Hamiltonian will be different. As was already pointed out, this is
because Eq.~(\ref{eq24}) requires a different truncation scheme. The
effective Hamiltonian is now produced by plugging the $S$ and $S'$
defined in Eq.~(\ref{eq29}) into Eq.~(\ref{eq15}). With naive
power-counting one can identify the potentially divergent terms.  At
the lower order of expansion, the divergent term is $\langle
0|[V_{A},S_{0}]|0\rangle $, the divergence in this term arises from a
double contraction of high-energy fields.

At this step the
contributions of the terms $[V_{B},S_{0}(S'_{0})]$ and
$[H_{1},S_{0}(S'_{0})]$ are zero, after projection on to the
high-frequency vacuum. There is one other divergent term, $\langle
0|[V_{C},S'_{0}]|0\rangle $, but this is harmless and will be cancelled
out by $\langle 0|[\big[H_{2},S_{0}],S'_{0}\big]|0\rangle$. We thus
obtain
\begin{eqnarray}
\delta H(\lambda)&=&-\frac{\lambda_{1}^{2}}{2(2\pi)^{3}}\int_{\mu}^{\Lambda} {\frac{d^{3}k}{\omega_{\theta}(k)(\omega_{N}(p'-k)+\omega_{\theta}(k))}}\big[\int d^{3}p' N^{\dag}(p')N(p')\big]\nonumber\\
&-&\frac{\lambda_{2}^{2}}{2(2\pi)^{3}}\int_{\mu}^{\Lambda}{ \frac{d^{3}k}{\omega_{\bar{\theta}}(k)(\omega_{V}(p'-k)+\omega_{\bar{\theta}}(k))}}\big[\int d^{3}p' V^{\dag}(p')V(p')\big].\
\end{eqnarray}
From this expression one can immediately deduce the renormalization
factors $Z_{m_{V}}$ and $Z_{m_{N}}$, we take $\omega_{O}\simeq |k|$
for $\mu \gg m_{O}$, therefore
\begin{eqnarray}
Z_{M_{V}}&=& 1+\frac{\lambda_{1}^{2}}{8\pi^{2}}(\Lambda-\mu),\nonumber\\
Z_{M_{N}}&=& 1+\frac{\lambda_{2}^{2}}{8\pi^{2}}(\Lambda-\mu).\
\end{eqnarray}
There is no mass renormalization for $\theta$ and $\bar{\theta}$ and
accordingly there are no vacuum polarization type diagrams. Thus $\theta
$ and $\bar{\theta}$ remain unrenormalized,
$Z_{\theta}=Z_{m_{\theta}}=Z_{\bar{\theta}}=Z_{m_{\bar{\theta}}}=1
$. The other contribution of $H^{\text{eff}}$ at one-loop which are
not zero after projecting on to vacuum come from
\begin{equation}
\delta H(\lambda)=-\langle 0|\big[[H_{1},S_{1}],S'_{0}\big]+\langle 0|\big[[H_{1},S_{1}],S'_{1}\big]|0\rangle.
\end{equation}
The divergent contribution emerges from the first terms, the leading
divergence of this expression is logarithmic which means that we can
neglect the difference between $p$ and $k-p'$ (for the divergent
contribution only). After evaluating a momentum integral we finally
get,
\begin{eqnarray}
\delta H&=&-\frac{\lambda_{1}^{2}}{16\pi^{2}}\ln \left[\frac{\Lambda}{\mu}\right]\int \omega_{V}(p')V^{\dag}(p')V(p')-\frac{\lambda_{2}^{2}}{16\pi^{2}}\ln\left[\frac{\Lambda}{\mu}\right]\int \omega_{N}(p')N^{\dag}(p')N(p')\nonumber\\
&-&\frac{\lambda_{1}^{2}+\lambda_{2}^{2}}{32\pi^{2}}\ln\left[\frac{\Lambda}{\mu}\right]\int \frac{\lambda_{1}}{((2\pi)^{3}\omega_{\theta}(k'))}V^{\dag}(p')N(p'-k')a(k')\nonumber\\
&-&\frac{\lambda_{1}^{2}+\lambda_{2}^{2}}{32\pi^{2}}\ln\left[\frac{\Lambda}{\mu}\right]\int \frac{\lambda_{2}}{((2\pi)^{3}\omega_{\bar{\theta}}(k'))}N^{\dag}(p')V(p'-k')b(k').\
\end{eqnarray}
From this expression we deduce the renormalization factor $Z_{\lambda_{1}},Z_{\lambda_{2}},Z_{V} $ and $Z_{N}$:
\begin{eqnarray}
&&Z_{V}^{2}=1+\frac{\lambda_{1}^{2}}{16\pi^{2}}\ln\frac{\Lambda}{\mu},\nonumber\\
&&Z_{N}^{2}=1+\frac{\lambda_{2}^{2}}{16\pi^{2}}\ln\frac{\Lambda}{\mu},\nonumber\\
&&Z_{\lambda_{1}}=Z_{\lambda_{2}}=1+\frac{\lambda_{1}^{2}+\lambda_{2}^{2}}{32\pi^{2}}\ln\frac{\Lambda}{\mu}.\
\end{eqnarray}
It is obvious from the equations above that one can define renormalized coupling constants in terms of the bare couplings and wave function renormalization
\begin{equation}
\lambda_{i}=\lambda_{i}^{0}/Z_{V}Z_{N}, \hspace{2cm}i=1,2.
\end{equation}
This definition corresponds for $\lambda_{1}=\lambda_{2}$ with the renormalization introduced to compute the $T$-matrix for the $N-\theta$ interaction \cite{19,24}. The one-loop $\beta$-function and anomalous dimension $\gamma$ are
\begin{eqnarray}
&&\beta_{\lambda_{i}}(\lambda_{1},\lambda_{2})=\frac{\partial \lambda_{i}}{\partial \ln M}|_{\Lambda}=\frac{\lambda_{i}}{32\pi^{2}}(\lambda_{1}^{2}+\lambda_{2}^{2}),\hspace{.5cm} i=1,2.\nonumber\\
&&\gamma_{V}=1/2\frac{\partial \ln Z_{V}}{\partial \ln M}|_{\Lambda}=-\frac{\lambda_{1}^{2}}{32\pi^{2}},\nonumber\\
&&\gamma_{N}=1/2\frac{\partial \ln Z_{N}}{\partial \ln M}|_{\Lambda}=-\frac{\lambda_{2}^{2}}{32\pi^{2}}.\label{g}\
\end{eqnarray}
Since the fixed points of the theory are the zero solutions of the
$\beta$-function, one immediately identifies the trivial solution
$\lambda_{1}=\lambda_{2}=0$ (we ignore the nonphysical imaginary
solution). It is now obvious from Eq.~(\ref{g}) that
$\gamma_{V}=\gamma_{N}=0$ only at trivial solutions of the
$\beta$-functions. This result is in correspondence with property that
for real field theories the $\gamma$-function is not zero unless at
trivial fixed point of the theory \cite{c}. Now to investigate the
behaviour of the theory at high momentum, we must compute the
momentum-dependent effective coupling constant $\lambda_{1}(k)$ and $
\lambda_{2}(k)$ by
\begin{eqnarray}
\left\{\begin{array}{ll}
& k\frac{d\lambda_{1}(k)}{dk}=\beta_{\lambda_{1}}(\lambda_{1}(k)\lambda_{2}(k)) ,\hspace{.51cm} \lambda_{1}(k)|_{k=1}=\lambda^{ph}_{1}\\
&\hspace{8cm},\\
&k\frac{d\lambda_{2}(k)}{dk}=\beta_{\lambda_{2}}(\lambda_{1}(k),\lambda_{2}(k)) ,\hspace{.51cm} \lambda_{2}(k)|_{k=1}=\lambda^{ph}_{2}\label{cou}\\
\end{array}
\right.
\end{eqnarray}
 where $\lambda_{1}^{ph}$ and $\lambda_{2}^{ph}$ are dimensionless
 physical renormalized coupling constants defined at the renormalization
 scale $k=1$. The coupled equations (\ref{cou}) can be solved by going to
 polar coordinates $r^{2}(k)=\lambda_{1}^{2}(k)+\lambda_{2}^{2}(k)$ and
 $ \theta(k)=\tan^{-1}\frac{\lambda_{1}(k)}{\lambda_{2}(k)}$,
\begin{eqnarray}
\lambda_{1}(k)&=&\frac{\bar{r}}{\sqrt{1-(16\pi^{2})^{-1}\bar{r}^{2}\ln k}}
\sin {\bar{\theta}},\nonumber\\
\lambda_{2}(k)&=&\frac{\bar{r}}{\sqrt{1-(16\pi^{2})^{-1}\bar{r}^{2}\ln k}}
\cos{ \bar{\theta}}, \label{so}\
\end{eqnarray}
where $\bar{r}$ and $\bar{\theta}$ denote the value at the renormalization
scale. The behaviour of the ELM in the deep-Euclidean region is
obtained by allowing $ k\rightarrow \infty$. From Eq.~(\ref{so}) one
observes that $\lambda_{1}(k)$ and $\lambda_{2}(k)$ in this region are
imaginary. This means that the effective Hamiltonian is non-hermitian
and the theory generates ghost states when the cut-off is removed. The
ghost state appears as a pole in $ V$ and $N$-propagators. Since a
theory is said to exhibit asymptotic freedom if (i)
$\frac{d\beta}{d\lambda}|_{\lambda(\infty)}<0$ (ultraviolet stability
at the fixed point $\lambda(\infty)$ and (ii)
$\lambda(\infty)=\lim_{k\rightarrow \infty}\lambda(k)=0$,
Eqs.~(\ref{so}) indicate that the ELM can not exhibit asymptotic
freedom at $D=4$.

\section{Conclusion}

In this chapter we have reviewed the merits and shortcomings of
several approaches for the construction of the effective field
theories in the Hamiltonian framework. 

We have outlined a new strategy to derive effective renormalized
operators. The formulation is not restricted to any quantization
scheme (e.g., equal time or light cone). The effective low-frequency
operator is obtained by the condition that it should exhibit
decoupling between the low- and high-frequency degrees of freedom. All
other irrelevant degrees of freedom like many-body states can be
systematically eliminated in the same way. We have shown that the
similarity transformation approach to renormalization can be
systematically classified. The non-hermitian formulation gives a very
simple description of decoupling, leading to a partial diagonalization
of the high-energy part.

The techniques proposed are known from the coupled cluster many-body
theory. We fully utilized Wilsonian Exact renormalization
group within the CCM formalism. Our approach invoke neither perturbation nor unitarity
transformation. It can be conceived as a topological
expansion in number of correlated excitation of the high-energy
modes. We showed that our formalism can be solved perturbatively. In
this way, it was revealed that diagonalization at first order in
coupling constant defines a correct low-energy effective Hamiltonian
which is valid up to the order $\lambda^{3}$. 

We showed that non-unitarity representation inherent in our
formulation is in favour of economic computation and does not produce
any non-hermiticity in the relevant terms. One can show that the
non-hermiticity of the effective Hamiltonian is controllable and might
appear in higher order which is beyond our approximation or in
irrelevant terms which can be ignored in renormalization group
sense. We argued that our formulation is free of any small-energy
denominator plaguing old-fashion perturbation theory. We showed that the
non-hermiticity of the coupled-cluster parametrization leads to the
compatibility of the formulation with the Hellmann-Feynman theorem and
it also induces a symplectic structure. More importantly, it provides
a simple framework for the renormalization of an arbitrary
operators. Notice that all these features are connected with each
other and one can not give up any of them without spoiling the
others. One may conceive that the non-hermiticity adds a auxiliary
(non-physical) sector to the physical phase space, thereby, it makes
the whole phase space geometrically meaningful and moreover it gives
enough room to keep the formulation to be conformed with the cluster
decomposition property and Poincar\'e invariance regardless of a
regularization method. There is a long tradition behind such
approaches, of course with different motivation, e.g., in the BRST
formulation, the phase space is enlarged by anti-commuting canonical
coordinates, another example is the bosonization of spin algebraic or
fermionic system, where one maps the original Hilbert space of the
system into a boson Hilbert space $\mathcal{H}^{B}$ which turns out to
be larger than original Hilbert space, in the sense that physically
realizable states in the original space map into a subspace of
$\mathcal{H}^{B}$. Interestingly, in this approach as well, the boson
Hamiltonian can be either Hermitian or non-Hermitian \cite{11}.

Notice that our RG method is non-perturbative although we have already
shown that perturbation expansion in coupling can be easily
implemented \cite{me,me1}. We successfully applied our RG formalism to
compute perturbatively the effective Hamiltonian for $\phi^{4}$ and
extended Lee theory up to two- and one-loop order, respectively. We
have employed a sharp cutoff, however this idealization should be
removed since generally it may lead to pathologies in renormalization,
since it induces non-locality and moreover potentially violates the
gauge symmetry.

One of the key features which has not yet been exploited is the
non-perturbative aspect of the method; it may well be able to obtain
effective degrees of freedom that are very different from the ones
occurs at the high-energy scale. This is a promising avenue for future
work. Another interesting question is that the connection between
our non-perturbative truncation scheme $SUB(\mathcal{N},\mathcal{M})$
and other non-perturbative scheme e. g., the large $N_{c}$
truncation. A systematic scheme which relates the large $N_{c}$ limit
with an approximate RG equation remains yet to be discovered.

\chapter{Renormalization problem in many-body system}
\section{Introduction} 
In this chapter we shall concentrate on the
renormalization problem in many-body theory of nuclear matter. As we
already pointed out, renormalization of many-body system in a
truncated (in number of particles) space is problematic, the so-called
Tamm-Dancoff problem. The question is how can one renormalize the
many-body problem equations obtained by non-perturbative approaches
(such as the CCM, Brueckner's reaction matrix (or $G$-matrix) theory
\cite{b-ex} and etc.,) in a truncated Fock space.

In the last chapter we showed that a renormalized effective interaction in small number of particles can be obtained
by imposing certain decoupling conditions between the model- and
excluded-spaces. In this sense, Feshbach formalism is in contrast with the effective interaction theory
since it is not derivable from such decoupling conditions. Notice that the
energy-dependence of the Feshbach formalism (and any Green function type
formulations, e.g., Schwinger-Dyson resummation, Faddeev approach)
emerges from the fact that the effects of the excluded Hilbert space
is taken into account by a ``quasi-potential'', while in the effective interaction approach
the latter is taken into account by imposing a certain decoupling
conditions. Therefore a given truncated Hilbert space becomes
independent of the remaining sectors and accordingly it can in principle be described
by an energy-independent prescription. 

In order to clarify the differences between an energy-dependent and an energy-independent formulation, here we investigate how can one
resolve the renormalization problem by fully utilizing the Feshbach projection operator technique
\cite{fesh} in the framework of the CCM. Therefore, we pursue an inverse of the EI approach. With a field
theoretical consideration, we show that the coupled-cluster formalism
by means of Feshbach projection technique leads
to a renormalized generalized Brueckner ($E$-dependent) theory.

The Feshbach projection technique was introduced to treat nuclear
reactions with many channels present. It was originally formulated
under the assumption that the number of elementary particles involved
is conserved, however clearly this is not that case for field
theory. An extension of the Feshbach formalism has been developed for
the pion-deuteron system \cite{20-ex} and general pion-nucleus
reactions
\cite{21-ex}. This technique bears some resemblance to Okubo's methods
\cite{oku}, which is consistent with meson field theory, but is
developed in terms of an effective Schr\"odinger equation so as to
remain in close contact with conventional nuclear physics.  This
approach was already pursued by Sch\"utte and Providencia \cite{13-e}
in the framework of the CCM for the Lee model. However, in the Lee model, because of an
inherent Tamm-Dancoff approximation (which limits the number of mesons
present at any instant), and exact solvability of the model, the issue
of renormalizibility of the nuclear matter properties is unclear. Here,
we follow their approach in an extended version of Lee model which is
not exactly solvable and does not display these trivialities. The
renormalization of the extended Lee model in the few-body sector was
already investigated in section 2.10.

Notice that the CCM as introduced by Sch\"utte and Providencia based
on Rayleigh-Ritz-Variational principle (e.g., see \cite{13-e}) is
different from the CCM introduced by Arponen and Bishop from the
standpoint of variational principle and the Hellman-Feynman theorem
\cite{ccm1,ccm3}. As we have already illustrated in the first two chapters, we believe that the latter has more advantages and is more suitable for a 
field theoretical application since it is naturally embedded in the
modern effective field theoretical framework. Having said that our
main goal in this chapter is to introduce other quantum many-body theory
techniques and challenge if they are adoptable for a field theoretical application.

\section{The extended Lee Model}
The extended Lee model (ELM) is a simple model connecting elementary
and composite particles \cite{19}. Although this model is not a chiral
model but it exhibits many field-theoretical features. We define four
kinds of particles. The $V$- and $N$- particles, two different types
of fermions, and the $\theta$ scalar boson and the $ \bar{\theta}$
anti-boson. We take for the Hamiltonian the expression:
\begin{eqnarray}
H^{0}&=&H_{0}^{0}+H^{0}_{I} ,\label{e1}\nonumber\\
H_{0}^{0}&=&\sum_{\alpha}E^{0}_{\alpha}V^{\dag}_{\alpha}V_{\alpha}+\sum_{\beta}E^{0}_{\beta}N_{\beta}^{\dag}N_{\beta}+\sum_{k}\omega_{k}(a_{k}^{\dag}a_{k}+b_{k}^{\dag}b_{k}),\nonumber\\
H^{0}_{I}&=&\sum_{\alpha\beta k}W^{0}_{\alpha\beta
k}V^{\dag}_{\alpha}N_{\beta}a_{k}+\sum_{\alpha\beta
k'}W^{0}_{\alpha\beta k'} N^{\dag}_{\beta}V_{\alpha}b_{k'}+ h.c.
\hspace{0.5 cm}.\
\end{eqnarray}
Here $a_{k},a^{\dag}_{k}$ and $b_{k},b^{\dag}_{k}$ are the
annihilation and creation operators which satisfy boson
commutation rules. The $V_{\alpha}^{\dag},V_{\alpha}$
and $N_{\beta}^{\dag},N_{\beta}$ define the fermion sector and
obey the usual anticommutator rules. The $\alpha,\beta,k$ and $
k'$ are abbreviations for all quantum numbers (such as momentum,
spin and isospin, etc). Within our formal investigation, we leave
open the specification of $E^{0}_{\alpha},E^{0}_{\beta}$ and
$\omega_{k}$. It can be taken as either a relativistic or
non-relativistic expression. The bare kinetic energies
$E_{\alpha}^{0}$ and $E_{\beta}^{0}$ are renormalized to
$E_{\alpha}$ and $E_{\beta}$ by the interaction. The matrices
$W^{0}_{\alpha\beta k}$ and $W^{0}_{\alpha\beta k'}$ are the bare
interaction strength  renormalizing to $W_{\alpha\beta k}$ and
$W_{\alpha\beta k'}$, respectively. The interaction strength is
defined by the kind of bosons exchanged (the scalar, pseudoscalar
or vector bosons). The exchange of higher spin $(J\geqslant 2)$
bosons, such as the f(1260), $\text{A}_{2}$(1310),
$\text{f}'$(1514) and g(1680) seems to have little influence on
the low-energy $NN$ data. The reason is that their poles are
located far away from the physical region. In other words, they
give rise to the contributions of the very short range which are
masked by the very strong repulsion coming from vector
meson-exchange. These contributions, however, are essentially
masked or parametrized by form factors necessary to regularize the
one-boson-exchange diagrams. The interaction term in $H$ describes
the process
\begin{eqnarray}
&&V \leftrightarrows N+\theta,\label{e3}\\
&&N \rightleftarrows V+\bar{\theta}.\label{e4}
\end{eqnarray}
The Lee model can be recovered if we decouple the anti-boson
$\bar{\theta}$, \(W^{0}_{\beta\alpha k'}\to 0\) (i. e. to remove
crossing symmetry) \cite{sch,22,23}.  In the Lee model the virtual
process Eq.~(\ref{e4}) is not included and thus the $N$-particle state
remains unrenormalized and the model becomes exactly solvable
\cite{sch}. We have already shown that the ELM can not exhibit
asymptotic freedom at space-time dimension $D$=4. It is well known
that such a model in four dimension exhibit a ghost state as the
cutoff is removed. It has been shown that the composite-particle
theory, the meson pair theory of Wentzel \cite{14-e}, can be obtained
as a strong-coupling limit of the ELM, in which limit the wave
function renormalization constant of the $V$-particle vanish
\cite{19}. One can find two operators commuting with the Hamiltonian,
\begin{eqnarray}
B&=&\sum_{\alpha}V_{\alpha}^{\dag}V_{\alpha}+\sum_{\beta}N_{\beta}^{\dag}N_{\beta},\label{e5}\\
Q&=&\sum_{\beta}N_{\beta}^{\dag}N_{\beta}+\sum_{k'}b_{k'}^{\dag}b_{k'}-\sum_{k}a^{\dag}_{k}a_{k}.\nonumber
\end{eqnarray}
Clearly $B$ is a baryon number operator and $Q$ is a charge
operator. In setting up the charge operator, we have assigned the
charges 1,0,-1 and 1 to the particles $N,V,\theta$ and
$\bar{\theta}$, respectively. The sectors of the model are labeled
by the eigenvalues $(b,q)$ of the operators $(B,Q)$. The most
trivial sectors are: $(0,0)={|o\rangle}$, the physical vacuum is
thus the same as the free particle vacuum for this model,
$(0,1)=\{b^{\dag}_{k'}|0\rangle=|k'\rangle\} $ and
$(0,-1)=\{a^{\dag}_{k}|0\rangle=|k\rangle\}$ which define respectively
anti-meson and meson states, these states stay unrenormalized. The
sectors $(1,0)$ and $(1,1)$ contain:
\begin{eqnarray}
&&(1,0)=\{V^{\dag}_{\alpha}|0\rangle=|\alpha\rangle,
N^{\dag}_{\beta}a^{\dag}_{k}|0\rangle=|\beta,k\rangle,|\alpha
kk'\rangle,...\}, \nonumber\\ &&(1,1)=\{|\beta\rangle, |\alpha
k'\rangle, |\beta kk'\rangle,...\}.\
\end{eqnarray}
In the Lee model $|\beta\rangle$ stays unrenormalized. This makes the
model exactly solvable, but this is not the case in ELM. The
renormalization of such a model was already studied in section
2.5. The one-loop renormalization of the interaction strength is
defined as
\begin{eqnarray}
&&W_{\alpha \beta k}(W_{\alpha\beta k'})=W^{0}_{\alpha \beta
k}(W^{0}_{\alpha\beta k'})/Z_{\alpha}Z_{\beta},
\label{new1}\nonumber\\
 &&Z^{2}_{\alpha}=1+\sum_{\beta
k}\frac{|W^{0}_{\alpha
\beta k}|^{2}}{(E_{\alpha}-E^{0}_{\beta}-\omega_{k})^{2}},\nonumber\\
&&Z^{2}_{\beta}=1+\sum_{\alpha k'}\frac{|W^{0}_{\alpha\beta
k'}|^{2}}{(E_{\beta}-E^{0}_{\alpha}-\omega_{k'})^{2}}.\
\end{eqnarray}
In one-loop order, the physical neutron state $|\psi_{\alpha}\rangle$
is a bound state in the $(1,0)$ sector with
$H|\psi_{\alpha}\rangle=E_{\alpha}|\psi_{\alpha}\rangle$, where
$E_{\alpha}$ is renormalized $V$-particle energy. We note that in the
limit that the coupling constant vanishes, the $V$-particle state
$|\psi_{\alpha}\rangle$ goes over into $|\alpha\rangle$. Therefore, the
simplest form of a bound state in the $(1,0)$ sector at one-loop order
can be written as 
\begin{equation}
|\psi_{\alpha}\rangle=|\alpha\rangle+\sum_{\beta,k} \phi(\beta,k)|\beta k\rangle,
\end{equation}
where the unknown coefficient
$\phi(\beta,k)$ is determined by requiring that
$|\psi_{\alpha}\rangle $ be an eigen-function of Hamiltonian with
eigenvalue $E_{\alpha}$. After straightforward calculations, one obtains 
\begin{eqnarray}
&&|\psi_{\alpha}\rangle=|\alpha\rangle+\sum_{\beta
k}\frac{W^{0*}_{\alpha \beta
k}}{E_{\alpha}-E^{0}_{\beta}-\omega_{k}}|\beta k\rangle,\label{n}\nonumber\\
&&E_{\alpha}=E^{0}_{\alpha}+h_{\alpha}(E_{\alpha}), \label{massv}\nonumber\\
&&h_{\alpha}(z)=\sum_{\beta k}\frac{|W^{0}_{\alpha \beta
k}|^{2}}{z-E^{0}_{\beta}-\omega_{k}}.\
\end{eqnarray}
The same argument can be applied for the sector $(1,1)$ and one can
obtain in lowest order, the physical proton state
$|\psi_{\beta}\rangle$ as a bound state in this sector which can
be shown to obey
\begin{eqnarray}
&&|\psi_{\beta}\rangle=|\beta\rangle+\sum_{\alpha
k'}\frac{W^{0*}_{\alpha\beta
k'}}{E_{\beta}-E^{0}_{\alpha}-\omega_{k'}}|\alpha
k'\rangle,\label{p}\nonumber\\
&&E_{\beta}=E^{0}_{\beta}+h_{\beta}(E_{\beta}),\label{massn}\nonumber\\
&&h_{\beta}(z)=\sum_{\alpha k'}\frac{|W^{0}_{\alpha\beta
k'}|^{2}}{z-E^{0}_{\alpha}-\omega_{k'}}.\
\end{eqnarray}
\begin{figure}[!tp]
\centerline{\includegraphics[width=14 cm]{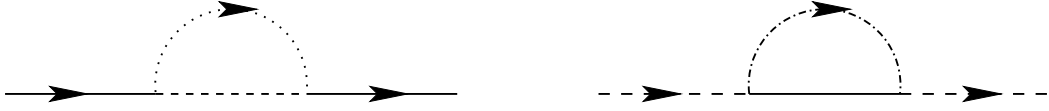}}
\caption{Diagrams in effective mass operators, $h(z)$. Solid lines and dashed lines denote $V$- and $N$-particles, respectively. 
Dotted and dash-doted lines denote $\theta$- and $\bar{\theta}$-particles.}
\end{figure}
The $h_{\alpha}(z)$ and $h_{\beta}(z)$ are the mass operators, and show the 
off-shell contribution to the self-energy, see Fig.~3.1. The mass
renormalization in the model is now performed by adding corresponding
terms as counter terms to the Hamiltonian (these terms will not change
the conservation of $B$ and $Q$ operators). We assume that the
parameters of the model are chosen in such a way that there exists one
bound state for each $\alpha$ and $\beta$ . The form factors contained
in $W_{\alpha \beta k}(W_{\beta\alpha k'})$ are assumed to make $
Z_{\alpha}>0$, $Z_{\beta}>0$ and consequently the coupling constant
renormalization finite.

\section{Nuclear matter equations}
We now wish to consider the binding energy problem of $H$ for the
sector $(2n,n)$ with $ n\to\infty$ so-called $N-V$ matter. We assume that the
non-interacting ground-state wave function of this system be the
Slater determinant built up by an equal number of $V$- and $N$- particle
up to a Fermi momentum $P_{F}$,
\begin{equation}
|\phi\rangle=\prod_{\alpha,\beta<P_{F}}V_{\alpha}^{\dag}N_{\beta}^{\dag}|0\rangle. \label{e6}
\end{equation}
We denote by $a(b)$ occupied $V$-particle ($N$-particle) states,
by $A(B)$ the unoccupied states and by $\alpha(\beta)$ either
states. We write the exact correlated ket ground state, in
a coupled cluster formulation, as
\begin{equation}
|\psi\rangle=e^{S+R}|\phi\rangle.     \label{e7}
\end{equation}
Here the cluster operator is separated into two parts, one, $S$,
without mesons which is obviously the same as in the CCM with a
phenomenological potential and $R$ with mesons contributions which
contains the additional parts originating from quantum field
theory. Here we define bra ground state as a hermitian conjugate of
ket ground state in every level of truncation. Using the symmetry
properties Eq.~(\ref{e5}) and momentum conservation the general form
of $R$ and $S$ are
\begin{eqnarray}
S&=&\sum_{i=1}^{n/2}\sum_{j=0}^{n/2}\hat{S}_{ij}, \label{e8}\nonumber\\
S_{ij}&=&\frac{1}{i!^{2}j!^{2}}\sum_{ABab}\langle A_{1}..A_{i}B_{1}..B_{j}|S_{ij}|a_{1}..a_{i}b_{1}..b_{j}\rangle_{A}\nonumber\\
&\times& V_{A_{1}}^{\dag}.. V_{A_{i}}^{\dag} N_{B_{1}}^{\dag}..N_{B_{j}}^{\dag}N_{b_{j}}..N_{b_{1}}V_{a_{i}}..V_{a_{1}},\nonumber\\
R&=&\sum_{k=1}^{n/2}\sum_{l=0}^{n/2}\sum_{m=1}^{k}\sum_{n=1}^{k'}\hat{R}_{klmn},\nonumber\\
R_{klmn}&=&\frac{1}{k!^2l!^2}\sum_{ABabKK'}\langle A_{1}..A_{k-m+n}B_{1}..B_{l+m-n}K_{1}..K_{m}K'_{n}..K'_{1}|R_{klmn}\nonumber\\
&\times&|a_{1}..a_{k}b_{1}..b_{l}\rangle_{A} V_{A_{1}}^{\dag}.. V_{A_{k-m+n}}^{\dag} N_{B_{1}}^{\dag}..N_{B_{l+m-n}}^{\dag}..N_{b_{l}}..N_{b_{1}}\nonumber\\
&\times&V_{a_{k}}..V_{a_{1}}a_{k_{1}}^{\dag}..a_{k_{m}}^{\dag}b_{k'_{1}}^{\dag}..b_{k'_{n}}^{\dag}.\
\end{eqnarray}
The subscript $A$ of the $R$- and $S$-amplitudes stands for
antisymmetrization of the states. For obtaining the binding energy of
$N$-$V$ matter one should solve the Schr\"odinger equation by making use
of the definitions in Eqs.~(\ref{e6})-(\ref{e8}), therefore
\begin{equation}
e^{-S}e^{R}He^{R}e^{S}|\phi\rangle=E|\phi\rangle.
\label{e9}
\end{equation}
Projecting this equation onto the complete orthonormal set of states
\begin{equation}
|\phi\rangle ,N_{B}^{\dag}V_{a}a_{k}^{\dag}|\phi\rangle,V_{A}^{\dag}N_{b}b_{k'}^{\dag}|\phi\rangle,..\hspace{0.2cm} \label{state},
\end{equation}
 one obtains a system of coupled integral equations of which the first
 determines the binding energy and others fix the corresponding
 amplitudes $\langle R\rangle$ and $\langle S\rangle$. These coupled
 integral equations from effective interaction view point are the 
 decoupling conditions and leads to
 energy-independent prescription. In order to derive the
 standard Brueckner theory we apply the Rayleigh-Ritz-Variational
 principle. This type of the CCM formulation was originally proposed by
 Providencia and Shakin \cite{shakin}. This is
 of course different from coupled-cluster
 theory introduced by Arponen and Bishop from the standpoint of variational
 principle and the Hellman-Feynman theorem \cite{ccm1,ccm3} (the so-called NCCM
 scheme, see chapter 1.2). 

One of our goals here is to show that in the consistent truncation scheme the
many-body problem does not require further renormalization and all
the bare parameters introduced in definition of the wave function can be fixed in the nucleon-nucleon
scattering. It is clear that Eq.~(\ref{e8}) needs to be truncated.
Aiming at a two-hole line and one-meson-exchange expansion of the
ground-state energy, we choose the following terms for $S$ and $R$
from Eq.~(\ref{e8}).
\begin{eqnarray}
S&=&\sum R_{ABab}N_{B}^{\dag}V_{A}^{\dag}V_{a}N_{b},
\label{e11}\nonumber\\
 R&=&\sum
R_{aBk}N_{B}^{\dag}V_{a}a_{k}^{\dag}+\sum
R_{Abk'}V_{A}^{\dag}N_{b}b_{k'}^{\dag}
+\sum 1/2 R_{BB'abk}N_{B}^{\dag}N_{B'}^{\dag}N_{b}V_{a}a_{k}^{\dag}\nonumber\\
&&+\sum 1/2
R_{AA'abk'}V_{A}^{\dag}V_{A'}^{\dag}V_{a}N_{b}b_{k'}^{\dag}=R_{1}+R_{2}+R_{3}+R_{4}.\
\end{eqnarray}
All other term vanish up to this order because of the symmetries
of the ELM Hamiltonian. Now we define cluster expansions of
expectation value of the Hamiltonian with respect to
$|\psi\rangle$, which is a straightforward generalization of the
standard expansion \cite{shakin,shakin2}. We neglect three-body and
higher-order cluster. The one-body, two-body, etc., correlated
wavefunction are defined as
\begin{eqnarray}
&&|\psi_{a}\rangle=e^{S+R}
V_{a}^{\dag}|0\rangle=(1+R_{1})V_{a}^{\dag}|0\rangle,
\label{e12}\nonumber\\
&&|\psi_{b}\rangle=e^{S+R} N_{b}^{\dag}|0\rangle=(1+R_{2})N_{b}^{\dag}|0\rangle,\nonumber\\
&&|\psi_{ab}\rangle=e^{S+R}
N_{b}^{\dag}V_{a}^{\dag}|0\rangle=(1+S+R)N_{b}^{\dag}V_{a}^{\dag}|0\rangle.\
\end{eqnarray}
\begin{figure}[!htp]
\centerline{\includegraphics[width=14 cm]{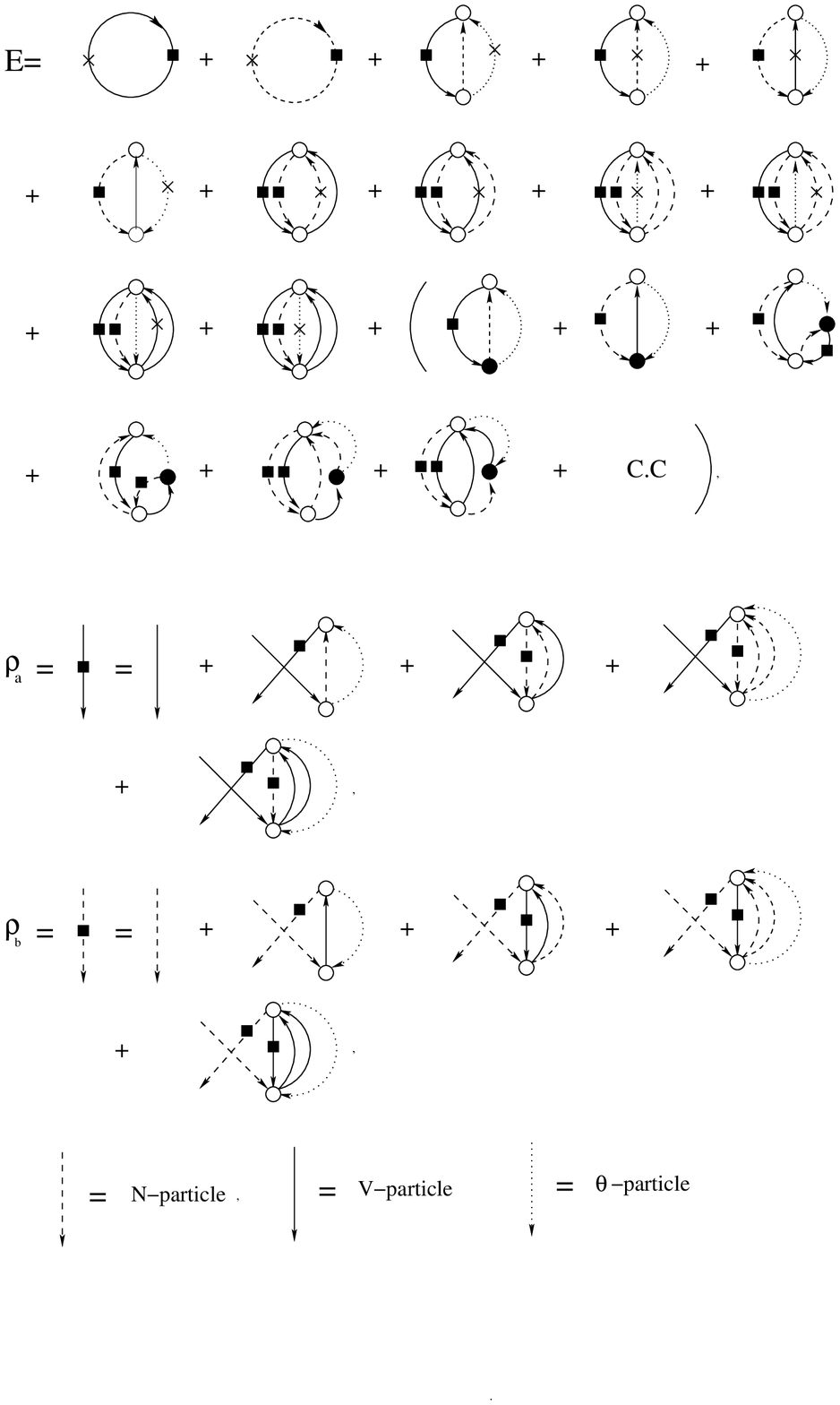}}
\caption{The symbols $\times$ and $\circ$ stand for kinetic energy and different parameters of wave function($C,C',D,F,F'$), respectively.  The symbol $\bullet$
denotes the interaction coupling $W$. The out-going arrows form
$\circ$ denote $V, N, \theta$-particle and in-coming arrows denote
$V,N$-hole and $\bar{\theta}$-particle. The abbreviation C.C implies
the complex conjugate terms. } 
\end{figure}
The expectation value of the Hamiltonian is then obtained by and
shown in Fig.~3.2
\begin{equation}
E=
\frac{\langle\psi|H|\psi\rangle}{\langle\psi|\psi\rangle}=\sum_{a}h_{a}\rho_{a}+\sum_{b}h_{b}\rho_{b}+\sum_{ab}h_{ab}\rho_{a}\rho_{b},
\label{e15}
\end{equation}
where the Hamiltonian cluster integrals have been introduced as
\begin{eqnarray}
&&h_{a}=\langle\psi_{a}|H|\psi_{a}\rangle,\nonumber\\
&&h_{b}=\langle\psi_{b}|H|\psi_{b}\rangle, \label{e14} \nonumber\\
&&h_{ab}=\langle\psi_{ab}|H|\psi_{ab}\rangle-h_{a}n_{b}-h_{b}n_{a},\
\end{eqnarray}
here $n_{a}$ and $n_{b}$ are the norm of $|\psi_{a}\rangle$ and
$|\psi_{b}\rangle$ respectively. We disregard the other cluster
integrals like $h_{aa'}$ because they would only get contributions
from two-mesons intermediate states. This makes sense in consistent
one-meson exchange approximation. According to the selective summation made in the expression Eq.~(\ref{e15}), the occupation numbers
$\rho_{a}$ and $\rho_{b}$ should satisfy the following algebraic equations:
\begin{eqnarray}
&&n_{a}\rho_{a}+\sum_{b}n_{ab}\rho_{a}\rho_{b}=1,  \label{e16}\nonumber\\
&&n_{b}\rho_{b}+\sum_{a}n_{ab}\rho_{b}\rho_{a}=1,\
\end{eqnarray}
with notation,
\begin{equation}
n_{ab}=\langle\psi_{ab}|\psi_{ab}\rangle-n_{a}n_{b}.
\end{equation}
The basic idea for the treatment of
$\langle\psi|H|\psi\rangle/\langle\psi|\psi\rangle $ is to expand
it term by term using the Wick-rule to keep track of all possible
contribution. The expansion of $\langle\psi|H|\psi\rangle$ and
$\langle\psi|\psi\rangle$ can therefore be characterized by a set
of suitable diagrams and one can check  that the division by
$\langle\psi|\psi\rangle$ cancels out all contribution from
disconnected diagrams (the linked-cluster theorem).
 Now we minimize the quantity $E$ with respect to
the constrains given by Eq.~(\ref{e16}) and denote it
$\overline{E}$, thus we multiply Eq.~(\ref{e16}) by the Lagrange
multipliers $\epsilon_{a}$ and  $\epsilon_{b}$ and subtract them
from Eq.~(\ref{e15}), then by using the definition Eq.~(\ref{e14})
we find 
\begin{eqnarray}
\overline{E}&=&\sum_{a}\rho_{a}(E_{a}^{0}-\epsilon_{a})+
\sum_{b}\rho_{b}(E_{b}^{0}-\epsilon_{b})+\sum_{a}\rho_{a}\langle
a|R_{1}^{\dag}(H_{0}^{0}-\epsilon_{a})R_{1} \nonumber\\
&&+R_{1}^{\dag}H^{0}_{I}+H^{0}_{I}R_{1}|a\rangle+\sum_{b}\rho_{b}\langle
b|R_{2}^{\dag}(H_{0}^{0}-\epsilon_{b})R_{2}+R_{2}^{\dag}H^{0}_{I}+H^{0}_{I}R_{2}|b\rangle\nonumber\\
&&+\sum_{ab}\rho_{b}\rho_{a}\langle
ba|S^{\dag}(H_{0}^{0}-\epsilon_{a}-\epsilon_{b})S+R_{3}^{\dag}(H_{0}^{0}-\epsilon_{a}-\epsilon_{b})R_{3}\nonumber\\
&&+R_{4}^{\dag}(H_{0}^{0}-\epsilon_{a}-\epsilon_{b})R_{4}
+S^{\dag}H^{0}_{I}R+R^{\dag}H^{0}_{I}S|ab\rangle.\label{e17}\
\end{eqnarray}
It is convenient to treat the $\rho_{a}$ and $\rho_{b}$ as
independent variables, therefore we minimize $\overline{E}$
respect to
$R^{*}_{aBk},R^{*}_{Abk'},S^{*}_{ABab},R^{*}_{BB'abk},R^{*}_{AA'abk},\rho_{a}$,$\rho_{b}$
 which respectively yields:
\begin{eqnarray}
&&R_{aBk}(E_{B}^{0}+\omega_{k}-\epsilon_{a})+W_{aBk}^{0*}+\sum_{Ab}\rho_{b}S_{ABab}W_{Abk}^{0*}=0,\label{e18}\\
&&R_{Abk'}(E_{A}^{0}+\omega_{k'}-\epsilon_{b})+W_{Abk'}^{0*}+\sum_{Ba}\rho_{a}S_{ABab}W_{aBk'}^{0*}=0,\label{e19}\\
&&\langle
AB|(H_{0}^{0}-\epsilon_{a}-\epsilon_{b})S+H^{0}_{I}R|ab\rangle=0,
\label{e20}\\
&&\langle BB'k|(H_{0}^{0}-\epsilon_{a}-\epsilon_{b})R_{3}+H^{0}_{I}S|ab\rangle=0,\label{e21}\\
&&\langle AA'k'|(H_{0}^{0}-\epsilon_{a}-\epsilon_{b})R_{4}+H^{0}_{I}S|ab\rangle=0,\label{e22}\\
&&\epsilon_{a}=E_{a}^{0}+\langle a|R_{1}^{\dag}(H_{0}^{0}-\epsilon_{a})R_{1}
+R_{1}^{\dag}H^{0}_{I}+H^{0}_{I}R_{1}|a\rangle +\sum_{b}\rho_{b}\langle ab|(R_{1}^{\dag}+R_{2}^{\dag})H^{0}_{I}S|ab\rangle,\nonumber\\
&&\epsilon_{b}=E_{b}^{0}+\langle
b|R_{2}^{\dag}(H_{0}^{0}-\epsilon_{b})R_{2}+R_{2}^{\dag}H^{0}_{I}+H^{0}_{I}R_{2}|b\rangle
+\sum_{a}\rho_{a}\langle
ab|(R_{1}^{\dag}+R_{2}^{\dag})H^{0}_{I}S|ab\rangle.\label{e23}\nonumber\\
\end{eqnarray}
  Equations (\ref{e18})-(\ref{e23}) represent a set of unrenormalized Variational coupled integral equations for $N-V$ matter, consistent with two-hole-line truncation.

\section{The renormalized nuclear matter equations}
The method that we will use to reformulate Eqs.~(\ref{e18})-(\ref{e23})
in order to obtain renormalized equations for $N-V$ matter is based on
the Feshbach projection operator formalism \cite{fesh}. Projection operator techniques have also been used to
analyze several sectors of the Lee model \cite{23-ex}.

The total energy of $N-V$ matter can be obtained by making use of
Eq.~(\ref{e17}) and Eqs.~(\ref{e20})-(\ref{e23}) and exploiting the
constrains Eq. (\ref{e16});
\begin{equation}
E=\sum_{a}\epsilon_{a}+\sum_{b}\epsilon_{b}-
\sum_{ab}\rho_{a}\rho_{b}\langle ab|(R_{1}^{\dag}+R_{2}^{\dag})H^{0}_{I}S|ab\rangle. \label{e24}
\end{equation}
We can now directly add mass counter terms
to the bare Hamiltonian in Eqs.~(\ref{e18}) and (\ref{e19}) to dress
masses. We expand the last term in the binding energy
Eq.~(\ref{e24}) in terms of the coupling constants and $S$. This form
will be used later to achieve full renormalization for the binding
energy since there is a subtle relation between $S$ and the
renormalized coupling constant,
\begin{eqnarray}
E&=&\sum_{a}\epsilon_{a}+\sum_{b}\epsilon_{b}-\sum_{abk'AB}\rho_{a}\rho_{b} \frac{W^{0*}_{aBk'}W^{0}_{Abk'}}{\epsilon_{b}-E_{A}-\omega_{k'}}S_{ABab}\label{eE}\\
&-&\sum_{abkAB}\rho_{a}\rho_{b} \frac{W^{0*}_{Abk}W^{0}_{aBk}}{\epsilon_{a}-E_{B}-\omega_{k}}S_{ABab}
-\sum_{aa'bk'ABB'}\rho_{a}\rho_{b}\rho_{a'}\frac{W^{0*}_{aBk'}W^{0}_{a'B'k'}}{\epsilon_{b}-E_{A}-\omega_{k'}}S^{*}_{AB'a'b}S_{ABab}\nonumber\\
&-&\sum_{abb'kAA'B}\rho_{a}\rho_{b}\rho_{b'}\frac{W^{0*}_{Abk}W^{0}_{A'b'k}}{\epsilon_{a}-E_{B}-\omega_{k}}S^{*}_{A'Bab'}S_{ABab}.\nonumber\
\end{eqnarray}
The last two terms in Eq.~(\ref{eE}) are a three-hole-line
contribution which we ignore, consistent with our aim for a
two-hole-line expansion. We introduce the projection operators
\begin{eqnarray}
&&Q=\sum|AB\rangle\langle AB|, \hspace{1.5cm}Q'=1/2\sum|BB'k\rangle\langle BB'k|,\label{e25}\\&&Q''=1/2\sum|AA'k'\rangle\langle AA'k'|. \nonumber
\end{eqnarray}
Inspired by the projection technique of feshbach, one can use the above definition to combine Eqs.~(\ref{e20})-(\ref{e22}), 
\begin{equation}
\Big[H_{0}^{0}-z+QH^{0}_{I}\Big(Q'\frac{1}{z-H_{0}^{0}}Q'+Q''\frac{1}{z-H_{0}^{0}}Q''\Big)H^{0}_{I}Q\Big]S|ab\rangle=-QH^{0}_{I}(R_{1}+R_{2})|ab\rangle, \label{e26}
\end{equation}
where $z=\epsilon_{a}+\epsilon_{b}$. We can show that the right hand
side of Eq.~(\ref{e26}) can be reduced by using Eqs.~(\ref{e18}) and
(\ref{e19}) 
\begin{eqnarray}
&&\langle AB|H^{0}_{I}R_{1}|ab\rangle=\sum_{k} W^{0}_{Abk}R_{aBk}\label{e27}\nonumber\\
&&=\sum_{k}\frac{W^{0}_{Abk}W^{0*}_{aBk}}{\epsilon_{a}-E_{B}-\omega_{k}} -\sum_{b'A'k}\rho_{b'}
\frac{S_{A'Bab}W^{0}_{Abk}W^{0*}_{A'b'k}}{\epsilon_{a}-E_{B}-\omega_{k}},\\
&&\langle AB|H^{0}_{I}R_{2}|ab\rangle=\sum_{k'}W^{0}_{aBk'}R_{Abk'}\label{e28}\nonumber\\
&&=\sum_{k'}\frac{W^{0}_{aBk'}W^{0*}_{Abk'}}{\epsilon_{b}-E_{A}-\omega_{k'}} -\sum_{a'B'k'}\rho_{a'}
\frac{S_{AB'ab}W^{0}_{aBk'}W^{0*}_{a'B'k'}}{\epsilon_{b}-E_{A}-\omega_{k'}}.\
\end{eqnarray}
The second term in Eqs.~(\ref{e27}) and (\ref{e28}) are a three-hole-line
contribution which are again ignoreable in our approximation. Let us
introduce the unrenormalized effective interaction $\overline U(z)$ 
\begin{eqnarray}
&&\langle \alpha\beta|\overline U(z)|\alpha^{'}\beta^{'}\rangle=
\langle\alpha\beta |
H^{0}_{I}\frac{1}{z-\overline H_{0}}H^{0}_{I}|\alpha^{'}\beta^{'}\rangle_{\text{linked}}
\label{e29}\nonumber\\
&&=-\sum_{k}\frac{W^{0*}_{\alpha\beta^{'}k}W^{0}_{\alpha^{'}\beta k}}{z-\epsilon_{\beta}-\epsilon_{\beta^{'}}-\omega_{k}}-\sum_{k'}\frac{W^{0}_{\alpha \beta^{'}k'}W^{0*}_{\alpha^{'}\beta k'}}{z-\epsilon_{\alpha}-\epsilon_{\alpha^{'}}-\omega_{k'}},\nonumber\\
&&\overline H_{0}=\sum_{\alpha}\epsilon_{\alpha}V^{\dag}_{\alpha}V_{\alpha}+\sum_{\beta}\epsilon_{\beta}V^{\dag}_{\beta}V_{\beta}+\sum_{k}\omega_{k}(a_{k}^{\dag}a_{k}+b_{k}^{\dag}b_{k}). \
\end{eqnarray}
The mass renormalization terms has been taken into account by using
renormalized masses in the propagators. (Notice that
$\epsilon_{A}=E_{A}$ and $\epsilon_{B}=E_{B}$). This definition
correspondences to the quasi-potential in a Lippmann-Schwinger
type equation if one replaces $H_{0}\to \overline H_{0}$. Having
made use of defined effective interaction $\overline U(z)$ one can show 
\begin{equation}
\langle AB|H^{0}_{I}(R_{1}+R_{2})|ab\rangle\simeq \langle AB|\overline U(z)|ab\rangle. \label{e30}
\end{equation}
This expression is consistent with the two-hole-line approximation.
Now we decompose the ``p-p'' and ``n-n'' interaction involved in
Eq.~(\ref{e26}) after adding mass counter terms in bare Hamiltonian,
\begin{equation}
QH^{0}_{I}\big(Q'\frac{1}{z-H_{0}}Q'+Q''\frac{1}{z-H_{0}}Q''\big)H^{0}_{I}Q=Q\overline U(z)Q+h(z)+\overline q(z), \label{e31}
\end{equation}
where the operator $h(z)$ and $\overline q(z)$ are defined via
Eqs.~(\ref{massv},\ref{massn}) as
\begin{eqnarray}
&&h(z)|AB\rangle=[h_{A}(z-E_{B})+h_{B}(z-E_{A})]|AB\rangle,\label{e33}\\
&&\overline q(z)|AB\rangle=-\Big(\sum_{bk}\frac{|W^{0}_{Abk}|^2}{z-E_{b}-E_{B}-\omega_{k}}+\sum_{ak'}\frac{|W^{0}_{aBk'}|^2}{z-E_{a}-E_{A}-\omega_{k'}}\Big)|AB\rangle.\nonumber\
\end{eqnarray}
\begin{figure}[!tp]
\centerline{\includegraphics[width=14 cm]{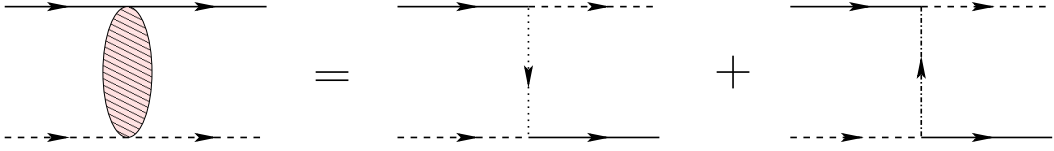}}
\caption{The effective two-body interaction, the lines are defined in Fig.~3.1.}
\end{figure}
In diagrammatic language Eq.~(\ref{e31}) contains self-energy diagrams
which contain unoccupied intermediate states because of the operators
$Q',Q''$, This is achieved by adding $\overline q(z)$ to $h(z)$.  We
now introduce the operator $B(z)$ to accomplish coupling constant
renormalization. Its structure emerges from renormalization of finite
sectors of ELM,
\begin{equation}
B(z)|\alpha\beta\rangle=Z^{2}_{\alpha}(z-E_{\beta})Z^{2}_{\beta}(z-E_{\alpha})|\alpha\beta\rangle, \label{e34}
\end{equation}
with notations,
\begin{eqnarray}
&&Z^{2}_{\alpha}(z)=1+\sum_{\beta k}\frac{|W^{0}_{\alpha \beta
k}|^2}{(E_{\alpha}-E^{0}_{\beta}-\omega_{k})(z-E^{0}_{\beta}-\omega_{k'})},\nonumber\\
&&Z^{2}_{\beta}(z)=1+\sum_{\alpha k'}\frac{|W^{0}_{\alpha \beta
k'}|^2}{(E_{\beta}-E^{0}_{\alpha}-\omega_{k'})(z-E^{0}_{\alpha}-\omega_{k'})},\
\end{eqnarray}
here $z=E_{\alpha}+E_{\beta}$. The benefit of $B(z)$ can be
manifested by following factorization property
\begin{equation}
Q(z-H_{0}^{0}-h(z))Q=Q(z-H_{0})B(z)Q. \label{fac}
\end{equation}
One may use $B(z)$ to show that $\overline U(z)$ is related to
renormalized effective $N-V$ potential
\begin{equation}
U(z)=B(z)^{1/2}\overline U(z) B(z)^{1/2}. \label{pot}
\end{equation}
Therefore, the renormalized effective two-body $N-V$ potential can be
readily found ( see Fig.~3.2)
\begin{eqnarray}
&&\langle \alpha\beta|U(z)|\alpha^{'}\beta^{'}\rangle=
-\sum_{k}\frac{r_{\alpha}(z-E_{\beta})r_{\beta}(z-E_{\alpha})W^{*}_{\alpha\beta^{'}k}W_{\alpha^{'}\beta k}r_{\alpha^{'}}(z-E_{\beta^{'}})r_{\beta^{'}}(z-E_{\alpha^{'}})}{z-\epsilon_{\beta}-\epsilon_{\beta^{'}}-\omega_{k}}\nonumber\\
&&-\sum_{k'}\frac{r_{\alpha}(z-E_{\beta})r_{\beta}(z-E_{\alpha})W_{\alpha \beta^{'}k'}W^{*}_{\alpha^{'}\beta k'}r_{\alpha^{'}}(z-E_{\beta^{'}})r_{\beta^{'}}(z-E_{\alpha^{'}})}{z-\epsilon_{\alpha}-\epsilon_{\alpha^{'}}-\omega_{k'}},
\end{eqnarray}
where we have introduced twice-subtracted ``dressing factor'' defining as
\begin{equation}
r_{\alpha}(z)=\frac{Z_{\alpha}(E_{\alpha})}{Z_{\alpha}(z)},
\hspace{2cm}
 r_{\beta}(z)=\frac{Z_{\beta}(E_{\beta})}{Z_{\beta}(z)}. \label{dress}
\end{equation}
Dressing factor appears independent of truncation made and describes
off shell correction of a self-energy of $V$ and $N$ particle. In Lee model $r_{\beta}(z)=1$, since $N$-particle stay
unrenormalized. Twice-subtracted dressing factors remain finite even
without form factor, due to the occurrence of a cubic energy
denominator, this can be verified by expanding Eq.~(\ref{dress}) using
relativistic expression for
$E_{\alpha},E_{\beta},\omega_{k},\omega_{k'}$ and interaction
strength.  Now one can rewrite Eq.~(\ref{e26}), by making use of
Eqs.~(\ref{e30}, \ref{e31}) and exploiting the factorization
property Eq.~(\ref{fac})
\begin{equation}
\big(H_{0}-z+q(z)+U(z)\big)B(z)^{1/2}S|ab\rangle=-QU(z)B(z)^{1/2}|ab\rangle,
\label{e35}
\end{equation}
where $q(z)$ is renormalized two-body operators obtained
from Eq. (\ref{e33})
\begin{eqnarray}
q(z)|AB\rangle
&&=-\Big(\sum_{bk}\frac{|W_{Abk}|^2
 r^{2}_{A}(z-E_{B})r^{2}_{B}(z-E_{A})}{z-E_{b}-E_{B}-\omega_{k}}\nonumber\\
&&+\sum_{ak'}\frac{|W_{aBk'}|^2 r^{2}_{A}(z-E_{B})r^{2}_{B}(z-E_{A})}{z-E_{a}-E_{A}-\omega_{k'}}\Big)|AB\rangle.\
\end{eqnarray}
It is observed the operator $B(z)$, transfers unrenormalized correlation function to renormalized one. This
is due to the simple structure of vertices renormalization in this
model. We now turn to Eq.~(\ref{e23}) for $\epsilon_{a}$ and
$\epsilon_{b}$, one may use Eqs.~(\ref{e18},\ref{e19}) to obtain 
\begin{eqnarray}
\epsilon_{a}&=&E^{0}_{a} +\sum_{Bk}R_{aBk}W^{0}_{aBk}+\sum_{bk'AB}
\rho_{b}R^{*}_{Abk'}W^{0*}_{aBk'}S_{ABab}\label{e36}\nonumber\\
&=&E^{0}_{a}+\sum_{Bk}\frac{|W^{0}_{aBk}|^2}{\epsilon_{a}-E^{0}_{B}-\omega_{k}}+\sum_{bkAB}\rho_{b}\frac{W^{0}_{aBk}W^{0*}_{Abk}}{\epsilon_{a}-E^{0}_{B}-\omega_{k}}S_{ABab}\nonumber\\
&+&\sum_{bk'AB}\rho_{b} \frac{W^{0*}_{aBk'}W^{0}_{Abk'}}{\epsilon_{b}-E^{0}_{A}-\omega_{k'}}S_{ABab}
+\sum_{a'bk'ABB'}\rho_{b}\rho_{a'}\frac{W^{0*}_{aBk'}W^{0}_{a'B'k'}}{\epsilon_{b}-E^{0}_{A}-\omega_{k'}}S^{*}_{AB'a'b}S_{ABab},\nonumber\\
\epsilon_{b}&=&E^{0}_{b} +\sum_{Ak'}R_{Abk'}W^{0}_{Abk'}+\sum_{akAB} \rho_{a}R^{*}_{aBk}W^{0*}_{bAk}S_{ABab} \nonumber\\
&=&E^{0}_{b}+\sum_{Ak'}\frac{|W^{0}_{Abk'}|^2}{\epsilon_{b}-E^{0}_{A}-\omega_{k'}}+\sum_{ak'AB}\rho_{a}\frac{W^{0}_{Abk'}W^{0*}_{aBk'}}{\epsilon_{b}-E^{0}_{A}-\omega_{k'}}S_{ABab}\nonumber\\
&+&\sum_{bkAB}\rho_{a} \frac{W^{0*}_{bAk}W^{0}_{aBk}}{\epsilon_{a}-E^{0}_{B}-\omega_{k}}S_{ABab}
+\sum_{ab'kA'AB}\rho_{a}\rho_{b'}\frac{W^{0*}_{Abk}W^{0}_{A'b'k}}{\epsilon_{a}-E^{0}_{B}-\omega_{k}}S^{*}_{A'Bab'}S_{ABab}.\nonumber\\
\end{eqnarray}
The last term in $\epsilon_{a}(\epsilon_{b})$ in Eq.~(\ref{e36}) are a
three-hole-line contribution. From Eqs.~(\ref{massv}) and
(\ref{massn}) it is obvious that the first two terms in
$\epsilon_{a}(\epsilon_{b})$ in Eq.~(\ref{e36}) are the renormalized
energy of $N(V)$ particles if we add the contribution of intermediate
occupied $a(b)$ states in self-energy. The corresponded terms should
be subtracted which contribute to the definition of
$\bar{\epsilon_{a}}(\bar{\epsilon_{b}})$. Having introduced mass
counter terms, one can eliminate $S$ by using  Eq.~(\ref{e35}) and obtain the
renormalized equation for $E$, $\epsilon_{a}$ and $\epsilon_{a}$,
\begin{eqnarray}
&&E=\sum_{a}\epsilon_{a}+\sum_{b}\epsilon_{b}-\sum_{ab}\rho_{b}\rho_{a}\langle ab|G(\epsilon_{a}+\epsilon_{b})|ab\rangle+\sum_{ab}\rho_{a}\rho_{b}\langle ab|U(\epsilon_{a}+\epsilon_{b})|ab\rangle,\label{E}\nonumber\\
&&\\
&&\epsilon_{a}=E_{a}+\bar{\epsilon_{a}}+\sum_{b} \rho_{b}\langle ab|G(\epsilon_{a}+\epsilon_{b})|ab\rangle,\label{e38}\\
&&\epsilon_{b}=E_{b}+\bar{\epsilon_{b}}+\sum_{a} \rho_{a}\langle ab|G(\epsilon_{a}+\epsilon_{b})|ab\rangle,\\
&&G(z)=U(z)+U(z)\frac{Q}{z-H_{0}-q(z)}G(z), \label{e39}\\
&&\bar{\epsilon_{a}}=-\sum_{bk}\frac{|W_{abk}|^2 r^{2}_{a}(\epsilon_{a})r^{2}_{b}(\epsilon_{b})}{\epsilon_{a}-E_{b}-\omega_{k}}-\sum_{b}\rho_{b}\langle ab|U(\epsilon_{a}+\epsilon_{b})|ab\rangle,\\
&&\bar{\epsilon_{b}}=-\sum_{ak'}\frac{|W_{abk'}|^2 r^{2}_{a}(\epsilon_{a})r^{2}_{b}(\epsilon_{b})}{\epsilon_{b}-E_{a}-\omega_{k'}}-\sum_{a}\rho_{a}\langle ab|U(\epsilon_{a}+\epsilon_{b})|ab\rangle,\
\end{eqnarray}
 where $\epsilon_{a}$ and $\epsilon_{b}$ are given by self-consistent equation of the Brueckner type and $ G(z)$ is the solution of a Brueckner-Bethe-Goldstone equation \cite{b-ex,24-ex}.
\section{Conclusion}
The correction to the standard many-body scheme emerges in this model
in different steps which are connected with the following physical
effects. We have to replace the $V^{OBE}$ by $ U(z)$, which accounts
for the renormalization effects of the boson propagating in the $N-V$
matter system. This can be understood since fermions feel the average
potential given by $\epsilon_{a}$ and $\epsilon_{b}$ during the time
when a boson is exchanged in $N-V$ matter \cite{24-ex}. In scattering
formalism ($T$-matrix) we have $z=E_{\alpha}+E_{\beta}$, the total
energy of $N-V$ system, whereas there is a medium effect in $N-V$
matter bringing the $z$-value to $ z=\epsilon_{a}+\epsilon_{b}$. The
energy denominator of the Brueckner Eq.~(\ref{e39}) contains the
correction $q(z)$ which is not present for standard $N-V$
scattering. This correction is due to the effect of the Pauli
principle on the self-energy of the fermions. The comparison between
many-body solution in the Lee model \cite{13-e} and the ELM reveals
following distinctions:\newline The simplest correction in the ELM is due
to the self-energy diagrams of $N$-particle and Pauli principle on
this diagrams which are manifested through the occurrence of the
corresponding dressing factor $r_{\beta}(z)$ and $q(z)$ respectively,
whereas these corrections in the Lee model do not appear because
$N$-particle remains unrenormalized . In the Lee model the lowest order
of $\epsilon_{b}$ would appear in a three-hole-line expansion while in
the ELM this term arises in two-hole-line contribution on an equal footing
with lowest order of $\epsilon_{a}$. 
\begin{figure}[!tp]
\centerline{\includegraphics[width=16 cm]{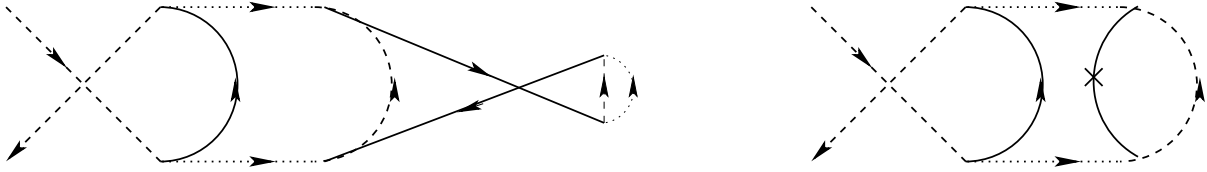}}
\caption{The symbol $\times$ stands for the renormalization counter term which is shown on the left land side.}
\end{figure}
The $\rho_{b}$
renormalization correction of the Lee model (see Fig.~3.4) goes to contribution of mass
renormalization of $N$-particle in the ELM. Therefore the renormalization
correction of occupation numbers gets contribution from higher order
which are not generated in our approximation.  It is notable that the
ELM due to possessing the crossing symmetry preserves the symmetry between $V$
and $N$ part of the renormalization correction, whereas this is not
the case for many-body problem within Lee model.

This presentation is conclusive that coupled-cluster theory without
the decoupling property in systematic truncation scheme leads to
derivation of generalized Brueckner ($E$-dependent) theory which
includes renormalization correction originating from medium effect. We
showed as well that a combination of the coupled-cluster theory and Feshbach
projection technique provides a powerful method to renormalize quantum many-body problem in a truncated Fock space.

\section{Some final remarks}
We have employed the coupled-cluster theory in various versions and
investigated the merits and short-comings of such techniques for field
theoretical applications. We showed that the CCM version introduced by
Arponen and Bishop can be easily adopted with Wilsonian
renormalization group and provides very strong tools for describing
(non)-perturbative phenomena (please see the conclusion of the last
chapter 2.11). On the other hand, the CCM version introduced by
Sch\"utte and Providencia can be equipped with
the Feshbach projection formalism and produces a renormalized generalized
Brueckner theory. One should bear in mind that in the latter, the
famous ``small-energy denominator'' problem is not avoidable and is
not at all clear how systematically high-energy modes can be
integrated out for a very complicated system.

In the rest of this work, we employ an effective QCD model as can
be constructed from the techniques of the first part of this
thesis and address with greater details, non-perturbative phenomenological phenomena such as
nucleon and diquark solutions, confinement and spontaneous chiral symmetry breaking.

\chapter{QCD properties and effective quark chiral models}
It is believed that the strong interactions are described by a quantum
field theory known as quantum chromodynamics (QCD)
\cite{ph4,qcd1}. QCD from many aspects is an unique theory. Quantum
electrodynamics (QED), and its expansion to the electroweak standard
model of particle physics, is also a quantum field theory. However,
QED breaks down at short distances and is not well-defined. QED is
renormalizable theory but it loses all his credibility as we approach
to Landau pole. On the other hand if the cutoff goes to infinity, QED
becomes trivial. QED is not the only theory with a Landau pole
problem, every theory which is not asymptotically free suffers from
this problem. The quantum field theory of gravity obtained from
general relativity suffers from nonrenormalizability. QCD is the
only known theory which is free from such problems. QCD only needs few
parameters to be defined completely, one universal coupling strength
and one mass for each kind of quark.

In following chapter we introduce part of QCD which
is relevant to nuclear physics.  We refrain from discussing all
details since it can be found in many quantum-field theory textbooks, see e.g., Refs.~\cite{ph4,qcd1}.
\section{QCD Symmetries} 
In this section we will introduce the
underlying symmetries of QCD. The Lagrangian of QCD is given by
\cite{ph4,qcd1},
\begin{equation}
\mathcal{L}=\bar{q}(i\gamma^{\mu}\partial_{\mu}-m^{0})q-\frac{1}{4}(F^{a}_{\mu\nu})^{2}+g\bar{q}\gamma^{\mu}A_{\mu}q, \label{qcd1}
\end{equation}
where $q$ is the quark field which is defined in the fundamental
representation of the color and flavor group, and the conjugate Dirac
field is defined as $\bar{q}=q^{\dag}\gamma^{0}$. The gluon field
matrix $A^{\mu}=A^{a}_{\mu}\lambda^{a}/2$ is defined in the
fundamental $SU(N_{c}=3)$ representation, $\lambda^{a}$ being the
generators of the gauge group which satisfies
$[\lambda^{a}/2,\lambda^{b}/2]=if^{abc}\lambda^{c}/2$. We define $g$ as strong
coupling constant. The field strength $F^{a}_{\mu\nu}$ is given by
\begin{equation}
F^{a}_{\mu\nu}=\partial_{\mu}A^{a}_{\nu}-\partial_{\nu}A^{a}_{\mu}+gf^{abc}A^{b}_{\mu}A^{c}_{\mu}.
\end{equation}
The non-Abelian nature of QCD is manifested by the quadratic term in the gauge field strength. The color and
flavour indices of the quark field are suppressed. $m^{0}$ is the current
quark mass which is not directly observable if QCD confines quarks. The current quark mass is color-independent and can be
brought diagonal in flavour space. There are six flavours of quarks,
each of which has a different mass. The three light quarks are called
up (u), down (d) and strange (s), while the three heavy quarks are
called charm (c), bottom (b) and top (t). The following values for the light current quark masses are found from the Particle Data group \cite{qcd-data},
\begin{equation}
m^{0}_{u}=2~~\text{to}~~ 8~~\text{MeV}, \hspace{1cm}  m^{0}_{d}=5~~\text{to}~~ 15~~\text{MeV},\hspace{1cm}  m^{0}_{s}=100~~\text{to}~~300~~\text{MeV}.
\end{equation}
Notice that the quark masses are renormalization-scheme dependent. The
above values are obtained in a subtraction scheme at a renormalization
scale $\mathcal{O}(1\text{GeV})$.  In addition to flavour, quarks carry
another quantum number known as colour. Each quark comes in three
colours, red, green and blue.

The Lagrangian  Eq.~(\ref{qcd1}) has a large classical symmetry: we have the local gauge symmetry $SU(N_{c})$ by construction,
\begin{eqnarray}
&&q\to U_{c}q, \hspace{2cm} \bar{q}\to\bar{q}U^{\dag}_{c}, \hspace{2cm} U_{c}(x)=\exp(i\theta^{a}(x)(\frac{\lambda^{a}}{2})_{c}),\nonumber\\
&& A_{\mu}\to U_{c}A_{\mu}U^{\dag}_{c}-\frac{1}{g}U_{c}i\partial_{\mu}U^{\dag}_{c}. \nonumber\
\end{eqnarray}
We have also global flavour symmetry which does not affect the gluon fields,
\begin{equation}
q\to U_{V}q, \hspace{2cm} \bar{q}\to\bar{q}U^{\dag}_{V},\hspace{2cm}  U_{V}=\exp(i\theta^{a}_{V}(\frac{\lambda^{a}}{2})_{F}). \label{qcd2}
\end{equation}
where $(\frac{\lambda^{a}}{2})_{F}$ denotes the generators of the
flavor group $U(N_{f})$ and $N_{f} $ denotes the number of
flavors. The above symmetry is referred to as vector flavor symmetry
$U_{V}(N_{f})$. When the generator matrix is taken unit matrix we have
$U_{V}(1)$ symmetry associated with conservation of baryon number. There is
another global symmetry which is exact at $m^{0}=0$, namely chiral
symmetry. This symmetry is very similar to vector flavor symmetry,
apart from an extra factor of $\gamma_{5}$ in the generator of
the transformation.
\begin{equation}
q\to U_{A}q, \hspace{2cm} \bar{q}\to\bar{q}U_{A}, \hspace{2cm} U_{A}=\exp\left(i\gamma_{5}\theta_{A}^{a}(\frac{\lambda^{a}}{2})_{F}\right).
\end{equation}
Notice that due to the factor $\gamma_{5}$ the quark field and its
conjugate partner are transformed by the same matrix in contrast
to vector transformation Eq.~(\ref{qcd2}). This transformation is
called the axial-vector transformation and can be combined with
the vector transformation to define a bigger symmetry at chiral
$m^{0}=0$ which is then called chiral symmetry $U_{V}(N_{f})\times
U_{A}(N_{f})$. One may alternatively define right- and left-handed 
quark fields by following transformation
\begin{equation}
q_{L}=\frac{1-\gamma_{5}}{2}q,
\hspace{2cm}q_{R}=\frac{1+\gamma_{5}}{2}q,
\end{equation}
The right- and left-handed massless fermions are eigenvalues of the
helicity or chirality (with eigenvalue $\pm 1$) and are not mixed
together. The chiral symmetry can be equivalently written as
$U_{L}(N_{f})\times U_{R}(N_{f})$.

It is believed that intermediate-energy hadronic physics, say over
range of energy MeV-GeV is adequately described by the dynamics of the
lowest-mass quarks, $u$ and $d$.  The overall classical symmetry of
the Lagrangian with $N_{f}=2$ becomes
\begin{equation}
SU(N_{c})_{\text{local}}\times(SU(2)_{L}\times SU(2)_{R}\times
U(1)_{V}\times U(1)_{A})_{\text{global}},
\end{equation}
Not all above-mentioned symmetries survive quantization. Particles
with opposite helicity are related by a parity transformation,
therefore in a chirally symmetric world the hadrons should come in
parity doublets. However, in a real life we do not observe such
degeneracy. Therefore, one can conclude that chiral symmetry is not
realized in the ground state and chiral symmetry is spontaneously
broken. The Goldstone theorem tell us that the spontaneous
breaking of a continuous global symmetry implies the existence of
associated massless spinless particles. This indeed
confirmed due to the existence of the light pseudoscalar mesons in
nature (pions, kaons and etas) as the corresponding Goldstone
bosons \cite{qcd2}. Moreover, the existence of quark condensate $\langle\bar{q}q\rangle$
implies that the $SU(N_{f})_{L}\times SU(N_{f})_{R}$ symmetry is
spontaneously broken down to $SU(N_{f})_{V}$. Therefore one may
conceive QCD quark condensate as an order parameter for
chiral symmetry breaking. The concept of
spontaneous broken chiral symmetry is the cornerstone in the
understanding of the low-energy hadronic spectrum.

 The $U(1)_{A}$ symmetry implies that all hadrons should come with
 opposite parity partners. However, this is not the case, therefore
 this symmetry must be broken somehow. If the spontaneous symmetry
 breaking mechanism works here, then one should observe Goldstone
 boson associated with $U(1)_{A}$, namely an $I=0$ pseudoscalar meson
 having roughly the same mass as the pion. Surprisingly there is no
 such Goldstone boson. This problem is sometime called $U(1)_{A}$
 puzzle. It turned out that the
 $U(1)_{A}$ symmetry is explicitly broken by quantum effects. This
 effect is known as the axial anomaly \cite{qcd3}. It was shown by 't Hoof that
 due to instanton effects, the $U(1)_{A}$ symmetry is not manifested
 in nature \cite{qcd3}.

Finally, at $m^{0}=0$, the QCD Lagrangian is
invariant under a scale transformation which is called
dilatational symmetry:
\begin{equation}
q(x)\to \epsilon^{3/2}q(\epsilon^{-1}x), \hspace{2cm} A^{a}_{\mu}(x)\to\epsilon A^{a}_{\mu}(\epsilon^{-1}x), \hspace{2cm} x_{\mu}\to \epsilon^{-1}x_{\mu}.
\end{equation}
This symmetry is again broken at quantum level due to the trace
anomaly \cite{qcd4}.

\section{Non-perturbative features of QCD}
In this section we shall recapitulate the most important features of
QCD which are not accessible perturbatively. Let us firstly elaborate
why perturbation theory in terms of coupling $g$ can not be used in
the low-energy regime of the theory. Having introduced the gauge fixing
term and an associated ghost term by means of the Faddeev-Popv procedure \cite{ph4,qcd1,qcd5},
one can carry out perturbation theory in terms of coupling. Due to the renormalization process, a
renormalization scale $\mu$ enters the algebra \cite{qcd6}. Therefore the running coupling is described by the RG equation
\begin{equation}
\frac{dg(\mu)}{d\ln \mu}=\beta(g).
\end{equation}
Where the coupling is small, the $\beta$ function can be computed perturbatively,
\begin{equation}
\beta(g)=-\frac{\beta_{0}}{(4\pi)^{2}}g^{3}-\frac{\beta_{1}}{(4\pi)^{4}}g^{5}+...,
\end{equation}
with 
\begin{equation}
\beta_{0}=11-\frac{2}{3}N_{f}, \hspace{2cm} \beta_{1}=102-\frac{38}{3}N_{f}.
\end{equation}
Therefore, one can readily calculate the effective running coupling
\begin{equation}
\alpha_{s}(\mu)=\frac{g^{2}(\mu)}{4\pi}=\frac{12\pi}{(33-2N_{f})\ln(\mu^{2}/\Lambda^{2})}\times[1-\frac{(918-114N_{f})\ln(\ln(\mu^{2}/\Lambda^{2}))}{(33-2N_{f})^{2}\ln(\mu^{2}/\Lambda^{2})}],
\end{equation}
where $\Lambda$ is a scale parameter of QCD and depends on the subtraction scheme and the number of active flavours,
\begin{equation}
\Lambda_{\overline{MS}}^{(5)}=(208^{+25}_{-23}) \text{MeV},
\end{equation}
where the symbol $\overline{MS}$ stands for minimal subtraction scheme
\cite{qcd1} and the superscripts indicate the number of active
flavours. This value is taken from an analysis of the various high
energy processes, see Ref.~\cite{qcd7}. The most striking feature of
the running coupling is that it decreases logarithmically as $\mu$
increases.  Therefore perturbation theory works very well for large
$\mu$.  This phenomenon is called asymptotic freedom
\cite{qcd8}.  However, if $\mu$ is near $\Lambda_{\overline{MS}}$,
perturbation theory does not work anymore and non-perturbative
phenomena enter the stage. Admittedly, there is no unambiguous method
available to connect small and large distances in QCD.

One of the most important non-perturbative features of QCD is dynamical
chiral symmetry breaking which is responsible for generation of a quark
mass from nothing\footnote{There is another very different way to
generate mass from vacuum, the so-called Casimir effect
\cite{qcd-new81} which originates from the response of vacuum in the
presence of non-perturbative boundary condition. The existence of boundary conditions in quantum
field theory is not always free of problems (see e. g.,
\cite{qcd-new82}).}. In order to show that this phenomenon is purely
non-perturbative, we employ the QCD gap equation \cite{asy},
\begin{equation}
S(p)^{-1}=(i\gamma .p+m^{0})+\int \frac{d^{4}q}{(2\pi)^{4}} g^{2}D_{\mu\nu}(p-q)\frac{\lambda^{a}}{2}\gamma_{\mu}S(q)\Gamma^{a}_{\nu}(p,q), \label{qcd-gap}
\end{equation}
where $m^{0}$ and $g$ are the current-quark bare mass and the coupling constant,
respectively. $D_{\mu\nu}(p-q)$ is the dressed-gluon
propagator and $\Gamma^{a}_{\nu}(p,q)$ is the dressed-quark-gluon vertex.
The general solution of the gap equation is a dressed-quark propagator of the form
\begin{equation}
S(p)=\frac{1}{i\gamma .pA(p^{2})+B(p^{2})}=\frac{Z(p^{2})}{i\gamma .p+M(p^{2})}. \label{qcd-s}
\end{equation}
One may now use the gap equation to work out the fermion self-energy perturbatively \cite{qcd9}. One obtains,
\begin{equation}
B(p^{2})=m^{0}\left(1-\frac{\alpha}{\pi}\ln (p^{2}/m^{2})+...\right).
\end{equation}
It is observed that at all orders of loop expansion terms are
proportional to the current-quark mass and consequently vanish as
$m^{0}\to 0$. Therefore, no mass (the quark mass is defined as a pole
of the dressed-quark propagator) is generated at current-quark mass
equal to zero, i.e. the dynamical chiral symmetry breaking is
impossible in perturbation theory and there is no mixing between left-
and right-handed quarks ``perturbatively''. Notice that apart from the
trivial solution $B(p^{2})=0$ at $m=0$, a non-trivial solution
$B(p^{2})\ne 0$ can indeed be found at the chiral point, albeit
accessible non-perturbatively. The renormalization effect is not
included in Eq.~(\ref{qcd-gap}), but it does not change the above
argument \cite{qcd9}. As we already mentioned, there is a close
relationship between the generation of the quark mass, $B(p^{2})\ne
0$, and the fact that $\langle\bar{q}q\rangle\ne 0$.  The quark
condensate in QCD is given by the trace of the full quark propagator
Eq.~(\ref{qcd-s}),
\begin{equation}
\langle\bar{q}q\rangle=-i\lim_{y\to x} \Tr S(x,y).
\end{equation}
Notice that since $\bar{q}q$ is a gauge invariant object, one may take any
gauge to obtain the dressed quark propagator which has a general
form as equation (\ref{qcd-s}). It is obvious when we have
$B(p^{2})= 0$, never does the quark condensate take place, simply
because of the identity $\Tr\gamma_{\mu}=0$. It has been shown in
Landau gauge that the dynamical quark mass, $M(p^{2})$ is large in
infrared, $M(0)\sim 0.5$ GeV, but is power-law suppressed in the
ultraviolet \cite{qcd10},
\begin{equation} \label{qcd-mass}
M(p^{2})
^{\text{large}-p^{2}}=\frac{2\pi^{2}\gamma_{m}}{3}\frac{-\langle\bar{q}q\rangle^{0}}{p^{2}\left(\frac{1}{2}\ln[\frac{p^{2}}{\Lambda_{QCD}}]\right)^{1-\gamma_{m}}},
\end{equation}
where $\gamma_{m}=12/(33-2N_{f})$ is the mass anomalous dimension and
$\langle\bar{q}q\rangle^{0}$ is the renormalization group invariant
vacuum quark condensate.  The dressed-quark mass-function
Eq.~(\ref{qcd-mass}) is a longstanding prediction of the
Dyson-Schwinger equation \cite{asy} and has been recently confirmed by
quenched lattice QCD \cite{qcd11}. It has been shown in many
non-perturbative approaches that the emergence of a dynamical quark mass
leads to the non-vanishing of quark condensate and vice versa, e.g.,
see chapter 6.

Another important non-perturbative feature of QCD is color confinement
\cite{qcd12}. Loosely speaking, confinement is defined as the absence of any
colored object in nature. But it is possible that there exists a
composite colored particle which can form colorless bound states with another
colored particle like quarks.  The color confinement is still not
properly understood, and a clear and indisputable mechanism
responsible for this effect yet remains to be discovered \footnote{The
Clay Foundation is offering $\$1$ million prize to anyone who can
provide a mathematical proof of confinement.}. Confinement
originates non-perturbatively, since it is associated with a linear
potential with a string tension $\sigma\propto
\Lambda^{2}e^{-\int\frac{dg}{\beta(g)}}$ which is obviously non-perturbative in
the coupling\footnote{Note that the string picture of quark confinement is not
free of flaws, since string breaking will occur once the potential
energy approaches the quark pair creation threshold.}.

One may wonder if there is a non-trivial solution for gap equation
$B(p^{2})\ne 0$ which gives rise to a pole of the quark propagator,
which would contradict QCD confinement since the quark is colored
\footnote{It is well-known that for confinement it is sufficient
that no colored Schwinger function possesses a spectral
representation. It is equivalent to say that all colored Schwinger
functions violate reflection positivity \cite{germany,asy}. This is one way of realization of QCD confinement. There
are in fact many different ways that the confinement can be realized such as monopole condensation effect, infrared
enhancement of the ghost propagator etc. For a review of this subject see Ref.\cite{qcd12}.}. Indeed this is one of
the subtle point in every QCD model and can not be easily resolved. In
principle, there will be a long-range force between massive quarks to
confine them and also a short range spin-spin interaction between
massive dressed quarks. The former will modify the low momentum part
of the propagator to remove the quark from being on-shell. Actually,
this describes a phenomenologically motivated picture of a constituent
quark model based on the dynamical symmetry breaking.  Having said
that it is very hard to incorporate the dynamical symmetry breaking
and the confinement into a QCD model.  In fact, many models
constructed to describe the low-energy properties of hadrons are
assumed to be only dominated by the quark flavor dynamics and
dynamical symmetry breaking and are indeed reliable only at
intermediate scales, between confinement scale few hundred MeV up to a
scale about 1 GeV.

\section{Effective low-energy quark interaction}
Physical hadrons are colorless objects and their properties seem to
be determined by the flavor dynamics which is induced by an effective QCD
interaction. The first step toward a construction of such
effective theory is to integrate out gluonic degrees of freedom, then by standard bozonization and hadronization methods \cite{qcd13} derive
the desired effective low-energy theories based on relevant degrees of
freedom. There have been many attempts to attack this difficult 
problem. Two very well established
methods are the global color model approach introduced by
Cahill and Roberts \cite{qcd14}, and the so-called field strength approach
introduced by Reinhardt and collaborators \cite{qcd15}. Our main goal in this section is
to give a taste of such approaches and focus only on the main themes
rather than details. First we rewrite the quark-gluon
interaction term in QCD Lagrangian Eq~(\ref{qcd1}) by rewriting 
\begin{equation}
\bar{q}\gamma^{\mu}A_{\mu}q=A^{a}_{\mu}J^{\mu}_{a}, \hspace{2cm}
J^{\mu}_{a}=\bar{q}\frac{\lambda^{a}}{2}\gamma^{\mu}q.
\end{equation}
The full quantum theory of QCD is solved by computing the
functional integral describing the vacuum-to-vacuum transition
amplitude,
\begin{equation}
Z_{QCD}=\int \mathcal{D}q\mathcal{D}\bar{q}\int \mathcal{D}
A^{a}_{\mu} e^{i\int d^{4}x \mathcal{L}_{QCD}}.
\end{equation}
In order to handle the gluonic part of the QCD functional integral, one
first has to define gauge inequivalent orbits by using a standard gauge fixing
procedure and the Faddeev-Popov method \cite{ph4,qcd1} (in what follows we assume that
this procedure has already been carried out). We now split the above
generating functional integral as
\begin{equation}
Z_{QCD}=\int \mathcal{D}q\mathcal{D}\bar{q} \exp\left(i\int d^{4}x
\bar{q}(i\gamma^{\mu}\partial_{\mu}-m^{0})q+\Gamma[J]\right),
\end{equation}
 where the gluon part of action is defined in
\begin{equation}
\Gamma[J]=\log \int \mathcal{D}A^{a}_{\mu}\exp\left(-\frac{1}{4} \int
F^{2}+g\int A^{a}_{\mu}J^{\mu}_{a}\right).
\end{equation}
If we could evaluate exactly the gluonic part of the functional integral
then we would be done. But unfortunately this integration can not be
handled unless we resort to some sort of systematic approximation
(this is due to the presence of cubic and quartic terms of $A^{a}_{\mu}$ in
the Lagrangian). One possibility to proceed is to expand the effective
action in powers of the quark current $J^{a}_{\mu}$ as suggested by
Cahill and Roberts \cite{qcd14},
\begin{eqnarray}
\Gamma[J]&=&\Gamma[J=0]+g\int
\Gamma^{(1)a}_{\mu}J^{\mu}_{a}dx_{1}+\frac{g^{2}}{2}\int
\Gamma^{2}(x_{1},x_{2})^{a_{1}a_{2}}_{\mu_{1}\mu_{2}}J^{\mu_{1}}_{a_{1}}J^{\mu_{2}}_{a_{2}}dx_{1}dx_{2}+...\nonumber\\
&+&\frac{g^{n}}{n!}\int
\Gamma^{(n)}(x_{1},x_{2},..x_{n})^{a_{1}..a_{n}}_{\mu_{1}...\mu_{n}}J^{\mu_{1}}_{a_{1}}...J^{\mu_{n}}_{a_{n}}
dx_{1}... dx_{n},\label{qcd-ex}\
\end{eqnarray}
where the coefficients
\begin{equation}
\Gamma^{(n)}(x_{1},x_{2},...,x_{n})^{a_{1}...a_{n}}_{\mu_{1}...
\mu_{n}}= \left(\frac{\partial^{n}\Gamma[J]}{\partial
J^{a_{1}}_{\mu_{1}}..
\partial J^{a_{n}}_{\mu_{n}}}\right)_{J=0},
\end{equation}
are defined as one-particle irreducible gluon correlation functions in
the absence of quarks, \begin{eqnarray}
\Gamma^{(1)}(x_{1})^{a_{1}}_{\mu_{1}}&=&\langle
A^{a_{1}}_{\mu_{1}}(x_{1})\rangle,\nonumber\\
\Gamma^{(2)}(x_{1},x_{2})^{a_{1},a_{2}}_{\mu_{1},\mu_{2}}&=&\langle
A^{a_{1}}_{\mu_{1}}(x_{1})A^{a_{2}}_{\mu_{2}}(x_{2})\rangle-\langle
A^{a_{1}}_{\mu_{1}}(x_{1})\rangle\langle
A^{a_{2}}_{\mu_{2}}(x_{2})\rangle. \nonumber\
\end{eqnarray}
In the above expression, the brackets denotes the functional average over the gluon field
\begin{equation}
\langle...\rangle=\frac{\int \mathcal{D}A...\exp\left(-\frac{1}{4}\int F^{2}\right)}{\int \mathcal{D}A \exp\left(-\frac{1}{4}\int F^{2}\right)}.
\end{equation}
The zeroth order term $\Gamma[J=0]$ does not depend on the quark field and
is therefore an irrelevant constant. The first order term
$\Gamma^{(1)}$ gives the expectation value of the gluon field and is zero
in the absence of external fields. The leading non-trivial term is
$\Gamma^{2}(x_{1},x_{2})^{a_{1}a_{2}}_{\mu_{1}\mu_{2}}=D^{ab}_{\mu\nu}(x-y)$
which is the exact gluon propagator and includes all gluon
self-interactions and gluon-ghost interactions. Notice that the quark
loops are incorporated through the quark current attached as legs to
the exact gluon propagator. The main approximation in this approach is to
ignore all terms $n\geq 3$. Note that none of the gluon correlation
functions is gauge or Lorentz invariant. While each term in the
expansion is separately invariant under Lorentz and ``global color
symmetry'', nevertheless the whole expansion in Eq.~(\ref{qcd-ex}) is invariant
under local gauge symmetry. We truncate Eq.~(\ref{qcd-ex}) up to $n=2$,
and with this simplification the QCD generating functional is approximated by
\begin{equation}
Z_{QCD}\approx \int \mathcal{D}q \mathcal{D}\bar{q}\exp(iS_{QFD}),
\end{equation}
where the $S_{QFD}$ is the induced non-local QCD action which describes the quark flavour dynamics,
\begin{equation}
S_{QFD}= \int \bar{q}(i\gamma^{\mu}\partial_{\mu}-m^{0})q+\frac{g^{2}}{2}\int\int
\bar{q}(x)\gamma_{\mu}\frac{\lambda^{a}}{2}q(x)D(x-y)\bar{q}(y)\gamma_{\mu}\frac{\lambda^{a}}{2}q(y).
\end{equation}
The exact form of $D(x)$ is not available at the moment.
Hence, the main phenomenology task is to simulate the simplest form of
gluon propagator in order to reproduce the confinement and asymptotic
freedom of quarks. Although there is no a priori reason to believe
that the effective interaction of quarks propagating in the QCD vacuum
should retain the form of a one gluon exchange interaction, it has
been proved that already this simple approximation reproduces most 
phenomenological models such as the instanton liquid model \cite{non1},
 Nambu-Jona-Lasinio (NJL) model \cite{njl}, various chiral bag or
topological-soliton models \cite{qcd14} and the Dyson-Schwinger equation approximation \cite{asy}, etc.
As an example, a useful starting point for low energy effective action is to employ a gluon propagator which is reduced to its crudest form,
\begin{equation}
D^{ab}_{\mu\nu}(x-y)=\frac{1}{\Lambda_{\chi}^{2}}\delta(x-y)g_{\mu\nu}\delta^{ab},
\end{equation}
where $\Lambda_{\chi}$ is a constant with dimension of
$(\text{energy})^{2}$. This choice leads to a local NJL type
model with interaction:
\begin{equation}
\mathcal{L}_{I}=-\frac{g^{2}}{2\Lambda_{\chi}^{2}}[ \bar{q}(x)\gamma_{\mu}\frac{\lambda^{a}}{2}q(x)][\bar{q}(x)\gamma^{\mu}\frac{\lambda^{a}}{2}q(x)]. \label{qcd-njl}
\end{equation}
This interaction describes a system of quarks interacting via a
two-body-force. The local form of the interaction of course causes
ultraviolet divergences, which introduces an energy scale
$\Lambda_{\chi}$, breaking the scale invariance of the classical
Yang-Mills Lagrangian (at zero current quark mass),  in an anomalous fashion. We have
sketched how an effective quark theory can be approximately obtained
from QCD by eliminating gluonic degrees of freedom. This is slightly
different from the RG approach discussed in the first part of the thesis in which
our main goal was to eliminate the high-energy degrees of
freedom. Having said that, both have the same foundation, namely
eliminating the irrelevant degrees of freedom and as we already discussed can be implemented
at the same time.   

Unfortunately, there is no economic way to derive
effective theories from QCD for a given system. Therefore, one may
write down the most general Lagrangian based on relevant degrees of
freedom, having imposed some general constraints such as symmetry
properties. This approach was introduced by Weinberg \cite{qcd16} and later
by Gasser and Leutwyler \cite{qcd17}. For an example, an effective chiral quark theory can be presented as
\begin{equation}
\mathcal{L}_{\text{eff}}(x)=\sum_{n}c_{n}\mathcal{O}_{n}(x)\left(\frac{1}{\Lambda}\right)^{\text{dim}\mathcal{O}_{n}-4},
\end{equation}
where $\mathcal{O}_{n}$ are the local chiral-invariant operators
consisting of quark fields and $c_{n}$ are dimensionless coupling
constants. The theory is only valid below the scale
$\Lambda$, and the momenta of the loop integrals are cut off at
$\Lambda$. One may now truncate the above series and by first obtaining the
coupling constant through experimental input, calculate other
quantities. This EFT approach has been discussed in detail in Ref.~\cite{qcd17}. 

\section{Fierz-transformation and the effective quark interaction}
The NJL model Lagrangian Eq.~(\ref{qcd-njl}) contains the color
octet flavor singlet currents of quark. Since hadrons are color
singlets, it is desirable to recast the Lagrangian in such to
act in color singlet channels. This can be accomplished by a Fierz
transformation, see e.g., Ref.~\cite{njl,njl-ma}, using the relation 
\begin{equation}
\left(\frac{\lambda^{a}}{2}\right)_{ij}\left(\frac{\lambda^{a}}{2}\right)_{kl}=\frac{1}{2}\left(1-\frac{1}{N^{2}_{c}}\right)\delta_{il}\delta_{kj}-\frac{1}{N_{c}}\left(\frac{\lambda^{a}}{2}\right)_{il}
\left(\frac{\lambda^{a}}{2}\right)_{kj},
\end{equation}
If we take the interaction in $\bar{q}q$ channel, then the first part
of the above expression creates color singlet mesons, while the second
part is color octet. However, the signs behind the color singlet and
the color octet are opposite, therefore if we choose the negative sign for
color singlet (in the Lagrangian level), i.e. an attractive
interaction, then the interaction for color octet will be repulsive
and consequently no bound state exist for this channel consistent with
nature. Moreover, at large $N_{c}$ the color octet $\bar{q}q$ can be
neglected. In the same fashion one can recast the interaction in the $qq$
channel. In order to make a color-singlet baryon out of three quarks, we
first couple two quarks in the fundamental triplet representation
$3_{c}$ which leads to either a sextet $6_{c}$ or an antitriplet
$\bar{3}_{c}$. But only an antitriplet can be combined with another
quark in $3_{c}$ to make a color singlet state. Notice as well that
$\bar{3}_{c}$ is antisymmetric and $6_{c}$ is symmetric in color
quantum numbers. We can now use the following Fierz transformation,
\begin{equation}
\left(\frac{\lambda^{a}}{2}\right)_{ij}\left(\frac{\lambda^{a}}{2}\right)_{kl}=\frac{1}{2}\left(1-\frac{1}{N^{2}_{c}}\right)\delta_{il}\delta_{kj}+\frac{1}{2N_{c}}\epsilon_{mik}\epsilon_{mlj}.
\end{equation}
It is obvious that the interaction in the $qq$ $\bar{3}_{c}$ channel becomes
attractive, therefore diquark formation in $\bar{3}_{c}$ is in
principle possible. In the same way, one can ``Fierz'' the flavor and
Dirac quantum numbers. For the meson channel we use
\begin{equation}
\delta_{ij}\delta_{kl}=2\left(\frac{\lambda^{0}}{2}\right)_{il}\left(\frac{\lambda^{0}}{2}\right)_{kj}+2\sum_{a=1}^{N^{2}_{f}-1}\left(\frac{\lambda^{a}}{2}\right)_{il}\left(\frac{\lambda^{a}}{2}\right)_{kj}.
\end{equation}
Notice that in contrast to color group $SU(N_{c})$ the flavor group is
$U(N_{f})$, hence the flavor generators includes
$\lambda^{0}/2=\openone/\sqrt{2N_{f}}$. One may immediately read off
from the above decomposition that for $N_{f}=3$, mesons occur as nonets
under flavor transformation. For the diquark channel we use
 \begin{equation}
\delta_{ij}\delta_{kl}=2\sum_{m=1}^{3}\left(\frac{\lambda_{s}^{m}}{2}\right)_{il}\left(\frac{\lambda_{s}^{m}}{2}\right)_{kj}+2\sum_{n=1}^{6}\left(\frac{\lambda_{a}^{n}}{2}\right)_{il}\left(\frac{\lambda_{a}^{n}}{2}\right)_{kj},
\end{equation}
where $\lambda^{m}_{s}$ denotes the symmetric generators of
$U(N_{f})$, i. e. $\lambda^{7},-\lambda^{5}$, and
$\lambda^{2}$. The antisymmetric part $\lambda^{n}_{a}$ stands for $\lambda^{0,1,3,4,6,8}$ .
Finally the Dirac indices can be rearranged for meson channels by making use of
\begin{equation}
(\gamma_{\mu})_{ij}(\gamma^{\mu})_{kl}=\delta_{il}\delta_{kj}+(i\gamma_{5})_{il}(i\gamma_{5})_{kj}-\frac{1}{2}\left((\gamma_{\mu})_{il}(\gamma^{\mu})_{kj}
+(\gamma_{\mu}\gamma_{5})_{il}(\gamma^{\mu}\gamma_{5})_{kj}\right).
\end{equation}
For the diquark channel we employ
\begin{equation}
(\gamma_{\mu})_{ij}(\gamma^{\mu})_{kl}=C_{ik}C_{lj}+(i\gamma_{5}C)_{ik}(Ci\gamma_{5})_{lj}-\frac{1}{2}\left((\gamma_{\mu}C)_{ik}(C\gamma^{\mu})_{lj}
+(\gamma_{\mu}\gamma_{5}C)_{ik}(C\gamma^{\mu}\gamma_{5})_{lj}\right),
\end{equation}
where $C=i\gamma^{2}\gamma^{0}$ is the charge conjugate matrix. Using
all the above transformations, one can now rewrite the gluon exchange
Lagrangian Eq.~(\ref{qcd-njl}) (for $N_{f}=N_{c}=3$) in the following compact form
\begin{eqnarray}
\mathcal{L}_{I}&=&-\frac{g^{2}}{2\Lambda_{\chi}^{2}}\left[ \bar{q}\gamma_{\mu}\frac{\lambda^{a}}{2}q\right]\left[\bar{q}\gamma^{\mu}\frac{\lambda^{a}}{2}q\right]=\mathcal{L}_{I}^{\bar{q}q}+\mathcal{L}^{qq}_{I},\nonumber\\
&=&
\frac{g^{2}}{3\Lambda_{\chi}^{2}}\{\left(\bar{q}\mathcal{M}_{\alpha}q\right)\left(\bar{q}\mathcal{M}^{\alpha}q\right)+\left(\bar{q}\Sigma_{\alpha}q^{c}\right)\left(\bar{q}^{c}\Sigma^{\alpha}q\right)\},\
\end{eqnarray}
where the charge conjugate spinor spinor is given  by $q^{c}=C\bar{q}^{T}$ and the other notations are defined as follows
\begin{eqnarray}
\mathcal{M}_{\alpha}&=&\openone_{c}\otimes\left(\frac{\lambda^{A}}{2}\right)_{F}\otimes\Gamma_{\alpha}, \hspace{2cm} A=0,...8,\nonumber\\
\Sigma_{\alpha}&=&\left(\frac{i\epsilon^{a}}{\sqrt{2}}\right)_{c}\otimes t^{A}_{F}\otimes\Gamma_{\alpha} \hspace{2cm} a=1,2,3; \hspace{1cm} t^{A}_{F}\in\{\lambda_{a},\lambda_{s}\},\nonumber\\
\Gamma_{\alpha}&\in&\{\openone,i\gamma_{5},\frac{i}{\sqrt{2}}\gamma_{\mu},\frac{i}{\sqrt{2}}\gamma_{\mu}\gamma_{5}\}.\
\end{eqnarray}
We will study with full detail the mesons, diquarks and the baryons structures within a non-local version of the above Lagrangian in the chapter 6.

Here some remarks are in order. First of all, the Fierz transformation
does not spoil chiral symmetry. The meson channel
$\mathcal{L}^{\bar{q}q}_{I}$ and diquark $\mathcal{L}^{qq}_{I}$ are
separately chirally invariant. The main difference between mesonic and
diquark interaction stems from the Pauli principle. The diquark $qq$
vertices are antisymmetric in color indices, hence the Pauli principle
requires a certain combinations of Dirac and flavor vertices. Notice
that for $N_{c}\neq 3$, one has to replace the factor $1/3$ with
$(N_{c}-1)/2N_{c}$ and $1/N_{c}$ for meson and diquark channels,
respectively. This indicates that in large $N_{c}$ limit, the colored
diquark channels are suppressed by a $1/N_{c}$ factor and only color
singlet mesons and baryons survive. This is in accordance with
Witten's conjecture that in the large-$N_{c}$ limit, QCD transforms
into a theory of weakly interacting mesons and baryons emerging as
solitons of this theory \cite{qcd18}. We will elaborate on this
conjecture in the next chapter.

\chapter{QCD inspired pictures for baryons}
\section{Introduction}
Despite all efforts to describe hadron physics
in terms of its underlying QCD theory, a unified and unambiguous
description of hadrons is still missing. This is due to the
complexities and the non-perturbative features of low-energy sector of
QCD which prohibits a straightforward computation of hadronic
properties in this regime. 
One well-established method toward understanding the hadronic physics is the use of QCD sum rules \cite{sol0}, which aim
to interpolate between the calculable high-energy behaviour of the
QCD and low-energy phenomenology. Although this approach has been reasonably 
successful phenomenologically, there are many uncertainties induced
by the choice and formulation of the phenomenological part, and the
result can exhibit a significant dependence on the mass scale at which
the matching is performed.  More importantly, the confinement
phenomenon and instanton effects are not incorporated in this formalism. 
Another method is to simulate QCD on a lattice of spacetime points. 
Some appreciable progress has been made in
ab initio calculations of low-lying baryon resonances using lattice QCD
simulation on a computer. This is limited by computational resources. This leads to pion masses of usually more than $400$ MeV
(thus, the chiral point is not accessible in lattice QCD) or to
simulations within the quenched approximation, where sea quark effects
are neglected. Moreover, there is yet incomplete understanding of
systematic errors, e.g., finite-size effects \footnote{For understanding
QCD phase structure, one should describe the important
non-perturbative nature of QCD and the hot/dense QCD in a unified
way. Thus far, lattice regularization have not be able to overcome the
issue associated with the fact that at non-zero chemical potential the
fermionic determinant is complex. On the other hand due to
computational difficulties, a single lattice of equally spaced points is
forced to span all distance scale which does not allow a sequence of
descriptions intermediate between the constituent quark model and QCD.
Full consideration of QCD phase structure is not
possible. Nevertheless, the effective field theories approach based on
simple models embodied essential ingredients of QCD at low-energy to
give a qualitative (and even sometimes quantitative)
prescription.}.

Therefore, any models which simplify the QCD dynamics and is capable of describing
part of the hadronic physics (which might not be accessible to ab initio methods, like lattice QCD) is greatly appreciated.   

Our main goal in the following is to study
baryon pictures in a relativistic framework by employing only quark
degrees of freedom. There are two distinct possibilities to build a
model for baryons, one is in the limit of a large number of colors
based on the picture of baryons as soliton, the second one is to describe baryons in term of bound state of a finite number of colors.
The key ingredient of the former will be introduced in section 5.2, and the latter will be described in sections 5.3,4.   

\section{Baryons as chiral solitons; Skyrme model}
The idea to describe baryons as solitons was first introduced by
Skyrme \cite{sol1} before the advent of the quarks and
gluons. However, this was not fully appreciated until the conjecture
of Witten \cite{qcd18}. It was 't Hooft's proposal that the inverse of
the number of colors $1/N_{c}$, can be treated as an effective
expansion parameter \cite{sol2}. Later, Witten pursued this idea and
showed that QCD is reduced to an effective theory of weakly
interacting mesons (with an effective four-meson vertex scaling like
$1/N_{c}$), and that baryons emerge as soliton solutions of this meson
field without any further reference to their quark content
\cite{qcd18}. Since then, this subject has been well studied. Here, we
intended to explore this approach in a selective way, following
Ref.~\cite{sol3} closely .

At low-energy one expects that the effective meson theory is a
type of non-linear $\sigma$ model with pions as the lighter
mesons. We shall focus on the massless two flavor model. One can combine the
four fields $(\boldsymbol{\pi},\sigma)$ into one unitary $2\times 2$ matrix $U$
(chiral field) with only one isovector field $\varphi(x,t)$,
\begin{equation}
U(x)=\exp\left(\frac{i}{f_{\pi}}\boldsymbol{\tau}.\varphi(x)\right), \label{sk-1}
\end{equation}
where the isovector $\boldsymbol{\tau}$ contains the pauli matrices.
In terms of $U(x)$ the non-linear $\sigma$ model is defined by the Lagrangian
\begin{equation}
\mathcal{L}^{(2)}=\frac{c^{2}}{4}\tr\left(\partial_{\mu}U\partial^{\mu}U^{\dag}\right), \label{sk-2}
\end{equation}
where $c$ is a constant. The elements of chiral $SU(2)\times SU(2)$ transform any chiral field as
\begin{equation}
U(x) \to LU(x)R^{\dag}, \label{sk-3}
\end{equation}
where $R$ and $L$ are arbitrary $SU(2)$ matrices. The Lagrangian
$\mathcal{L}^{(2)}$ is invariant under such transformations. The
vacuum configuration ($\varphi=0$ i.e. $ U=1$) is only invariant
under the coset $L=R$ which reflects the spontaneous breaking of
chiral symmetry. One can readily construct the Noether current
associated with the symmetry transformation Eq.~(\ref{sk-3}). The
vector and axial-vector current correspond to $R=L$ and
$R^{\dag}=L$, respectively. The matrix element of the axial-vector
current between the vacuum and a one pion state relates the unknown
coefficient $c$ in Eq.~(\ref{sk-2}) to the pion decay constant. It
turns out that $c=f_{\pi}=93$ MeV \cite{sol3,sol3-2}.

Now, we want to find a solitonic solution for our meson theory. A
fundamental requirement for a solitonic solution of the equation of
motion is finiteness of energy. The static soliton configuration
$U(r)$ represents mappings $U: \mathcal{R}^{3}\to SU(2)$. A necessary
condition is to require the boundary condition
\begin{equation}
\lim_{r\to\infty} U(r)=1.
\end{equation}
This means that spatial infinity is mapped to one point in flavor
space, i.e. $\mathcal{R}^{3}$ is compactified to the hypersphere
$S^{3}$.
\begin{equation}
U: S^{3}\to S^{3},
\end{equation}
which induces a topological invariant, the winding numbers. The associated topological current has been given by Skyrme \cite{sol1}
\begin{equation}
B^{\mu}=\frac{1}{24\pi^{2}}\epsilon^{\mu\nu\rho\lambda}\Tr[(U^{\dag}\partial_{\nu}U)(U^{\dag}\partial_{\rho}U)(U^{\dag}\partial_{\lambda}U)],
\end{equation}
with conservation law $\partial^{\mu}B_{\mu}=0$. The integral over the zero component defines the topological charge
\begin{equation}
B=\int B_{0} d^{3}x \label{sk-wi},
\end{equation}
which as we will show, introduces the winding number $n$. A static
configuration of the soliton is given by the spherically symmetric hedgehog Ansatz
\begin{equation}
U(r)=\exp\left(i\bold{\tau}.\hat{r}\Theta(r)\right). \label{sk-an}
\end{equation}
In order to understand the geometrical meaning of the above Ansatz, we use this expression to compute the topological charge from Eq.~(\ref{sk-wi}).
\begin{eqnarray}
B&=& \frac{1}{24\pi^{2}}\epsilon^{ijk}\int \Tr[(U^{\dag}\partial_{i}U)(U^{\dag}\partial_{j}U)(U^{\dag}\partial_{k}U)]d^{3}x,\nonumber\\
&=&\frac{1}{2\pi^{2}}\int \frac{\sin^{2}(\Theta)}{r^{2}}\partial_{r}\Theta~ d^{3}x=\frac{2}{\pi}\int_{\Theta(0)}^{\Theta(\infty)} \sin^{2}(\Theta)d\Theta=n,\
\end{eqnarray}
where we imposed the boundary conditions $\Theta(0)=-n\pi$ and
$\Theta(\infty)=0$. Solitons with different winding numbers are
topologically distinct, and there thus exists no continuous
deformation connecting solitons of different winding number. It has
been conjectured by Skyrme \cite{sol1} that the topological current
$B^{\mu}$ can be related to the baryon current and the winding number with the
baryon number.

One may scale the spatial coordinate $r$ in $U(r)$ by $\lambda r$
i.e. $U\to U(\lambda r)$, this leads to scaling the potential
part of Lagrangian Eq.~(\ref{sk-2}): $\mathcal{L}^{(2)}\to
\frac{1}{\lambda}\mathcal{L}^{(2)}$. Therefore the minimal energy
is only obtained for $\lambda\to \infty$, e. i., no stable
solitons can be found and the soliton collapse to zero size.
However, if the Lagrangian contains a term containing products of four
spatial derivatives (but only quadratic in time derivative for sake
of quantization), it will scale as $\mathcal{L}^{(2)}\to
\frac{1}{\lambda}\mathcal{L}^{(2)}+\lambda\mathcal{L}^{(4)}$ which
stabilizes at $\lambda=\mathcal{L}^{(2)}/\mathcal{L}^{(4)}$. A
possible form of $\mathcal{L}^{(4)}$ is given by
\begin{equation}
\mathcal{L}^{(4)}=\frac{1}{32e^{2}}\left(Tr[(U^{\dag}\partial_{\mu}U),(U^{\dag}\partial_{\nu}U)][ (U^{\dag}\partial^{\mu}U),(U^{\dag}\partial^{\nu}U)]\right),
\end{equation}
where $e$ is an extra parameter which determines the size of the
particle. 

A natural question is how baryons get their half
integer spin and isospin within soliton picture since the pion field
possesses spin zero and isospin one. An immediate answer is
quantization. We refrain to go through details here and concentrate
only on main points. The time dependent soliton solution should be
obtained as a first step toward quantization. However, such solutions
are hardly available and one needs to resort to an approximation. A
reasonable Ansatz for such a time-dependent solution is given by \cite{sol3}
\begin{equation}
U(r,t)=A(t)U_{0}(r)A^{\dag}(t),
\end{equation}
where $A(t)\in SU(2)$ is referred to as the collective rotation and
contains collective coordinates (it can be parametrized in terms of
Euler angles). This Ansatz does not change the potential energy of the
hedgehog. The time derivative in the Lagrangian, produces terms which
represent the rotational energy of a rotating skyrmions. Having used
hedgehog properties and a suitable definition of angular velocities as
canonical variables, one can show that the absolute values of spin and
isospin are equal,  $T^{2}=J^{2}$, and the energy eigenvalues of the
system forms a rotational spectrum $E_{J}=\frac{1}{2\theta}J(J+1)+M $,
where $M$ is the static energy of the soliton and $\theta$ denotes the
moment of inertia. One can immediately observe that the quantum numbers of the low energy baryons
(e.g. the nucleon with $J=T=1/2$ and delta $J=T=3/2$ ) are consistent 
with this picture.  One of the obvious shortcomings of this
presentation is that it does not give any reason in favor or against
half-integer or integer values of $J$ for the quantized skyrmions. We
will argue later that the fermionic character of the quantized baryon
can be revealed in a model with $SU(3)\times SU(3)$ symmetry by inclusion of the Wess-Zumino-Witten term.

An extension of the theory to three flavors with chiral $SU(3)\times
SU(3)$ symmetry is essential if one wants to consider the baryon octet
and decuplet, see for details Ref.~ \cite{sol3,sol3-2}. We assume the generic form of the underlying
Lagrangian remains unchanged  $\mathcal{L}=\mathcal{L}^{(2)}+\mathcal{L}^{(4)}$. However, the chiral field $U$ needs to be increased to a
$SU(3)$ field with the mesons fields $\phi^{a}, a=1,...,8$, which in
addition to the pions, contains the kaons and the octet component of
the $\eta$
\begin{equation}
U=\exp\left(i\sum_{a=1}^{8} \frac{\phi^{a}}{f_{a}}\lambda^{a} \right),
\end{equation}
where $\lambda^{a}$ denotes the Gell-Mann matrices. The decay
constants $f_{a}$ are defined through the gradient expansion of the
axial-vector current, analogous to the case with only two flavors. In a
similar fashion to the two flavor case, one can obtain solitonic solutions of
the theory and quantization can be done by introducing the collective
rotations.

In order to link the effective
meson theory to QCD, one should firstly consider if all symmetries of
the Lagrangian $\mathcal{L}$ are in accordance with QCD. Witten \cite{sol4}
observed that the Lagrangian Eq.~(\ref{sk-2}) possesses an extra discrete
symmetry that is not a symmetry of QCD. Under parity transformation
$P$, the pseudoscalar meson fields as described by QCD should obey
$P\boldsymbol{\pi}(x,t)=-\boldsymbol{\pi}(-x,t)$. In our meson theory this means,
\begin{equation}
P: \boldsymbol{x}\to -\boldsymbol{x},\hspace{0.5cm} t\to t,\hspace{0.5cm} U\to U^{\dag}. \label{sk-pa}
\end{equation}
But it is obvious that the Lagrangian Eq.~(\ref{sk-2}) is invariant under
$\boldsymbol{x}\to -\boldsymbol{x}$ and $U\to U^{\dag}$,
separately. In order to break this unwanted symmetry, one needs to add
some extra term to the meson action. Unfortunately, there exists no
local term in four spacetime dimension which can be added to the
Lagrangian, so as to get rid of this separate symmetry. However, it is very easy to look for such extra
term by considering the equation of motion. Witten \cite{sol4} suggested that the simplest term (with lowest possible
number of derivatives) which needs to be added are as follows
\begin{equation}
\frac{f_{\pi}^{2}}{8}\partial^{\mu}C_{\mu}+\lambda \epsilon^{\mu\nu\rho\sigma}C_{\mu}C_{\nu}C_{\rho}C_{\sigma}=0,\label{swi-1}
\end{equation}
where $C_{\mu}=U^{\dag}\partial_{\mu}U$. The new term is odd under
$\boldsymbol{x}\to -\boldsymbol{x}$ while the first and similar
higher-order terms are even. However, the new term is even under
transformation $U\to U^{\dag}$ while the first term is odd. Therefore,
Eq.~(\ref{swi-1}) is invariant only under the combined action of $P$
Eq.~(\ref{sk-pa}). The problem now is that the four-dimensional action
corresponding to the new extra term can not be written in a
chirally invariant form. However, this action can be rewritten in such a form in five dimensions. Therefore, we extend the coordinate
space to five-dimensional manifold $M_{5}$ in such a way that our conventional
four-dimensional spacetime $M_{4}$ is the boundary of a five-dimensional
volume i.e. $\partial M_{5}=M_{4}$. Therefore, one can write
\begin{equation}
\Gamma=\lambda\int_{M_{5}}\epsilon^{ijklm}\Tr(C_{i}C_{j}C_{k}C_{l}C_{m})d^{5}x. \label{sk-wzw}
\end{equation}
This action indeed leads to the equations of motion Eq.~(\ref{swi-1})
where written in four-dimensional spacetime by means of
Stokes'theorem. Witten in his remarkable paper \cite{sol4}, argued that the coefficient $\lambda$ in above
equation must be integer multiple of a normalization factor,
\begin{equation}
\lambda=n\frac{-i}{240\pi^{2}}.
\end{equation}
This is comprehensible, since the path-integral formulation requires
the action to be changed by a multiple of $2\pi$ when going from
$M_{5}$ to its complement, which has the identical boundary with
opposite orientation. This is indeed in the same spirit of Dirac's
quantization of a magnetic monopole. The physical meaning of the
integer $n$ is fixed through a connection to the Wess-Zumino action
\footnote{ The Wess-Zumino action was introduced to account for anomalies
which occur through the renormalization of fermion loops in quantum field theories where
pseudoscalar mesons are coupled to fermions} \cite{sol5}. Witten
included the photon fields in a gauge invariant way which generates a
vertex for the decay $\pi^{0}\to \gamma\gamma$ with
\begin{equation}
\frac{-n}{96\pi^{2}f_{\pi}}\pi^{0}\epsilon_{\mu\nu\rho\sigma}F^{\mu\nu}F^{\rho\sigma},
\end{equation}
where $F^{\mu\nu}$ is the field strength tensor of the photon. One can
immediately compare this result with the well-known triangle anomaly
in QCD and find that $n=N_{C}$. In this way, the effects of anomalies
in QCD are correctly reproduced by the action Eq.~(\ref{sk-wzw}) which
is called Wess-Zumino-Witten (WZW) term. Notice that if one considers
an adiabatic $2\pi$ rotation of the soliton, the WZW term produces a
contribution $N_{c}\pi$ to the action while other terms do not
contribute. Therefore, the soliton acquires a phase $(-1)^{N_{c}}$ for
such a rotation as required for fermions with $N_{c}$ odd or bosons
with $N_{c}$ even.  It is interesting that for two flavours, this
conclusion can not be made since the WZW term is zero. Therefore, for
three flavors, the WZW term provides some hint about the statistical
difference between baryons and mesons (fermions or bosons).  The fact
that the Skyrme Lagrangian needs to be augmented by the WZW term
indicates that the underlying physics of the Skyrme is a model of
quarks and gluons which possess QCD properties.

As we discused in the last
chapter, one of the simplest but viable quark model at the present is
the NJL model. The solitonic solutions of NJL models have been
extensively investigated \cite{sol3-2}. Notice that, in contrast to Skyrme-like
models with infinite energy barriers separating sectors with different
winding numbers, chiral quark models, such as the NJL models have
finite energy barriers separating the different sectors, and they give
rise to so-called non-topological solitons.

Despite all appealing features of soliton models, there are some
shortcomings: It is well-known that solitonic models are not very
accurate, e.g., in leading order of $N_{c}$, the quasi-classical soliton
configuration with quantized collective variables can produce 
baryonic observables with errors of about $20\%-30\%$, and corrections due to
mesonic fluctuations seem to be very important. Another point of
weakness is that exotic states are very controversial in these models\footnote{As
Cohen \cite{sol6} argued the main reason is that the rigid-rotor quantization is
not valid for such states. In other words, the assumption that the
collective motion is orthogonal to vibrational motion is only true for
non-exotic motion, but the Wess-Zumino term induces mixing at leading
order between collective and vibrational motion with exotic quantum
numbers. Recent discovery of Pentaquark $\theta^{+}$ which was already
predicted based on soliton model have brought a lot of activity on this
subject.}.

\section{Bag models}
In 1974 MIT group \cite{qcd7} developed a new picture of hadrons based on the simple assumption
that the physical vacuum prohibits free quarks and gluons, but instead creates
bubbles of hadronic size in which quarks and gluons may propagate
ordinarily. This idea has been employed in various models: a hybrid
bag model where the nucleon consists of a quark bag surrounded by a meson cloud, the little bag model where in contrast to the hybrid bag,
pions are not allowed to propagate inside the bag, a cloudy bag model
where the mesons are constrained to the chiral circle and are allowed
inside bag, and finally the chiral bag model where the constrained
mesons outside bag are described by the Skyrme Lagrangian. In all these models, there is
one extra parameter in the model, the bag radius, which provides some hint
about the quark and pion distributions. A review of various bag models can be found in Ref.~\cite{sol8}.

Chiral bag models seem to interpolate between two different aspects of QCD,
the long range-low energy (non-perturbative) and the small distance
(perturbative) behaviours. This idea was proposed by Rho et al \cite{sol9}. The very small volume $V$ represents the
perturbative domain of QCD containing quarks and gluons only, as opposed
to its complementary piece containing the confined phase of QCD with color-singlet objects such as mesons,
\begin{eqnarray}
\mathcal{L}&=&\mathcal{L}_{\text{bag}}\theta_{V}+\mathcal{L}_{\text{meson}}(1-\theta_{V})+\mathcal{L}_{\text{boundary}}\delta_{V},\nonumber\\
\mathcal{L}_{\text{bag}}&=&-i\bar{q}\gamma_{\mu}\partial^{\mu}q,\nonumber\\\
\mathcal{L}_{\text{boundary}}&=&-\bar{q}\exp\left(-\gamma_{5}\hat{x}.\hat{\tau}\Theta(r)\right)q,\nonumber\
\end{eqnarray}
where inside the bag we have massless quark fields $q$ and outside we
have chiral meson fields $U$ which obeys the Skyrme Lagrangian (for
simplicity the hedgehog configurations
$U=\exp\left(i\tau.\hat{x}\Theta(r)\right)$ with spherical bags of radius $R$ are assumed ). The boundary
term cause the full Lagrangian to be invariant under a combined chiral
symmetry;
\begin{equation}
q'\to\exp\left(i\gamma_{5}\alpha.\tau\right)q, \hspace{2cm} U'\to \exp\left(-i\alpha.\tau\right)U\exp\left(-i\alpha.\tau\right).
\end{equation}
We already identified the topological charge carried by the meson
field as baryon number. On the other hand, the quarks inside the bag
each carry one third amount of baryon charge, therefore, it seems
puzzling that the total baryon number is not integral.  Goldstone and
Jaffe \cite{sol10} proved that a conjecture of Rho {\em et al}
\cite{sol9} that baryon number in the hybrid model remains one is
indeed right. The crucial observation they made is that the charge of
the vacuum baryon number, inside the bag is changed due to boundary
effects. This shows that the baryon number remains one regardless of
the profile of the Skyrme fields and the size of bag. It was later
proved that the total energy from bag, chiral fields and vacuum
(Casimir energy) is insensitive to the bag size as well
\cite{sol11}. Therefore, it is tantalising to assume that observables
should not depend on the details of the bag (e.g., its size)
also. This statement is called ``The Cheshire Cat Principle (CCP)''
\cite{sol12}. Topological quantities, like the baryon number satisfy
an exact CCP in (3+1) dimensions while non-topological observables
such as masses, static properties and also non-static properties
satisfy it approximately well \cite{sol12}.

The main problem with various bag models is that they are not fully
covariant and possible large modification of observables due to
quantum fluctuations make model prediction less reliable.

\section{Diquark-quark picture; Relativistic Faddeev approach}
In the previous sections, we reviewed the basic foundations of two well
established picture of baryons, motivated from QCD properties. Two
main shortcomings within these approaches, namely the uncertainty of
computed quantities and lack of covariance, make these not viable for
phenomenological usage in the intermediate-energy regime where a fully
covariant formulation is required.  A new generation of continuous
beam facilities such as CEBAF at TJNAF, ELSA, COSY, MAMI, etc, which
are designed to explore the intermediate energy between
non-perturbative and perturbative regime of QCD, needs accurate
covariant formulation to describe the forthcoming data. The
diquark-quark picture of baryons based on a relativistic Faddeev
approach, is a framework for such a fully covariant approach. This
approach has been extensively employed during the last decade, and is
phenomenologically very successful
\cite{njln1,njln2,njln3,njln4,o1,o2,o3,o4}. In this picture, a bound
state of a baryon \footnote{There is another very old-fashioned approach
to obtain baryon bound state which was invented in the early sixties,
the constituent quark models or quark potential models. In these
models one starts with simple potential e.g, the hyperfine type
interaction and employs 3-particle Schr\"odinger (or Dirac ) equation
to obtain the spectrum. This approach is not covariant and does not
incorporate the minimal field theoretical aspect of QCD, like quarks
degree of freedom. We refrain from discussing this approach here.} is
obtained as a pole of the three-quark correlation by summing over
infinitely many interaction graphs. This process is very similar to
obtaining a two-body bound state which leads to the Bethe-Salpeter
equation.

In the context of local quantum field theory, the few-body problem
seems to be ill-defined, since any restriction on degrees of freedom
(e.g., particle numbers) may spoil the Lorentz invariance (e.g., on the
equal-time quantization, the boosts generators involve interactions
and change the number of particles, therefore limiting the number of particles is against the Lorentz invariance). Nevertheless, we know from a
phenomenological point of view that a fixed number of constituent
quarks might be enough to describe baryons. Therefore, we introduce
the notion of baryon wave functions as matrix elements of three quark
operators between the physical vacuum $|\Omega\rangle$ and a (nucleon)
bound state $|P_{N}\rangle$;
$\Psi\sim\langle\Omega|T(qqq)|P_{N}\rangle$. Having said that, QCD
vacuum is non-perturbative and indeed possesses non-trivial
condensates, and thus the wave function contains sea quark and gluonic
parts. However, it may be reasonable to ignore all other operator
matrix elements which involve an arbitrary number of quarks and gluons
as irrelevant in comparison to the dominant amplitude $\Psi$. One can
now solve the three-body problem by means of the Green's function
formulation of quantum field theory. As we will show, the
non-perturbative feature of the vacuum can then be effectively
incorporated into the formalism in a systematic fashion.

In the following we introduce an approximation scheme based on
the relativistic Faddeev approach to simplify the three-quark problem
in form of a diquark and quark interacting via quark exchange. For
simplicity, we use a formal presentation; all color, flavor and
Dirac indices are implicit in the single particle labels. We use
a Euclidean metric in momentum space. We denote a dressed single quark propagator by $S_{i}$, with
\begin{equation}
(2\pi)^{4}\delta^{4}(k_{i}-p_{i})S_{i}(k_{i};p_{i})=
\int d^{4}x_{k_{i}}d^{4}x_{p_{i}}e^{i(k_{i}.x_{k_{i}}-p_{i}.y_{p_{i}})}\langle 0|Tq_{i}\left(x_{i})q_{j}(x_{j}\right)|0\rangle.
\end{equation}
In our definition of Green's functions and bound state matrix
elements, we always take out one $\delta$-function corresponding to conservation of energy-momentum.
We define the full quark
six-point function (or the three-quark correlation function) as
\begin{eqnarray}
&&(2\pi)^{4}\delta^{4}\left(\sum_{i=1}^{3}(k_{i}-p_{i})\right)G(k_{i};p_{i})=\nonumber\\
&&\int\Pi_{i=1}^{3} d^{4}x_{k_{i}}d^{4}y_{p_{i}} \exp\left(i\sum_{i=1}^{3}(k_{i}.x_{k_{i}}-p_{i}.y_{p_{i}})\right)\langle 0|T\Pi_{i=1}^{3}q(x_{k_{i}})\bar{q}(y_{p_{i}})|0 \rangle.\
\end{eqnarray}
The three-quark correlation function satisfies the Dyson equation,
\begin{equation}
G=G_{0}+G_{0}\otimes K\otimes G, \label{Fa-1}
\end{equation}
where $G_{0}$ denotes the disconnected three dressed quark propagator and $K$ stands for the three-quark
scattering kernel containing all two and three-body irreducible
diagrams. The symbol "$\otimes$" denotes summation and integration
over all internal and dummy indices. A bound state of mass $M$
with wave function $\Psi$ emerges as a pole of the three-quark
correlation function,
\begin{equation}
G(k_{i},p_{i})\sim
\frac{\Psi(k_{1},k_{2},k_{3})\Psi(p_{1},p_{2},p_{3})}{P^{2}+M^{2}}, \label{bound-n}
\end{equation}
where $P=p_{1}+p_{2}+p_{3}$ and we defined $\Psi$ as a three-body
wave function which represents the transition matrix element between
the vacuum and a bound state with mass $M$,
\begin{equation}
(2\pi)^{4}\delta^{4}\left(\sum_{i=1}^{3}(p_{i}-P)\right)\Psi(p_{1},p_{2},p_{3})=\int \Pi_{i=1}^{3} d^{4}x_{i} \exp\left(i\sum_{i=1}^{3}p_{i}.x_{i}\right) 
\langle 0|\Pi_{i=1}^{3}q_{i}(x_{i})|P\rangle. 
\end{equation}
We now substitute the bound state parametrization Eq.~(\ref{bound-n})
into Eq.~(\ref{Fa-1}) and compare the residues. This leads to the
homogeneous bound state equation,
\begin{equation}
\Psi=G_{0}\otimes K\otimes \Psi, \hspace{1cm}
\longleftrightarrow\hspace{1cm} G^{-1}\otimes \Psi=0. \label{fa-2}
\end{equation}
Solving this equation exactly is almost impossible since neither
the detail of all two- and three-particle irreducible graphs appearing in $K$,
nor the full dressed quark propagator contained in $G_{0}$ are
known. It is well known that the problem becomes more tractable
if one employs the Faddeev approximation, by discarding all
three-body irreducible graphs from the interaction kernel $K$. In
this way one can write the kernel as a sum of three two-body
interaction kernels,
\begin{equation}
K=K_{1}+K_{2}+K_{3}, \label{fa-3}
\end{equation}
where $K_{i}$ with $i=1,2,3$ refers to the interactions of quark
pairs $(jk)$ with a spectator quark $(i)$. The two-quark propagators
$g_{i}$ satisfy their own Dyson equation with kernel $K_{i}$,
\begin{equation}
g_{i}=G_{0}+G_{0}\otimes K_{i}\otimes g_{i}, \label{fa-4}
\end{equation}
where $g_{i}$ and $K_{i}$ are defined in three-body space, since
$G_{0}$ is defined in three-body space. Hence, the former contains a
factor $S_{i}$ (the propagator of the spectator quark), and the latter
contains a factor $S^{-1}_{i}$ i.e., $K_{i}=k_{qq}\otimes
S^{-1}_{i}$. One may associate a disconnected scattering amplitude to
every spectator quark $(i)$, i.e. $T_{i}=t_{i}\otimes S^{-1}_{i}$,
where $t_{i}$ describes the scattering between the quarks $(j)$ and $(k)$
in two-quark subspace. The matrix $T_{i}$ is obtained by amputating
all incoming and outgoing quark legs from the connected part of $g_{i}$,
\begin{equation}
g_{i}=G_{0}+G_{0}\otimes T_{i}\otimes G_{0}. \label{fa-5}
\end{equation}
By combining the two previous equations, the Dyson equation for $T_{i}$ can be found as
\begin{equation}
T_{i}=K_{i}+K_{i}\otimes G_{0}\otimes T_{i}.
\end{equation}
Now we define Faddeev components $\Psi_{i}$ via Eqs.~(\ref{fa-2},\ref{fa-3})
\begin{equation}
\Psi_{i}=G_{0}\otimes K_{i}\otimes \Psi, \label{fa-6}
\end{equation}
where we have $\Psi=\sum \Psi_{i}$. We rewrite the Eqs.~(\ref{fa-5})
as $g_{i}\otimes G_{0}^{-1}=1+G_{0}\otimes T_{i}$ and plug this
expression into Eq.~(\ref{fa-6}) and make use of Eq.~(\ref{fa-4}).  We then find the well-known Faddeev bound state equations,
\begin{equation}
\Psi_{i}=G_{0}\otimes T_{i}\otimes (\psi_{j}+\Psi_{k})=S_{j}S_{k}\otimes t_{i}\otimes (\Psi_{j}+\Psi_{k}). \label{fa-7}
\end{equation}
We have shown that the complicated three-quark problem can be
systematically simplified by employing the full two-quark correlation
function $t_{i}$, instead of the kernel $K$. In this way the eight
dimensional Eq.~(\ref{fa-2}) is reduced to a set of coupled
four-dimensional equations (\ref{fa-7}). A further simplification of
this equation can be achieved by approximating the full two-quark correlation $t_{i}$ as a sum of separable correlations,
\begin{equation}
t_{i}(k_{1},k_{2};p_{1},p_{2})=\sum_{a} \chi_{i}^{a}(k_{1},k_{2})D^{a}_{i}(k_{1}+k_{2})\bar{\chi}_{i}^{a}(p_{1},p_{2}), \label{fa-sep}
\end{equation}
where the function $\chi_{i}^{a}$ is the vertex function of
two-quark with a diquark and $\bar{\chi}_{j}^{a}$ denotes its
complex conjugate. The index ``$a$'' denotes the different
channels of the diquarks, and $D^{a}_{i}$ is the corresponding
diquark propagator. Note that separability implies that $t_{i}$ does not
depend on any of the scalar products $k_{i}.p_{j}$. The diquarks parametrization to some extend contains the unknown
non-perturbative physics within the baryon structure. A natural
Ansatz for the Faddeev component $\Psi_{i}$ is given by
\begin{equation}
\Psi^{\alpha\beta\gamma}_{i}(p_{i},p_{j},p_{k})=\sum_{a} S^{\alpha\alpha'}_{i} S^{\beta\beta'}_{j}S^{\gamma\gamma'}_{k}\chi_{i,\beta'\gamma'}^{a}(p_{i},p_{k})D^{a}_{i}(p_{j}+p_{k})\Phi^{a}_{i\alpha'}(p_{i},p_{j}+p_{k}), \label{fa-phi}
\end{equation}
where summation over repeated indices is assumed. The Greek
multi-indices $\alpha,\beta,..$ denote color, flavor and Dirac
indices, and $i,j,k$ indicate the type of quarks. The quark $(i)$ and diquark labels
$(jk)$ are fixed. The quantity $\Phi$ is the baryon-quark-diquark vertex
function and depends only on the relative momentum between the
momentum of the spectator quark, $p_{i}$ and the momentum of the
diquark quasi-particle, $P_{j}+P_{k}$. In a relativistic formulation
of a few-body system there is no unique definition of the momentum
transfered between individual particles. Therefore, we introduce a new
parameter $\eta$ which parametrises this ambiguity and shows the
distribution of the total momentum within the system (diquark and
quark). We define a relative momentum between the quark $(i)$ and the
diquark consisting of the quarks $(jk)$ by
\begin{equation}
p=(1-\eta)p_{i}-\eta(p_{j}+p_{k})=p_{i}-\eta P,
\end{equation}
where $P=p_{1}+p_{2}+p_{3}$. The physical properties of the baryons
will indeed not depend on $\eta$. One can employ the
definition of (\ref{fa-phi}) to rewrite the Faddeev equation
(\ref{fa-7}) in terms of a vertex function $\Phi$,
\begin{equation}
\Phi^{a}_{i,\alpha}=\sum_{b}[\bar{\chi}^{a}_{i,\beta\gamma}S^{\gamma\gamma'}_{k}\chi^{b}_{j,\gamma'\alpha}][D^{b}S^{\beta\beta'}_{j}\Phi^{b}_{j\beta'}]+(j\longleftrightarrow k). \label{fa-beth}
\end{equation}
In the above derivation we assumed that the quark-diquark vertex function
is antisymmetric under the exchange of the quark labels
($\chi^{a}_{i,\beta\gamma}=-\chi^{a}_{i,\gamma\beta}$) as a
consequence of the Pauli exclusion principle. Eq.~(\ref{fa-beth})
resembles a Bethe-Salpeter equation; the first bracket can be
conceived as a quark-diquark interaction kernel since it contains the
exchange of a single quark between a quark and a diquark, the second
term couples the interaction kernel to baryon-quark-diquark vertex via
diquarks and quarks propagator.
\begin{figure}[!htp]
       \centerline{\includegraphics[width=11 cm]
                                   {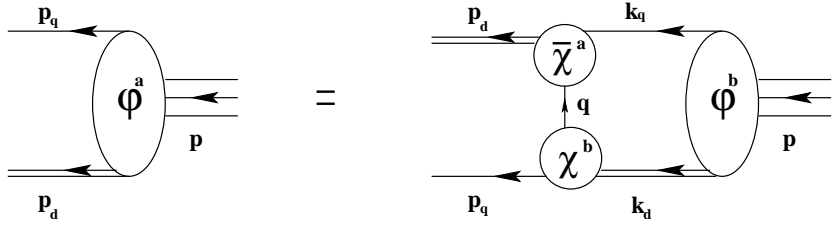}}
\caption{The coupled Bethe-Salpeter equations for the vertex function $\Phi$.}
\end{figure}
For identical quarks the
antisymmetrization of the vertex functions is essential, we can now sum
over the type of particle and drop the index like $(i)$. Therefore, equation (\ref{fa-beth}) can be rewritten as,
\begin{eqnarray}
\Phi^{a}_{\alpha}(p,P)&=&\sum_{b}\int\frac{d^{4}k}{(2\pi)^{4}}K_{\alpha\gamma}(k,p,P)G_{0,\gamma\beta}(k,P)\Phi^{b}_{\beta}(k,P),\nonumber\\
G_{0,\gamma\beta}(k,P)&=&2S_{\gamma\beta}(k_{q})D^{b}(k_{d}),\nonumber\\
K_{\alpha\gamma}(k,p,P)&=&\bar{\chi}^{a}_{\beta'\gamma}(k,q)S_{\beta'\gamma'}\chi^{b}_{\alpha\gamma'}(q,p_{q}), \
\end{eqnarray}
where we define the spectator quark and diquark momenta as $k_{q}(p_{q})=\eta
P+k(p)$, $k_{d}(p_{d})=(1-\eta)P-k(p)$, respectively. Momentum
conservation fixes the momentum of exchanged quark $q=-k-p+(1-2\eta)P$, see fig.~5.1. 
The factor of two in the above presentation originates from the summation
over the type of quarks and can be absorbed into the definition of diquark
propagator.

In conclusion, we managed to recast the three-body Faddeev
equation in the form of an effective two-body Bethe-Salpeter equation
between the diquarks and the quarks, having summed over the ladder-type quark
exchange diagrams between the quarks and the diquarks. The only assumptions we
have made are; 1) we neglect all three-particle irreducible
contributions 2) we model the connected two-body correlation as a sum
of separable terms which are identified by diquark channels, see
Eq.~(\ref{fa-sep}). In order to obtain an equation for physical
baryons, we have to project Eq.~(\ref{fa-beth}) onto
the baryon quantum numbers. This will be carried out for
a model in the next section.

\chapter{Baryons structure in a non-local quark confining NJL model}
\section{Introduction}
The NJL model is a successful (low-energy) phenomenological field theory
inspired by QCD \cite{njl}. The model is constructed to obey the basic
symmetries of QCD in the quark sector, but unlike the case of low-energy QCD,
quarks are not confined. The basic ingredient of the model, apart
from the standard bilinear Lagrangian in the quark fields, is a
zero-range interaction containing four fermion fields.  This means
that the model is not renormalizable. If we make the standard
one over the number of colours ($1/N_C$) loop expansion, already at
one-loop level an ultraviolet subtraction (usually implemented by a
cut-off) supplemented with a regularisation method is required.  The
value of the cut-off can be related to the scale of physical processes
not included in the model, and thus determines its range of validity.
Consequently, processes involving a large momentum transfer, such as
anomalous decay, can not be described by the model. At higher orders
in the loop expansion, which are necessary for calculating mesonic
(baryonic) fluctuations \cite{njl-cut,cut-b}, one needs extra cut-off
parameters. It is hard to determine these parameters from independent
physics, and thus to build a viable phenomenology. A similar problem
appears in the diquark-quark picture of baryons where an additional
cut-off parameter is required to regularise the diquark-quark loops
\cite{cut-b}. It has been shown that a renormalizable extension of the
NJL model (at least to one-loop level) \cite{ren} can be constructed
by matching the NJL-contact interaction at low energy with a one-gluon
exchange type interaction above the Landau pole. Another way to cope
with the non-renormalisability of the model is to embed this model
into a renormalizable theory such as the linear $\sigma$-model and apply a
renormalisation group approach \cite{p,p-1}. However, all these
approaches add extra complexity which make them difficult to apply
for anything but very simple problems.

Another drawback of the model is the absence of confinement, which
makes it questionable for the description of few-quark states
and for quark matter.  If energetically allowed, the mesons of the
model can decay into free quark-antiquark pairs, and the presence
unphysical channel is another limitation of the applicability of NJL
model. At the same time, it is also known that the NJL model exhibits
a zero-temperature phase transition at unrealistically low baryon
density
\cite{njl-ma}. This problem is caused by the formation
of unphysical coloured diquark states. These may be explicitly excluded at
zero density by a projection onto the physical channels, but dominate
the behaviour at finite density.  The model is not able to describe nuclear
matter, even in the low-density regime \cite{w}. 

We do not know how to implement colour confinement in the model and,
anyway, the exact confining mechanism of QCD is still unknown. In the
context of an effective quark theory, a slightly different mechanism
of ``quark confinement'' can be described by a quark propagator which
vanishes due to infra-red singularities \cite{rho2} or when it does
not produce any poles corresponding to asymptotic quark states
\cite{rho1,asy}. Another realisation of quark confinement
can be found in Ref.~
\cite{rho}. It has been shown that a non-local covariant extension of
the NJL model inspired by the instanton liquid model
\cite{non1} can lead to quark confinement for acceptable values of
the parameters
\cite{pb}. Here the quark propagator
has no real pole, and consequently quarks do not appear as asymptotic
states. The quark propagator has infinitely many pairs of
complex poles corresponding to quarks which have a finite lifetime. This phenomenon
was also noticed in Schwinger-Dyson equation studies in QED and QCD
\cite{sde-p,sde-new,sde-qe}. 

We can simply accept the appearance of
these poles as an artifact of the naive truncation scheme
involved. However, it has been recently suggested that it might be a
genuine feature of the full theory, and be connected with the
underlying confinement mechanism \cite{sde-new,sde-qe}. For example,
it has been shown by Maris that if we remove the confining potential
in QED in 2+1D the mass singularities are located almost on the
time axis, and if there is a confining potential, the mass-like
singularities move from the time axis to complex momenta
\cite{sde-qe}. In this chapter, we study this kind of confinement from
another viewpoint. We show that when we have quark confinement in the
non-local NJL model, the baryons become more compact, compared to a
situation where we have only real poles for quark propagator.

There are several other advantages of the non-local version of the
model over the local NJL model: the dynamical quark mass is
momentum-dependent and also found in lattice simulations of QCD
\cite{lat-m}. More importantly, the
non-locality regularises the model preserving anomalies \cite{pb}, and the regulator makes the theory finite to
all orders in the $1/N_{c}$ expansion, and leads to small
next-to-leading order corrections \cite{pb-2}. As a result, the
non-local version of the NJL model may have more predictive power.

The instanton-liquid model is only one way to motivate such a model
\cite{non1}.  Many effective field theories constructed by the
Wilsonian renormalisation group approach lead to a non-locality, at
least as irrelevant terms in the renormalisation group
sense. Non-locality also emerges naturally in the Schwinger-Dyson
resummation
\cite{asy}. 
Considerable work has been done on these nonlocal NJL models including
applications to the mesonic sector \cite{pb,a1}, phase transitions
at finite temperature and densities \cite{a2}, and the study of chiral
solitons \cite{a3}.

In this chapter we present our first results
from a calculation of the relativistic Faddeev equation for a
non-local NJL model \cite{new-a3}, based on the covariant diquark-quark picture of baryons
\cite{njln1,njln2,njln3,njln4,o1,o2,o3,o4}. 
Such an approach has been extensively employed to study baryons in the
local NJL model, see, e.g., Refs.~\cite{njln1,njln2,njln3,njln4}.
We include both scalar and the axial-vector diquark correlations. We
do not assume a special form for the interaction Lagrangian,
but we rather treat the coupling in the diquark channels as free
parameters and consider the range of coupling strengths which lead to
a reasonable description of the nucleon. We construct diquark and nucleon 
solutions and study the possible implications of the quark
confinement in the solutions. Due to the separability of the non-local
interaction, the Faddeev equations can be reduced to a set of
effective Bethe-Salpeter equations. This makes it possible to adopt
the numerical method developed for such problems in Refs.~\cite{o1,o2,o3,o4}.


\section{A non-local NJL model\label{sec:model}}
We consider a non-local NJL model Lagrangian with $SU(2)_{f}\times
SU(3)_{c}$ symmetry.
\begin{equation}
\mathcal{L}=\bar{\psi}(i\slashit\partial-m_{c})\psi+\mathcal{L}_{I},
\end{equation}
where $m_{c}$ is the current quark mass of the $u$ and $d$ quarks
and $\mathcal{L}_{I}$ is a chirally invariant non-local
interaction Lagrangian. Here we restrict the interaction terms to
four-quark interaction vertices.

There exist several versions of such non-local NJL models. Regardless
of what version is chosen, by a Fierz transformation one can rewrite
the interaction in either the $q\bar{q}$ or $qq$ channels, and we
therefore use the interaction strengths in those channels as
independent parameters. For simplicity we truncate the mesonic
channels to the scalar ($0^{+},T=0$) and pseudoscalar ($0^{-},T=1$)
ones. The $qq$ interaction is truncated to the scalar ($0^{+},T=0$)
and axial vector ($1^{+},T=1$) colour $\overline{3}$ $qq$ channels (the colour $6$ channels do not contribute to the colourless
three-quark state considered here). We parametrise the relevant part
of interaction Lagrangian as
\begin{eqnarray}\label{n1-rp}
\mathcal{L}_{I}&=&\frac{1}{2}g_{\pi} j_{\alpha}(x)j_{\alpha}(x)+g_{s}\overline{J}_{s}(x)J_{s}(x)+g_{a}\overline{J}_{a}(x)J_{a}(x),\nonumber\\
j_{\alpha}(x)&=&\int
d^{4}x_{1}d^{4}x_{3}f(x-x_{3})f(x_{1}-x)\overline{\psi}(x_{1})\Gamma_{\alpha}\psi(x_{3}),\nonumber\\
\overline{J}_{s}(x)&=&\int
d^{4}x_{1}d^{4}x_{3}f(x-x_{3})f(x_{1}-x)\overline{\psi}(x_{1})\big[\gamma_{5}C\tau_{2}\beta^{A}\big]\overline{\psi}^{T}(x_{3}),\nonumber\\
J_{s}(x)&=&\int
d^{4}x_{2}d^{4}x_{4}f(x-x_{4})f(x_{2}-x)\psi^{T}(x_{2})\big[C^{-1}\gamma_{5}\tau_{2}\beta^{A}\big]\psi(x_{4}).\nonumber\\
\overline{J}_{a}(x)&=&\int
d^{4}x_{1}d^{4}x_{3}f(x-x_{3})f(x_{1}-x)\overline{\psi}(x_{1})\big[\gamma_{\mu}C\tau_{i}\tau_{2}\beta^{A}\big]\overline{\psi}^{T}(x_{3}),\nonumber\\
J_{a}(x)&=&\int
d^{4}x_{2}d^{4}x_{4}f(x-x_{4})f(x_{2}-x)\psi^{T}(x_{2})\big[C^{-1}\gamma^{\mu}\tau_{2}\tau_{i}\beta^{A}\big]\psi(x_{4}),\
\end{eqnarray}
where $\Gamma_{\alpha}=(1,i\gamma_{5}\tau)$. The matrices
$\beta^{A}=\sqrt{3/2} \lambda^{A}(A=2, 5, 7)$ project onto the colour
$\overline{3}$ channel with normalisation
$\tr(\beta^{A}\beta^{A'})=3\delta^{AA'}$ and the ${\tau_{i}}$'s are
flavour $SU(2)$ matrices with $\tr
(\tau_{i}\tau_{j})=2\delta_{ij}$. The object $C=i\gamma_{2}\gamma_{5}$
is the charge conjugation matrix.

Since we do not restrict ourselves to specific choice of interaction,
we shall treat the couplings $g_{s}$, $g_{a}$ and $g_{\pi}$ as
independent parameters. We assume $g_{\pi,s,a}>0$, which leads to
attraction in the given channels (and repulsion in
the $q\bar{q}$ colour octet and $qq$ colour antisextet channels). The
coupling parameter $g_{\pi}$ is responsible for the pions and their
isoscalar partner $\sigma$. The coupling strengths $g_{s}$ and $g_{a}$
specify the behaviour in the scalar and axial-vector diquark channel,
respectively. 

For simplicity, we assume the form factor $f(x-x_{i})$
to be local in momentum space, since it leads to a separable
interaction
\begin{equation}
f(x-x_{i})=\int \frac{d^{4}p}{(2\pi)^{4}} e^{-i(x-x_{i})\cdot p}f(p). \label{forfa}
\end{equation}
It is exactly this separability that is also present in the instanton
liquid model \cite{sep-i}.  The dressed quark propagator $S(k)$ is now
constructed by means of a Schwinger-Dyson equation (SDE) in the
rainbow-ladder approximation.  Thus the dynamical constituent quark
mass, arising from spontaneously broken chiral symmetry, is obtained
in Hartree approximation\footnote{Notice that as we demonstrated in
section 4.4, the exchange diagrams (for four-fermion interactions) can always be cast in form of
direct diagram via a Fierz transformation. This means, the
Hartree-Fock approximation is equivalent to the Hartree approximation
with properly redefining coupling constants. Therefore, Hartree approximation
is as good as Hartree-Fock one as long as the interaction
terms in Lagrangian are not fixed by some underlying theory.} (the
symbol $\Tr$ denotes a trace over flavour, colour and Dirac indices
and $\tr_{D}$ denotes a trace over Dirac indices only)
\begin{equation}
M(p)=m_{c}+ig_{\pi}f^{2}(p)\int \frac{d^{4}k}{(2\pi)^{4}} \Tr
[S(k)] f^{2}(k), \label{n3-gap}
\end{equation}
where
\begin{equation}
S^{-1}(k)=\slashit{k}-M(k), \label{n3}
\end{equation}
one can simplify this equation by writing $M(p)$ in the form
\begin{equation}
M(p)=m_{c}+(M(0)-m_{c})f^{2}(p).  \label{nI}
\end{equation}
The non-linear equation can then be solved iteratively for 
$M(0)$.

Following Ref.~\cite{pb}, we choose the form factor to be Gaussian in
Euclidean space, $f(p_{E})=\exp(-p_{E}^{2}/\Lambda^{2})$, where
$\Lambda$ is a cutoff of the theory. If one assumes that $\Lambda$ is
related to the average inverse size of instantons $1/\bar{\rho}$, then
its value parametrises aspects of the non-perturbative properties of
the QCD vacuum \cite{non1}. This choice respects Poincar\'e invariance
and for certain values of the parameters it leads to quark, but not
colour, confinement.
For values of $M(0)$ satisfying
\begin{equation} \label{n4}
\frac{M(0)-m_{c}}{\sqrt{m^{2}_{c}+\Lambda^{2}}-m_{c}} >
\frac{1}{2}\exp\left(-\frac{(\sqrt{m^{2}_{c}+\Lambda^{2}}+m_{c})^{2}}{2\Lambda^{2}}\right)
\end{equation}
the dressed quark propagator has no poles at real $p^{2}$ in Minkowski
space ($p^{2}+M^{2}(p^{2})\neq 0$).  The propagator has infinitely
many pairs of complex poles, both for confining and non-confining
parameter sets. This is a feature of these models and due care should
be taken in handling such poles, which can not be associated with
asymptotic states if the theory is to satisfy unitarity.  One should
note that the positions of these poles depend on the details of the
chosen form factor and the cut-off, hence one may regard them as a
pathology of the regularisation scheme.  Since the choice of the
cut-off is closely related to the truncation of the mesonic channels,
(for example, if one allows mixing of channels, the cut-off and the
positions of poles will change. In Fig.~ 6.5 we have shown the
positions of the first poles of the quark propagator for various
cutoff. We have examined that in the presence of $\pi a_{1}$ mixing, these
positions will change, but it follows very similar trend). Even though the
confinement in this model has no direct connection to the special
properties of the pion, there is an indirect connection through the
determination of the parameters from the pionic properties.
\begin{figure}[!tp]
       \centerline{\includegraphics[width=15 cm] {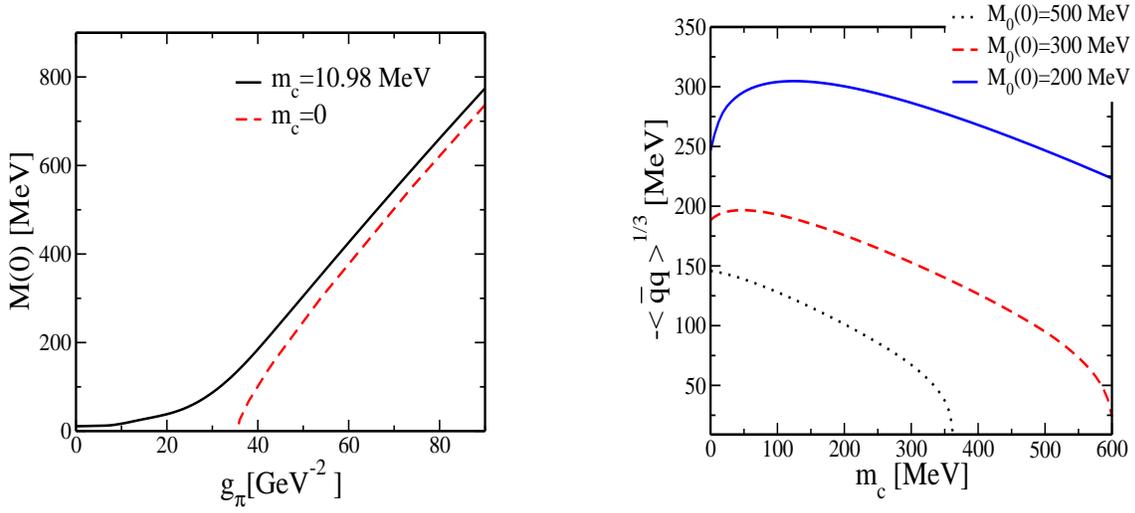}}
\caption{Quark condensate for various sets of parameters as a function of current quark mass (right). The dynamical quark mass at zero momentum as a
       function of coupling (left) for parameter set: $M_{0}(0)=300$ MeV,
       $\Lambda=860.58$ MeV $m_{c}=10.98$, the mass generation occur at critical coupling $g^{cr}_{\pi}=27.0$ [GeV$^{-2}$].}
\end{figure}
From the gap equation Eq.~(\ref{n3-gap}) one can show that dynamical
symmetry breaking occurs for $1/g_{\pi}-1/g^{cr}_{\pi}<0$, where
$1/g^{cr}_{\pi}=\frac{N_{c}N_{f}\Lambda^{2}}{12\pi^{2}}$, $N_{c}$ and
$N_{f}$ are the number of colours and  flavours, respectively.  
For
$g_{\pi}>g_{\pi}^{cr}$ fermions become massive and the vacuum
accommodates a non-vanishing condensate $\langle
\bar{\psi}\psi\rangle$, and consequently there exists a massless
Nambu-Goldstone boson, see Fig~6.1.

\section{Meson channel\label{sec:meson}}
The quark-antiquark $T$-matrix in the pseudoscalar channel can be
solved by using the Bethe-Salpeter equation in the random phase
approximation (RPA), as shown in Fig.~\ref{fig:meson}, see
Ref.~\cite{pb}.
\begin{figure}
\centerline{\includegraphics[clip,width=10 cm]{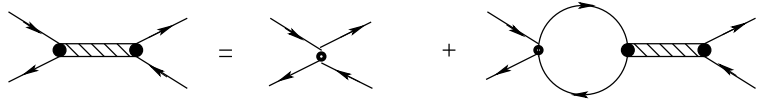}}
\caption{A graphical representation of the Bethe-Salpeter equation for the 
$\bar{q}q$ $T$-matrix in RPA approximation. The solid lines denote the
dressed quark propagators Eq.~(\ref{n3}) and shaded boxes denote meson
propagators.\label{fig:meson}}
\end{figure}
\begin{eqnarray}
T(p_{1},p_{2},p_{3},p_{4})&=&f(p_{1})f(p_{2})\big[i\gamma_{5}\tau_{i}\big]\frac{g_{\pi}i}{1+g_{\pi}J_{\pi}(q^{2})}\big[i\gamma_{5}\tau_{i}\big]
f(p_{3})f(p_{4})\label{pi1}\nonumber\\
&&\times\delta(p_{1}+p_{2}-p_{3}-p_{4}),\
\end{eqnarray}
where 
\begin{eqnarray}
 J_{\pi}(q^{2})&=& i \Tr \int
 \frac{d^{4}k}{(2\pi)^{4}}f^{2}(k)\gamma_{5}\tau_{i}S(k)\gamma_{5}\tau_{i}S(q+k)f^{2}(q+k),\nonumber\\
&=&6i\int\frac{d^{4}k}{(2\pi)^{4}}\tr_{D}[\gamma_{5}S(k)\gamma_{5}S(k+q)]f^{2}(k)f^{2}(q+k),\label{pi2}\
\end{eqnarray}
where $q$ denotes the total momentum of the $\bar{q}q$ pair. The pion
mass $m_{\pi}$ corresponds to the pole of $T$-matrix. One 
immediately finds that $m_{\pi}=0$ if the current quark mass $m_{c}$
is zero, in accordance with Goldstone's theorem. The residue of the
$T$-matrix at this pole has the form
\begin{equation}
V^{\pi}(p_{1}, p_{2})=ig_{\pi qq}[\openone_{c}\otimes \tau^{a}\otimes\gamma_{5}]f(p_{1})f(p_{2}),    \label{pi3}      
\end{equation}
where $g_{\pi qq}$ is the pion-quark-antiquark coupling
constant and is related to the corresponding loop integral $J_\pi$ by
\begin{equation}
g^{-2}_{\pi qq}=\left.\frac{dJ_{\pi}}{dq^{2}}\right|_{q^{2}=m^{2}_{\pi}}.
\end{equation}
Notice that $Z=g^{2}_{\pi qq}$ can be regarded as a pion wavefunction
renormalisation constant. For space-time dimension $D=4$, one can show
$Z^{-1}\varpropto \Lambda^{2}$ (for the local NJL model we have
$Z^{-1}\varpropto \Ln\Lambda$), therefore in the continuum limit
$\Lambda\to\infty$ we have $Z=0$ which is precisely the compositeness
condition \cite{win}. In this extreme limit pions become pointlike. The
cutoff for spacetime $D=4$ can be removed only at the expense of making
the theory trivial in the continuum limit. It has been shown that for
four-fermion theories the renormalisability, nontriviality and
compositeness are intimately related \cite{p-1,si}.
\begin{figure}
\centerline{\includegraphics[clip,width=10 cm]{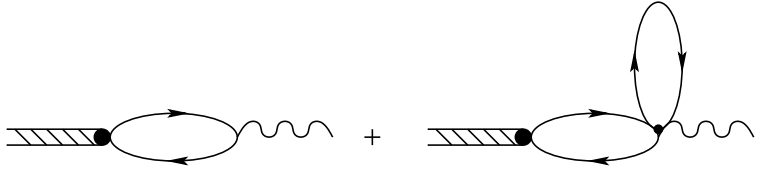}}
\caption{One-pion-to-vacuum matrix element in RPA, contributing to the weak pion decay. The lines are as defined in Fig.~\ref{fig:meson}. The wavy line denotes a weak decay.\label{fig:fpi}}
\end{figure}

The pion decay constant $f_{\pi}$ is obtained from the coupling of the
pion to the axial-vector current. Notice that due to the non-locality
the axial-vector current is modified \cite{pb,non-j} and consequently
the one-pion-to-vacuum matrix element gets the additional contribution
shown in Fig.~\ref{fig:fpi}. This new term is essential in order to
maintain Gell-Mann-Oakes-Renner relation \cite{pb} and makes a
significant contribution. The pion decay constant is
given by
\begin{eqnarray}
f_{\pi}&=&\frac{ig_{\pi \bar{q}q}}{m_{\pi}^{2}}\int
\frac{d^{4}k}{(2\pi)^{4}}\Tr[\slashit{q}\gamma_{5}\frac{\tau_{a}}{2}(S(p_{-}))\gamma_{5}\tau_{a}(S(p_{+}))]f(p_{-})f(p_{+})\nonumber\\
&+&\frac{ig_{\pi}}{2m^{2}_{\pi}}\int
\frac{d^{4}k}{(2\pi)^{4}}\Tr[S(k)]\int
\frac{d^{4}k}{(2\pi)^{4}}\Tr[V^{\pi}(p_{-},p_{+})S(p_{-})\gamma_{5}\tau_{a}S(p_{+})]\nonumber\\
&&\times
[f^{2}(k)\left(f^{2}(p_{+})+f^{2}(p_{-})\right)-f(p_{+})f(p_{-})f(k)\left(f(k+q)+f(k-q)\right)], \label{fpi}\
\end{eqnarray}
where $V_{\pi}(p_{-},p_{+})$ is defined in Eq.~(\ref{pi3}), with notation $p_{\pm}=p\pm\frac{1}{2}q$.    
\begin{table}
\begin{center}
\caption{The parameters for the sets $A$ and $B$,
fitted to $f_{\pi}=92.4$ MeV and $m_{\pi}=139.6$ MeV. Resulting values of the
dynamical quark mass $M(0)$ are also shown.\label{tab:modpar}}
\begin{tabular}{rllclc}
\hline
\hline
Parameter &set A & set B  \\
\hline
$M(0)$ (MeV) & 297.9& 351.6  \\
$M_{0}(0)$ (MeV) & 250 & 300 \\
$m_{c}$ (MeV) & 7.9 & 11.13 \\
$\Lambda$ (MeV) & 1046.8 &847.8 \\
$g_{\pi} (\text{GeV}^{-2})$ &31.6 & 55.80 \\
\hline
\hline
\end{tabular}
\end{center}
\end{table}

\begin{table}
\begin{center}
\caption{The first two sets of  poles of the quark propagator (in  magnitude) in
the Minkowski frame. \label{tab:quarkpoles}}
\begin{tabular}{rllc}
\hline
\hline
set A  & set B \\
\hline
$\pm 391$ MeV&$\pm 408 \pm 238i$ MeV\\
$\pm 675$ MeV& $\pm 1575\pm 307i$ MeV\\
\hline
\hline
\end{tabular}
\end{center}
\end{table}
\begin{figure}[!tp]
       \centerline{\includegraphics[width=14 cm] {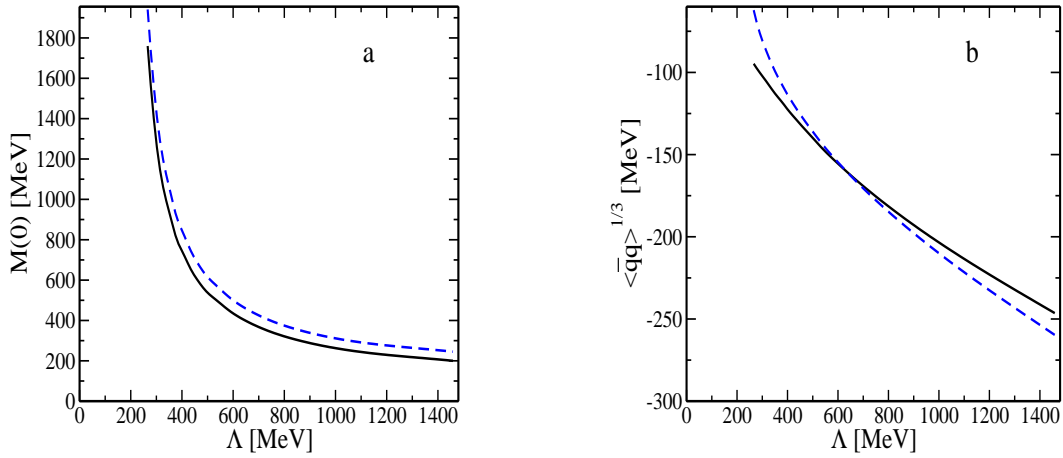}}
       \caption{The dynamical quark mass at zero momentum (a) and quark condensate
       (b) with respect to the cutoff for fixed $f_{\pi}= 93$ MeV and $m_{\pi}=140$ MeV. The solid line denotes
       the result when $m_{c}$ is not zero and the dashed line denotes the chiral limit $m_{c}=0$.  } 
\end{figure}
Our model contains five parameters: the current quark mass $m_{c}$,
the cutoff ($\Lambda$), the coupling constants $g_{\pi}$ , $g_{s}$ and
$g_{a}$. We first fix $g_{\pi}$ and the current quark mass $m_{c}$ for
arbitrary values of $\Lambda$ by fitting $f_{\pi}$ and $m_{\pi}$ to
their empirical values. In this way, we can consider the entire parameter space 
of the model. The corresponding solution of gap
equation are shown in Fig.~6.4. In the left panel the constituent mass
$M$ at zero momentum is shown as a function of the cutoff. It is
obvious that for very small cutoff, there is no solution for the gap
equation. On the right panel of Fig.~6.4, we show the corresponding
values of the quark condensate.  The quark condensate $\langle
\bar{\psi}\psi\rangle=i\Tr S(0)$ is closely related to the gap
equation Eq~(\ref{n3-gap}). In the latter there appears an extra form
factor inside the loop integral. The quark condensate with non-zero
current-quark mass is quadratically divergent and is regulated by a
subtraction of its perturbative value.  These values can fall
within the limits extracted from QCD sum rules $190\text{MeV}\lesssim
-\langle \bar{q}q\rangle^{1/3}\lesssim 260\text{MeV}$ at a
renormalisation scale of $1$ GeV \cite{qcdsum} and lattice calculation
\cite{lat-con}, having in mind that QCD condensate is a renormalised
and scale-dependent quantity. In right-hand side of Fig.~6.1 we show
the quark condensate with respect to current quark mass for various
parameter sets, it is seen that for large coupling (or large $M(0)$),
the magnitude of quark condensate decreases as the current quark mass
increases. This feature is consistent with the behaviours of lattice
and sum rule results. 

We analyse two sets of parameters, see Table \ref{tab:modpar}. The set $A$ is a non-confining parameter set,
while set $B$ leads to quark confinement (i.e., it satisfies the
condition Eq.~(\ref{n4}). The quark condensate in the chiral limit is
$-(207 \text{ MeV})^{3}$ and $-(186\text{ MeV})^{3}$ for sets A and B,
respectively. At non-zero current quark mass one obtains $-(215
\text{MeV})^{3}$ and $-(191 \text{MeV})^{3}$ for sets A and B
respectively. The position of the quark poles are given in Table
\ref{tab:quarkpoles} for two sets of parameters. In Fig.~6.5 we show
the position of the first pole of quark propagator with respect to
various cutoff, which indicates that for large cutoff, we have only
real poles and we have indeed complex poles for reasonable range of
cutoff. The real part of the first pole of dressed quark propagator
can be considered in much the same as the quark mass in the ordinary
NJL model. Since we do not believe in on-shell quarks or quark
resonances, this is also a measure for a limit on the validity of the
theory.  The real part of the first quark propagator pole $\mqr$ is
larger than the constituent quark mass at zero momentum $M(0)$, as can
be seen in Table \ref{tab:modpar}.

\begin{figure}[!tp]
\vspace{1cm}
       \centerline{\includegraphics[width=11 cm] {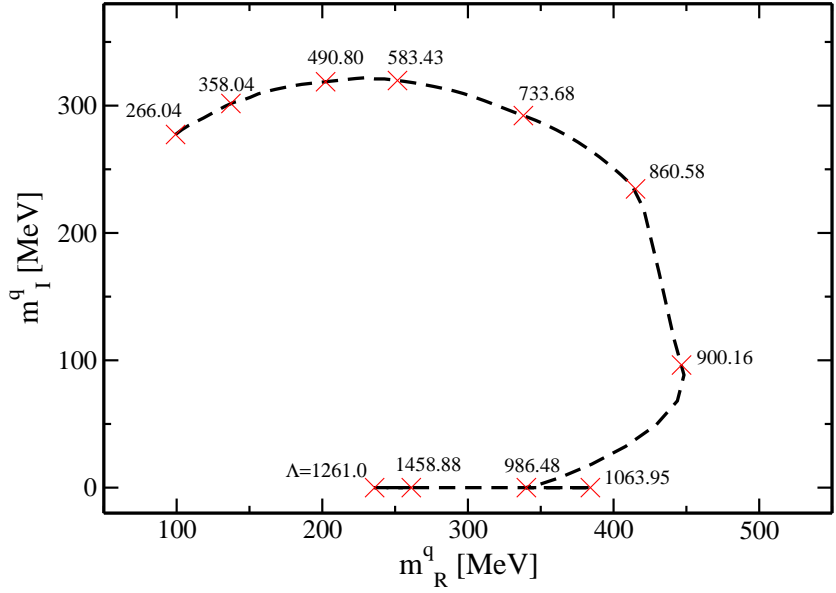}}
\caption{the position of the first poles of quark propagator in complex plane for various cutoff $\Lambda$ (MeV) 
for fixed $f_{\pi}= 93$ MeV and $m_{\pi}=140$ MeV. For every point,
there is a complex conjugate partner.  The imaginary part and real
part denote $m^{q}_{I}$ and $m^{q}_{R}$, respectively. Some values of
cutoff are given on the plot.}
\end{figure}
As we will see the mass
$\mqr$ appears as an important parameter in diquark and
nucleon solution rather than the constituent quark mass. The same
feature has been seen in the studies of the soliton in this model, where
$\mqr$ determines the stability of the soliton
\cite{a3}. In contrast to the
local NJL model, here the dynamical quark mass Eq.~(\ref{nI}) is
momentum dependent and follows a trend similar to that estimated from
lattice simulations \cite{lat-m}. Although this is less fundamental
since one is free to choose the form factor, nevertheless the quark
mass is a gauge dependent object and is not directly observable.

The parameters $g_{s}$ and $g_{a}$ are yet to be determined, we shall
treat them as free parameters, which allows us to analyse baryon
solutions in terms of a complete set of couplings. The
coupling-constant dependence appears through the ratios
$r_{s}=g_{s}/g_{\pi}$ and $r_{a}=g_{a}/g_{\pi}$.

\section{Diquark channels\label{sec:diquark}}
In the rainbow-ladder approximation the scalar $qq$ $T$-matrix can be
calculated from a very similar diagram to that shown in
Fig.~\ref{fig:meson} (the only change is that the anti-quark must be
replaced by a quark with opposite momentum). It can be written as
\begin{eqnarray}
T(p_{1},p_{2},p_{3},p_{4})&=&f(p_{1})f(p_{2})\big[\gamma_{5}C\tau_{2}\beta^{A}\big]\tau(q)\big[C^{-1}\gamma_{5}\tau_{2}\beta^{A}\big]f(p_{3})f(p_{4})\label{tm1}\nonumber\\
&&\times\delta(p_{1}+p_{2}-p_{3}-p_{4}),\
\end{eqnarray}
with
\begin{equation}
\tau(q)=\frac{2g_{s}i}{1+g_{s}J_{s}(q^{2})},\label{n5-0}
\end{equation}
where $q=p_{1}+p_{2}=p_{3}+p_{4}$ is the total momentum of the $qq$
pair, and
\begin{eqnarray}
 J_{s}(q^{2})&=& i \Tr \int
 \frac{d^{4}k}{(2\pi)^{4}}f^{2}(-k)\big[\gamma_{5}C\tau_{2}\beta^{A}\big]S(-k)^{T}\big[C^{-1}\gamma_{5}\tau_{2}\beta^{A}\big]S(q+k)f^{2}(q+k),\nonumber\\
&=&6i\int\frac{d^{4}k}{(2\pi)^{4}}\tr_{D}[\gamma_{5}S(k)\gamma_{5}S(k+q)]f^{2}(k)f^{2}(q+k)\label{l1}.
\end{eqnarray}
In the above equation the quark propagators $S(k)$ are the solution of
the rainbow SDE Eq.~(\ref{n3}).  The denominator of Eq.~(\ref{n5-0})
is the same as in the expression for the pion channel,
Eq.~(\ref{pi1}), if $g_{s}=g_{\pi}$. One may thus conclude that at
$r_{s}=1$ the diquark and pion are degenerate. This puts an upper
limit to the choice of $r_{s}$, since diquarks should not condense in
vacuum.  

One can approximate $\tau(q)$ by an effective diquark
exchange between the external quarks, and parametrise $\tau(q)$ around
the pole as
\begin{equation}
\tau(q)\equiv 2ig^{2}_{dsqq}V^{s}(q)D(q), \hspace{2cm} D^{-1}(q)=q^{2}-M^{2}_{ds},\label{n5}
\end{equation}
where $M_{ds}$ is the scalar diquark mass, defined as the position of
the pole of $\tau(q)$.
The
strength of the on-shell coupling of scalar diquark to quarks,
$g_{dsqq}$ is related to the polarisation operator $J_{s}$ by
\begin{equation}
g^{-2}_{dsqq}=\frac{dJ_{s}}{dq^{2}}|_{q^{2}=M^{2}_{ds}},\label{co}
\end{equation}
and $V^{s}(q)$ is the ratio between
the exact $T$-matrix and on-shell approximation. It is obvious that we
should have $V^{s}(q)|_{q^{2}=M^{2}_{ds}}=1$. 
\begin{figure}
\centerline{\includegraphics[clip,width=7 cm]{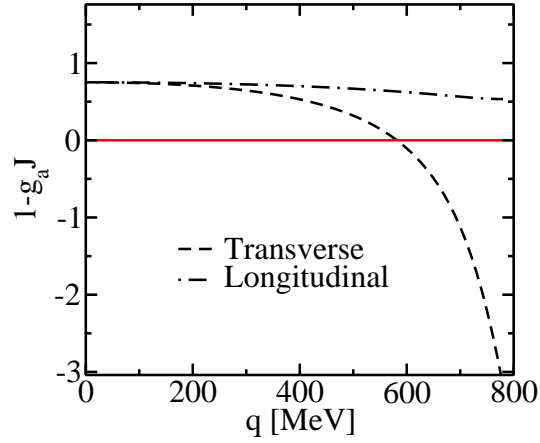}}
\caption{The denominator of the diquark $T$ matrix for the longitudinal and 
transverse axial vector channel. for parameter set A at $r_{a}=0.44$.
Note that there is no longitudinal pole.
\label{fig:avT}}
\end{figure}

Here, there is no mixing between the axial-vector diquark channel with
others, therefore in the same way one can write the axial-vector
diquark $T$-matrix in a similar form
\begin{eqnarray}
T(p_{1},p_{2},p_{3},p_{4})&=&f(p_{1})f(p_{2})\big[\gamma_{\mu}C\tau_{i}\tau_{2}\beta^{A}\big]\tau^{\mu\nu}(q)\big[C^{-1}\gamma_{\nu}\tau_{2}\tau_{i}\beta^{A}\big]f(p_{3})f(p_{4})\label{tm2}\nonumber\\
&&\times\delta(p_{1}+p_{2}-p_{3}-p_{4}),\
\end{eqnarray}
with
\begin{equation}
\tau^{\mu\nu}(q)=2g_{a}i\Big[\frac{g^{\mu\nu}-q^{\mu}q^{\nu}/q^{2}}{1+g_{a}J^{T}_{a}(q^{2})}+\frac{q^{\mu}q^{\nu}/q^{2}}{1+g_{a}J^{L}_{a}(q^{2})}\Big].
\end{equation}
Here we prefer to decompose the axial polarisation tensor into
longitudinal and transverse channels as well,
\begin{eqnarray}
J^{\mu\nu}_{a}(q^{2})&=& i \Tr \int
\frac{d^{4}k}{(2\pi)^{4}}f^{2}(-k)\big[\gamma^{\mu}C\tau_{i}\tau_{2}\beta^{A}\big]S(-k)^{T}\big[C^{-1}\gamma^{\nu}\tau_{2}\tau_{i}\beta^{A}\big]S(q+k)f^{2}(q+k),\nonumber\\
&=&6i\int\frac{d^{4}k}{(2\pi)^{4}}\tr_{D}[\gamma^{\mu}S(k)\gamma^{\nu}S(k+q)]f^{2}(k)f^{2}(q+k)\nonumber\\
&=&J^{T}_{a}(q^{2})(g^{\mu\nu}-q^{\mu}q^{\nu}/q^{2})+J^{L}_{a}(q^{2})q^{\mu}q^{\nu}/q^{2}.\label{lon}\
\end{eqnarray}
We find that the longitudinal channel does not produce a pole (see
Fig.~\ref{fig:avT}), and thus the bound axial-vector diquark solution
corresponds to a pole of the transverse $T$-matrix. The transverse
component of $\tau^{\mu\nu}(q)$ matrix is now approximated by $M_{da}$
as,
\begin{equation} 
\tau^{\mu\nu}(q)\equiv 2ig^{2}_{daqq}V^{a}(q)D^{\mu\nu}(q), \hspace{2cm} D^{\mu\nu}(q)=\frac{g^{\mu\nu}-q^{\mu}q^{\nu}/q^{2}}{q^{2}-M^{2}_{da}},\label{aloop}
\end{equation}
where $V^{a}(q)$ includes the off-shell contribution to the
$T$-matrix. The  coupling constant $g_{daqq}$ is related to the
residue at the pole of the $T$-matrix,
\begin{equation}
g^{-2}_{daqq}=\frac{dJ^{T}_{a}}{dq^{2}}|_{q^{2}=M^{2}_{da}}.\label{coa}
\end{equation}

\begin{figure}
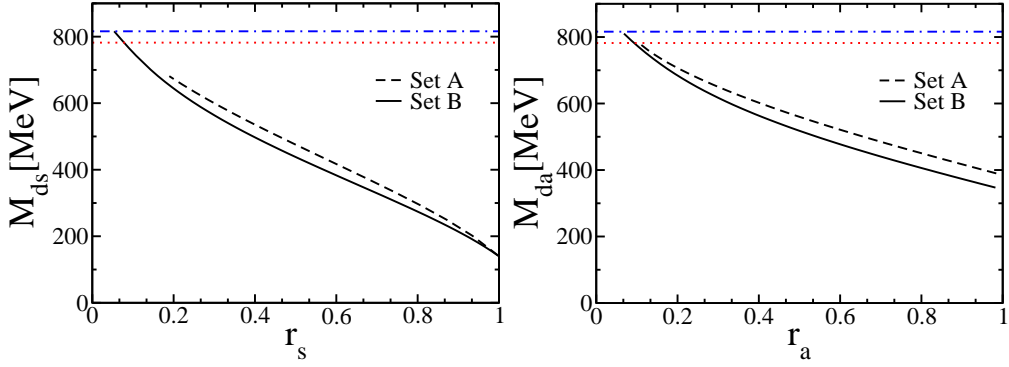

\vspace{.5cm}
\includegraphics[height=.23\textheight]{plotsab.eps}
\includegraphics[height=.23\textheight]{plotab.eps}
\caption{The scalar and axial-vector diquark mass 
as a function of $r_{s}$ and $r_{a}$, respectively, for both parameter
sets. The dotted and broken-dotted lines lines denote the quark-quark
pseudo-threshold for set A and B, respectively. \label{fig:samass}}
\end{figure}

\begin{figure}
     \centerline{\includegraphics[clip,width=7 cm] {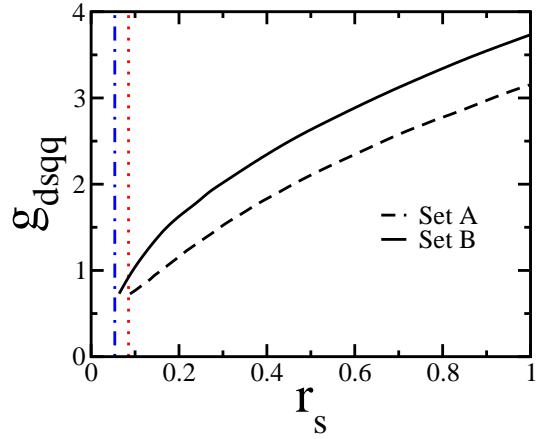}}
%
\caption{The scalar diquark-quark-quark coupling as a function of $r_{s}$.
The dotted and broken-dotted lines indicate the quark-quark pseudo-threshold for set A and B, respectively.\label{fig:gsdqq}}
\end{figure}

\subsection{Diquark Solution}
The loop integrations in Eq.~(\ref{l1}, \ref{lon}) are evaluated in
Euclidean space\footnote{We work in Euclidean space with metric
$g^{\mu\nu}=\delta^{\mu\nu}$ and a hermitian basis of Dirac matrices
$\{\gamma_{\mu},\gamma_{\nu}\}=2\delta_{\mu\nu}$, with a standard
transcription rules from Minkowski to Euclidean momentum space:
$k^{0}\to ik_{4} $, $\vec{k}^{M}\to -\vec{k}^{E}$}. For the current
model, the usual analytic continuation of amplitudes from Euclidean to
Minkowski space can not be used. This is due to the fact that quark
propagators of the model contain many poles at complex energies
leading to opening of a threshold for decay of a diquark (or meson)
into other unphysical states.  Any theory of this type need to be
equipped with an alternative continuation prescription consistent with
unitarity and macrocausality. Let us define a fictitious two-body
threshold as twice $\mqr$. For a confining parameter set, each quark
propagator has a pair of complex-conjugate poles. Above the two-body
pseudo-threshold $q^{2}<-4(\mqr)^{2}$, where $q$ is meson (diquark)
momentum, the first pair of complex poles of the quark propagator has
a chance to cross the real axis. According to the Cutkosky
prescription \cite{cutk}, if one is to preserve the unitarity and the
microcausality, the integration contour should be pinched at that
point. In this way, one can ensure that there is no spurious
$\bar{q}q$ (or $qq$) production threshold, for energies below the next
pseudo-threshold, i.e. twice the real part of the second pole of the
quark propagator. Note that it has been shown \cite{u} that the
removal of the $\bar{q}q$ pseudo-threshold is closely related to the
existence of complex poles in the form of complex-conjugate pairs.
Since there is no unique analytical continuation method available for
such problems, any method must be regarded as a part of the model
assumptions \cite{pb,a1,u}. Here, we follow the method used in Ref.~\cite{pb}.

We use the parameter sets determined in the mesonic sector shown in
table \ref{tab:modpar}. Our numerical computation is valid below the
first $qq$ pseudo-threshold. Note that the longitudinal polarisability
$J^{L}_{a}(q)$ defined in Eq.~(\ref{lon}) does not vanish here.
However this term can probably be ignored since it does produce any
poles in the $T$-matrix, and moreover there is no conserved current
associated to this channel. We find that for a wide range of $r_{s}$
and $r_{a}$, for all parameter sets, a bound scalar and axial-vector
diquark exist (the results for additional sets can be found in
\cite{me}). This is in contrast to the normal NJL model where a bound
axial-vector diquark exists only for very strong interaction
\cite{njln3}. The diquark masses for various values of $r_{s}$ and
$r_{a}$ are plotted in Fig.~\ref{fig:samass}. As already pointed out,
the scalar diquark mass is equal to the pion mass at $r_{s}=1$. It is
obvious from Fig.~\ref{fig:samass} that for $r_{s}=r_{a}$ the axial-vector diquark is heavier than the
scalar diquark, and consequently is rather loosely bound. For very
small $r_{s}$ and $r_{a}$ one finds
no bound state of either diquark. This kind of diquark confinement is
due to the screening effect of the ultraviolet cutoff and can not be
associated with confinement in QCD which originates from the infrared
divergence of the gluon and ghost propagators. Having said that, it is
possible that real diquark confinement may arise beyond the ladder
approximation
\cite{new}. There, in order to preserve Goldstone's theorem at every
order, we must include additional terms in the interaction. Although
these new terms should have minimal impact on the solutions for the
colour-singlet meson channels, they can provide a 
repulsive contribution to the colour-antitriplet diquark channels
which removes the asymptotic-diquark solutions from the spectrum. This
would indicate that diquark confinement is an independent phenomenon
and is not related to the particular realisation of quark
confinement.   
\begin{figure}[!tp]
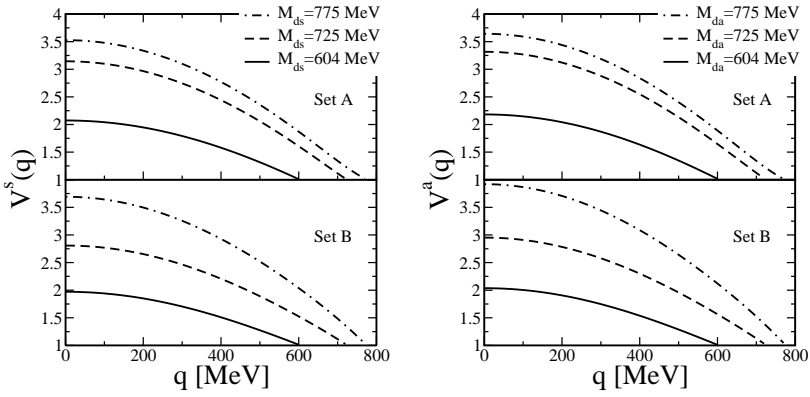

\begin{tabular}{cc}
\includegraphics[clip,height=.25\textheight]{plotoffshell-as.eps}&
\includegraphics[clip,height=.25\textheight]{plotoffshell-a.eps}
\end{tabular}
\caption{The ratio of the on-shell approximation compared to the 
exact diquark $T$-matrix for the various scalar and axial diquark
masses.\label{fig:offshellT}}
\end{figure}

One should note that the nucleon bound state in the diquark-quark
picture does not require asymptotic-diquark states since the diquark
state is merely an intermediate device which simplifies the three-body
problem. Nevertheless, evidence for correlated diquark states in
baryons is found in deep-inelastic lepton scatterings and in hyperon weak decays \cite{lep2}. 
At the same time, diquarks appear as bound 
states in many phenomenological models. It is puzzling that diquarks
are even seen in lattice calculations
\cite{lat-diq1,lat-diq2}. In contrast to our perception of QCD colour
confinement, the corresponding spectral functions for these supposedly
confined objects in the colour anti-triplet channel are very similar to
mesonic ones \cite{lat-diq2}.

In Fig.~\ref{fig:gsdqq}, we show the scalar diquark-quark-quark coupling
defined in Eq.~(\ref{co}) with respect to various scalar diquark couplings. A pronounced change in behaviour
around the quark-quark pseudo-threshold is observed in the confining set $B$,
and this seems to justify our emphasis on the pseudo-threshold defined by
twice the real part of the quark pole.

Next we study the off-shell behaviour of the diquark $T$-matrix. In
Fig.~\ref{fig:offshellT} we show the discrepancy between the exact
$T$-matrix and the on-shell approximation $V^{s,a}(q)$. At the pole
we have by definition that
$V^{s,a}(q)|_{q^{2}=M^{2}_{s,a}}=1$.
We see  that elsewhere
the off-shell contribution is very important due to the non-locality
of our model. We find that the bigger the diquark mass is, the bigger
the off-shell contribution.  The off-shell behaviour of the
scalar and the axial-vector channel for both parameter sets $A$ and
$B$ are rather similar.

\section{Three-body sector\label{sec:TB}}
In order to make three-body problem tractable, we discard any
three-particle irreducible graphs (this is sometimes called the
Faddeev approximation).  The relativistic Faddeev equation can be then
written as an effective two-body BS equation for a quark and a
diquark due to the locality of the form factor in momentum space (see
Eq.~(\ref{forfa})) and accordingly the separability of the two-body
interaction in momentum-space. We adopt the formulation developed by the
T\"ubingen group \cite{o1,o2,o3} to solve the resulting BS
equation. In the following we work in momentum space with Euclidean
metric. The BS wave function for the octet baryons can be presented in
terms of scalar and axialvector diquarks correlations,
\begin{equation}
\psi (p,P) u (P,s) =\left(\begin{array}{c}\psi^5 (p,P) \\ \psi^\mu(p,P) \end{array}\right)u(P,s), \label{bsn}
\end{equation}
where $u(P,s)$ is a basis of positive-energy Dirac spinors of spin $s$
in the rest frame. The parameters $p=(1-\eta)p_{i}-\eta(p_{j}+p_{k})$
and $P=p_{i}+p_{j}+p_{k}$ are the relative and total momenta in the
quark-diquark pair, respectively. The Mandelstam parameter $\eta$
describes how the total momentum of the nucleon $P$ is distributed
between quark and diquark.


One may alternatively define the vertex function associated with
$\psi (p,P)$ by amputating the external quark and diquark propagators (the legs) from the wave function;
\begin{equation}
\phi (p,P) =S^{-1}(p_{q})\tilde{D}^{-1}(p_{d})\left(\begin{array}{c}\psi^5 (p,P) \\ \psi^\nu(p,P) \end{array}\right), \label{psi}
\end{equation}
with 
\begin{equation}
\tilde{D}^{-1}(p_{d})=\left(\begin{array}{cc}D^{-1}(p_{d})&0\\0&(D^{\mu\nu}(p_{d}))^{-1},\end{array}\right) \label{di-pp}
\end{equation}
where $D(p), D^{\mu\nu}(p)$ and $S(p)$ are Euclidean versions of the
diquark and quark propagators which are obtained by the standard
transcription rules from the expressions in Minkowski space,
Eqs.~(\ref{n5},\ref{aloop}) and Eq.~(\ref{n3}), respectively. The
spectator quark momentum $p_{q}$  and the diquark momentum
$p_{d}$ are given by
\begin{eqnarray}
p_{q}&=&\eta P+p, \label{pk}\\
p_{d}&=&(1-\eta)P-p,\label{pkd}\
\end{eqnarray}
with similar expressions for $k_{q,d}$, where we replace $p$ by $k$
on the right-hand side.  In the ladder approximation, the coupled
system of BS equations for octet baryon wave functions and their
vertex functions takes the compact form,
\begin{equation}
\phi (p,P)=\int \frac{d^{4}k}{(2\pi)^{4}} K^{BS}(p,k;P) \psi(k,P),\label{n7}\\
\end{equation}
where $K^{BS}(p,k;P)$ denotes the kernel of the nucleon BS equation
representing the exchange quark within the diquark with the spectator quark (see Fig.~5.1), and in the colour
singlet and isospin $\frac{1}{2}$ channel we find (see Ref.~\cite{njln3})
\begin{equation}
K^{BS}(p,k;P)=-3\left(\begin{array}{cc}\chi^{5}(p_{1},k_{d})S^{T}(q)\bar{\chi}^{5}(p_{2},p_{d})&-\sqrt{3}\chi^{\alpha}(p_{1},k_{d})S^{T}(q)\bar{\chi}^{5}(p_{2},p_{d})\\-\sqrt{3}\chi^{5}(p_{1},k_{d})S^{T}(q)\bar{\chi}^{\mu}(p_{2},p_{d})&
-\chi^{\alpha}(p_{1},k_{d})S^{T}(q)\bar{\chi}^{\mu}(p_{2},p_{d})\end{array}\right),\label{n9}
\end{equation}
where $\chi$ and $\chi^{\mu}$
(and their adjoint $\bar{\chi}$ and $\bar{\chi}^{\mu}$) stand for the
Dirac  structures of the scalar and the axial-vector diquark-quark-quark
vertices and can be read off  immediately from Eqs.~(\ref{tm1}, \ref{n5})
and Eqs.~(\ref{tm2}, \ref{aloop}), respectively. Therefore we have 
\begin{eqnarray}
\chi^{5}(p_{1},k_{d})&=&g_{dsqq}(\gamma^{5}C)\sqrt{2 V^{s}(k_{d})}f(p_{1}+(1-\sigma)k_{d})f(-p_{1}+\sigma k_{d}),\nonumber\\
\chi^{\mu}(p_{1},k_{d})&=&g_{daqq}(\gamma^{\mu}C)\sqrt{2 V^{a}(k_{d})}f(p_{1}+(1-\sigma)k_{d})f(-p_{1}+\sigma k_{d}).\label{d-v}\
\end{eqnarray}
We have used an improved on-shell approximation for the contribution
of diquark $T$-matrix occurring in the Faddeev equations.  Instead of
the exact diquark $T$-matrices we use the on-shell approximation with
a correction of their off-shell contribution through $V^{s,a}(p)$.
What is neglected is then the contribution to the $T$-matrix beyond
the pseudo-threshold.  As we will see this approximation is sufficient
to obtain a three-body bound state. In order to evaluate the structure
of the diquark $T$-matrix completely, one normally employs the dispersion
relation, however, this is not applicable here, due to non-analyticity
of the diquark $T-$ matrix. Notice, as we already pointed out for the
confining set B, we do not have $qq$ continuum, however, there exists
many complex poles beyond the pseudo-threshold which might be ignored,
provided that they lie well above the energies of interest.

The relative momentum of quarks in the diquark vertices $\chi$ and
$\chi^{\mu}$ are defined as $ p_{1}=p+k/2-(1-3\eta)P/2$ and
$p_{2}=-k-p/2+(1-3\eta)P/2$, respectively. The momentum $k_{d}$ of the
incoming diquark and the momentum $p_{d}$ of the outgoing diquark are
defined in Eq.~(\ref{pkd}) (see Fig. 5.1). The momentum of
the exchanged quark is fixed by momentum conservation at $q=-p-k+(1-2\eta)P$.

It is interesting to note the non-locality of the diquark-quark-quark
vertices naturally provides a sufficient regularisation of the
ultraviolet divergence in the diquark-quark loop. In the expressions
for the momenta we have introduced two independent Mandelstam
parameters $\eta, \sigma$, which can take any value in $[0,1]$. They
parametrise different definitions of the relative momentum within the
quark-diquark ($\eta$) or the quark-quark system
($\sigma$). Observables should not depend on these parameters if the
formulation is Lorentz covariant. This means that for every BS
solution $\psi(p,P;\eta_{1},\sigma_{1})$ there exists a equivalent
family of solutions. This provides a stringent check on calculations,
see the next section for details.


We now constrain the Faddeev amplitude to describe a state of
positive energy, positive parity and spin $s=1/2$. The  parity
condition can be immediately reduced to a
condition for the BS wave function:
\begin{equation}
\mathcal{P}\left(\begin{array}{c}\psi^{5}(p,P)\\ \psi^{\mu}(p,P)\end{array}\right)=\left(\begin{array}{c}\gamma^{4}\psi^{5}(\bar{p},\bar{P})\gamma^{4}\\ \gamma^{4}\Lambda^{\mu\nu}_{\mathcal{P}}\psi^{\nu}(\bar{p},\bar{P})\gamma^{4} \end{array}\right)=\left(\begin{array}{c}\psi^{5}(p,P)\\ -\psi^{\mu}(p,P)\end{array}\right), 
\end{equation}
where we define $\bar{p}=\Lambda_{\mathcal{P}}p$ and $
\bar{P}=\Lambda_{\mathcal{P}}P$, with
$\Lambda^{\mu\nu}_{\mathcal{P}}=\text{diag}(-1,-1,-1,1)$. In order to
ensure the positive energy condition, we project the BS wave function
with the positive-energy projector
\(\Lambda^{+}=(1+\hat{\slashP})\), where the hat
denotes a unit four vector (in rest frame we have
$\hat{P}=P/iM$). Now we expand the BS wave function $\psi(p,P)$ in
Dirac space $\Gamma\in\{\textbf{1}, \gamma_{5},\gamma^{\mu},
\gamma_{5}\gamma^{\mu},\sigma^{\mu\nu}\}$. The above-mentioned
conditions reduce the number of independent component from sixteen to
eight, two for the scalar diquark channel, $S_{i},
(i=1,2)$ and six for the axial-diquark channel, $A_{i},(i=1,...6)$.
The most general form of the BS wave function is given by
\begin{eqnarray}
\psi^{5}(p,P)&=&\left(S_{1}-i\hat{\slashp}_{T}S_{2}\right)\Lambda^{+},\nonumber\\
\psi^{\mu}(p,P)&=&\Big(i\hat{P}^{\mu}\hat{\slashp}_{T}A_{1}+\hat{P}^{\mu}A_{2}-\hat{p}^{\mu}_{T}\hat{\slashp}_{T}A_{3}+
i\hat{p}^{\mu}_{T}A_{4}+\left(\hat{p}^{\mu}_{T}\hat{\slashp}_{T}-\gamma^{\mu}_{T}\right)A_{5}\nonumber\\
&&-(i\gamma^{\mu}_{T}\hat{\slashp}_{T}+i\hat{p}^{\mu}_{T})A_{6}\Big)\gamma_{5}\Lambda^{+}.
\label{main}\
\end{eqnarray}
Here we write
\(\gamma^{\mu}_{T}=\gamma^{\mu}-\hat{\slashP}\hat{P}^{\mu}\). 
The subscript $T$ denotes the component of a four-vector transverse to
the nucleon momentum, $p_{T}=p-\hat{P}(p.\hat{P})$. In the same way,
one can expand the vertex function $\phi$ in Dirac space, and since
the same constraints apply to the vertex function, we
obtain an expansion quite similar to Eq.~(\ref{main}), with new unknown
coefficients $\mathbb{S}_{i}$ and $\mathbb{A}_{i}$ which are
substituted the coefficients $S_{i}$ and $A_{i}$, respectively.  The
unknown scalar function $S_{i} (\mathbb{S}_{i})$ and $A_{i}
(\mathbb{A}_{i})$ depend on the two scalars which can be built from
the nucleon momentum $P$ and relative momentum $p$,
$z=\hat{P}.\hat{p}=\cos\omega$ (the cosine of the four-dimensional
azimuthal angle of $p^{\mu}$) and $p^{2}$.

In the nucleon rest frame, one can rewrite the Faddeev amplitude in
terms of tri-spinors each possessing definite orbital angular momentum
and spin \cite{o1}. It turns out that these tri-spinors can be written
as linear combinations of the eight components defined in
Eq.~(\ref{main}).  Thus from knowledge of $S_{i}$ and $A_{i}$, a full
partial wave decomposition can be immediately obtained
\cite{o1}. Notice that although the diquarks are not pointlike objects here,
they do not carry orbital angular momentum i. e. $L^{2}\chi^{5,\mu}(q)=0$. This is due to the fact that the
off-shell contribution $V^{s,a}(q)$ is a function of scalar $q^{2}$.
Moreover, the form factor in our model Lagrangian is also
scalar, hence the total momentum dependent part of the
diquark-quark-quark vertices are scalar functions and carry no
orbital angular momentum. Therefore, the  partial wave
decomposition obtained in Ref.~\cite{o1} for pointlike diquarks can
be used here. Note that no such partial wave decomposition can be found if
one uses the BS vertex function $\phi^{5,\mu}$  since the
axial-vector diquark propagator mixes the space component of the
vertex function and time component of the axial-vector diquark.

\subsection{Numerical method for the coupled BS equations}
For solving the BS equations we use the algorithm introduced by Oettel
{\em et al} \cite{o4}. The efficiency of this algorithm has already been
reported in several publications, see for example
Refs.~\cite{o1,o2,o3}.  We will focus here only on the key ingredients
of this method. The momentum dependence of quark mass in
our model increases the complexity of the computation significantly.

As usual, we work in the rest frame of the nucleon $P=(0,iM_{N})$. In
this frame we are free to chose the spatial part of the relative
momentum $p$ parallel to the third axis. Thus the momenta
$p$ and $k$ are given by
\begin{eqnarray} 
p^{\mu}&=&|p|(0,0,\sqrt{1-z^{2}},z), \nonumber\\ 
k^{\mu}&=&|k|(\sin\theta'\sin\phi'\sqrt{1-z'^{2}},\sin\theta'\cos\phi'\sqrt{1-z'^{2}},\cos\theta\sqrt{1-z'^{2}},z'),\
\end{eqnarray}
where we write $z=\cos\omega$ and $z'=\cos\omega'$. 
The wave function Eq.~(\ref{main}) consists of $2\times2$-blocks in Dirac space can be simplified to
\begin{eqnarray}
\psi^{5}(p,P)&=&\left(\begin{array}{cc}
S_{1}(p^{2}, z) & 0 \\
\sigma_{3}\sqrt{1-z^{2}}S_{2}(p^{2},z) &0
\end{array}\right),
\hspace{1.2cm}
\psi^{4}(p,P)=\left(\begin{array}{cc}
\sigma_{3}\sqrt{1-z^{2}}A_{1}(p^{2},z) & 0 \\
A_{2}(p^{2},z) & 0\end{array}\right),   \nonumber\\
\psi^{3}(p,P)&=&\left(\begin{array}{cc}
i\sigma_{3}A_{3}(p^{2},z) & 0 \\
i\sqrt{1-z^{2}}A_{4}(p^{2},z) & 0\end{array}\right),
\hspace{1.5cm}
\psi^{2}(p,P)=\left(\begin{array}{cc}
i\sigma_{2}A_{5}(p^{2},z) & 0 \\
-\sigma_{1}\sqrt{1-z^{2}}A_{6}(p^{2},z) & 0\end{array}\right),\nonumber\\
\psi^{1}(p,P)&=&\left(\begin{array}{cc}
i\sigma_{1}A_{5}(p^{2},z) & 0 \\
\sigma_{2}\sqrt{1-z^{2}}A_{6}(p^{2},z) & 0\end{array}\right).\label{rest-bs}\
\end{eqnarray}
The great advantage of this representation is that the scalar and the
axial-vector components are decoupled. Therefore the BS equation
decomposes into two sets of coupled equations, two for the scalar
diquark channel and six for the axial diquark
channel.  We expand the vertex (wave) functions in terms of Chebyshev
polynomials of the first kind, which are closely related to the
expansion into hyperspherical harmonics. This decomposition turns
out to be very efficient for such problems \cite{o1,o2,o3,o4}. Explicitly, 
\begin{eqnarray}
F^{\psi}_{i}(p^{2},z)&=&\sum^{n_{max}}_{n=0}i^{n}F^{\psi(n)}_{i}(p^{2})T_{n}(z),\label{che}\nonumber\\
F^{\phi}_{i}(p^{2},z)&=&\sum^{m_{max}}_{m=0}i^{n}F^{\phi(m)}_{i}(p^{2})T_{m}(z),\
\end{eqnarray}
where $T_{n}(z)$ is the Chebyshev polynomial of the first kind. 
We use a generic notation, the functions $F^{\psi}_{i}$( and $F^{\phi}_{i}$) substituting the function $S_{i},~ A_{i}$ (and
$\mathbb{S}_{i},~\mathbb{A}_{i}$),
\begin{eqnarray}
&& S_{1,2}\to F^{\psi}_{1,2}, \hspace{2cm}  A_{1...6}\to F^{\psi}_{3...8}, \nonumber\\
&& \mathbb{S}_{1,2}\to F^{\phi}_{1,2}, \hspace{2cm}  \mathbb{A}_{1...6}\to F^{\phi}_{3...8} .
\end{eqnarray}
We truncate the Chebyshev
expansions involved in $F^{\psi}_{i}$ and $F^{\phi}_{i}$ at
different orders $n_{max}$ and $m_{max}$, respectively. We also expand
the quark and diquark propagators into Chebyshev polynomials. In this
way one can separate the $\hat P\cdot \hat p$ and $\hat P \cdot \hat
k$ dependence in Eqs.~(\ref{n7}--\ref{psi}). Using the  
orthogonality relation between the Chebyshev polynomials, one can
reduce the four dimensional integral equation into a system of coupled
one-dimensional equations. Therefore one can rewrite Eqs.~(\ref{psi},
\ref{n7}) in the matrix form
\begin{eqnarray}
F^{\psi(n)}_{i}(p^{2})&=&\sum_{j=1}^{8}\sum_{m=0}^{m_{max}}g_{ij}^{nm}(p^{2})F^{\phi(m)}_{j}(p^{2}),\nonumber\\
F^{\phi(m)}_{i}(p^{2})&=&\sum_{j=1}^{8}\sum_{n=0}^{n_{max}}\int^{\infty}_{0}d|k| |k|^{3} H_{ij}^{mn}(k^{2},p^{2})F^{\psi(n)}_{j}(k^{2}).\label{bs-matrix}\
\end{eqnarray}
Here $g_{ij}^{nm}$ and $H_{ij}^{mn}$ are the matrix elements of the
propagator and the quark exchange matrices, respectively. The indices
$n, m$ give the Chebyshev moments and $i, j$ denote the individual channels.
To solve  Eq.~(\ref{bs-matrix}), we first rewrite it in the 
form of linear eigenvalue problem. Schematically
\begin{equation}
\lambda(P^{2})\varphi=K(P^{2})\varphi,
\end{equation}   
with the constraint that $\lambda(P^{2})=1$ at $P^{2}=-M^{2}_{N}$. This
can be used to determine the nucleon mass $M_{N}$ iteratively.

\begin{figure}

       \centerline{\includegraphics[clip,width=7cm]
                                   {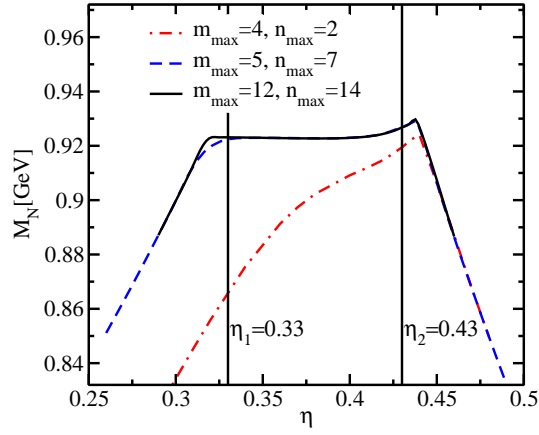}}
\caption{The dependence of the nucleon mass on the Mandelstam parameter $\eta$
for a few values of the cut-off on the expansion. Here we use set B,
with $M_{ds}=725 $ MeV and $M_{da}=630$ MeV. The variables $\eta_1$
and $\eta_2$ denote the position of the
singularities.\label{fig:Mandelstam}}
\end{figure}

As already pointed out the BS solution should be independent of the
Mandelstam parameters $\eta,\sigma$. As can be seen in Fig.~\ref{fig:Mandelstam}, there is indeed a large plateau for the $\eta$ dependence if we use
a high cut-off on the Chebyshev moments. The limitations on the
size of this area of stability can be understood by considering where
the calculation contains singularities due to quark and diquark poles,
\begin{eqnarray}
\eta&\in&\left[1-\frac{M_{ds}}{M_{N}}, \frac{\mqr }{M_{N}}\right], \hspace{2cm}\text{if} \hspace{1cm} M_{ds}<M_{da},\nonumber\\
\eta&\in&\left[1-\frac{M_{da}}{M_{N}}, \frac{\mqr }{M_{N}}\right],  \hspace{2cm}\text{if} \hspace{1cm} M_{da}<M_{ds}.\label{sin}
\end{eqnarray}
A similar plateau
has been found in other applications \cite{o1,o2,o3}. [Other complex poles
lies out side the minimal region given in Eq.~(\ref{sin}).] 
The
singularities in the quark-exchange  propagator put another constraint
on the acceptable range of $\eta$;
$\eta>\frac{1}{2}(1-\frac{\mqr }{M_{N}})$. 
No such constraint exists for
 $\sigma$, which relates to the relative momentum
between two quarks. To simplify the algebra 
we take $\sigma=1/2$.

In what follows we use a momentum mesh of $60\times 60$ for $p,k$,
mapped in a non-linear way to a finite interval. In the non-singular
regime of Mandelstam parameter $\eta$ Eq.~(\ref{sin}), the Faddeev
solution is almost independent of the upper limit on the Chebyshev
expansion, and for $m_{max}=10, n_{max}=12$, see the
Fig.~\ref{fig:Mandelstam}, this seems to be satisfied. This limit is
some higher than the reported values for simple models \cite{o1,o2,o3,o4}.

\subsection{Nucleon Solution}
In order to understand the role of the axial diquark in nucleon
solution, we first consider the choice $r_{a}=0$. For this case we
find that the non-confining set A can not generate a three-body bound
state in this model without the inclusion of the off-shell
contribution. 
For the confining set B one also has to enhance the
diquark-quark-quark coupling $g_{dsqq}$ by a factor of about $1.73$
over the value defined in Eq.~(\ref{co}) (as we will show, this extra
factor is not necessary when the axial-vector is included).  The
situation is even more severe in the on-shell treatment of the local
NJL model, since one needs to include the full $qq$ continuum
contribution in order to find a three-body bound state when the
axial-vector diquark channel is not taken into account \cite{njln1}.

\begin{figure}
 \centerline{\includegraphics[clip,width=7 cm]{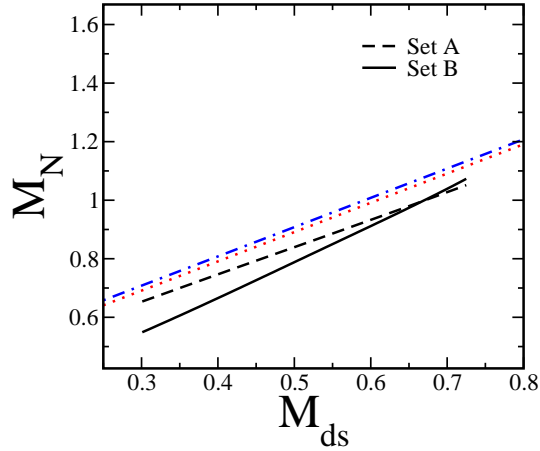}}
\caption{The nucleon mass without inclusion of the axial diquark channel. 
The dotted lines indicate the diquark-quark threshold. All values are
given in GeV.\label{fig:Nmass_scal}}
\end{figure}

As can be seen from Fig.~\ref{fig:samass} a decrease in $r_{s}$ leads
to a larger diquark mass, and an increase in the off-shell
contribution to the $qq$ $T$-matrix (see Fig.~\ref{fig:offshellT}). It
is this off-shell correction is need for a bound nucleon.

The nucleon result is shown in Fig.~\ref{fig:Nmass_scal}. We also show
a fictitious diquark-quark threshold defined as $M_{ds}+\mqr$. The
nucleon mass can be seen to depend roughly linearly on the scalar
diquark mass. A similar behaviour is also seen in the local NJL model
\cite{njln2}. Increasing the diquark mass (or decreasing $r_{s}$)
increases the nucleon mass, i.e. the scalar diquark channel is
attractive.  In order to obtain a nucleon mass of $940 $ MeV, we need
diquark mases of $608$ MeV and $623$ MeV for set A and B,
respectively. The corresponding nucleon binding energy measured from
the diquark-quark threshold are $ 56$ MeV and $ 91$ MeV for set A and
B, respectively, compared to the binding of the diquarks (relative to
the $qq$ pseudo threshold) of about $174$ and $193$ MeV for set A and
B, respectively. Such diquark clustering within the nucleon is also
observed in the local NJL model
\cite{njln2}, and is qualitatively in agreement with a instanton
model \cite{i-b} and lattice simulations
\cite{lat-bin}. The nucleon solutions for sets A and B behave
rather differently with respect to the diquark-quark threshold, see
Fig.~\ref{fig:Nbind_scal}. This indicates that for non-confining set
A, the nucleon solution is rather sensitive to the diquark-quark
threshold and tends not avoid it. However, for confining set, since
there is no well-defined threshold, this tendency is absent and as we
approach to the diquark-quark threshold, the nucleon binding energy
decreases and can approach to zero.

\begin{figure}
       \centerline{\includegraphics[clip,width=7 cm]
                                   {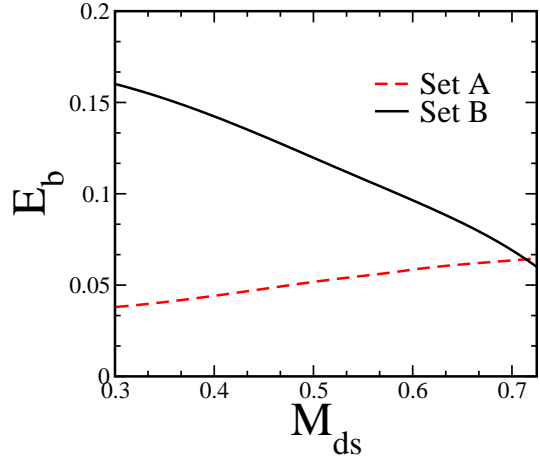}}
\caption{
The nucleon binding energy 
measured from the diquark-quark threshold with respect to scalar diquark
mass. Only scalar diquarks are included in the calculation
 The smooth behaviour of 
solid line indicates the absence of threshold effects for the
confining case. \label{fig:Nbind_scal}}
\end{figure}

\begin{figure}
\centerline{\includegraphics[clip,width=10cm]{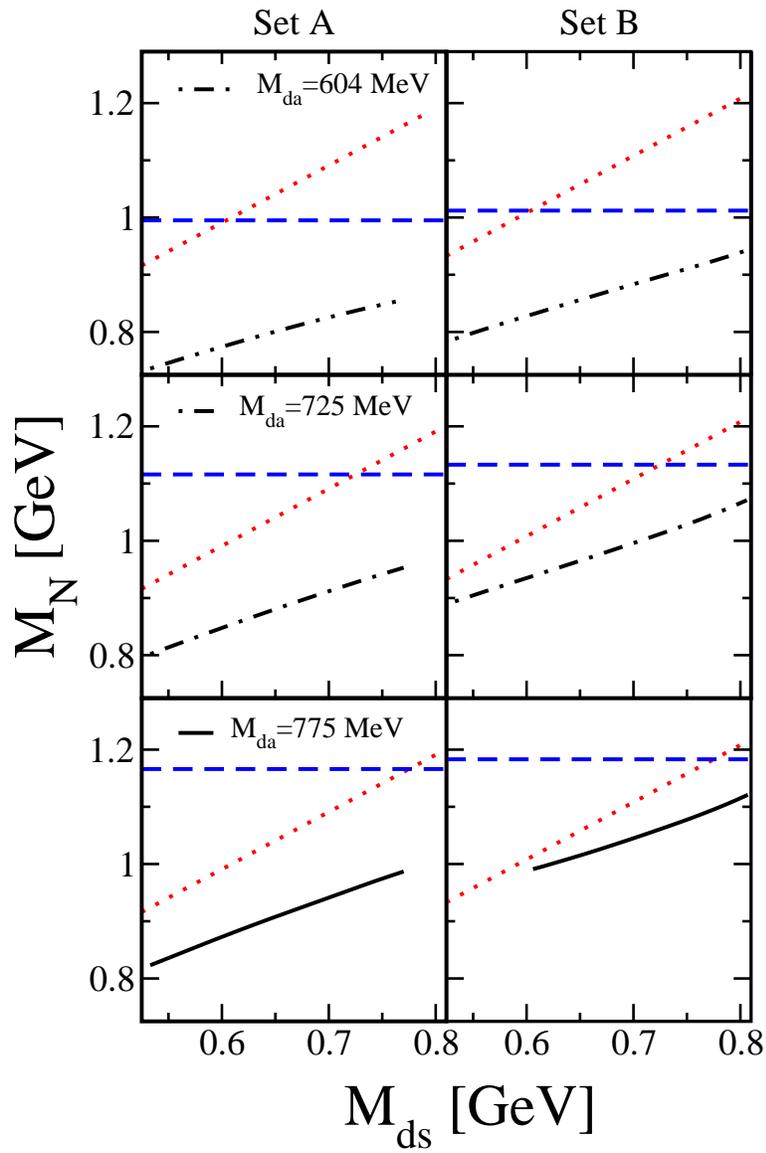}}
\caption{The nucleon mass as a function of the scalar diquark mass.
The dotted lines are the scalar diquark-quark threshold, the broken lines
are the axial vector diquark-quark threshold. All values are given in GeV.
\label{fig:Nmass_full1}}  
\end{figure}

\begin{figure}
       \centerline{\includegraphics[clip,width=7 cm]
                                  {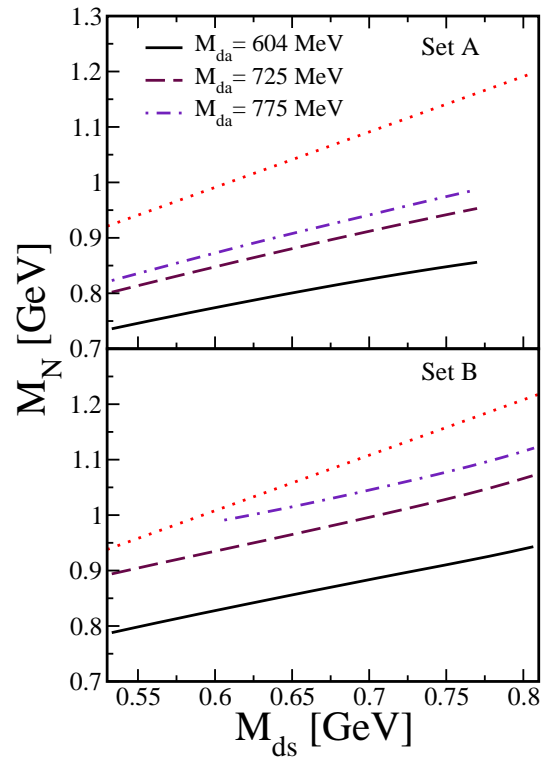}}
\caption{\label{fig:Nmass_full2} 
The nucleon mass as a function of the scalar 
diquark mass for various axial vector diquark masses for both parameter sets. The scalar diquark-quark threshold are shown by the doted lines.}
\end{figure}

Next we investigate the effect of the axial-vector diquark channel on
nucleon solution. We find that the axial-vector diquark channel
contributes considerably to the nucleon mass\footnote{Note however that
neglecting $\pi N$-loops may lead to a quantitative overestimate
of the axial-vector diquark role in the nucleon
\cite{pion-n}.} and takes away the need for the artificial enhancement of the
coupling strength for set B. In Figs.~\ref{fig:Nmass_full1},
\ref{fig:Nmass_full2} we show the nucleon mass as a function of
the scalar and axial-vector diquark mass.  Similar to the scalar diquark
channel, we define the axial-vector diquark-quark threshold as
$M_{da}+\mqr$. We see that as one increase the axial-vector diquark
(and scalar diquark) masses, the $qq$ interaction is weakened and
consequently the nucleon mass is increases. Therefore the contribution
of the axial-vector channel to the nucleon mass is also attractive.

\begin{figure}
       \centerline{\includegraphics[clip,width=7 cm]
                                   {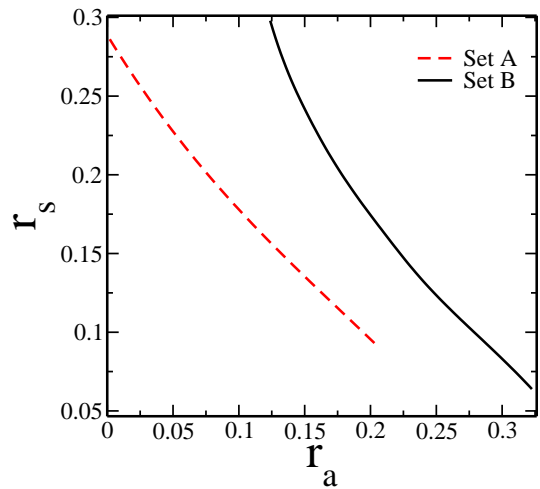}}
\caption{Range of parameters $(r_{s},r_{a})$ where we find a nucleon mass of 
$940$ MeV.\label{fig:Nmass_phys}}
\end{figure}

In Fig.~\ref{fig:Nmass_phys} we plot the parameter space of the
interaction Lagrangian with variable $r_{s}$ and $r_{a}$ which leads
to the nucleon mass $M_{N}=0.940$ GeV. The trend of this plot for the
non-confining set A is very similar to the one obtained in the local
NJL model \cite{njln3} (although we use a different parameter set) and
roughly depends linearly on the ratios $r_{s}$ and $r_{a}$,
\begin{equation}
M_{N}[0.940 \text{GeV}]=-r_{s}-0.94r_{a}+1.2.
\end{equation}
Therefore, any interaction Lagrangian with $r_{s}$ and $r_{a}$ which
satisfies the relation $r_{s}+0.94r_{a}=0.30$ gives a nucleon mass at
about the experimental value. This relation shows how interaction is
shared between scalar and axial-vector channels. If the scalar diquark
interaction $r_{s}$ is less than $0.14$, we need the axial-vector
interaction to be stronger than the scalar diquark channel
$r_{a}>r_{s}$ in order to get the experimental value of nucleon
mass. For set B, as we approach to $r_{a}=0$, the curve bends upward,
reflecting the fact that we have no bound state with only scalar
diquark channel.
In Fig.~\ref{fig:Nmass_phys} we see for the confining set B that the
interaction is again shared between the scalar and the axial-vector
diquark and for small $r_{s}<0.19$ one needs a dominant axial-vector
diquark channel $r_{a}>r_{s}$. It is obvious that the axial-vector
diquark channel is much more important in the confining than the
non-confining phase of model. 
\begin{table}
\caption{Diquark masses and coupling of diquarks to quarks
obtained for $M_{N}=0.940$. All masses are given in MeV.
 $E_{ds}(E_{da})$ denote the binding energy of diquarks in the
 nucleon, $E^{ds}_{N}(E^{da}_{N})$ denote the binding energy of the
 nucleon measured from scalar (axial) diquark mass.\label{tab:mdiquark}}
\centering
\begin{tabular}{ccccccc}
\hline
\hline
\multicolumn{1}{c}{}& \multicolumn{3}{c}{Set A}& \multicolumn{3}{c}{Set B}  \\
&Set A1 & Set A2 & Set A3 & Set B1 &Set B2 &Set B3\\
\hline
$M_{ds}$& 775     & 748   &698       & 802        &705  &609        \\
$g_{dsqq}$&0.74    &0.83   &1.04       & 0.73       &  1.31    &1.79       \\
$r_{s}$ &0.09      &0.12  &0.17       & 0.06         &  0.14   &0.24        \\
$E_{ds}$& 7 &34  &84       & 14   & 111   &207          \\
$E^{ds}_{N}$&226  &199 &149        & 270   & 173   &77           \\
$M_{da}$&705      &725 &775       & 604       & 660  &725       \\
$g_{daqq}$&1.08     &0.98  &0.79         & 1.99      & 1.67   &1.28      \\
$r_{a}$&0.20        &0.17  &0.11           & 0.32        & 0.23   &0.15     \\
$E_{da}$&77  &57  &7     & 212  & 156   &91       \\
$E^{da}_{N}$&156  &176 &226          & 72  & 123 &193       \\
$p_{\bot}^{\text{RMS}}$&194.88 &181.51 &163.70 &283.99& 232.86 &  209.95 \\
\hline
\hline
\end{tabular}
\end{table}



\begin{figure}[!htb]
\begin{tabular}{cc}
\includegraphics[height=.22\textheight]{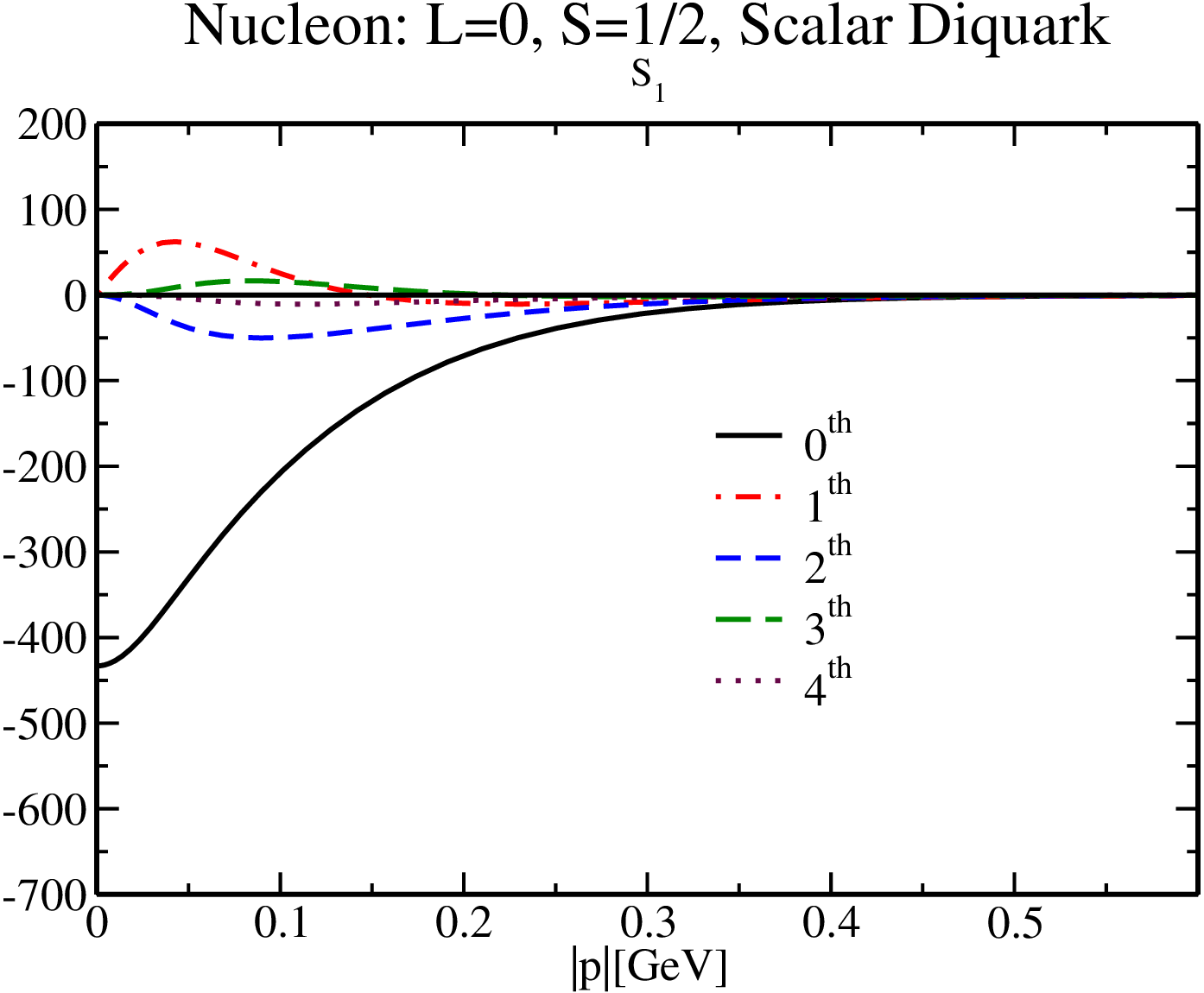}&
\hspace{0.5 cm}
\includegraphics[height=.22\textheight]{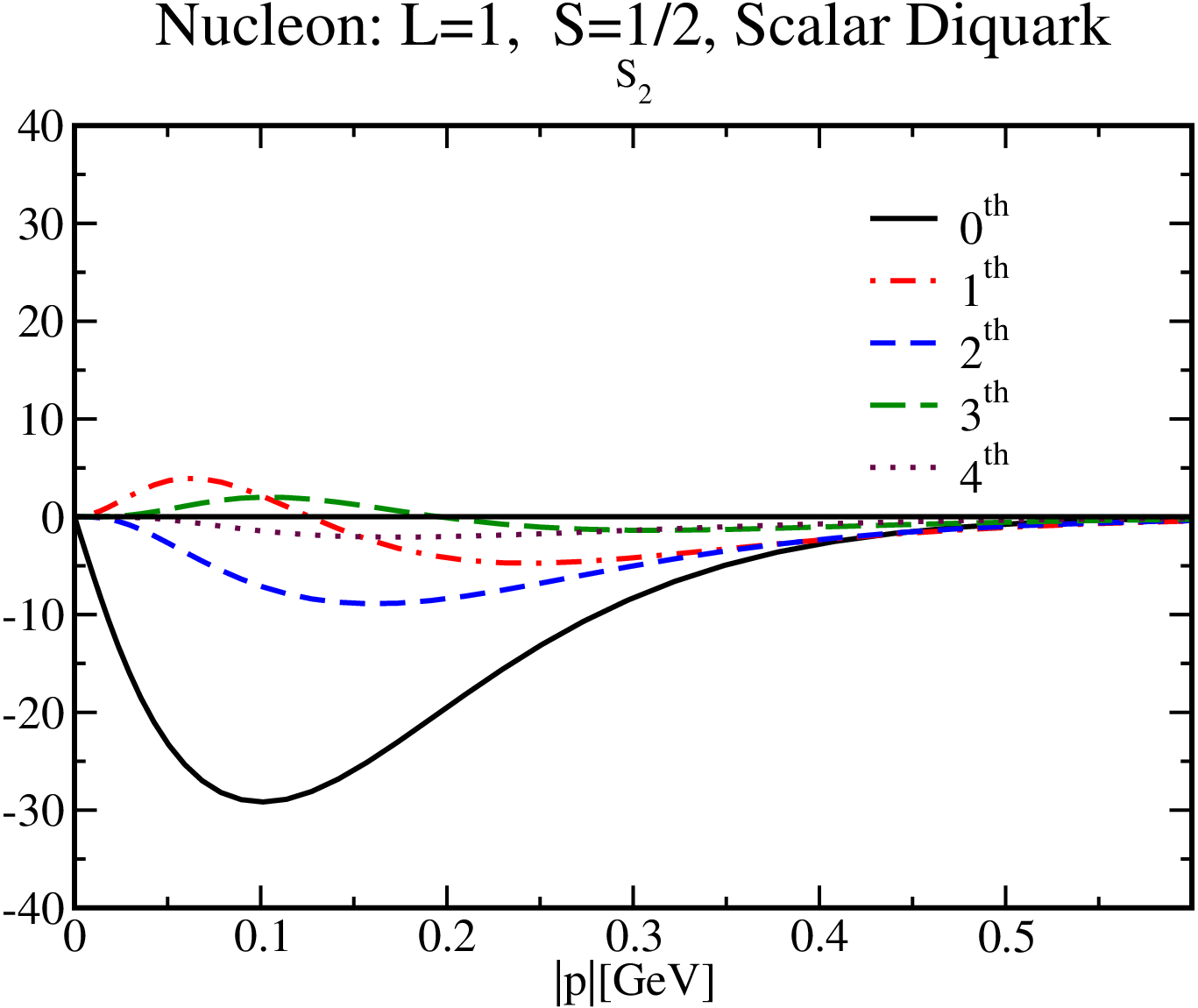}
\vspace{.20 cm}
\end{tabular}
\begin{tabular}{cc}
\includegraphics[height=.22\textheight]{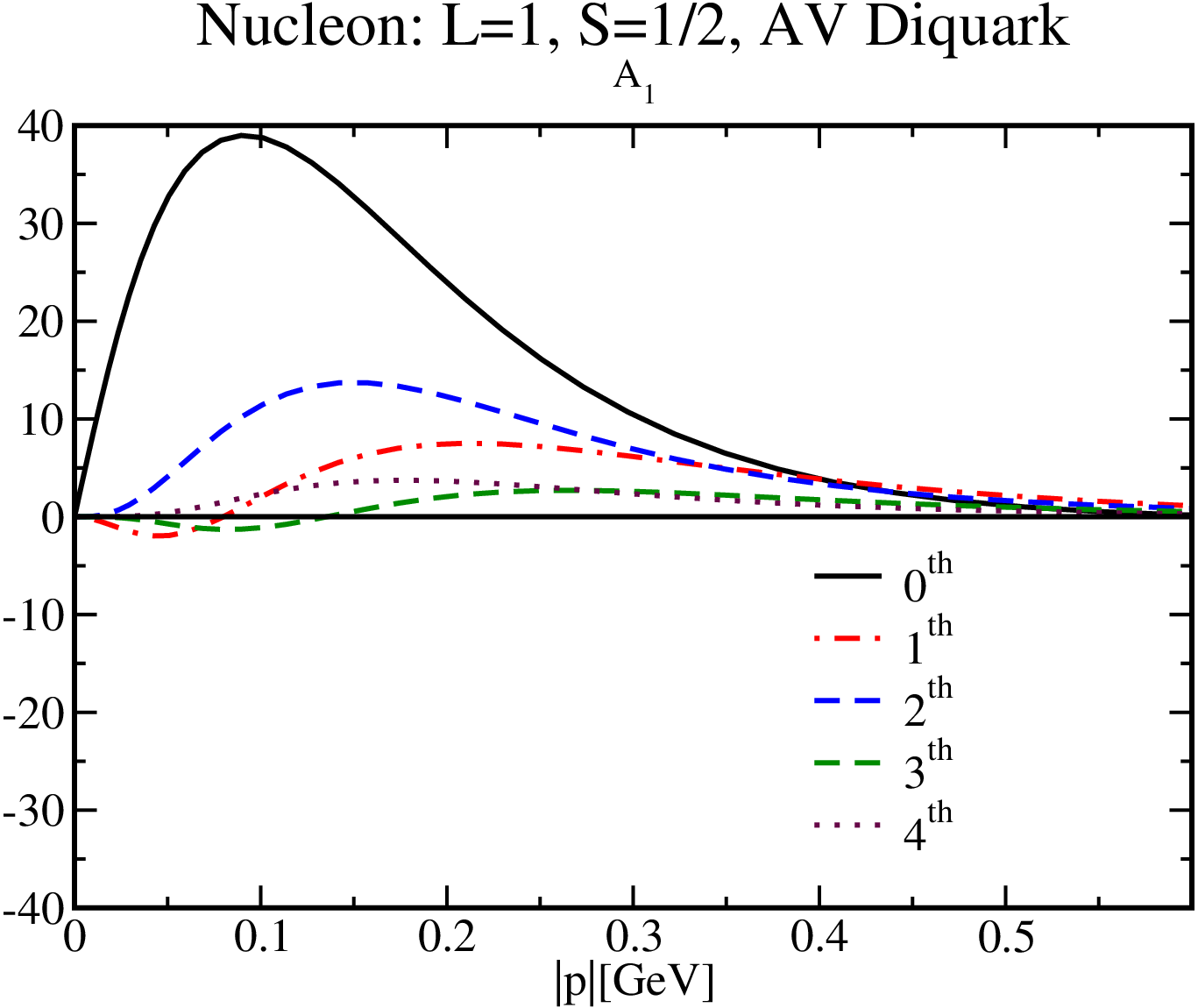}&
\hspace{0.5 cm}
\includegraphics[height=.22\textheight]{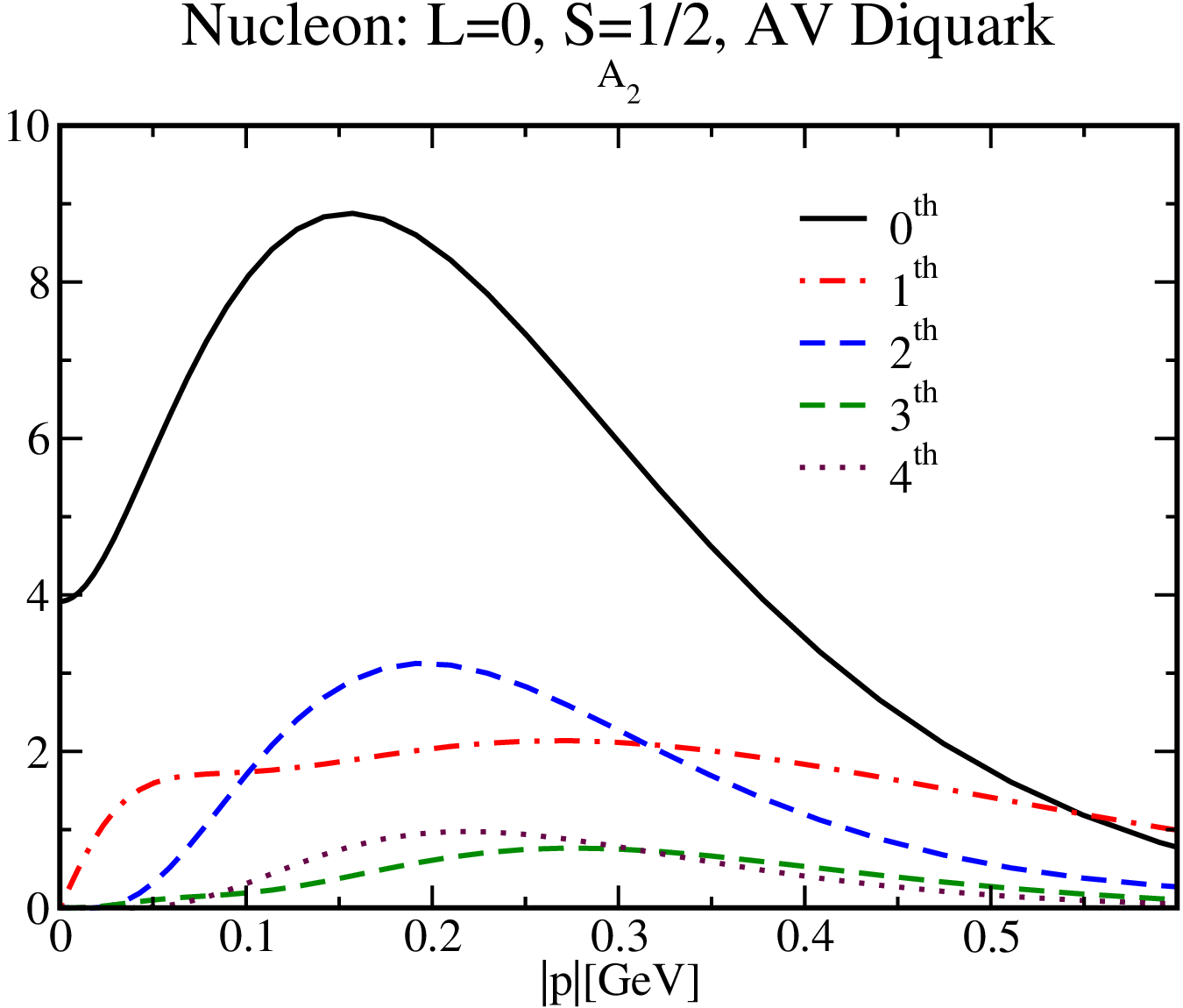}
\vspace{.20 cm}
\end{tabular}
\begin{tabular}{cc}
\includegraphics[height=.22\textheight]{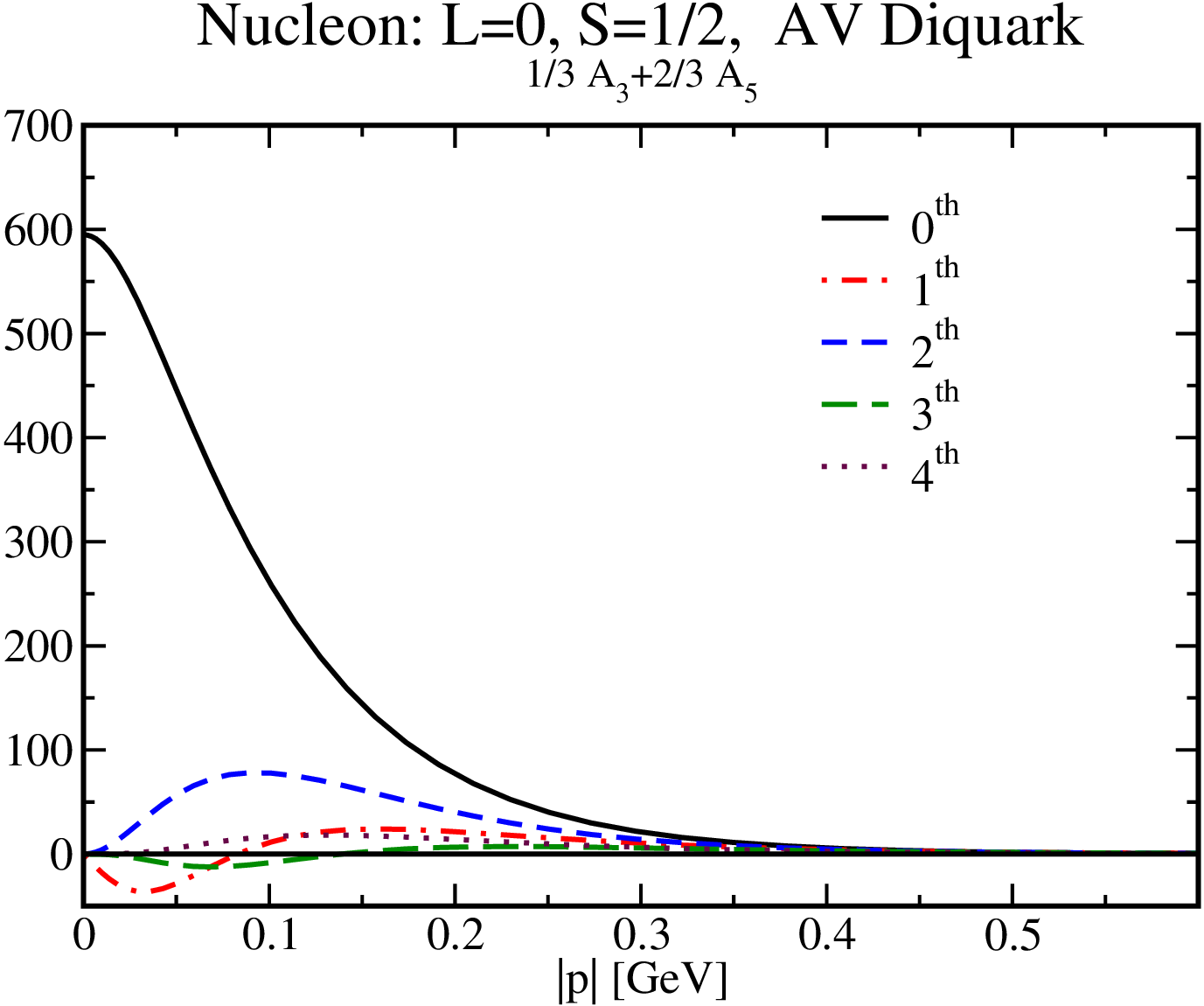}&
\hspace{0.5 cm}
\includegraphics[height=.22\textheight]{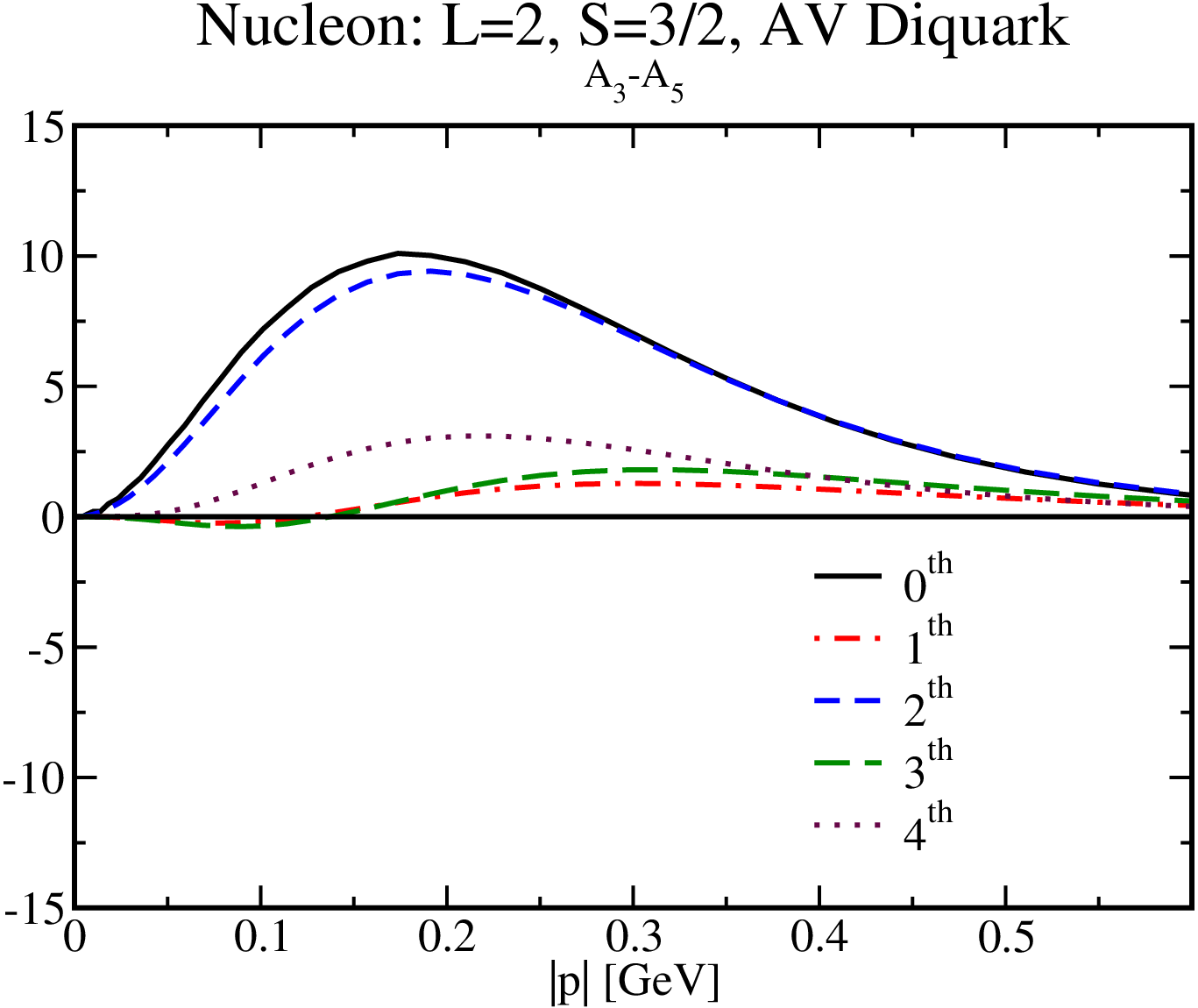} 
\vspace{.20 cm}
\end{tabular}
\begin{tabular}{cc}
\includegraphics[height=.22\textheight]{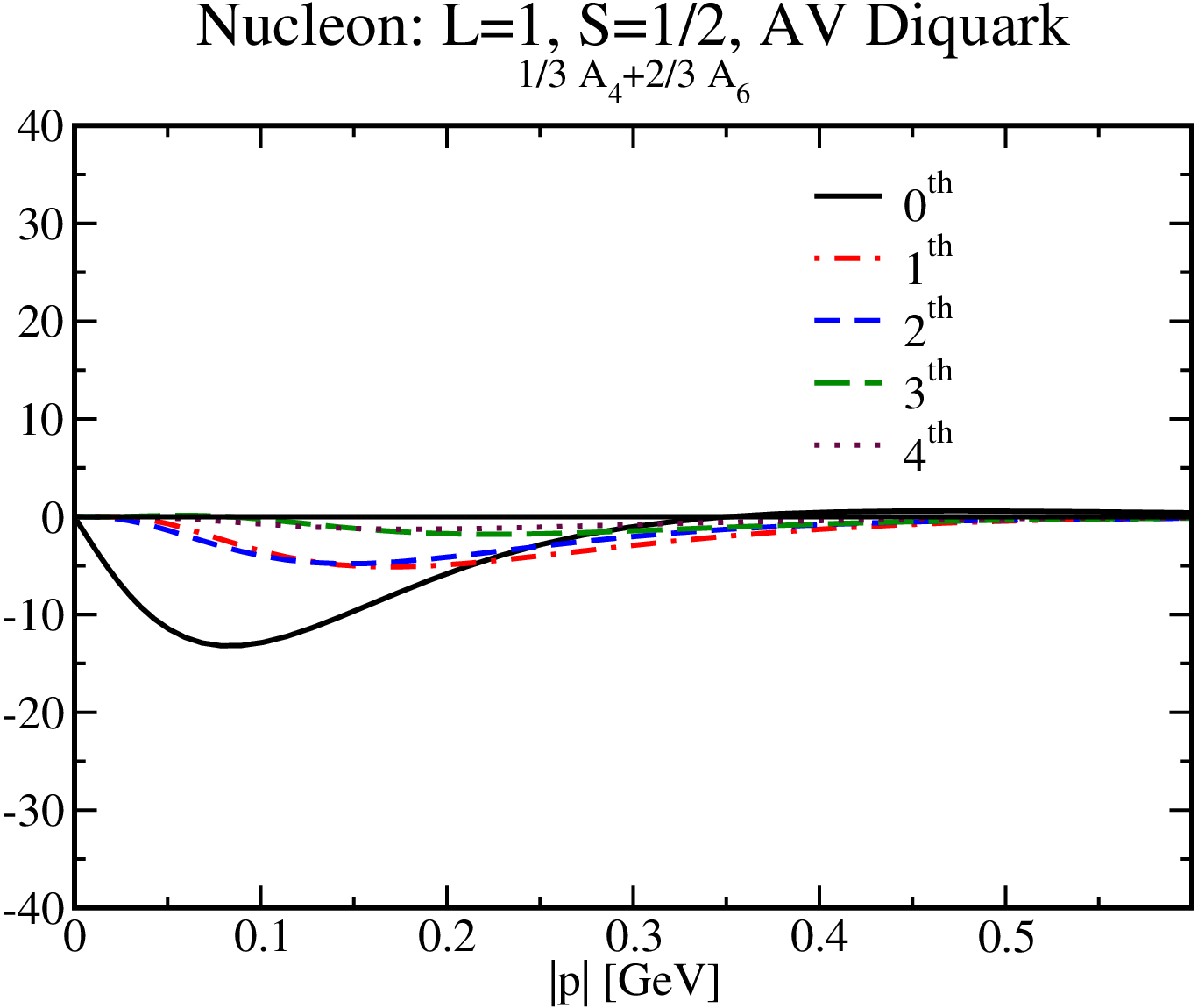}&
\hspace{0.5 cm}
\includegraphics[height=.22\textheight]{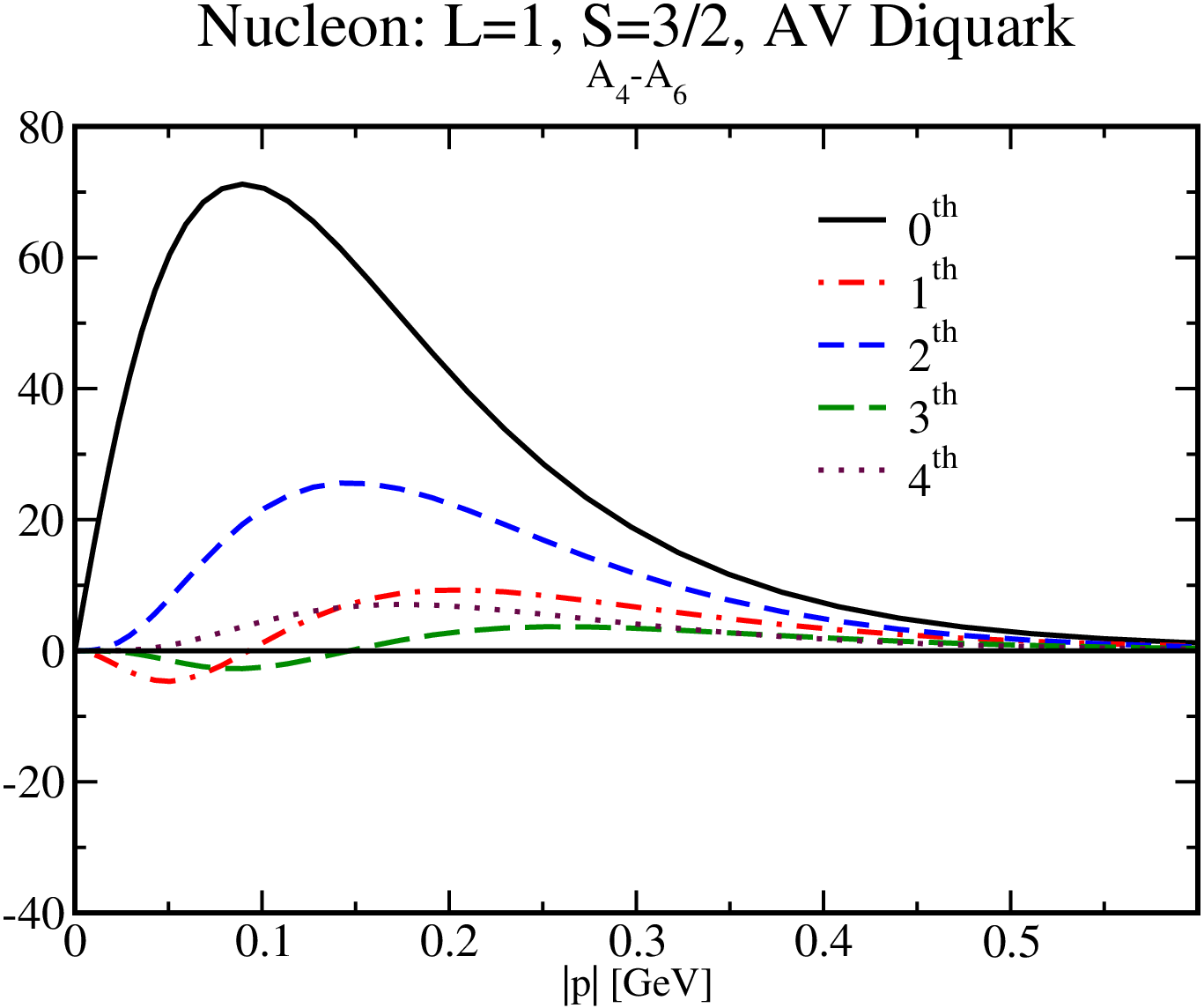}
\end{tabular}
\caption{Different Chebyshev moments (labeled by $n^{th}$) of scalar and axialvector (AV) diquark amplitudes of the nucleon BS wave function given by Set A1.}
\end{figure}

\newpage
\begin{figure}[!htb]
\begin{tabular}{cc}
\includegraphics[height=.22\textheight]{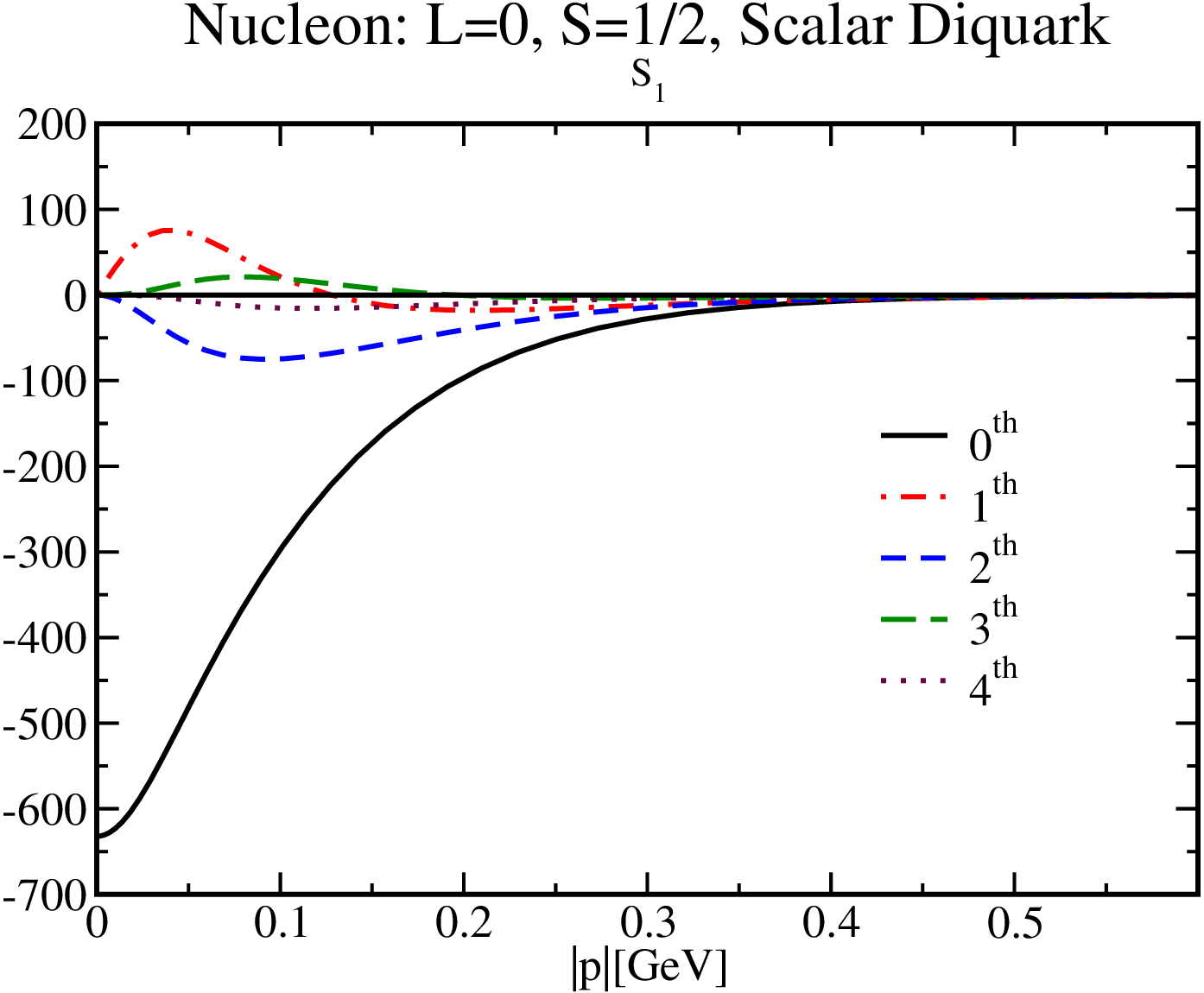}&
\hspace{0.5 cm}
\includegraphics[height=.22\textheight]{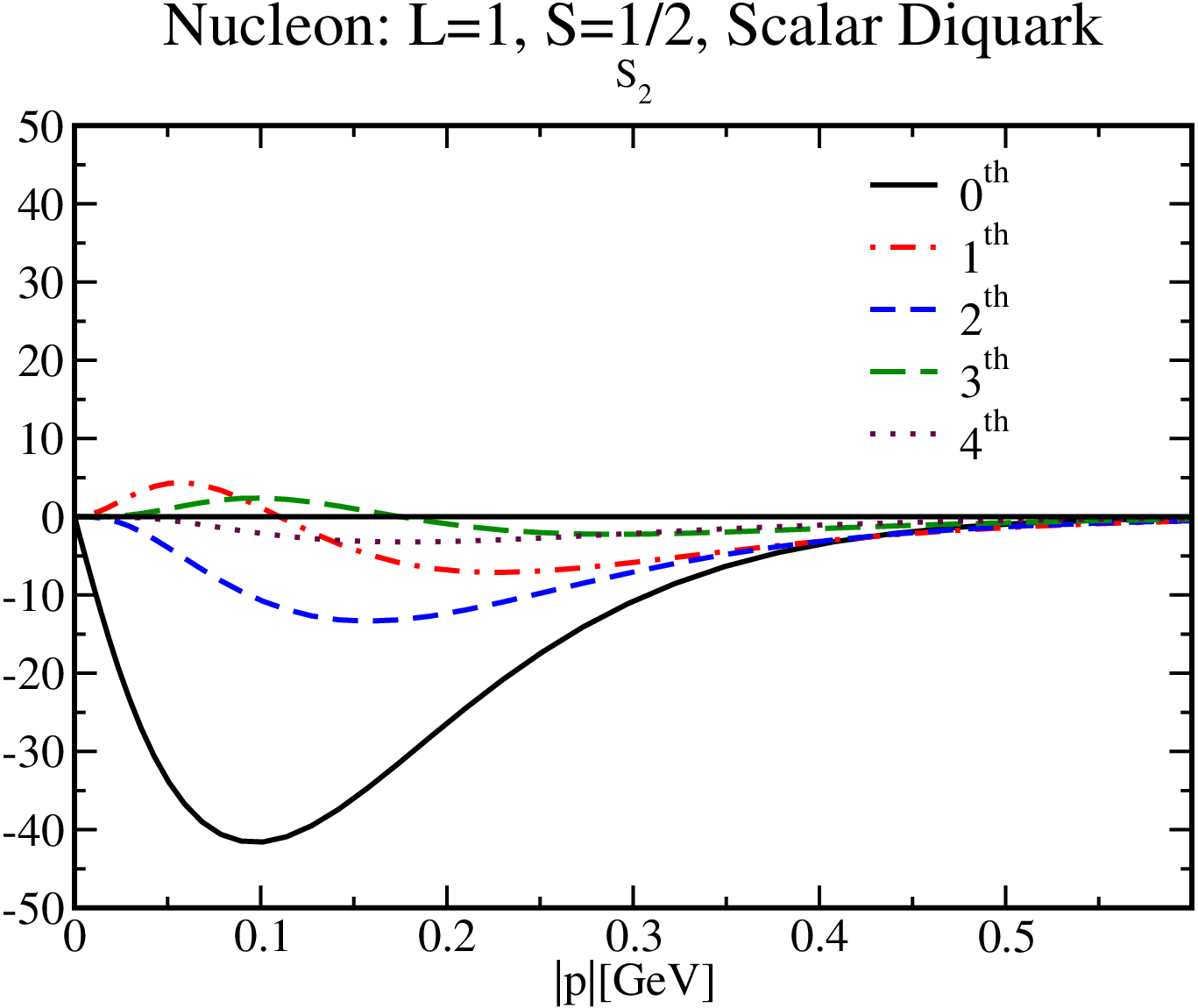}
\vspace{.20 cm}
\end{tabular}
\begin{tabular}{cc}
\includegraphics[height=.22\textheight]{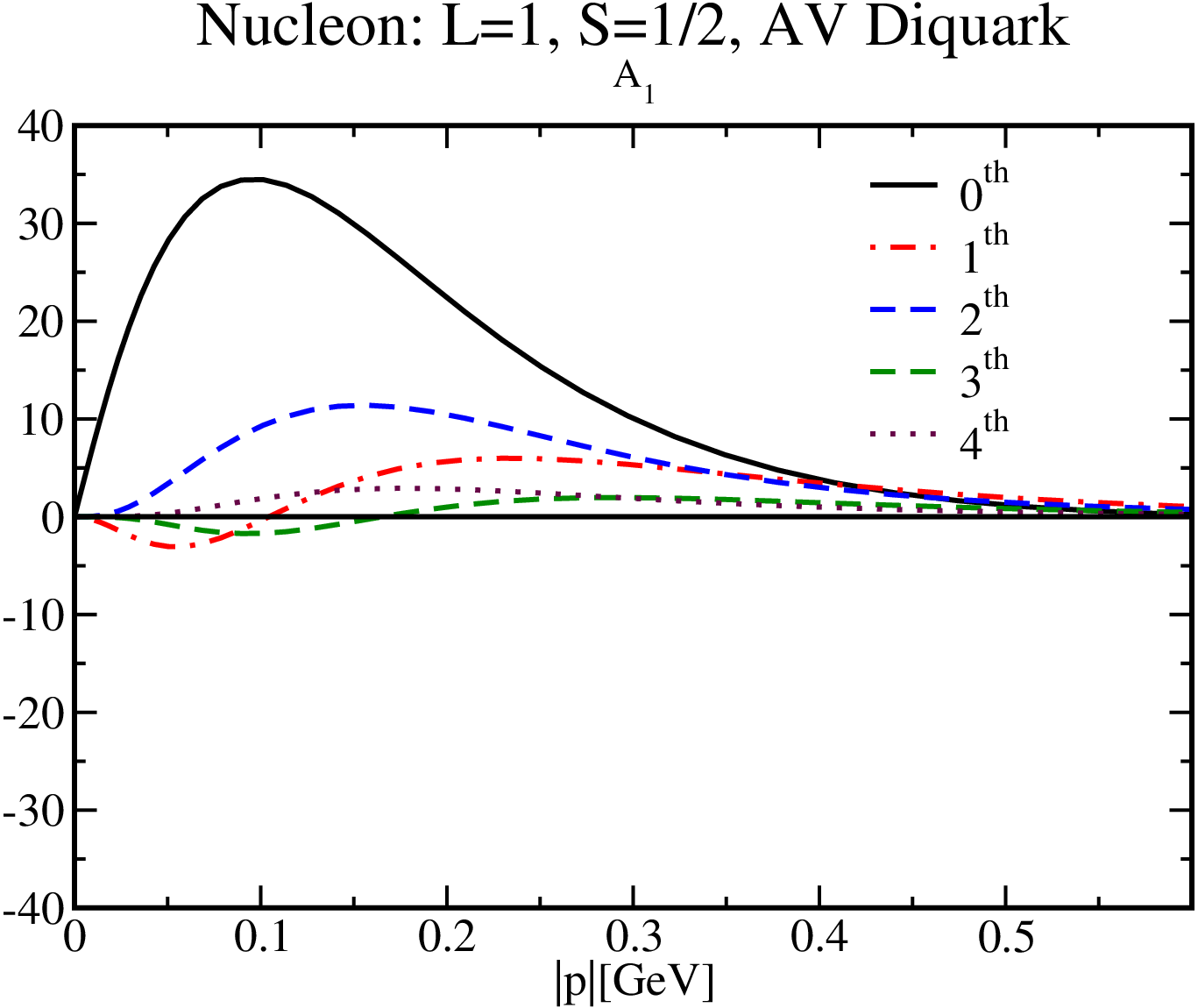}&
\hspace{0.5 cm}
\includegraphics[height=.22\textheight]{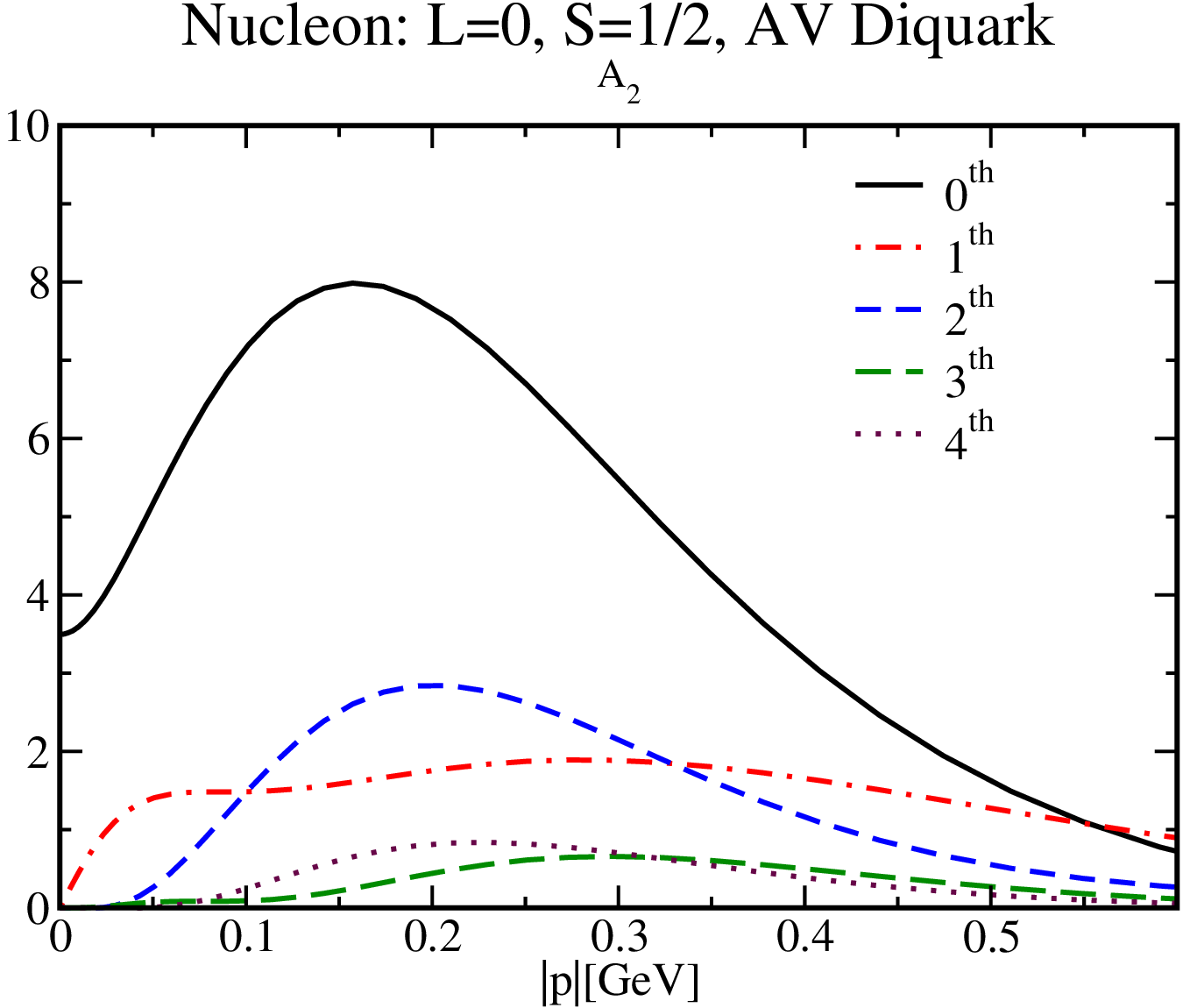}
\vspace{.20 cm}
\end{tabular}
\begin{tabular}{cc}
\includegraphics[height=.22\textheight]{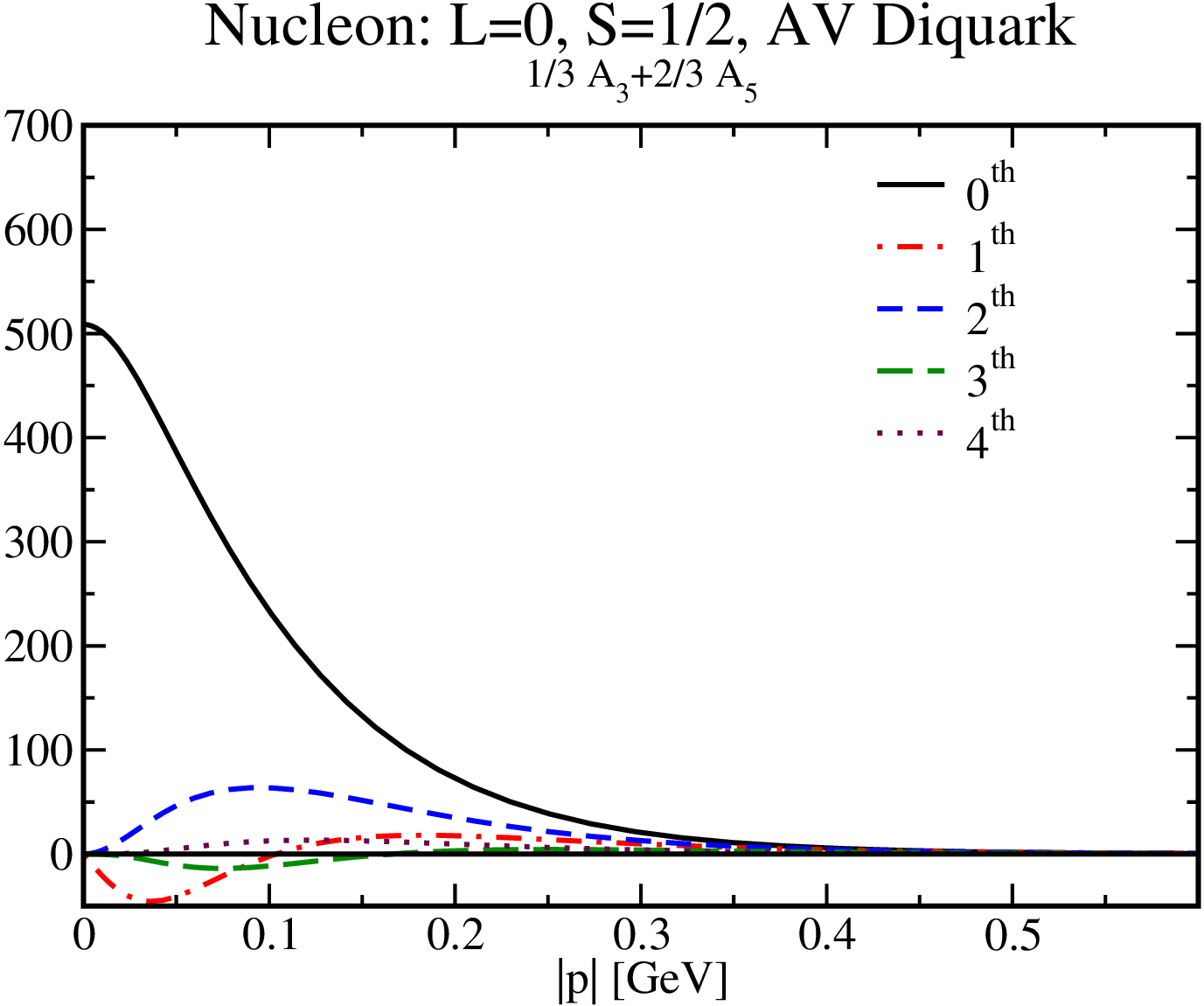}&
\hspace{0.5 cm}
\includegraphics[height=.22\textheight]{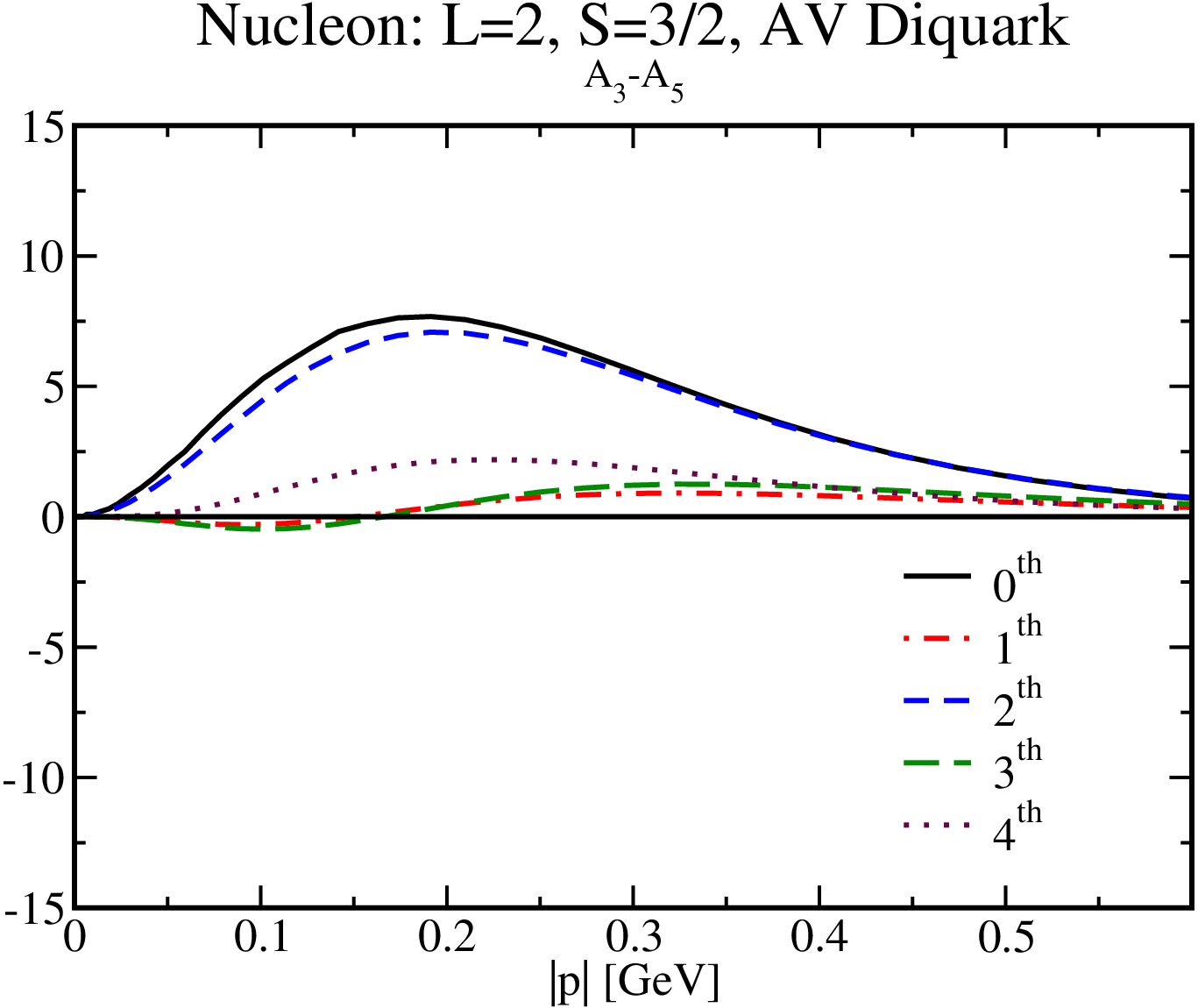} 
\vspace{.20 cm}
\end{tabular}
\begin{tabular}{cc}
\includegraphics[height=.22\textheight]{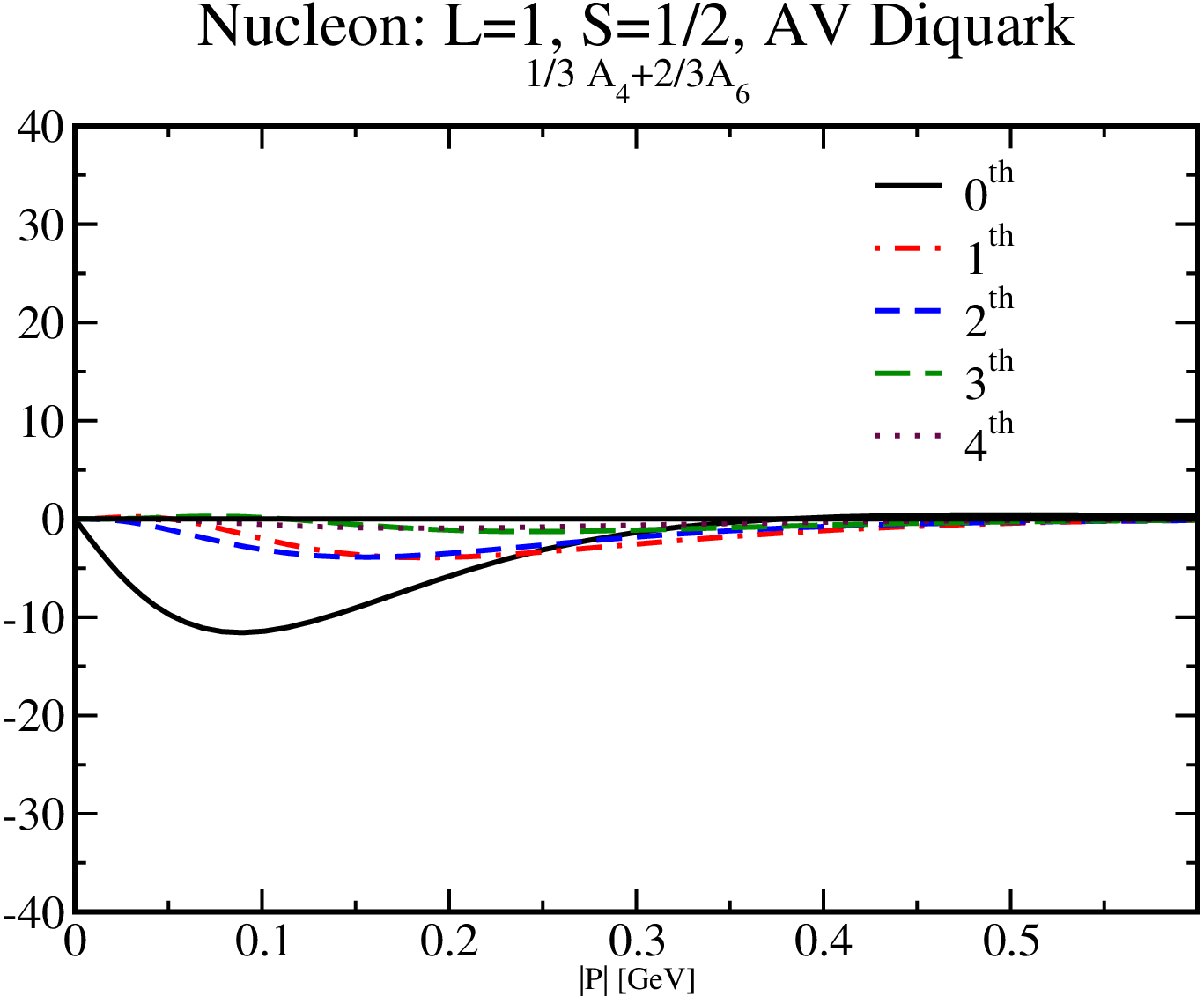}&
\hspace{0.5 cm}
\includegraphics[height=.22\textheight]{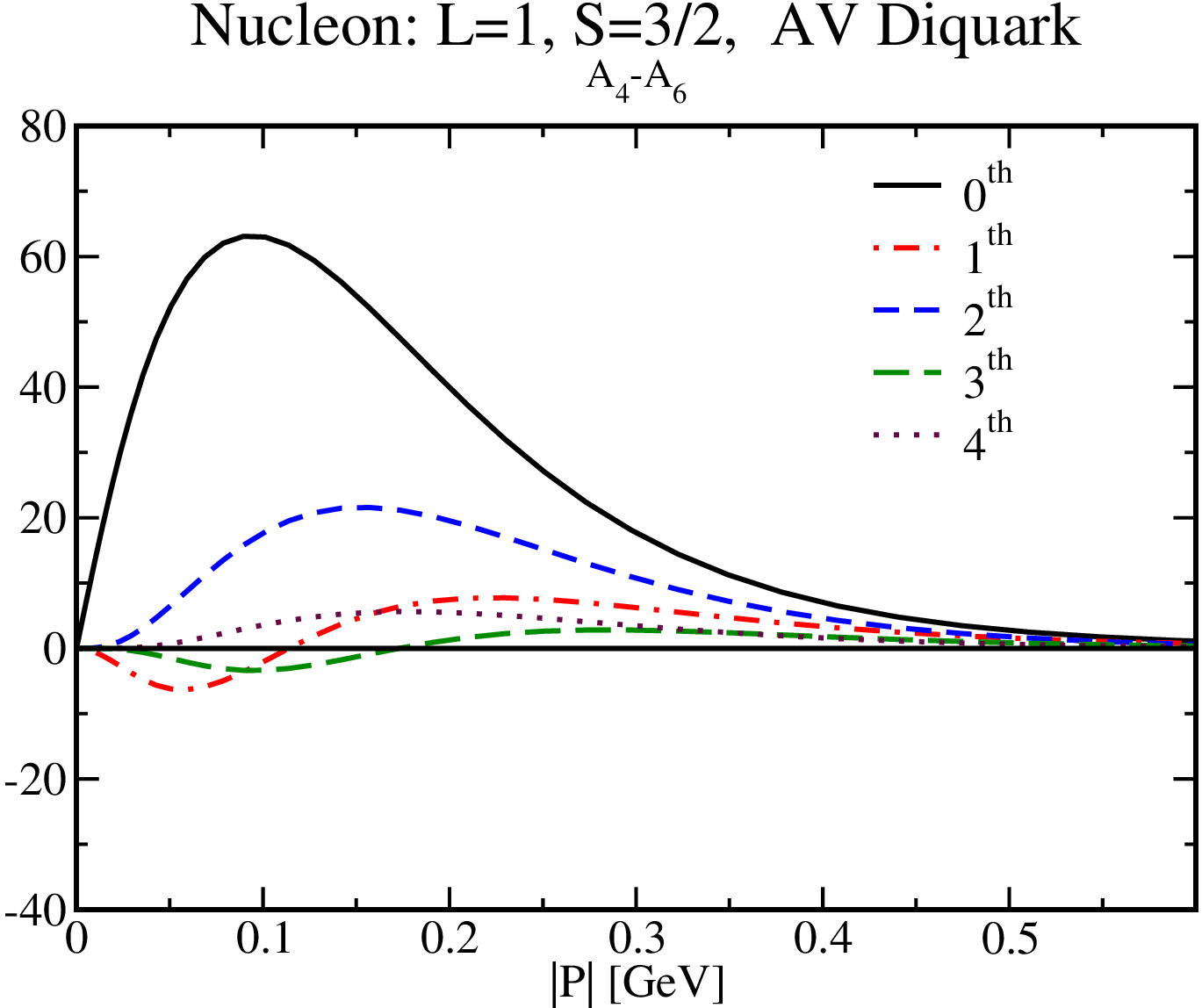}
\end{tabular}
\caption{Different Chebyshev moments (labeled by $n^{th}$) of scalar and axialvector diquark amplitudes of the nucleon wave function given by Set A2.}
\end{figure}

\newpage
\begin{figure}[!htb]
\begin{tabular}{cc}
\includegraphics[height=.22\textheight]{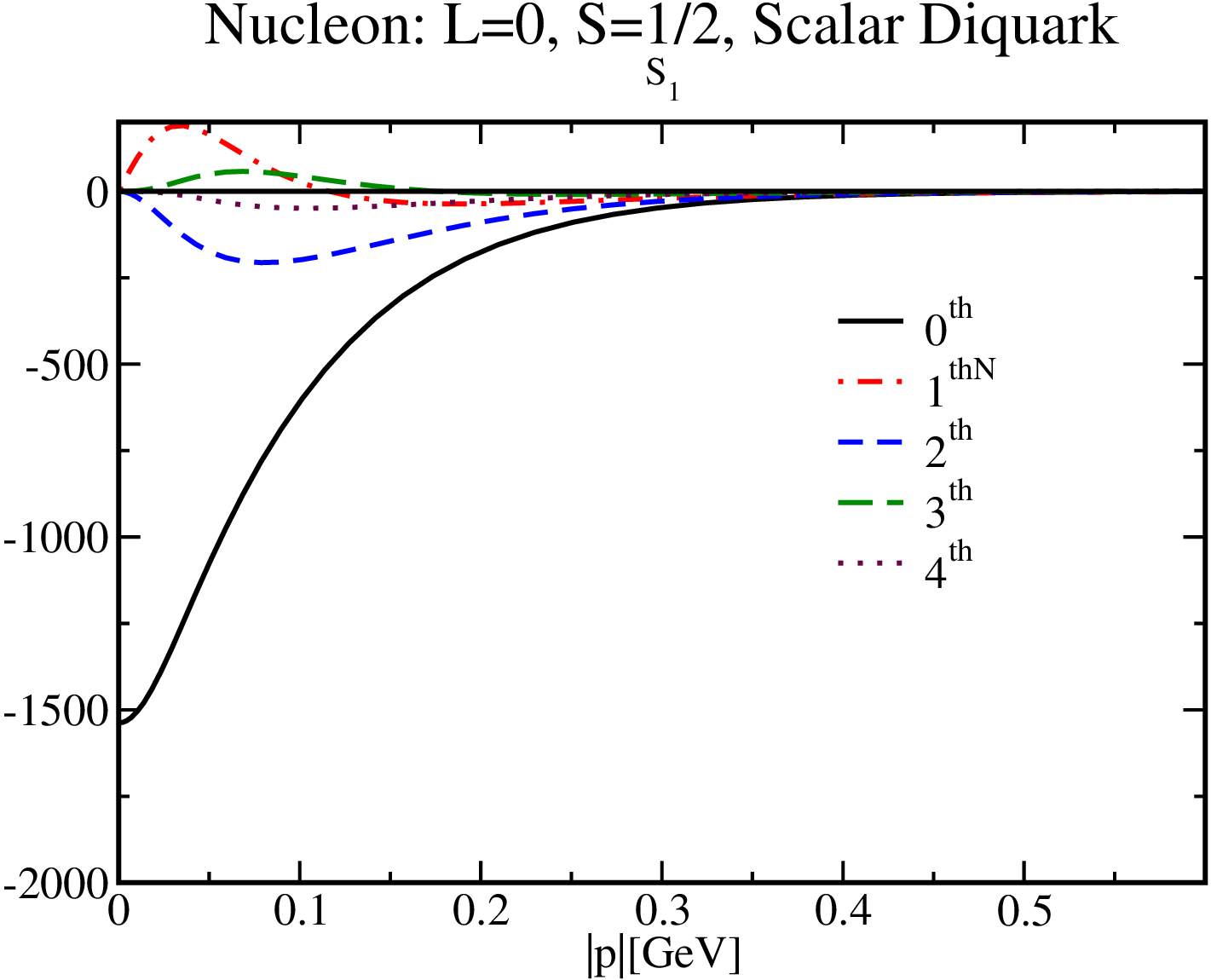}&
\hspace{0.5 cm}
\includegraphics[height=.22\textheight]{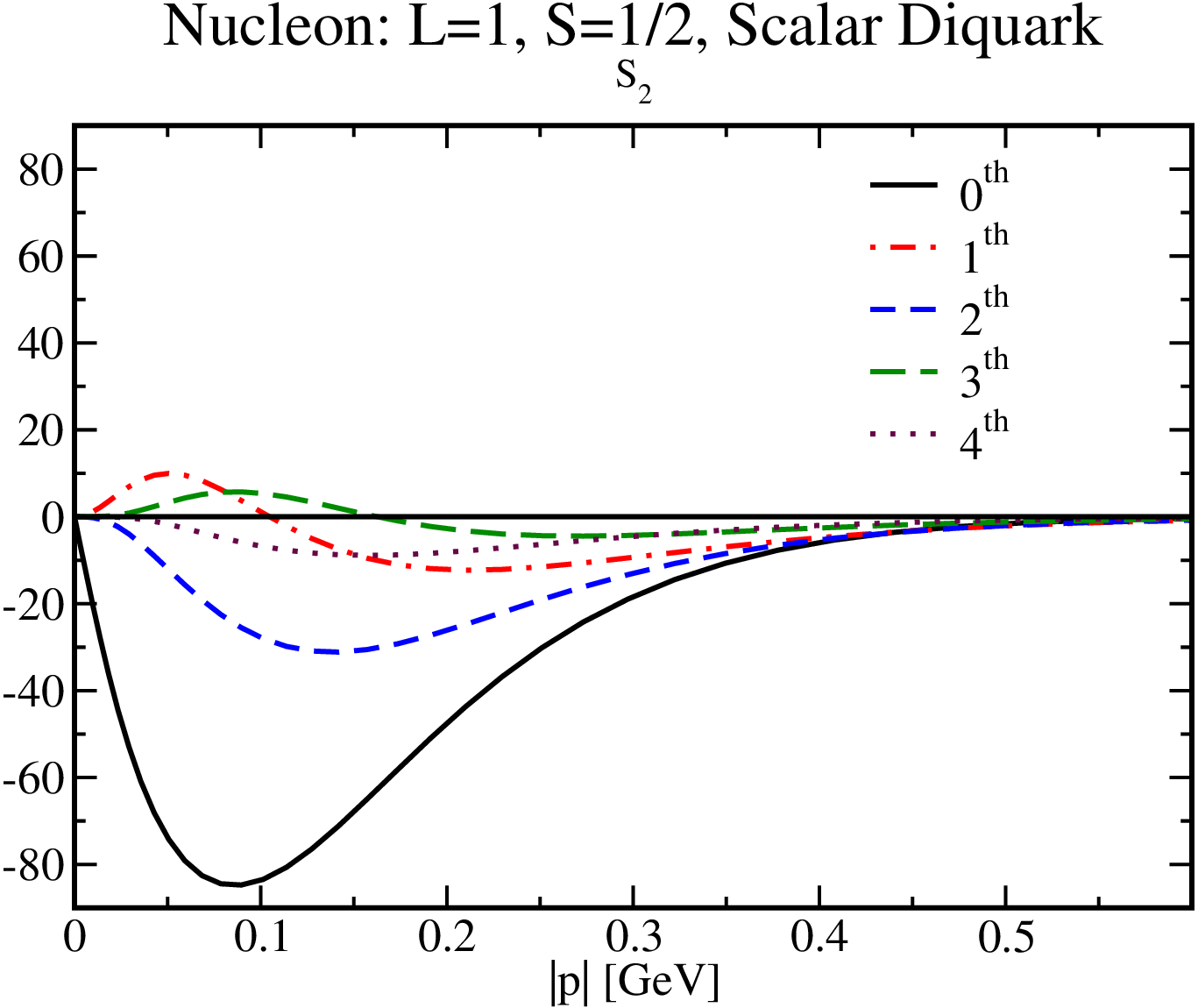}
\vspace{.20 cm}
\end{tabular}
\begin{tabular}{cc}
\includegraphics[height=.22\textheight]{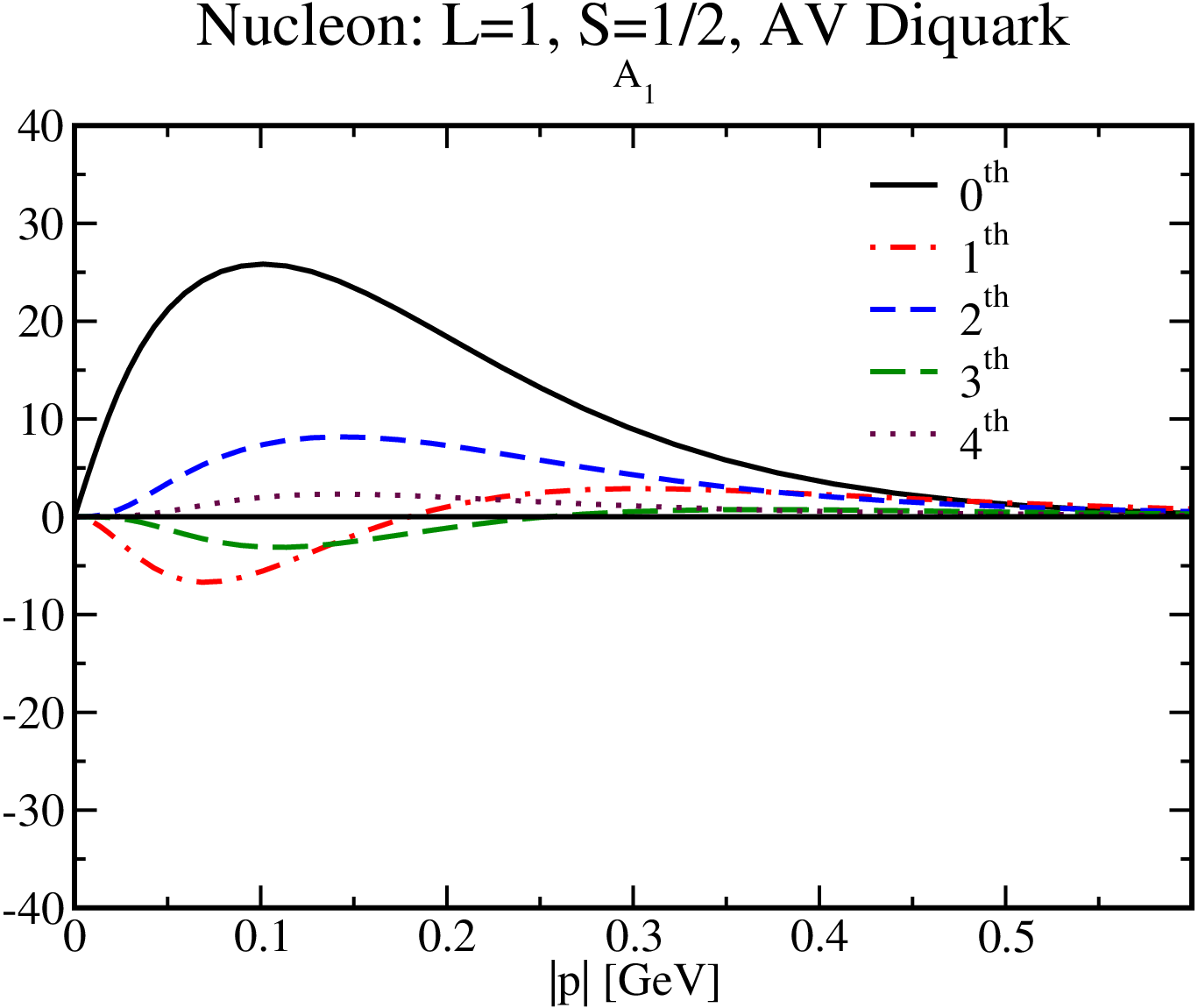}&
\hspace{0.5 cm}
\includegraphics[height=.22\textheight]{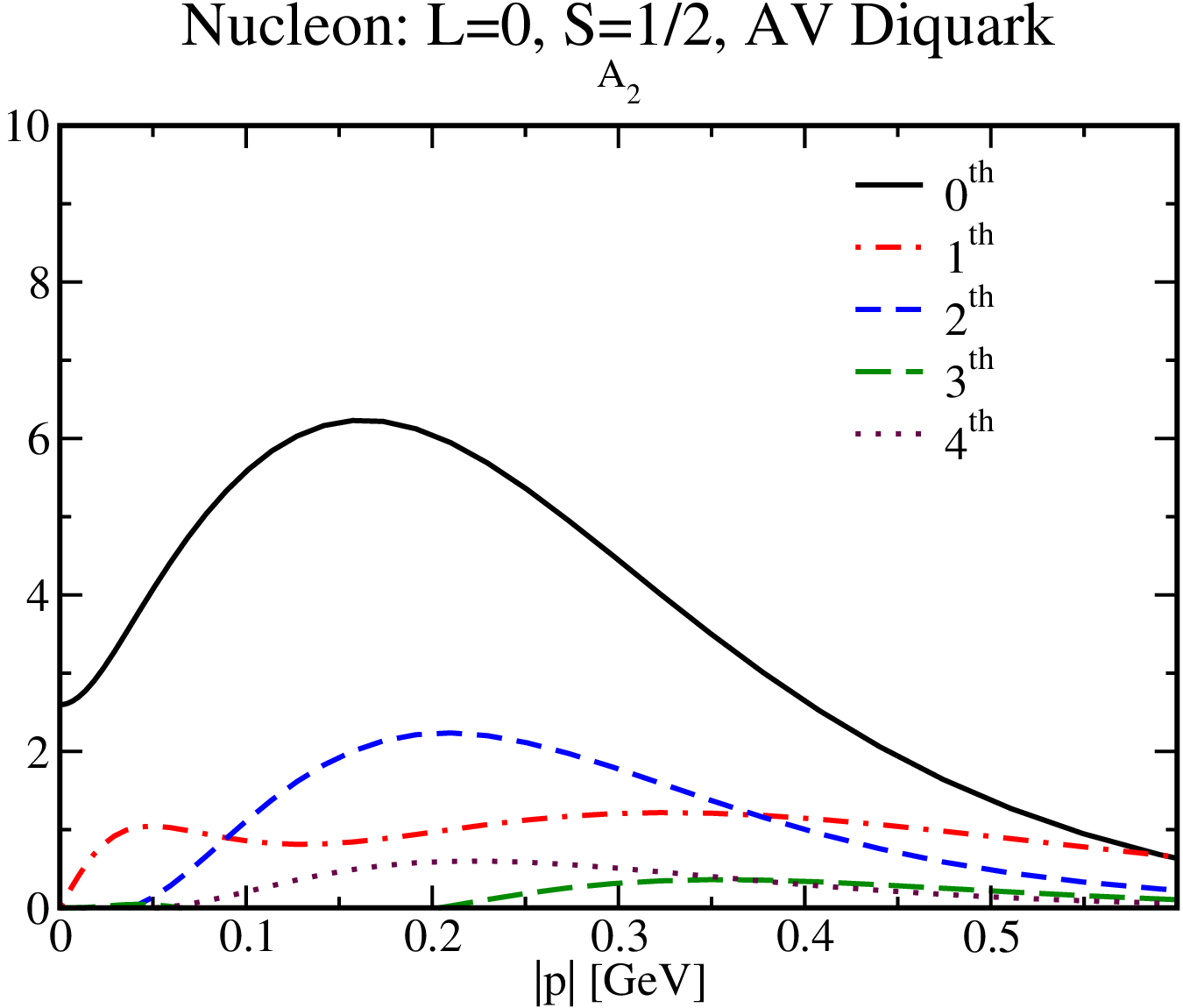}
\vspace{.20 cm}
\end{tabular}
\begin{tabular}{cc}
\includegraphics[height=.22\textheight]{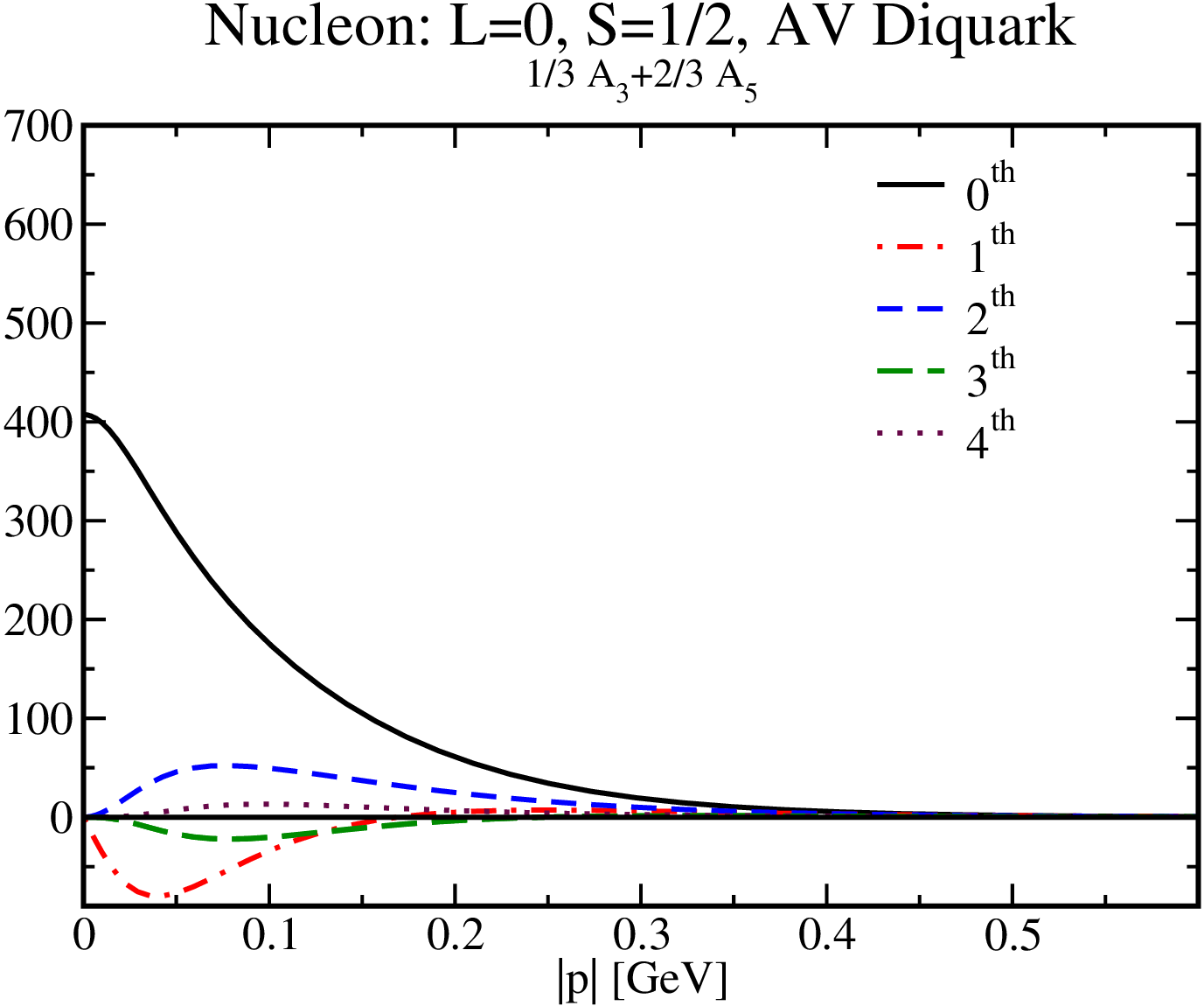}&
\hspace{0.5 cm}
\includegraphics[height=.22\textheight]{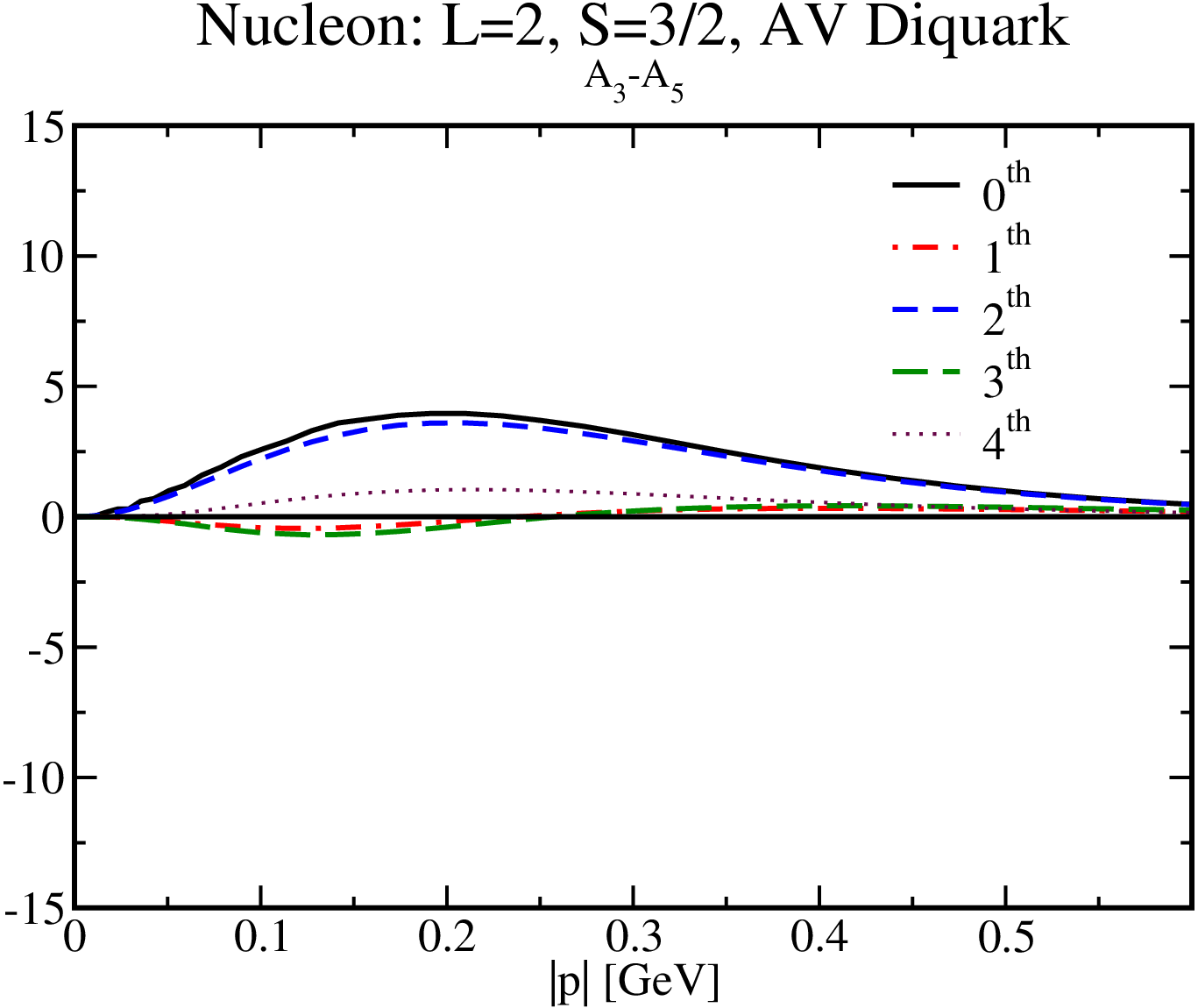} 
\vspace{.20 cm}
\end{tabular}
\begin{tabular}{cc}
\includegraphics[height=.22\textheight]{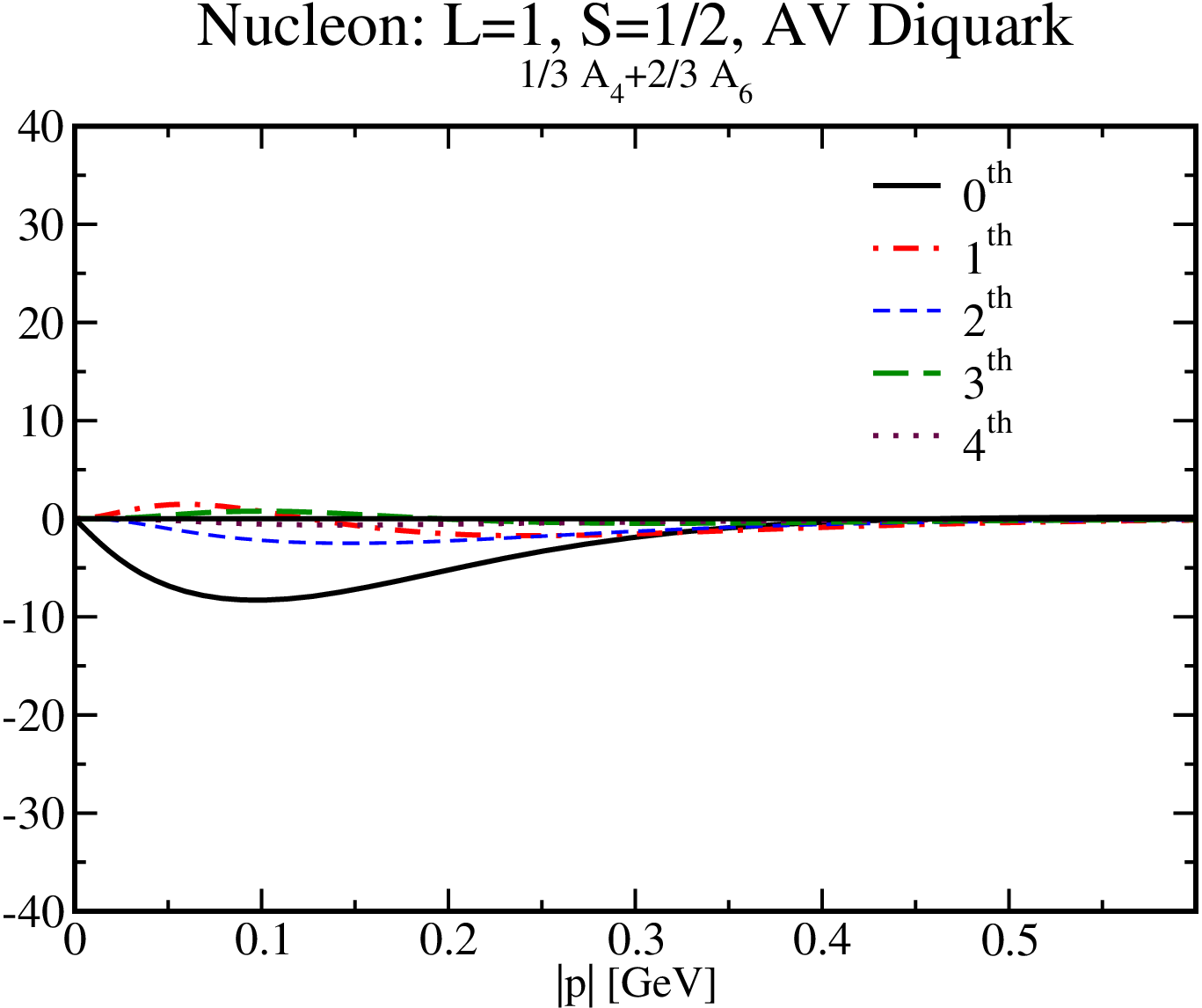}&
\hspace{0.5 cm}
\includegraphics[height=.22\textheight]{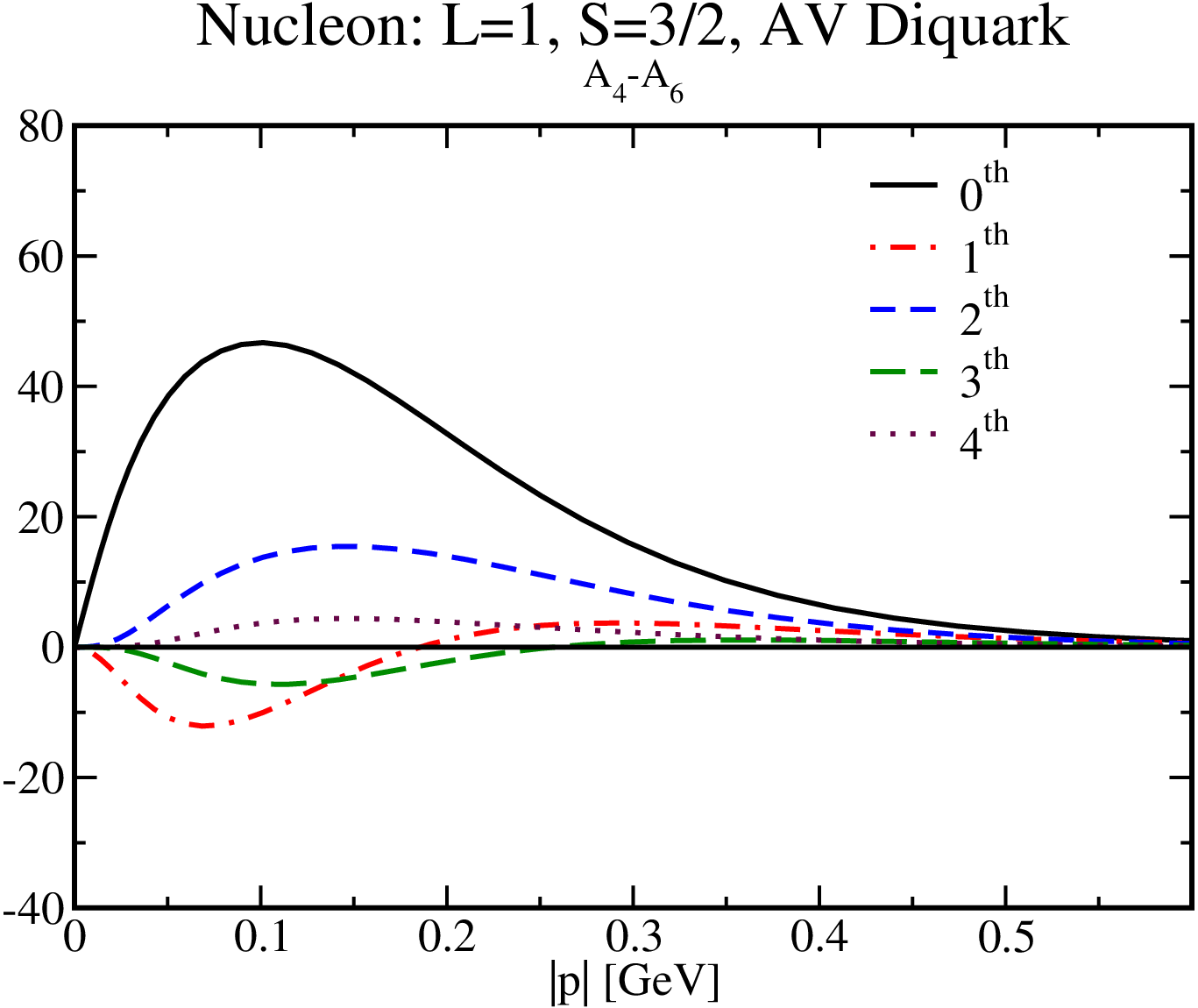}
\end{tabular}
\caption{Different Chebyshev moments (labeled by $n^{th}$) of scalar and axialvector diquark amplitudes of the nucleon wave function given by Set A3.}
\end{figure}

\newpage
\begin{figure}[!htb]
\begin{tabular}{cc}
\includegraphics[height=.22\textheight]{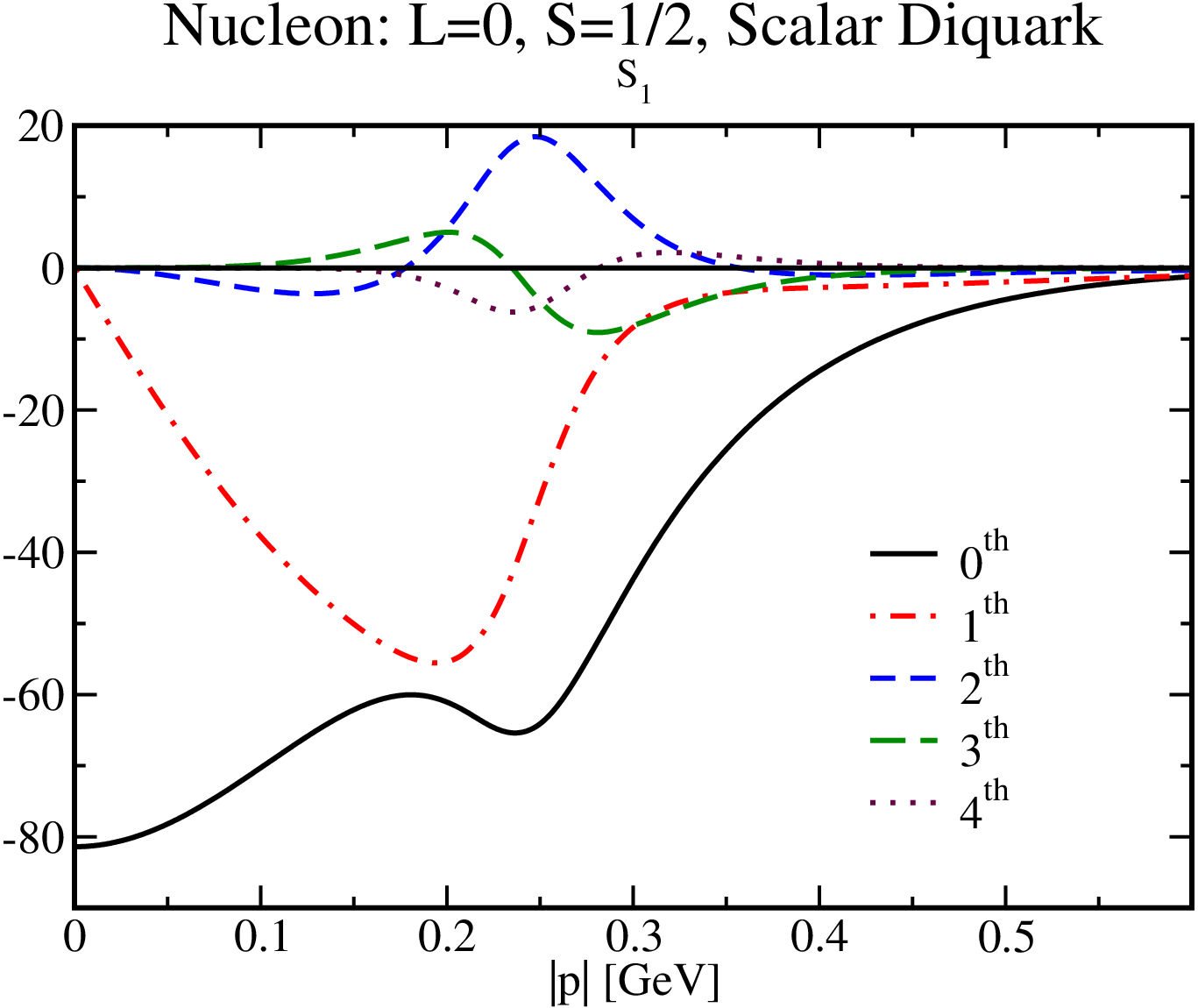}&
\hspace{0.5 cm}
\includegraphics[height=.22\textheight]{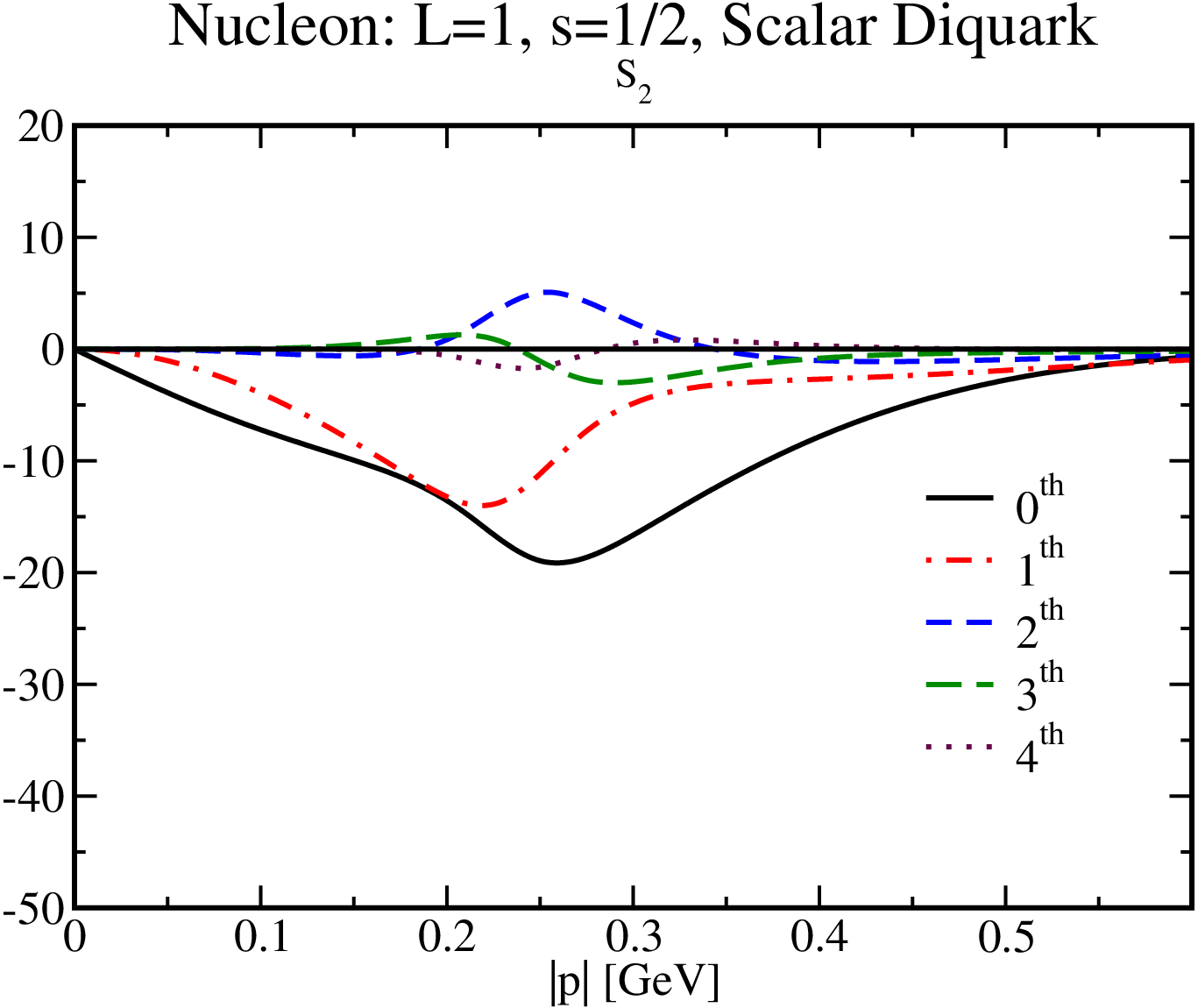}
\vspace{.20 cm}
\end{tabular}
\begin{tabular}{cc}
\includegraphics[height=.22\textheight]{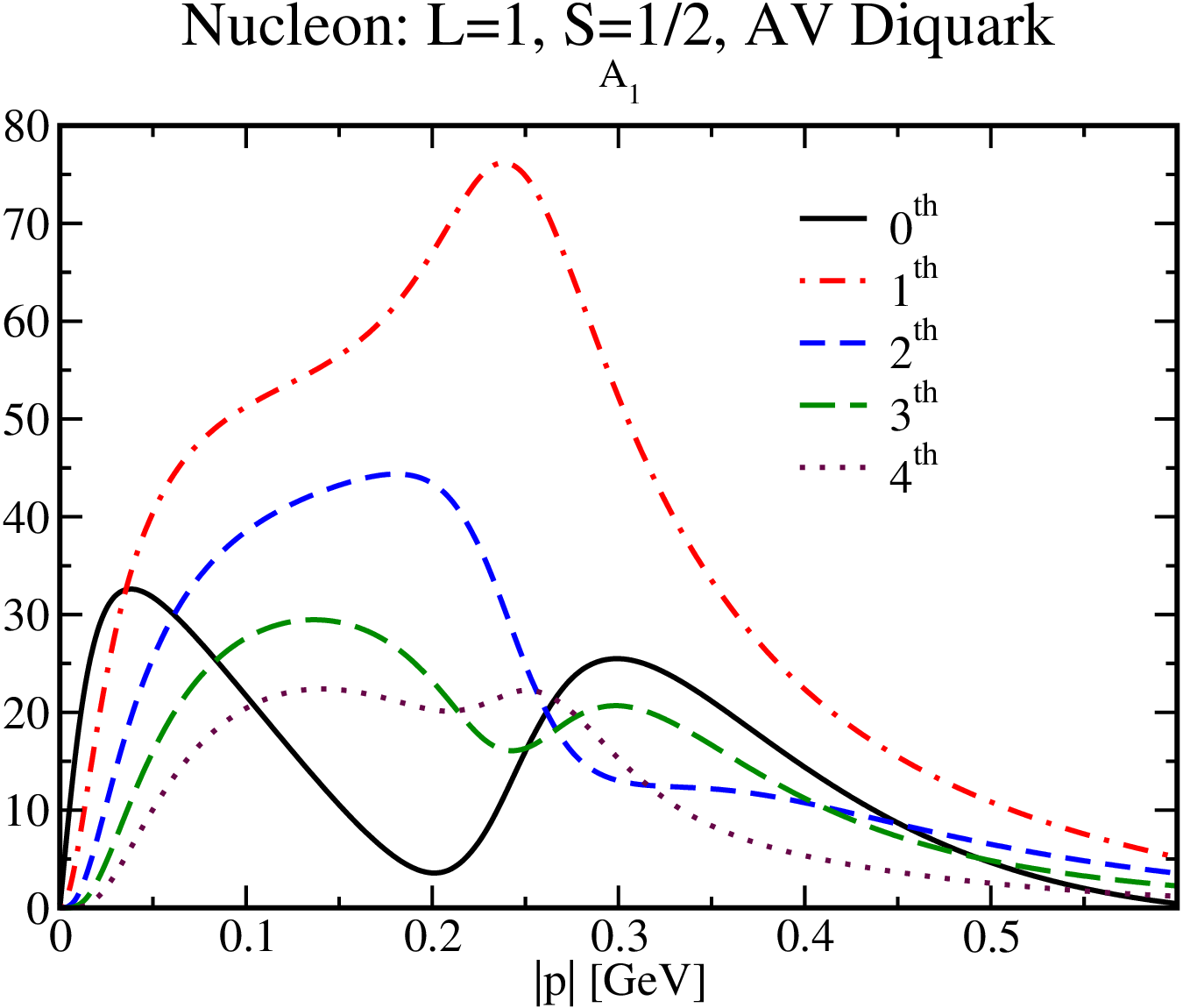}&
\hspace{0.5 cm}
\includegraphics[height=.22\textheight]{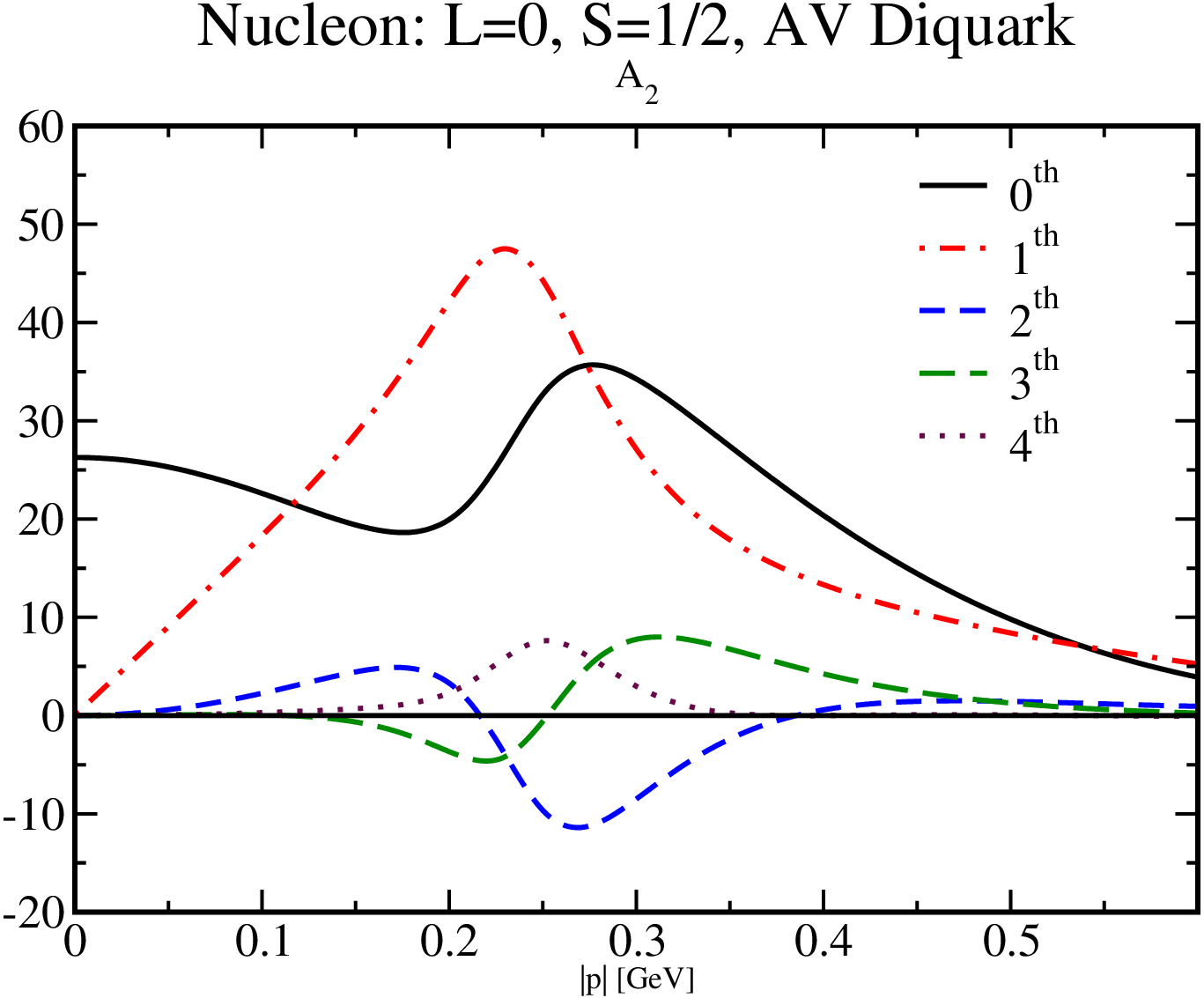}
\vspace{.20 cm}
\end{tabular}
\begin{tabular}{cc}
\includegraphics[height=.22\textheight]{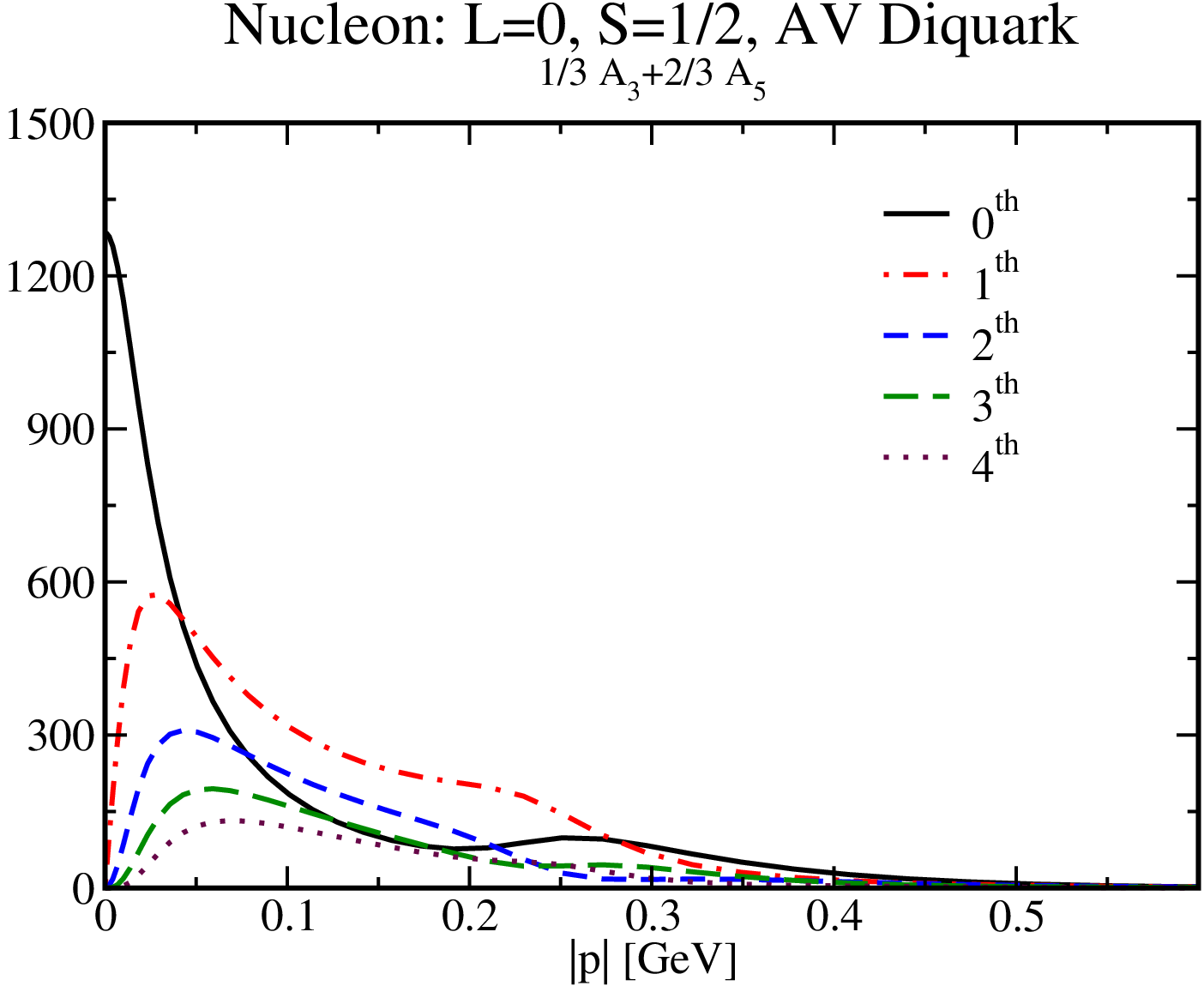}&
\hspace{0.5 cm}
\includegraphics[height=.22\textheight]{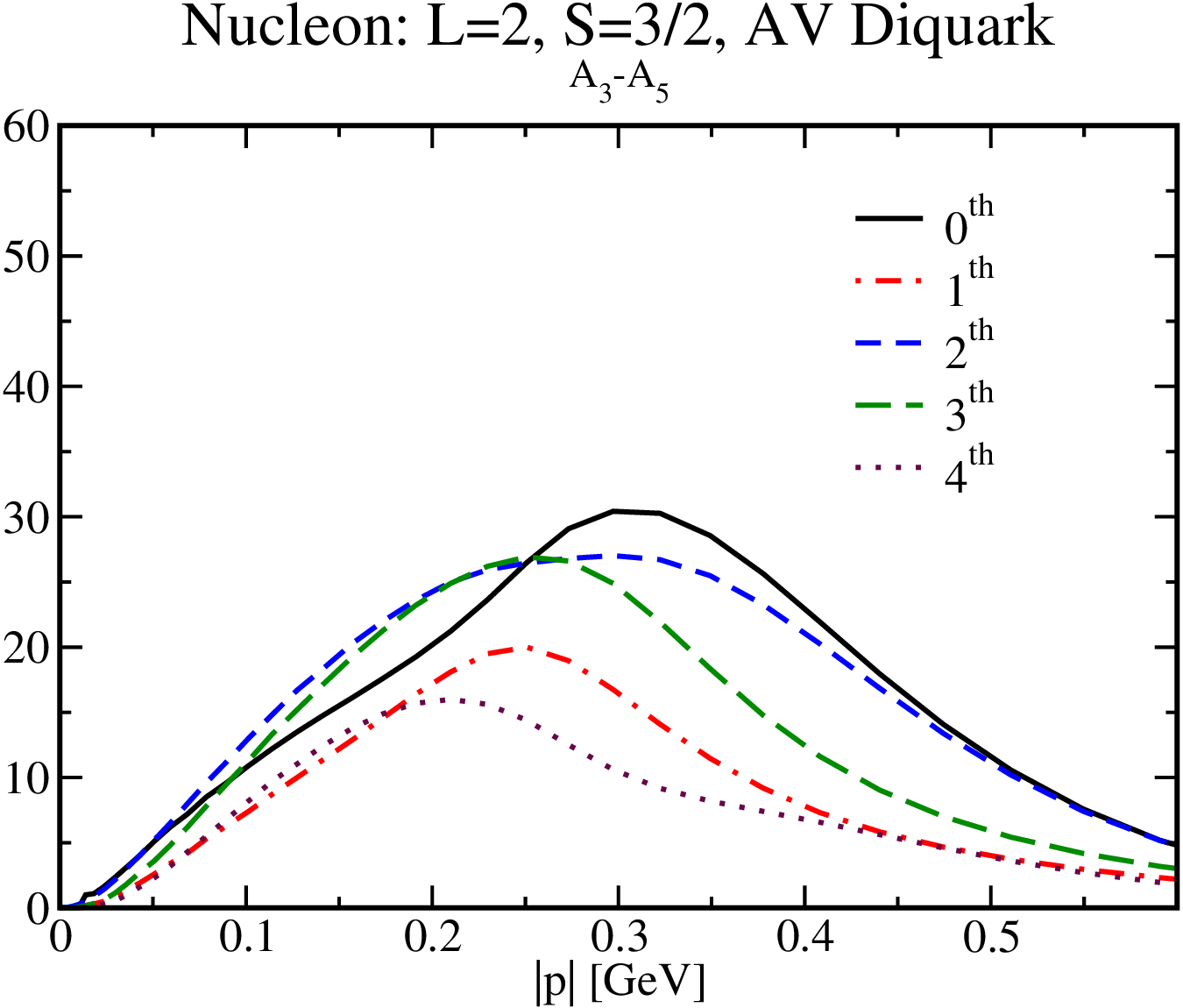} 
\vspace{.20 cm}
\end{tabular}
\begin{tabular}{cc}
\includegraphics[height=.22\textheight]{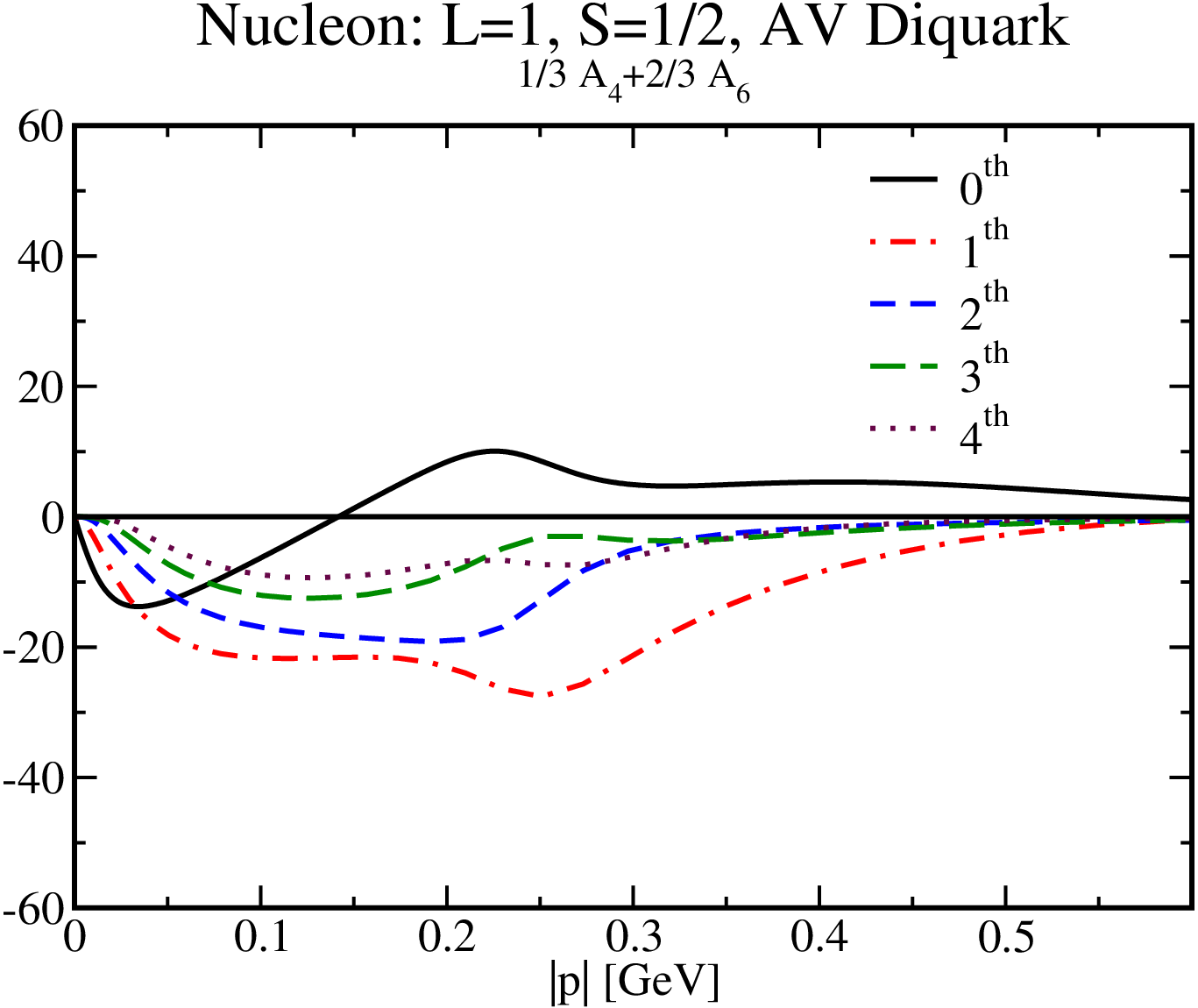}&
\hspace{0.5 cm}
\includegraphics[height=.22\textheight]{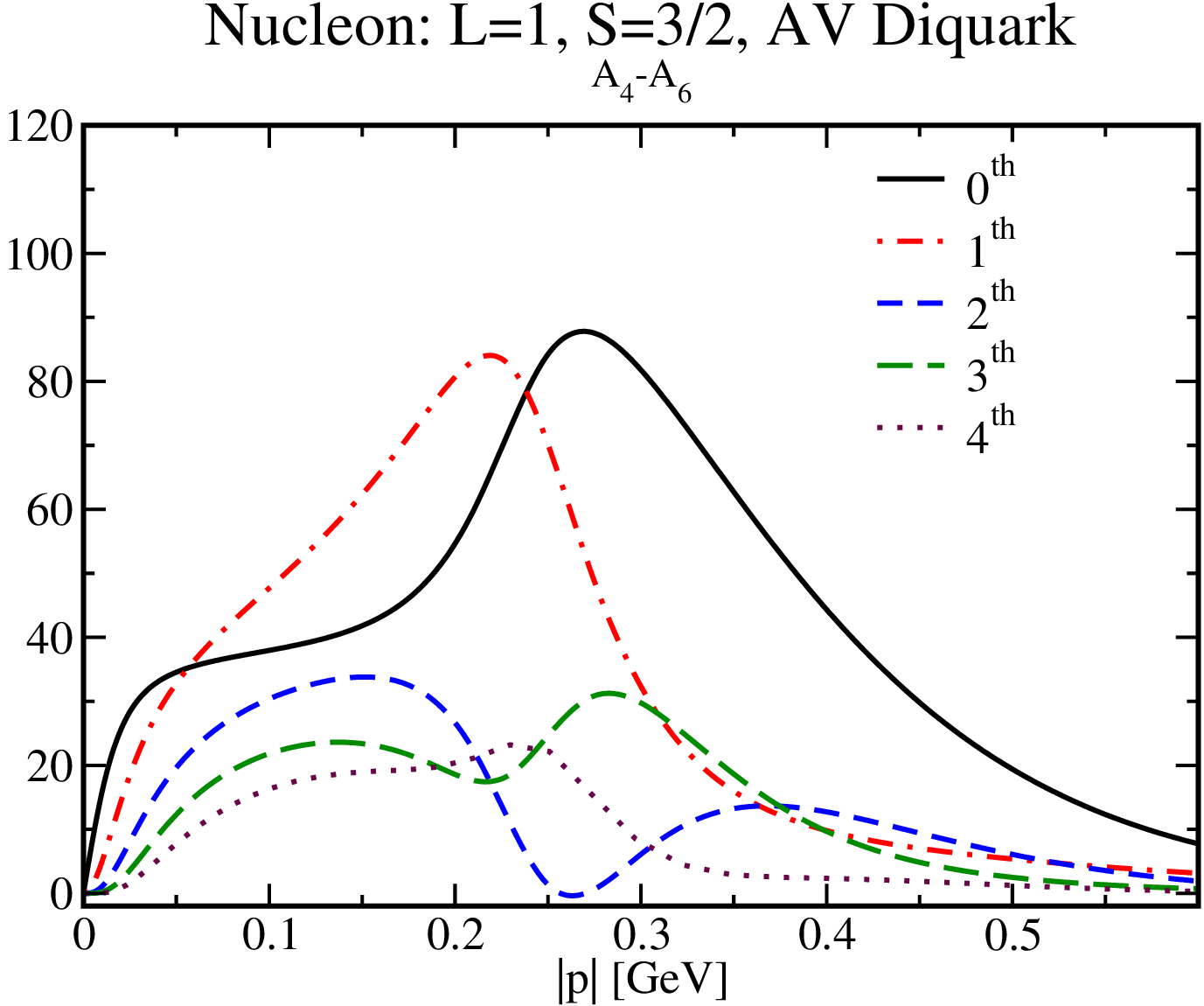}
\end{tabular}
\caption{Different Chebyshev moments (labeled by $n^{th}$) of scalar and axialvector (AV) diquark amplitudes of the nucleon BS wave function given by Set B1.}
\end{figure}

\newpage
\begin{figure}[!htb]
\begin{tabular}{cc}
\includegraphics[height=.22\textheight]{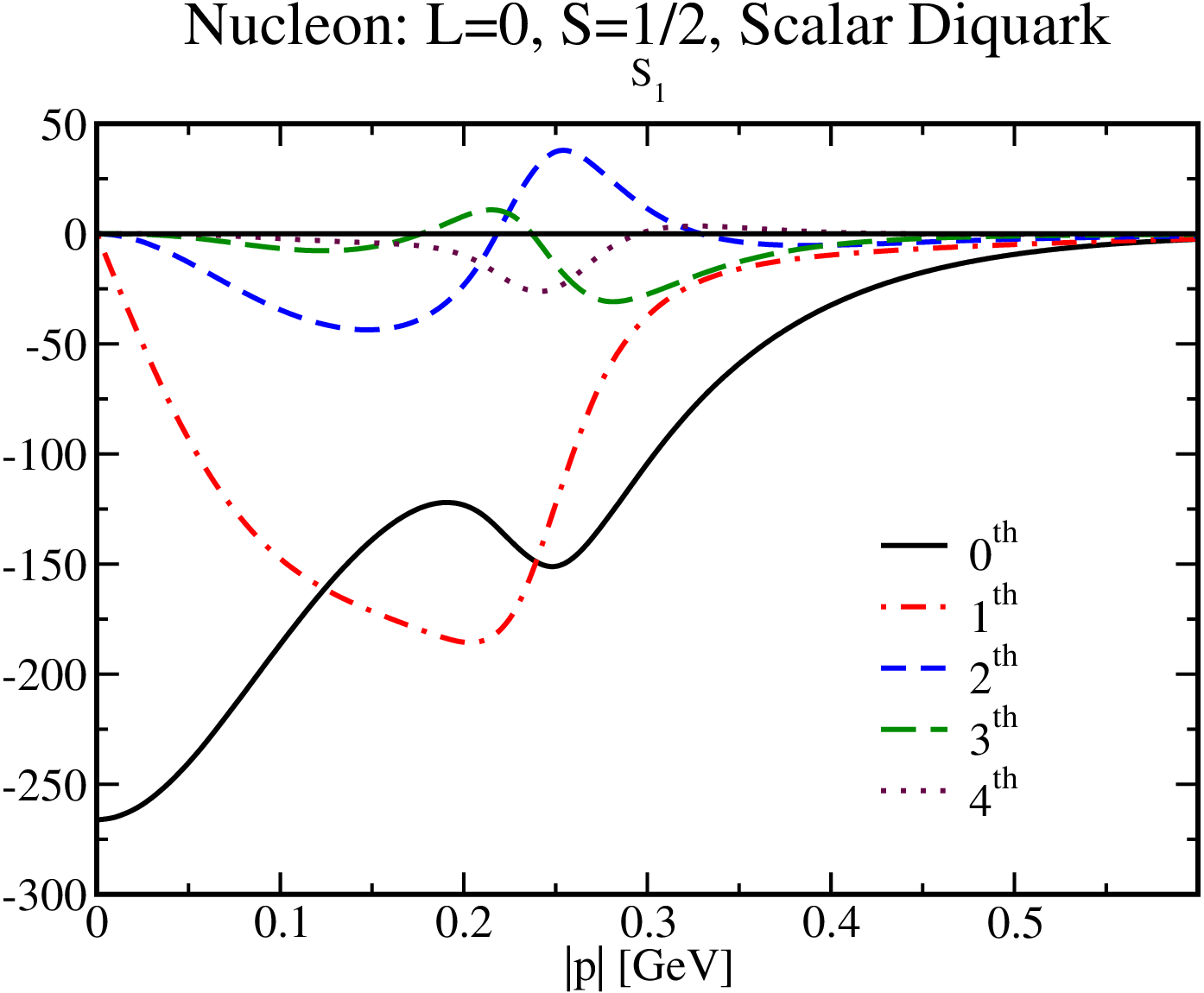}&
\hspace{0.5 cm}
\includegraphics[height=.22\textheight]{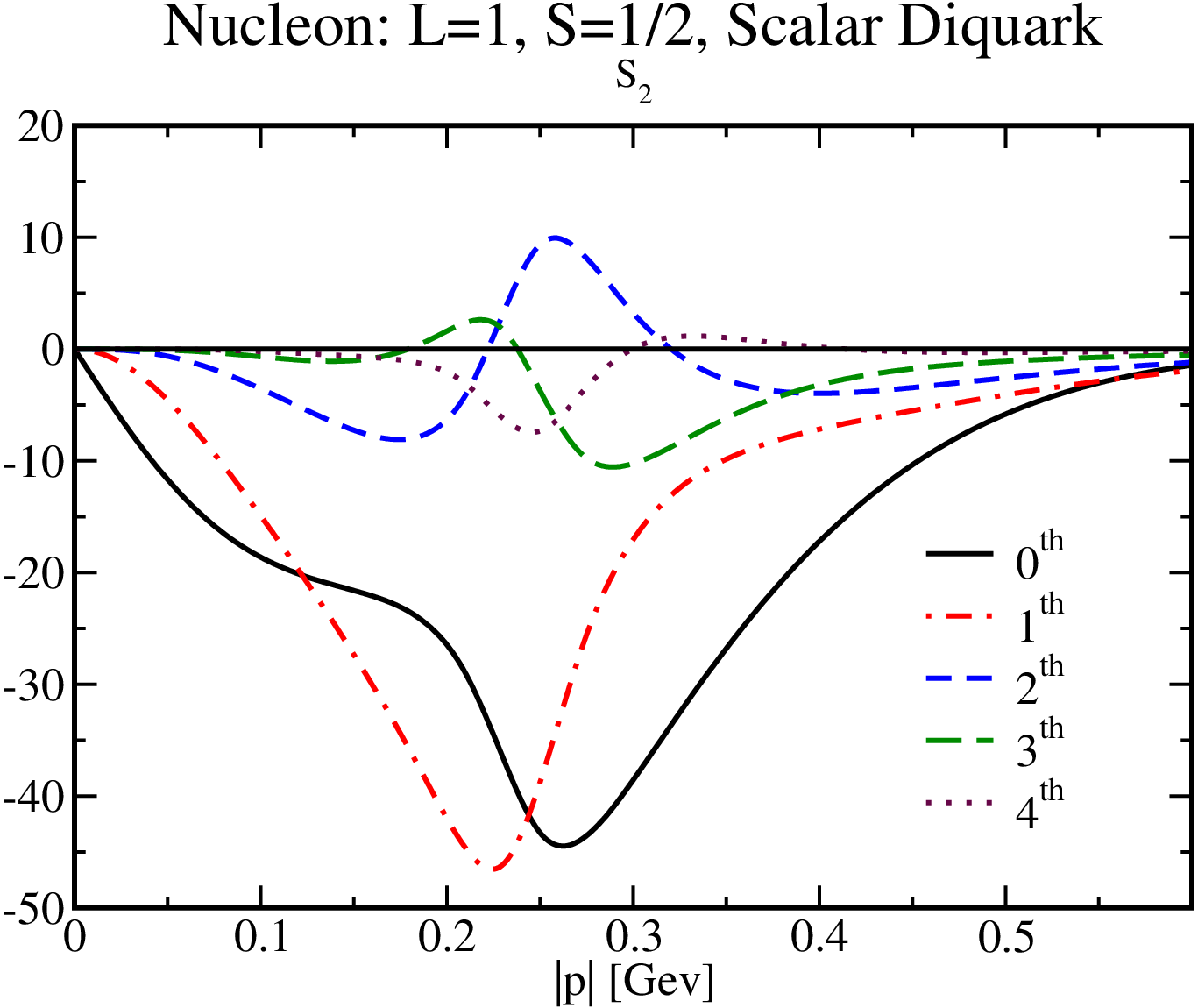}
\vspace{.20 cm}
\end{tabular}
\begin{tabular}{cc}
\includegraphics[height=.22\textheight]{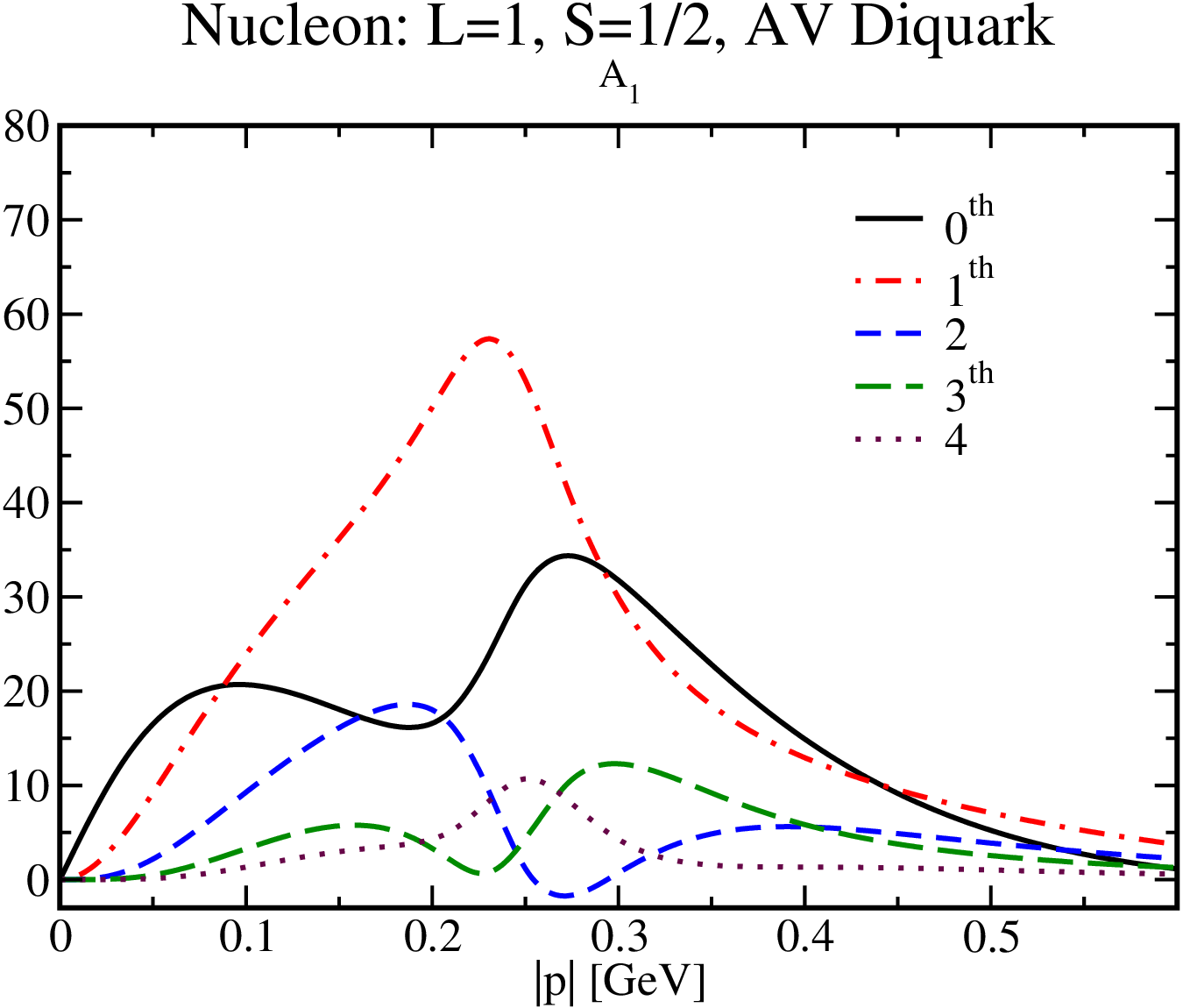}&
\hspace{0.5 cm}
\includegraphics[height=.22\textheight]{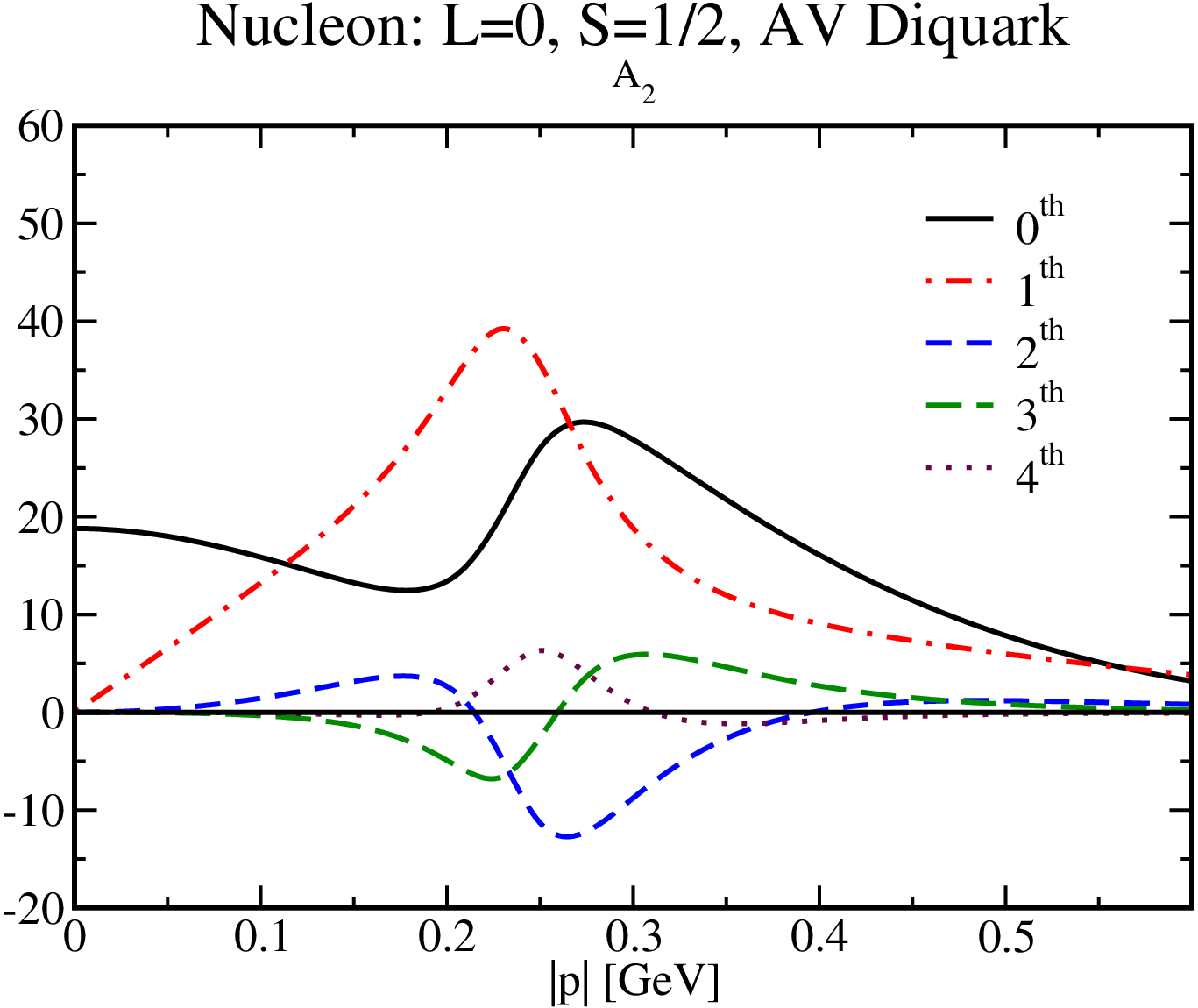}
\vspace{.20 cm}
\end{tabular}
\begin{tabular}{cc}
\includegraphics[height=.22\textheight]{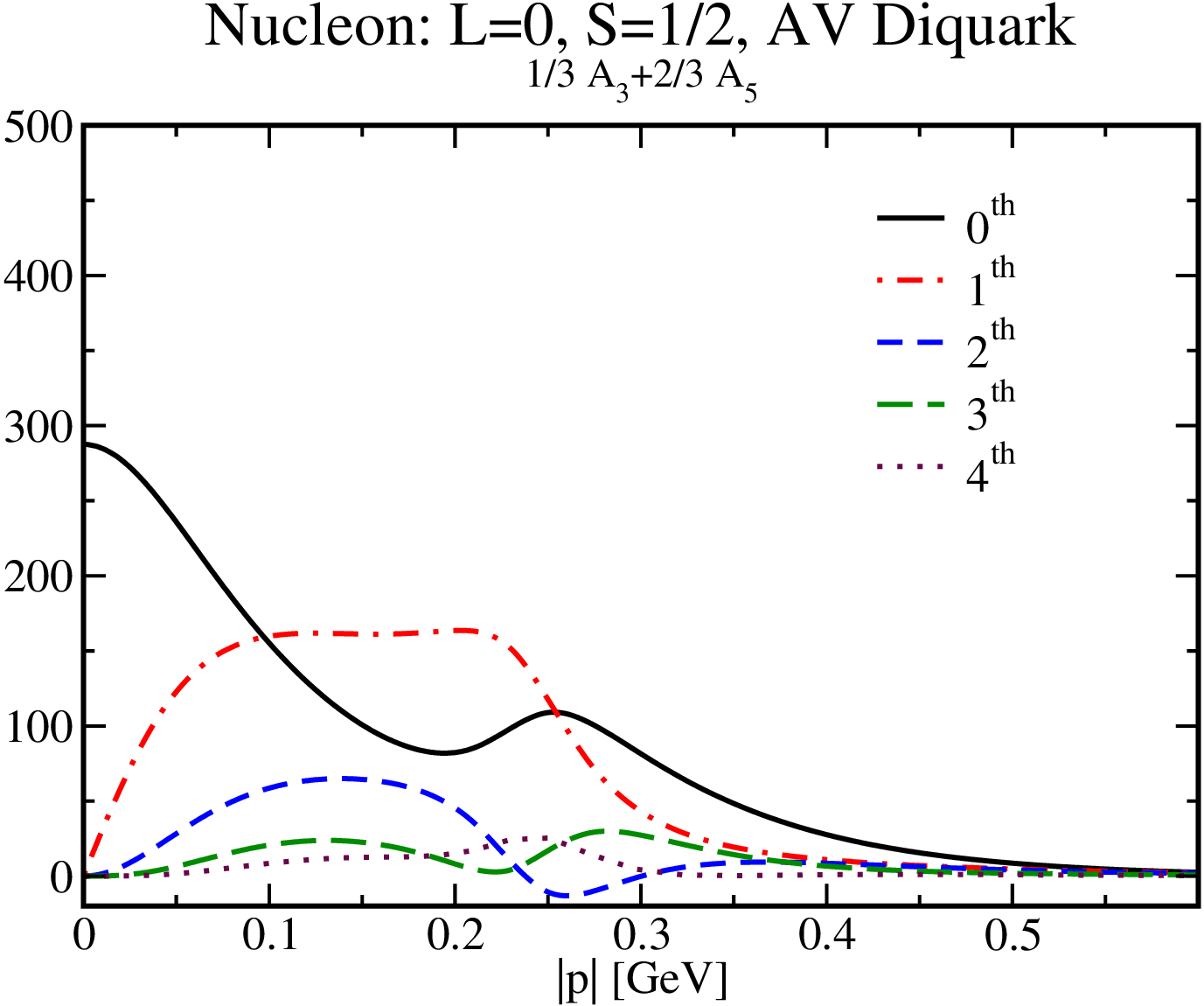}&
\hspace{0.5 cm}
\includegraphics[height=.22\textheight]{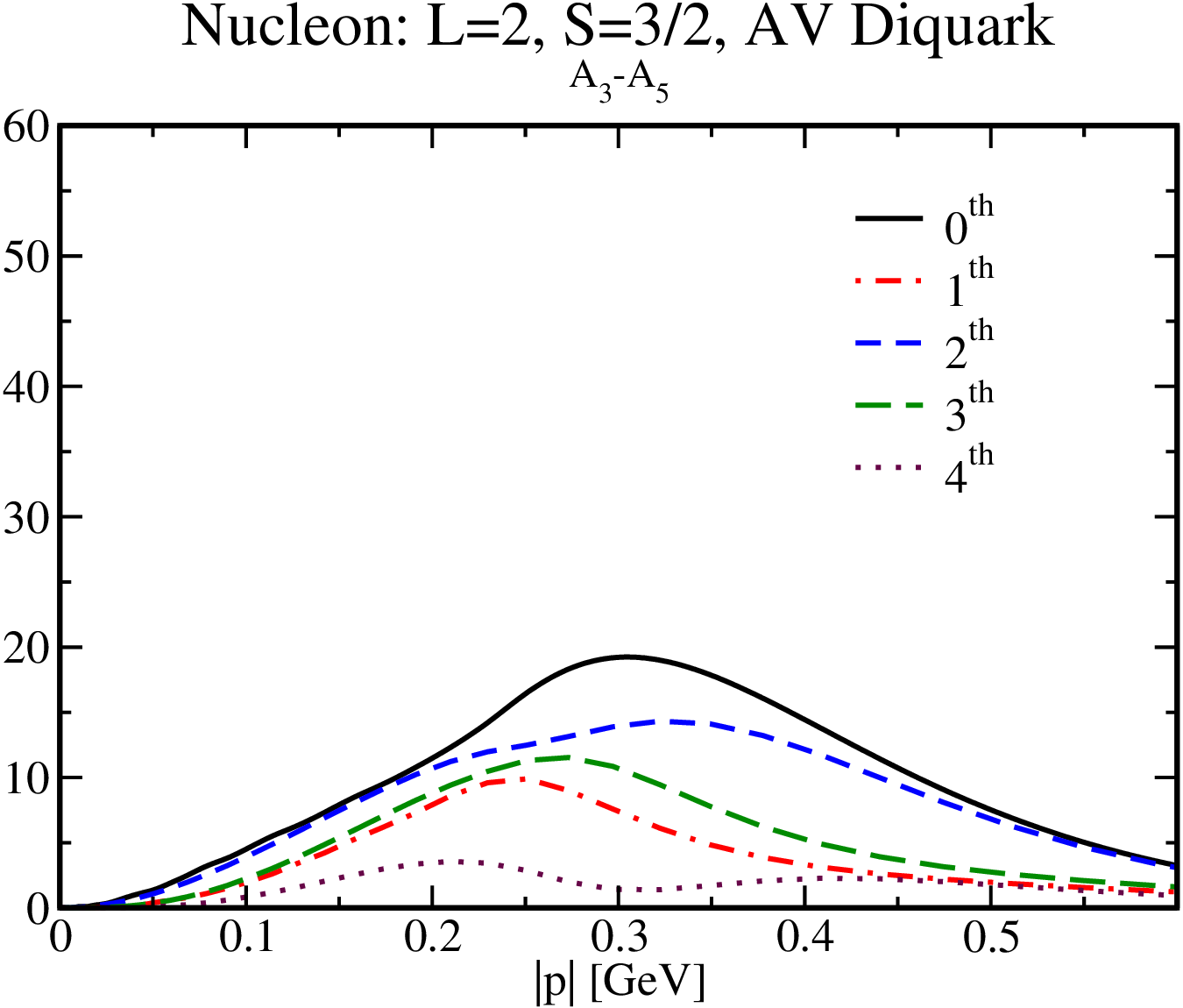} 
\vspace{.20 cm}
\end{tabular}
\begin{tabular}{cc}
\includegraphics[height=.22\textheight]{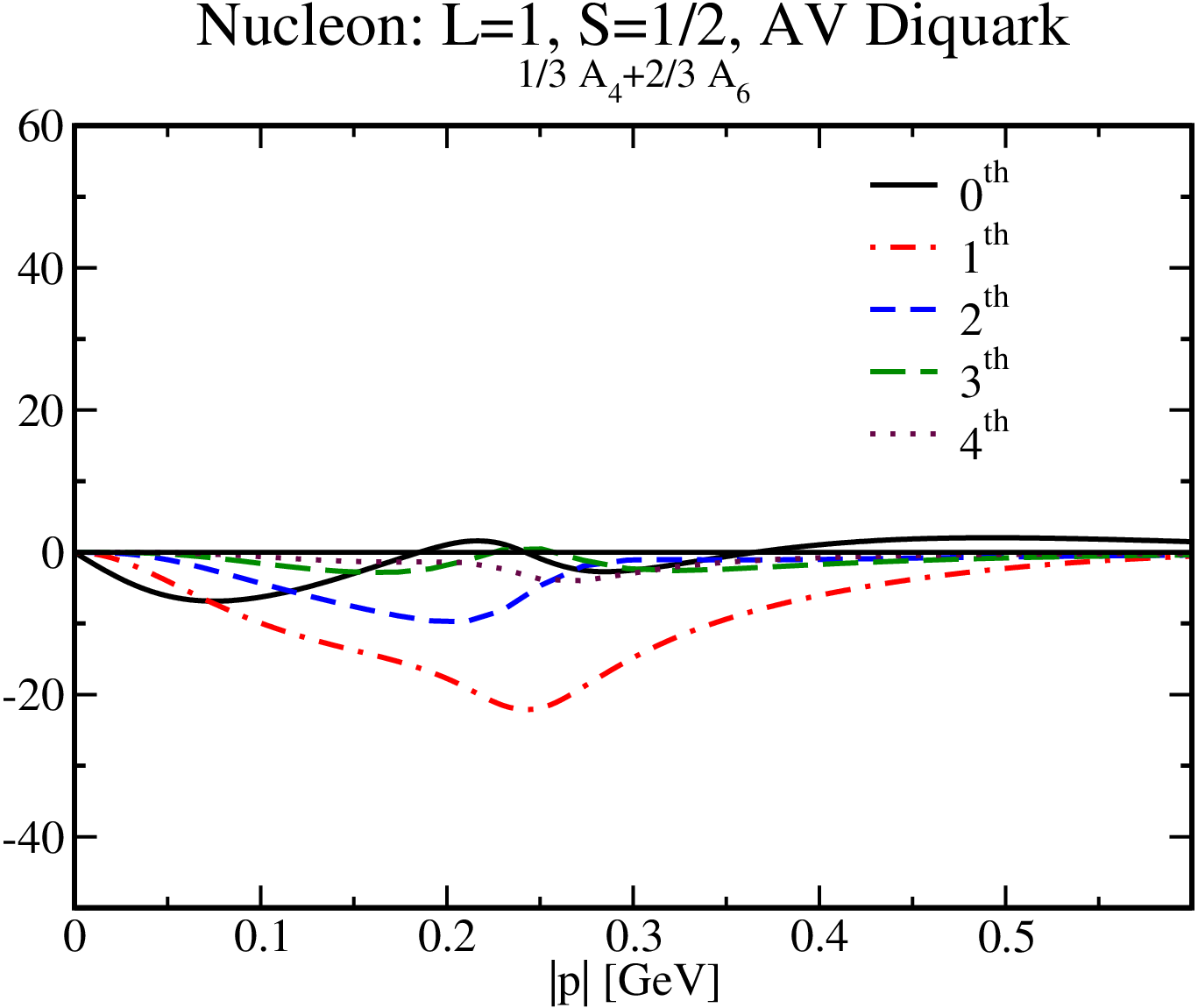}&
\hspace{0.5 cm}
\includegraphics[height=.22\textheight]{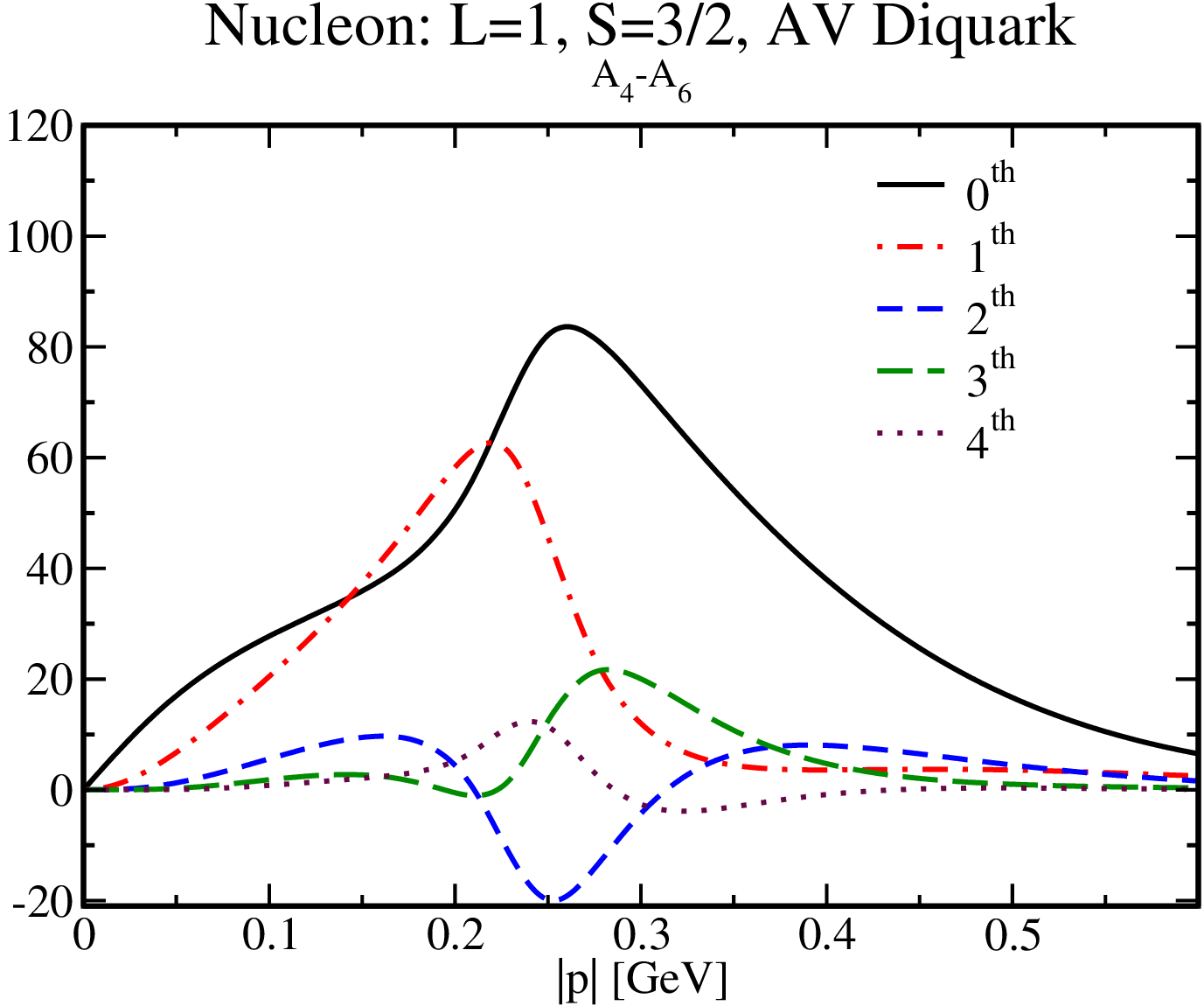}
\end{tabular}
\caption{Different Chebyshev moments (labeled by $n^{th}$) of scalar and axialvector diquark amplitudes of the nucleon wave function given by Set B2.}
\end{figure}

\newpage
\begin{figure}[!htb]
\begin{tabular}{cc}
\includegraphics[height=.22\textheight]{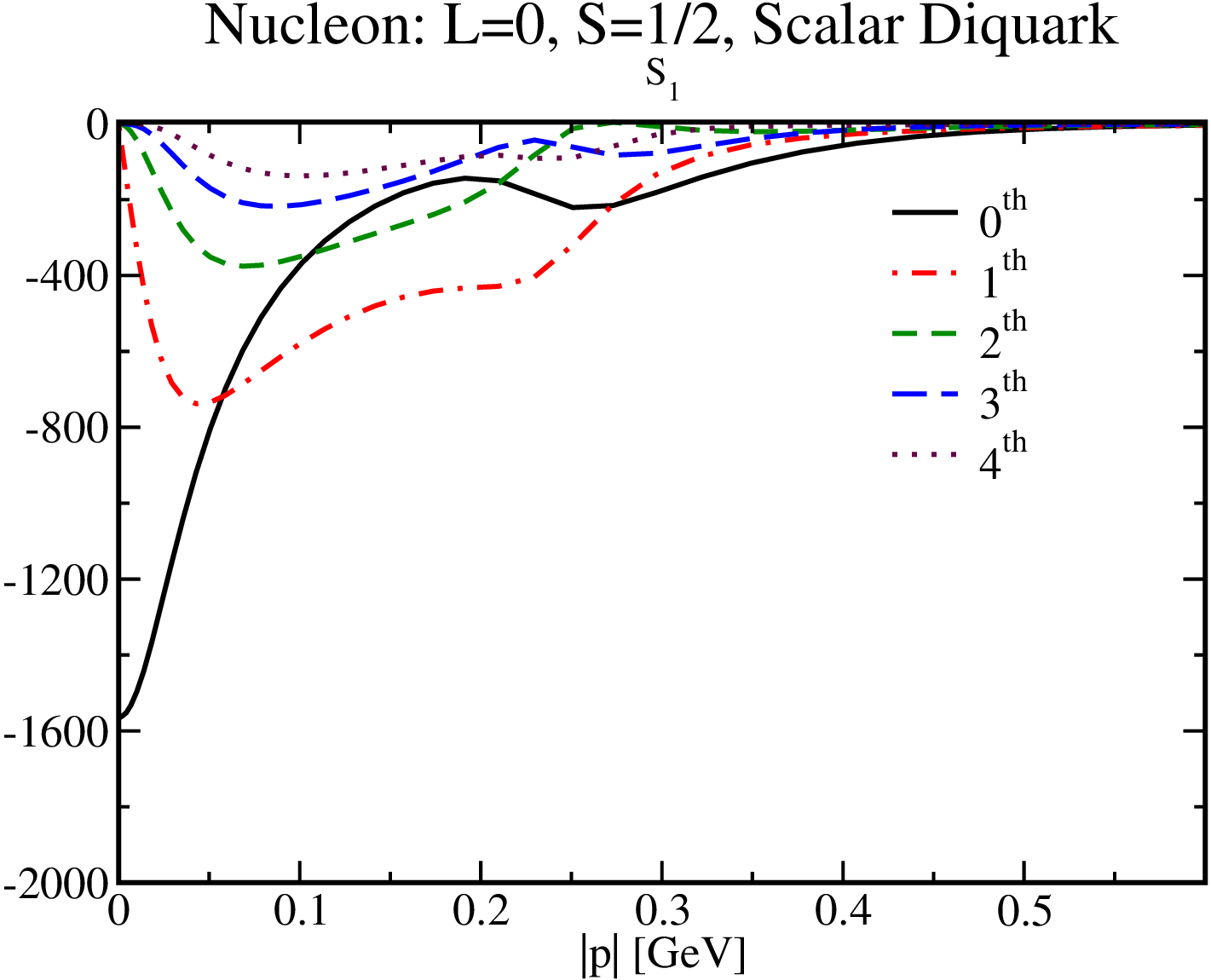}&
\hspace{0.5 cm}
\includegraphics[height=.22\textheight]{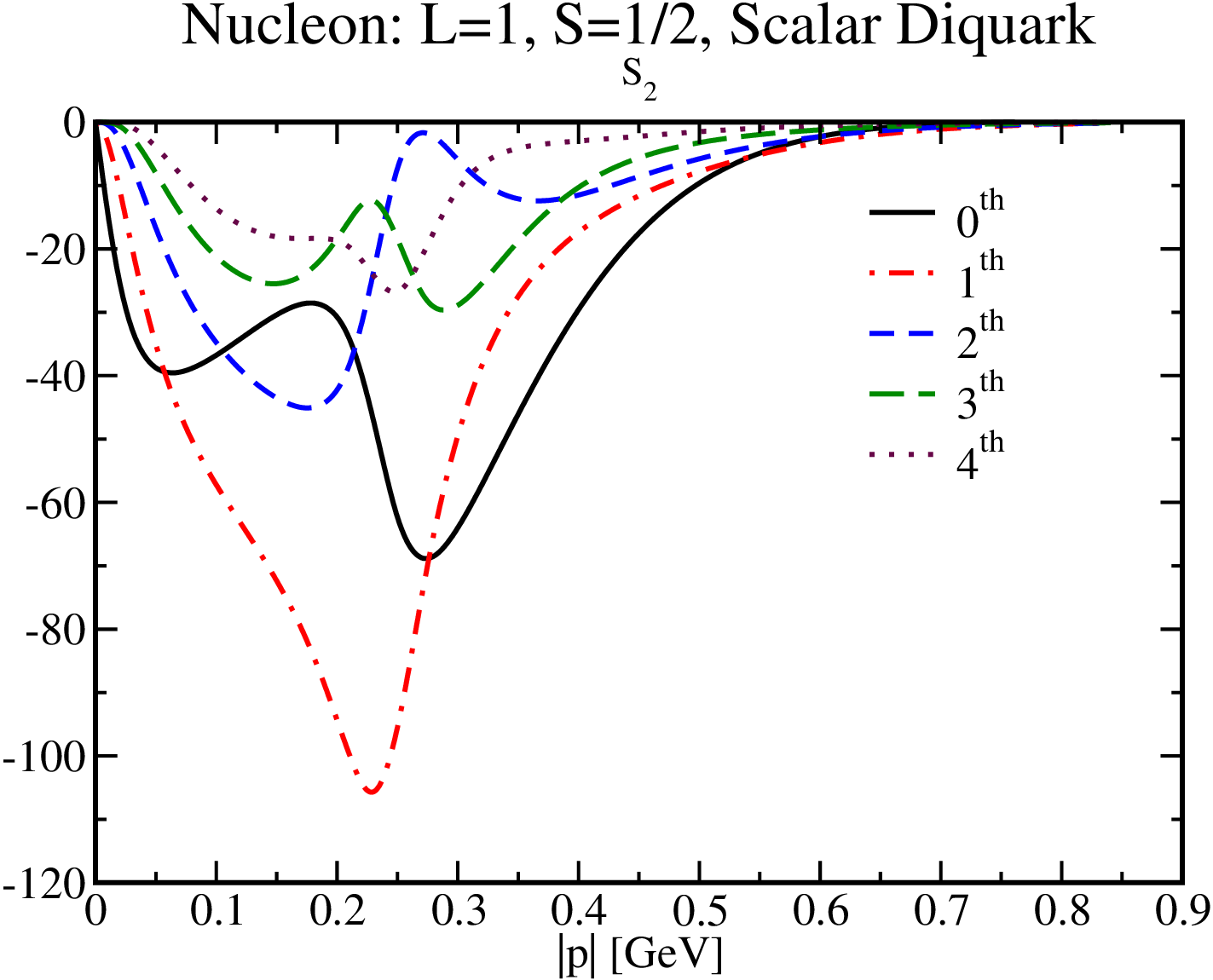}
\vspace{.20 cm}
\end{tabular}
\begin{tabular}{cc}
\includegraphics[height=.22\textheight]{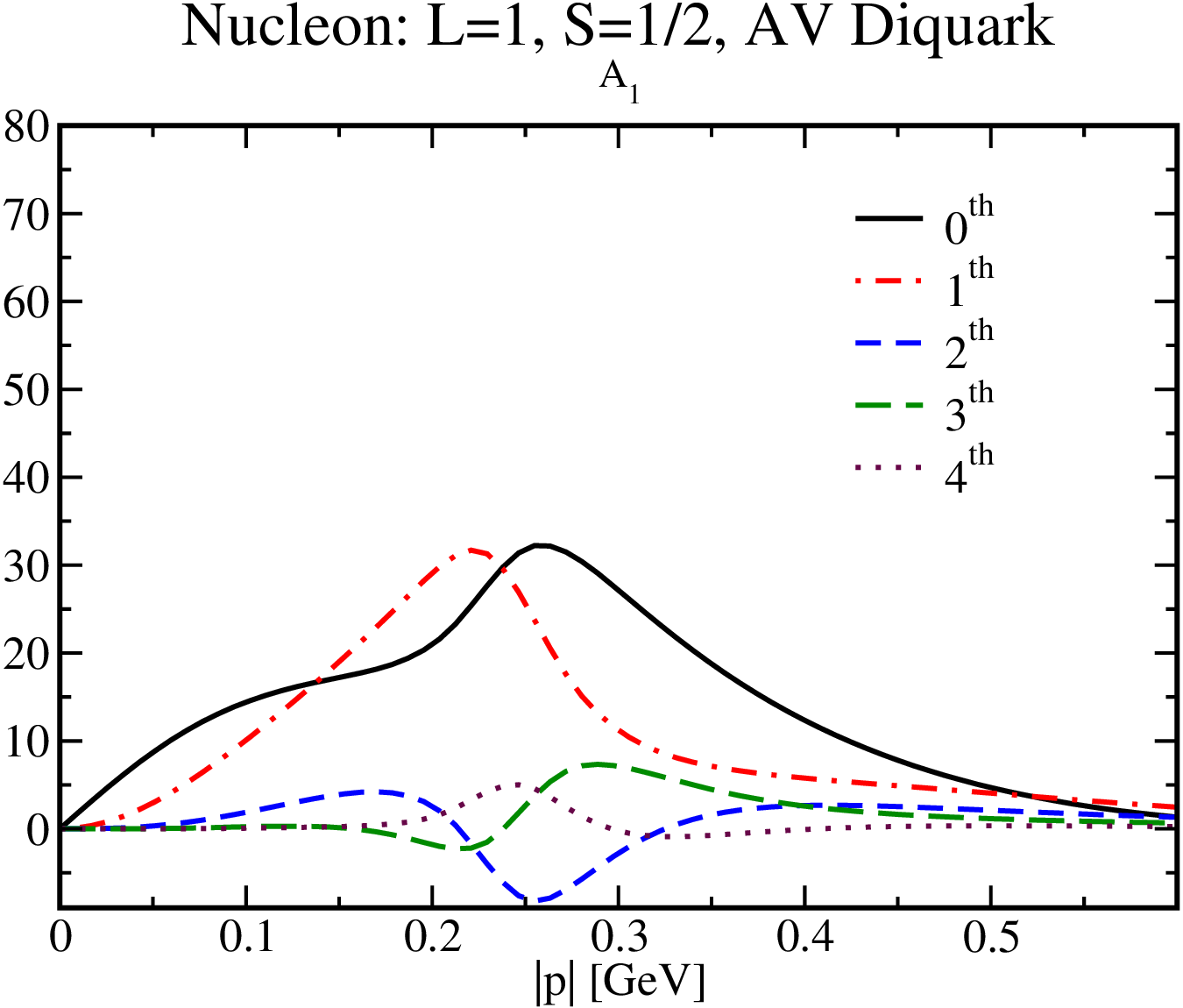}&
\hspace{0.5 cm}
\includegraphics[height=.22\textheight]{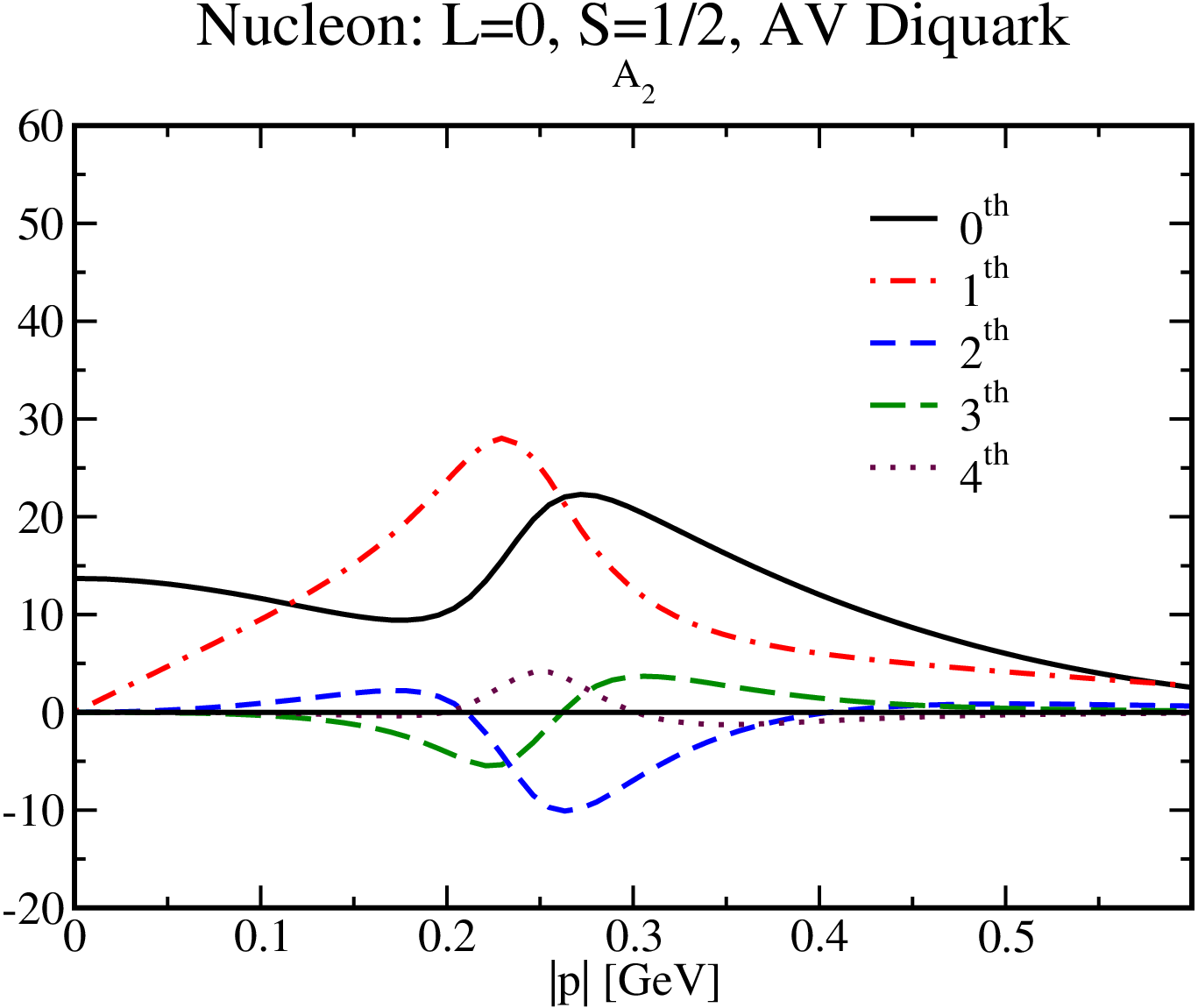}
\vspace{.20 cm}
\end{tabular}
\begin{tabular}{cc}
\includegraphics[height=.22\textheight]{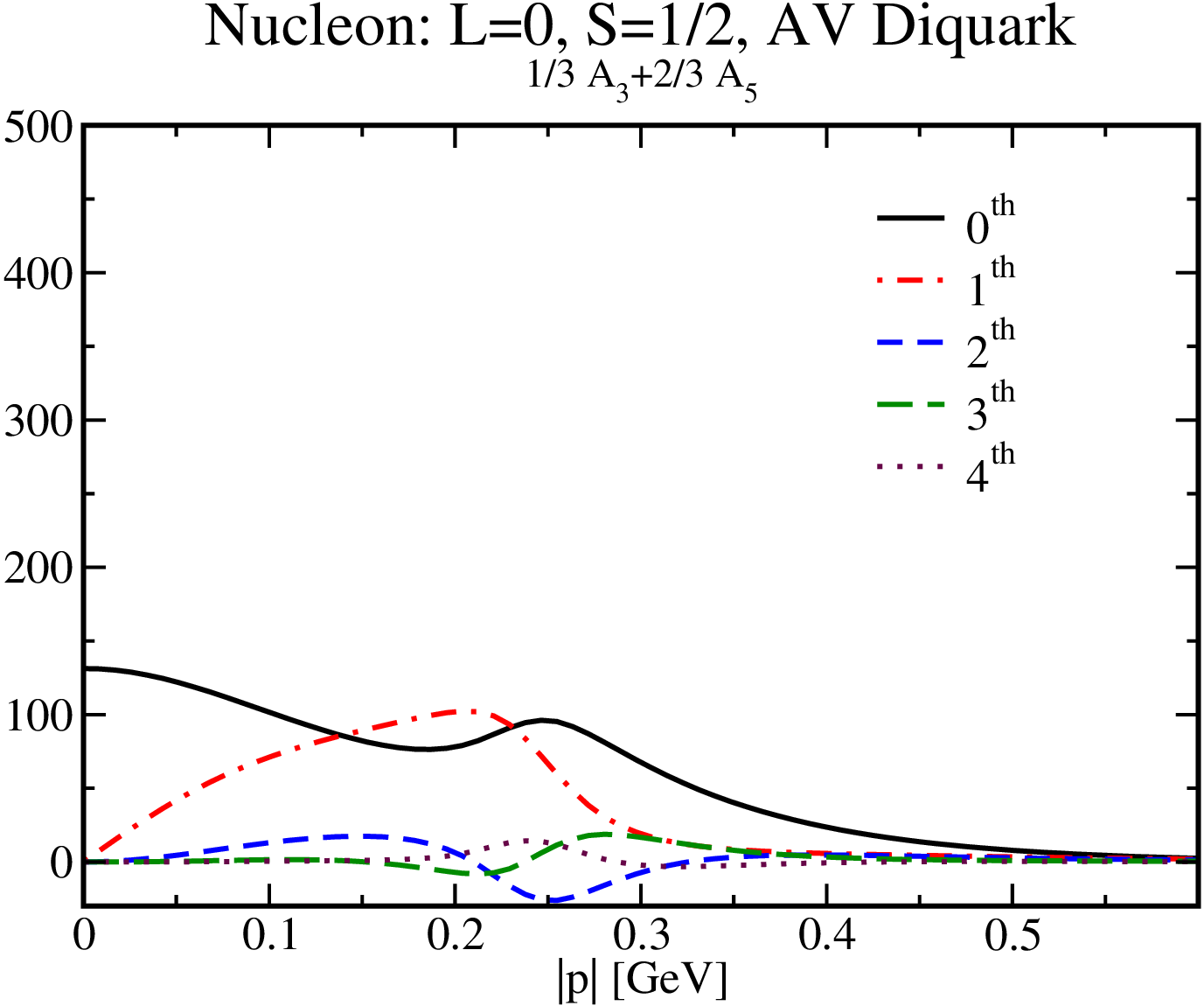}&
\hspace{0.5 cm}
\includegraphics[height=.22\textheight]{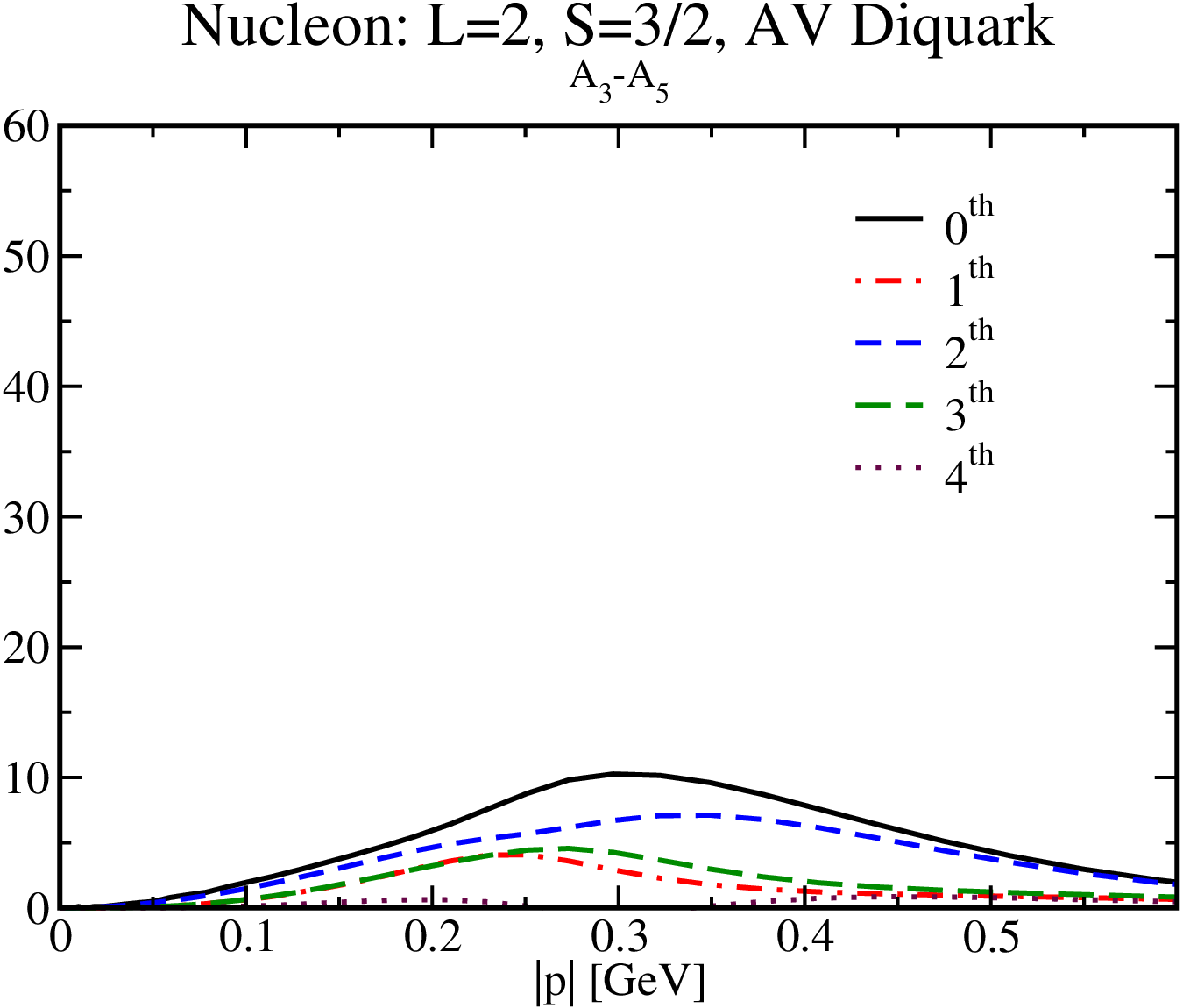} 
\vspace{.20 cm}
\end{tabular}
\begin{tabular}{cc}
\includegraphics[height=.22\textheight]{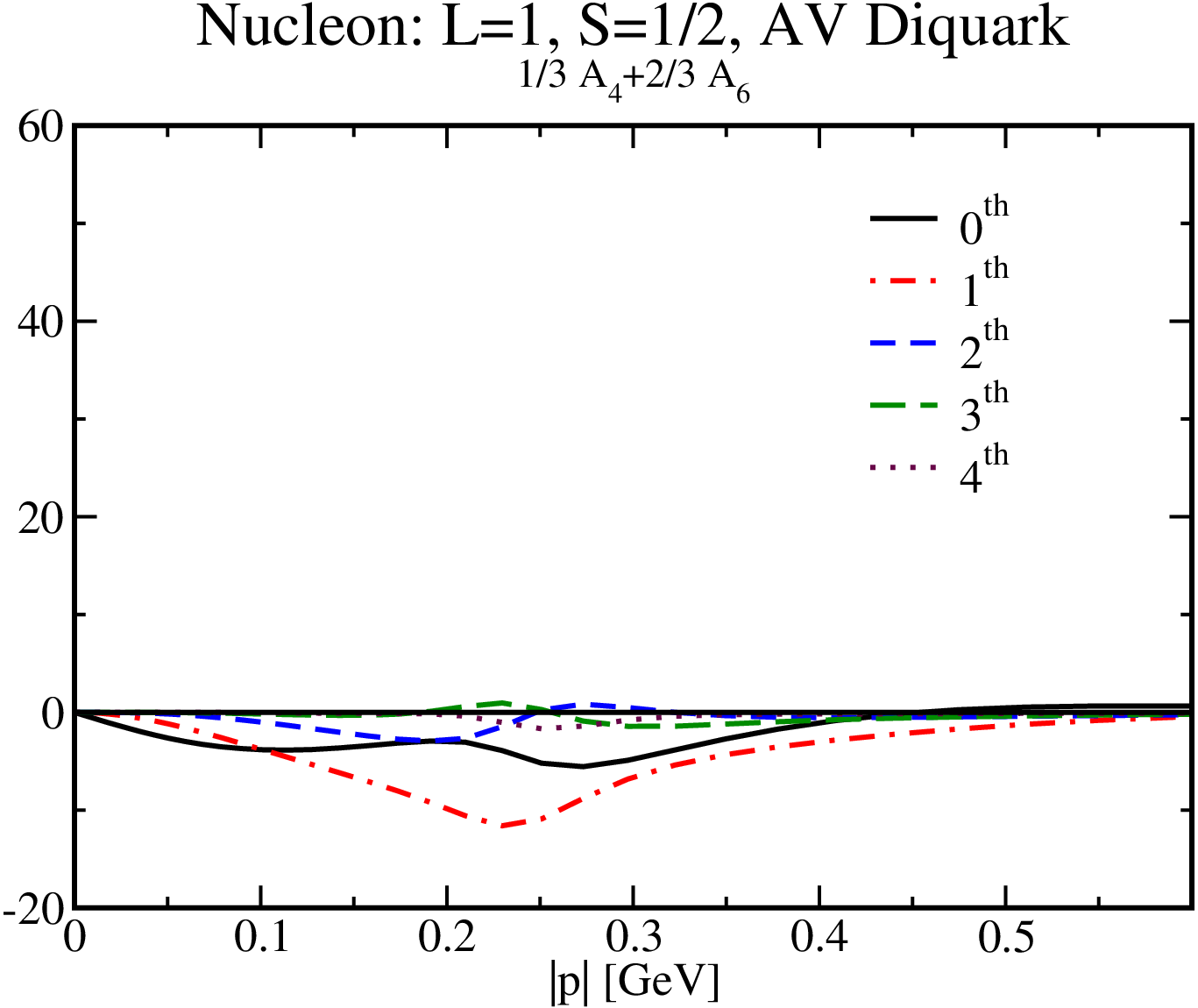}&
\hspace{0.5 cm}
\includegraphics[height=.22\textheight]{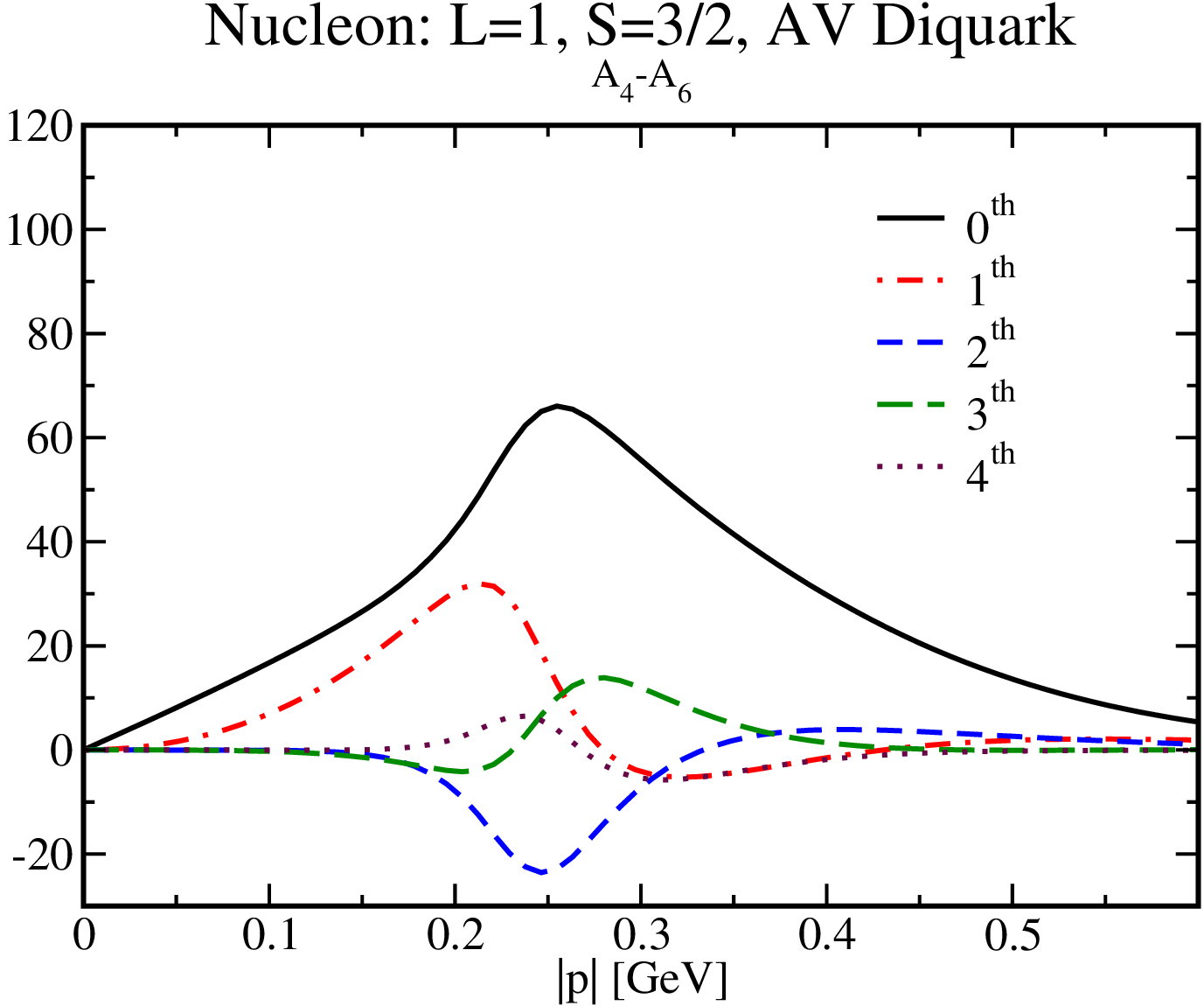}
\end{tabular}
\caption{Different Chebyshev moments (labeled by $n^{th}$) of scalar and axialvector diquark amplitudes of the nucleon wave function given by Set B3.}
\end{figure}

In order to study the implications of the quark confinement for the
description of the nucleon, we compare in Table \ref{tab:mdiquark} three
representative cases for both the non-confining and confining
parameter sets, which all give nucleon mass about $940$ MeV.  The
first three columns contain results for set $A$, and the last three
columns for the confining set $B$.

Given the definition of diquark-quark thresholds, in the presence of
both scalar and axial-vector diquark channels, the diquarks in the
nucleon can be found much more loosely bound, although one obtains a
very strongly bound nucleon solution near its experimental value, see
table \ref{tab:mdiquark}. Next we study the nucleon BS wave function
for the various sets given in Table \ref{tab:mdiquark}. The nucleon
wave and vertex function are not physical observables, but rather they
suggest how observables in this model will behave. In Figs.~6.16-21
we show the leading Chebyshev moments of the scalar functions of the
nucleon BS wave function for various sets (A1-3 and B1-3) which
describes the strengths of the quark-diquark partial waves with $S$ as
a total quark-diquark spin and $L$ as a total orbital angular
momentum. They are normalised to $F_{1}^{\phi(0)}(p_{1})=1$, where
$p_{1}$ is the first point of the momentum mesh. It is seen that the
contribution of higher moments are considerably small, indicating a
rapid convergence of the wave function amplitudes in terms of
Chebyshev polynomials. In the confining case Figs.~6.19-21
there is a clear interference which is not present in the
non-confining Figs.~6.16-19. Therefore in the confining case,
all wave function amplitudes are somehow shifted to higher relative
four-momenta between diquark and quark. In order to understand the
role of this interference we obtain a mass density for the various
channels. This density is defined as
\begin{equation}
\rho(p_{\bot},P)=\int dp_{4}\psi^{\dag}(p_{\bot},p_{4},P)\tilde{D}^{-1}(p_{d})\psi(p_{\bot},p_{4},P)  
\end{equation}
where $p_{\bot}$ stands for space component of relative momentum $p$
and $\tilde{D}^{-1}(p_{d})$ defined in Eq.~(\ref{di-pp}). This definition 
corresponds to a diagram occurring for the calculation of the isoscalar
quark condensate in the impulse approximation
\cite{njln4}. In the above definition
of the density function, we have integrated over the time component of
the relative momentum. In this way the density function becomes very
similar to its counterpart in Minkowski space. Although the above
definition of density is not unique, it does provide a useful
measure of the spatial extent of the wave function (we have examined
the possibility of taking matrix elements of other operators between
the BS wave function, since this does not lead to any significant
effect, the results are not presented here). The results are
plotted in Figs.~\ref{fig:bdensA} and \ref{fig:bdensB}.

It is noticeable that in various sets, the $s$-wave is the dominant
contribution to the ground state. The relative importance of the
scalar and the axial diquark amplitude in the nucleon changes with the
strength of the diquark-quark couplings $g_{dsqq}(g_{daqq})$ and
accordingly with $r_{s}(r_{a})$. We see that in the confining sets,
the nucleon density extends to higher relative momentum between the
diquark and the quark. This indicates a more compact nucleon in the
confining case. In order to find a qualitative estimation of the
confinement effect in our model, we calculate $p_{{\bot}}
^{\text{RMS}}=(\langle p^{2}_{\bot}\rangle-\langle p_{\bot}
\rangle^{2})^{1/2}$, the results can be found in table \ref{tab:mdiquark}. This can be
related to the mean-square radius of the nucleon, if we assume
minimal uncertainty. We see in the both confining and non-confining cases
 a decrease in $p_{{\bot}}^{\text{RMS}}$ with weakening
axial-vector diquark interaction (and consequently increasing the
scalar diquark interaction strength). If we compare $p_{{\bot}}
^{\text{RMS}}$ for the two sets $A2, B2$, which have
very similar interaction parameters $r_{s}(r_{a})$, an increase about
$25\%$ is found. This effect can not only be associated with the
non-locality of our interaction, since that is present in both confining and
non-confining cases.
\begin{figure}
\centerline{\includegraphics[width=16 cm]{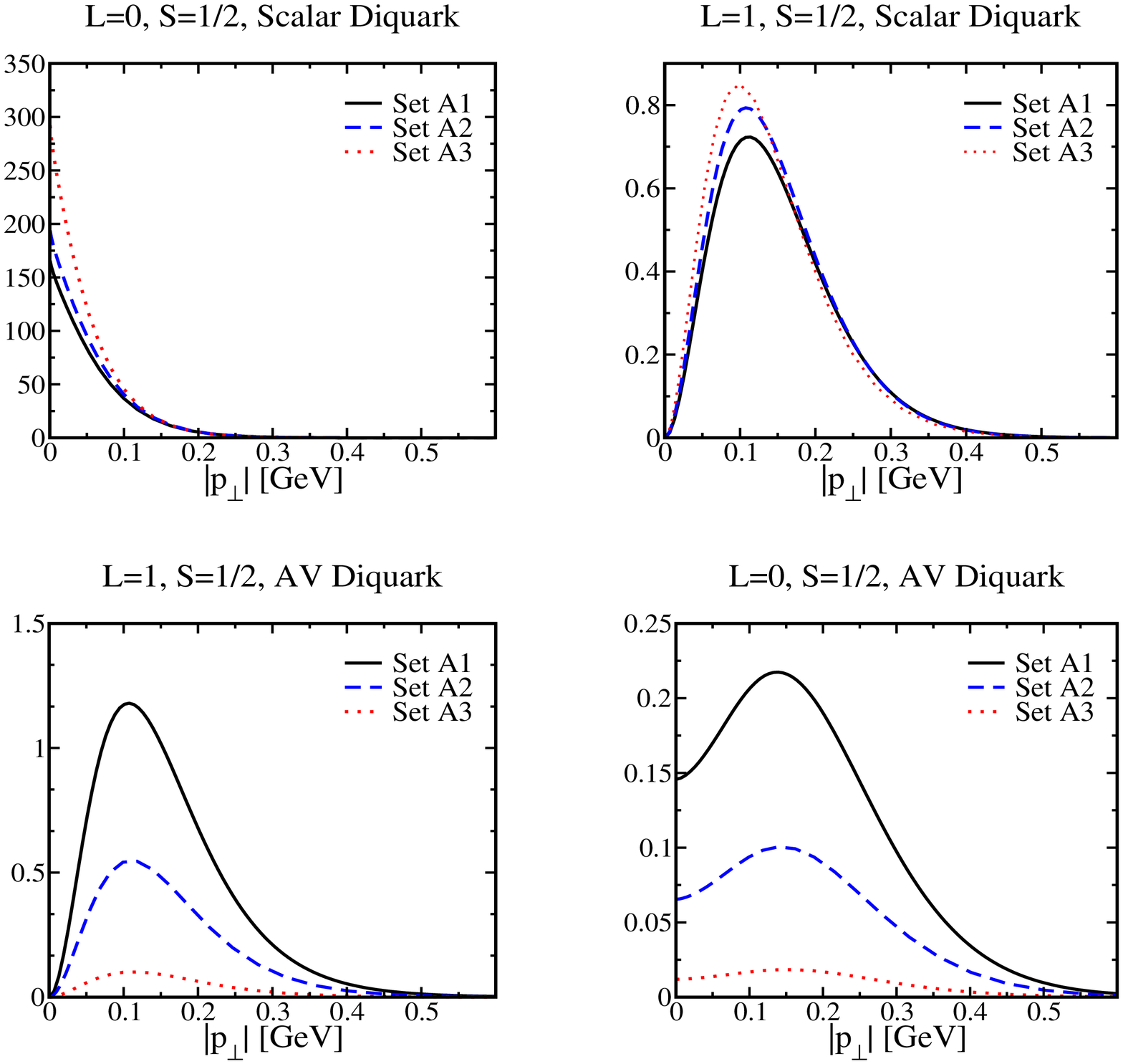}}
\caption{Shows the nucleon density ($M=0.940$ GeV, Set $A$) with respect to relative
momentum between diquark and quark for different set of $A1, A2$ and $A3$.
\label{fig:bdensA}}
\end{figure}

\begin{figure}
\centerline{\includegraphics[width=16 cm]{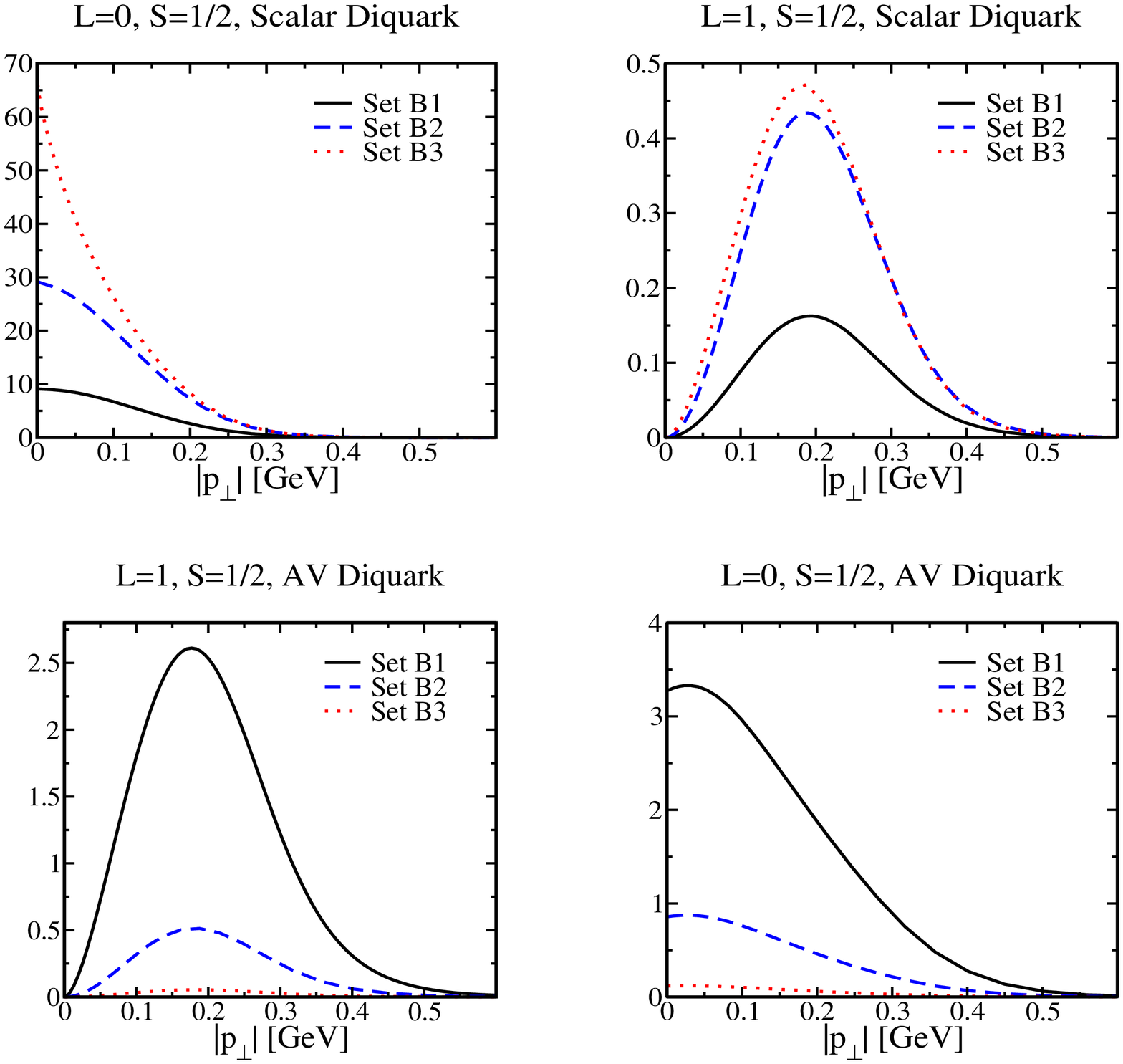}}
\caption{Shows the nucleon density ($M=0.940$ GeV, Set $B$) with respect to relative
momentum between diquark and quark for different set of $B1, B2$ and $B3$. \label{fig:bdensB}}
\end{figure}

\section{Summary and Outlook\label{sec:conc}}
In this chapter we investigated the two- and three-quark problems in a
non-local NJL model. We have truncated the diquark sector to the bound
scalar and the axial-vector channels. We have solved the relativistic
Faddeev equation for this model and have studied the behaviour of the
nucleon solutions with respect to various scalar and the axial-vector
interactions. We have also investigated a possible implication of the
quark confinement of our model in the diquark and the baryon sector.

Although the model is quark confining, it is not diquark confining (at
least in the rainbow-ladder approximation). A bound diquark can be
found in both scalar and the axial-vector channel for a wide range of
couplings. We have found that the off-shell contribution to the
diquark $T$-matrix is crucial for the calculation of the structure of
the nucleon: without its inclusion the attraction in the diquark
channels is too weak to form a three-body bound state. We have also
found that both the scalar and the axial-vector contribute
attractively to the nucleon mass. The role of axial-vector channel is
much more important in the confining phase of model. The nucleon in
this model is strongly bound although the diquarks within nucleon
are loosely bound. The confining aspects of the model are more
obvious in three-body, rather than the two-body sector. By
investigating the nucleon wave function we showed that  quark
confinement leads to a more compact nucleon. The size of nucleon is
reduced by about $25\%$ in confining phase.

For both confining and non-confining phases, an increase in the scalar
diquark channel interaction $r_{s}$ leads to a lower nucleon mass, see
Figs.~\ref{fig:Nmass_full1} and \ref{fig:Nmass_full2}, but the mass of
the $\Delta$ remains unchanged since it does not contain scalar
diquarks. In the standard NJL model where the axial-vector diquark
does not contribute significantly to the nucleon binding
\cite{cut-b,av-i,njl-av}, the difference between the $\Delta$ and
nucleon mass is directly related to the scalar diquark interaction. In
the current model where the axial-vector diquark makes a larger
contribution to the nucleon mass, therefore a detailed calculation for
the delta states is needed to understand the mechanism behind the
$\Delta$-$N$ mass difference.  In the standard NJL model this leads to
a contradiction, since for an axial-vector interaction which gives a
reasonable description of nucleon properties, both the axial-vector
diquark and more importantly the $\Delta$ are unbound
\cite{njl-av}. The crucial role of the axial-vector
diquark correlation in the non-local NJL model, especially in the
confining phase of the model, indicate that this model might do better.

In order to understand the implications of this model in baryonic
sector fully one should investigate properties of the nucleon such as
the axial vector coupling constant, the magnetic moment, etc. On the
other hand, the role of quark confinement in this model can be better
clarified by investigating quark and nuclear matter in this
model. One of the long standing problem in four-Fermi chiral quark
models is the fact that quark/nuclear  matter does not saturate \cite{under1},
mainly due to the strongly attractive quark interactions responsible
for the spontaneous chiral symmetry breaking. Recently, Bentz and
Thomas \cite{under2} have shown that a sufficient strong repulsive
contribution can arise if confinement effects are incorporated, albeit
in the cost of introducing a new parameter into model. It was shown
that such repulsive contribution can lead to saturation of nuclear
matter equation of state. It is indeed of interest to investigate the
stability of nuclear matter within this quark confining model.  Such problems can be studied based on the Faddeev
approach, see e. g., Ref.~\cite{w}.

\cleardoublepage

\end{document}